\PassOptionsToPackage{unicode}{hyperref}
\PassOptionsToPackage{hyphens}{url}
\PassOptionsToPackage{dvipsnames,svgnames,x11names}{xcolor}
\documentclass[11pt,a4paper]{article}
\usepackage[scale=0.8]{geometry}
\usepackage{rotating}
\usepackage{amsmath,amssymb}
\usepackage{iftex}
\ifPDFTeX
  \usepackage[T1]{fontenc}
  \usepackage[utf8]{inputenc}
  \usepackage{textcomp} 
\else 
  \usepackage{unicode-math}
  \defaultfontfeatures{Scale=MatchLowercase}
  \defaultfontfeatures[\rmfamily]{Ligatures=TeX,Scale=1}
\fi
\usepackage{lmodern}
\ifPDFTeX\else  
\fi
\IfFileExists{upquote.sty}{\usepackage{upquote}}{}
\IfFileExists{microtype.sty}{
  \usepackage[]{microtype}
  \UseMicrotypeSet[protrusion]{basicmath} 
}{}

\usepackage{xcolor}
\makeatletter
\ifx\paragraph\undefined\else
  \let\oldparagraph\paragraph
  \renewcommand{\paragraph}{
    \@ifstar
      \xxxParagraphStar
      \xxxParagraphNoStar
  }
  \newcommand{\xxxParagraphStar}[1]{\oldparagraph*{#1}\mbox{}}
  \newcommand{\xxxParagraphNoStar}[1]{\oldparagraph{#1}\mbox{}}
\fi
\ifx\subparagraph\undefined\else
  \let\oldsubparagraph\subparagraph
  \renewcommand{\subparagraph}{
    \@ifstar
      \xxxSubParagraphStar
      \xxxSubParagraphNoStar
  }
  \newcommand{\xxxSubParagraphStar}[1]{\oldsubparagraph*{#1}\mbox{}}
  \newcommand{\xxxSubParagraphNoStar}[1]{\oldsubparagraph{#1}\mbox{}}
\fi
\makeatother

\usepackage{longtable,booktabs,array}
\usepackage{calc} 
\usepackage{etoolbox}
\makeatletter
\patchcmd\longtable{\par}{\if@noskipsec\mbox{}\fi\par}{}{}
\makeatother
\IfFileExists{footnotehyper.sty}{\usepackage{footnotehyper}}{\usepackage{footnote}}
\makesavenoteenv{longtable}
\usepackage{graphicx}
\makeatletter
\def\maxwidth{\ifdim\Gin@nat@width>\linewidth\linewidth\else\Gin@nat@width\fi}
\def\maxheight{\ifdim\Gin@nat@height>\textheight\textheight\else\Gin@nat@height\fi}
\makeatother
\setkeys{Gin}{width=\maxwidth,height=\maxheight,keepaspectratio}
\makeatletter
\def\fps@figure{htbp}
\makeatother

\makeatletter
\@ifpackageloaded{caption}{}{\usepackage{caption}}
\AtBeginDocument{%
\ifdefined\contentsname
  \renewcommand*\contentsname{Table of contents}
\else
  \newcommand\contentsname{Table of contents}
\fi
\ifdefined\listfigurename
  \renewcommand*\listfigurename{List of Figures}
\else
  \newcommand\listfigurename{List of Figures}
\fi
\ifdefined\listtablename
  \renewcommand*\listtablename{List of Tables}
\else
  \newcommand\listtablename{List of Tables}
\fi
\ifdefined\figurename
  \renewcommand*\figurename{Figure}
\else
  \newcommand\figurename{Figure}
\fi
\ifdefined\tablename
  \renewcommand*\tablename{Table}
\else
  \newcommand\tablename{Table}
\fi
}
\@ifpackageloaded{float}{}{\usepackage{float}}
\floatstyle{ruled}
\@ifundefined{c@chapter}{\newfloat{codelisting}{h}{lop}}{\newfloat{codelisting}{h}{lop}[chapter]}
\floatname{codelisting}{Listing}

\makeatother
\makeatletter
\@ifpackageloaded{caption}{}{\usepackage{caption}}
\@ifpackageloaded{subcaption}{}{\usepackage{subcaption}}
\makeatother

\ifLuaTeX
  \usepackage{selnolig}  
\fi
\usepackage[round,sectionbib]{natbib}
\usepackage{bookmark}

\IfFileExists{xurl.sty}{\usepackage{xurl}}{} 
\usepackage{mathabx}
\usepackage{authblk}
\usepackage{enumerate}
\usepackage{threeparttable}
\usepackage{pdflscape}
\usepackage{newtxtext}

\newtheorem{theorem}{Theorem}
\newtheorem{assumption}{Assumption}

\newtheorem{passumption}{Assumption}

\usepackage{siunitx}
\newcolumntype{d}[1]{S[table-format=#1]}

\title{Quasi-Maximum Likelihood Estimation for a Genuinely Unbalanced Dynamic Network Panel Data Model}

\author[a,\textdagger]{Zhijian Wang}
\author[a,b,\textdagger]{Tuo Liu}
\author[b,a,c,*]{Xingbai Xu}
\affil[a]{Department of Statistics and Data Science, School of Economics, Xiamen University}
\affil[b]{MOE Key Lab of Econometrics and Fujian Key Lab of Statistics, Wang Yanan Institute for Studies in Economics}
\affil[c]{Paula and Gregory Chow Institute for Studies in Economics, Xiamen University}

\date{}
\hypersetup{pdftitle={Quasi-Maximum Likelihood Estimation for a Genuinely Unbalanced Dynamic Network Panel Data Model},pdfauthor={Zhijian Wang; Tuo Liu; Xingbai Xu},
pdfkeywords={Dynamic Network Panel Data Model; Quasi-maximum Likelihood Estimation; Spatial Dynamic Panel Data Model; Unbalanced Panel Data Model},colorlinks=true,linkcolor=blue,citecolor=blue,urlcolor=blue}
\allowdisplaybreaks[3]
\begin{document}
\maketitle

\begingroup
\renewcommand{\thefootnote}{}
\footnotetext{\textdagger\ Zhijian Wang and Tuo Liu contributed equally to this work.}
\footnotetext{* Corresponding author: \href{mailto:xuxingbai@xmu.edu.cn}{xuxingbai@xmu.edu.cn}.}
\endgroup
\begin{abstract}
This paper develops quasi-maximum likelihood estimation for genuinely unbalanced dynamic network panel data models. We first study a baseline one-way fixed effect specification and then extend the framework to two-way fixed effects. The model accommodates contemporaneous and lagged network spillovers, temporal dependence, and a listing effect that operates when a unit first appears in the panel. For the baseline specification, we establish consistency and asymptotic normality of the QMLE, characterize the incidental-parameter bias, and propose a bias-corrected estimator. We further provide a consistent estimator of the asymptotic covariance matrix, to enable feasible inference. Under the two-way fixed effect extension, we derive the corresponding asymptotic distribution, bias correction, and covariance matrix estimator, accounting for the additional bias component induced by the time fixed effects. Monte Carlo simulations examine the finite sample performance of the proposed estimators and inference procedure. An empirical application to Airbnb listings from New Zealand and New York City reveals region-specific patterns in spatial and temporal price transmission, highlighting the importance of modeling genuine unbalancedness in dynamic network settings.
\end{abstract}
\begin{quote}\small\textbf{Keywords:} Dynamic Network Panel Data Model; Quasi-maximum Likelihood Estimation; Spatial Dynamic Panel Data Model; Unbalanced Panel Data Model\end{quote}

\section{Introduction}
Panel data are widely used in social science and economics because they simultaneously capture individual heterogeneity and temporal dynamics. Compared with cross-sectional or time-series data, panel data provide richer information for analyzing both long-term relationships and short-term adjustments among variables. When units are spatially or network-connected---such as regions in an economic system, firms in a financial market, or individuals in a social platform---their outcomes often exhibit interdependence across both space and time. This has motivated the development of spatial dynamic panel data models, which integrate network interactions with temporal evolution. These models have been applied to a wide range of contexts, such as regional economics \citep{Amidi2020}, environmental systems \citep{repec:eee:eneeco:v:104:y:2021:i:c:s0140988321004680}, financial markets \citep{repec:spr:jospat:v:3:y:2022:i:1:d:10.1007_s43071-022-00025-8}, and social networks \citep{Han02012021}.

\begin{figure}[h]
    \centering
    \begin{subfigure}[b]{0.45\textwidth} 
        \centering
        \includegraphics[width=\textwidth]{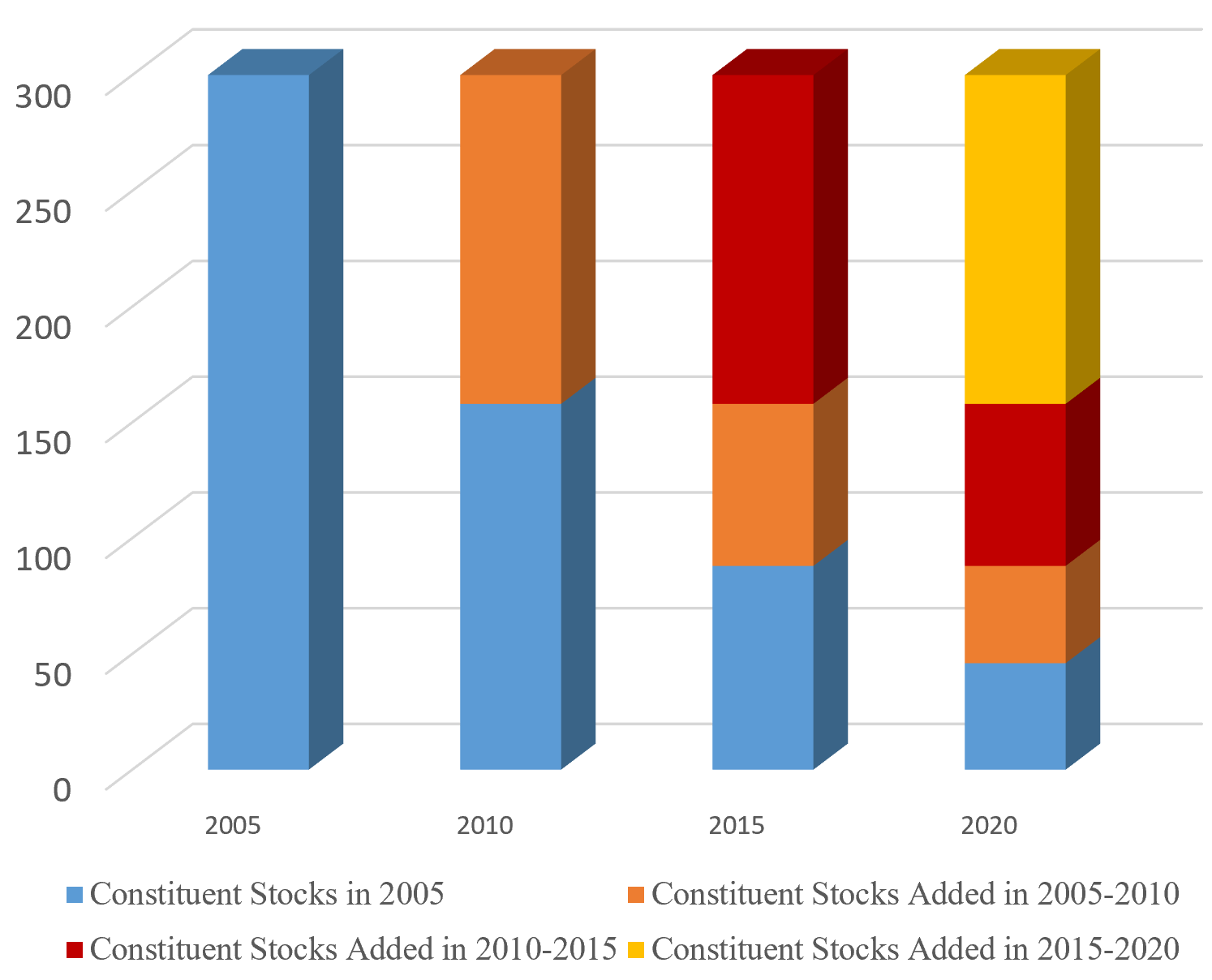}
        \caption{CSI 300}
    \end{subfigure}
    \begin{subfigure}[b]{0.5\textwidth}
        \centering
        \includegraphics[width=\textwidth]{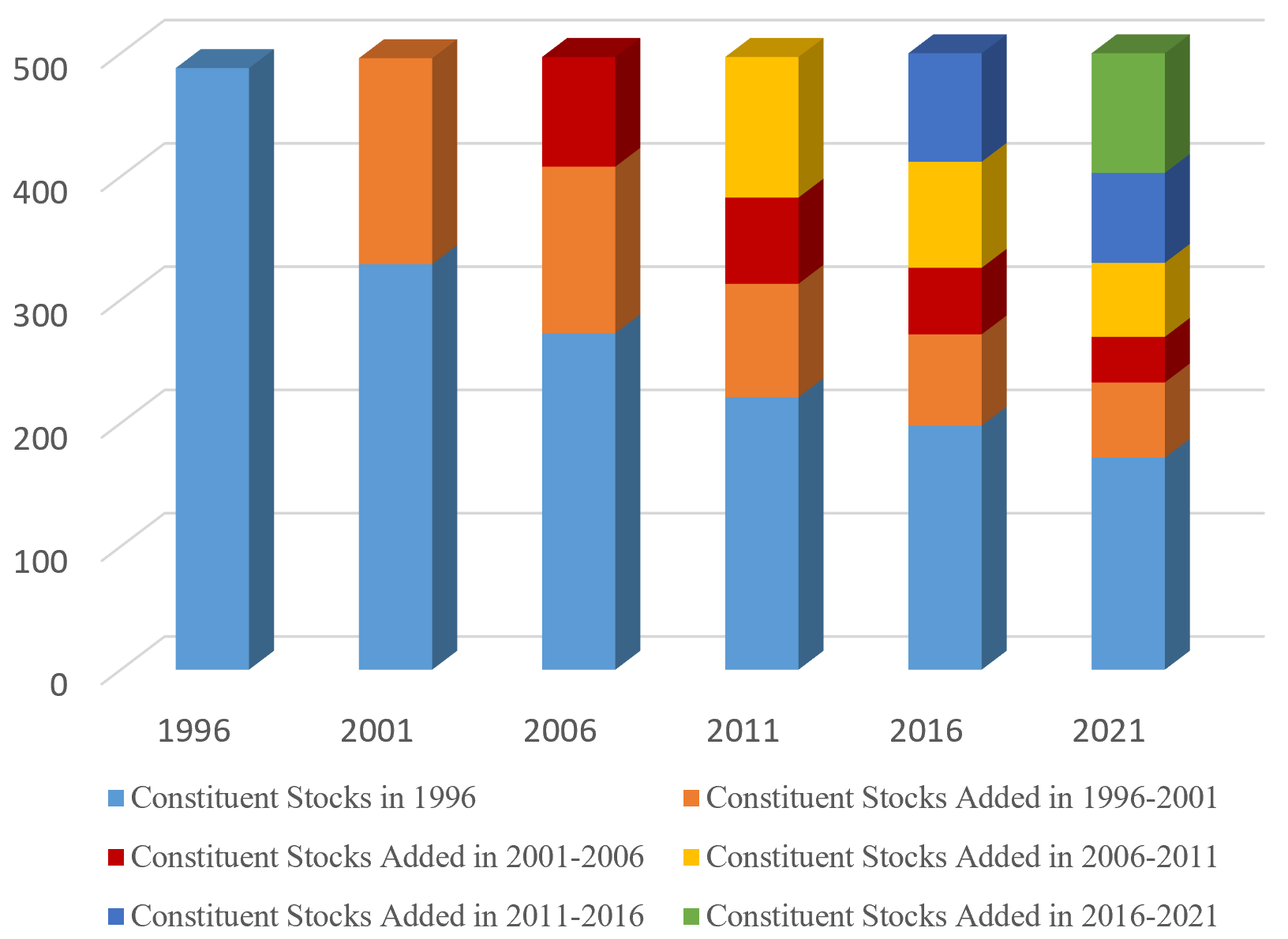}
        \caption{S\&P 500}
    \end{subfigure}
\caption{Historical retention status of CSI 300 and S\&P 500 constituent stocks}
\label{fig_CSI}
\end{figure}

Unbalanced dynamic network panel data are commonly observed in real-world contexts such as regional rental markets, financial stock markets, and social networks. For instance, in stock markets, firms may enter or exit over time, as illustrated in Figure ~\ref{fig_CSI}. The figure shows the historical retention status of CSI 300 and S\&P 500 constituent stocks and indicates that a substantial share of constituents is replaced approximately every five years. Critically, these firms are interconnected through business partnerships or ownership ties, giving rise to a complex and evolving network structure. Together, the continuous turnover of units and the network links define an unbalanced dynamic network panel. Similarly, in social network studies, individuals exhibit heterogeneous temporal activity, and relationships evolve as users join or leave the network, which also constitutes a prototypical example of an unbalanced dynamic network panel.

Despite their empirical prevalence, theoretical work on dynamic spatial or network panels has overwhelmingly assumed balanced panels. \cite{KAPOOR200797} study the estimation of static spatial panel data models. 
\cite{YU2008118} and \cite{LeeSDPD} propose quasi-maximum likelihood estimation (QMLE) for spatial dynamic panel data models. \cite{LEE2014174} consider the generalized method of moments (GMM) estimation for spatial dynamic panel data. \cite{lee_qml_2012} and \cite{QU2017173} extend the previous spatial dynamic panel data to scenarios with time-varying or endogenous spatial weight matrices.
\cite{pmid32094164} propose a least squares estimators based on the eigendecomposition of a spatial weight matrix method for general spatial dynamic panel data models.
On this basis, \cite{HOU2024105605} extend this framework to directed networks and develop a variable selection approach for high-dimensional spatial dynamic panel data models. Another related model is the network autoregressive model. See \cite{10.1214/16-AOS1476} and \cite{ZHU2019345}. 
While these studies collectively enrich the methodological toolkit for dynamic spatial and network panel models, they all focus on balanced panel data settings.

In the field of unbalanced panel data, existing studies have focused on specific subtopics but lack integration with spatial dynamic structures. \cite{c50faaf4-3ca5-360a-914d-60311fe2980c}, \cite{1ade2690-c518-3924-afd9-60d7e947837b}, and \cite{HONORE2004293} address censored unbalanced panel data models, while \cite{WOOLDRIDGE1995115} and \cite{d0ceac3d-09ee-347a-9a49-b8b357c03f08} investigate unbalanced panels with sample selection bias. \cite{su_estimation_2025} develop estimation and inference methods for unbalanced panels with interactive fixed effects, and \cite{SASAKI2017320} study dynamic panel data with irregular time intervals. For spatial panel data, \cite{Meng21042026} examine a static genuinely unbalanced spatial panel data model, while \cite{Meng25072025} focus on static spatial panel data models with missing observations. None provide a unified framework for estimation and inference in unbalanced dynamic network panels.

This paper bridges this gap. We propose a quasi-maximum likelihood estimator (QMLE) for an unbalanced dynamic network panel model that captures network interactions and temporal dynamics simultaneously, while explicitly accommodating unit entry and exit.
We focus on panels that are \textit{genuinely} unbalanced \citep{Meng21042026}, \textbf{meaning that the unbalancedness is driven by the entry and exit of units rather than by missing observations}. This specification allows us to avoid modeling the missing-data process.
Our main contributions and results are as follows.

\begin{enumerate}
\item We propose a baseline one-way fixed effect (OWFE) model for genuinely unbalanced dynamic network panels. The model incorporates contemporaneous network spillovers, lagged network and temporal effects, and a common listing
effect in a unit's first active period. For this model, we develop the QMLE, establish consistency and asymptotic normality, characterize the incidental-parameter bias, and derive a bias-corrected estimator with a consistent asymptotic covariance matrix estimator.

\item Since the size and elements of the network weight matrices vary over time, our setting differs from balanced spatial panel models. We establish new linear--quadratic form results that extend the existing theory from balanced spatial dynamic panels to genuinely unbalanced dynamic network panels. These results may also be useful for future studies of unbalanced network panel data.

\item We further extend the baseline framework to a two-way fixed effect (TWFE) specification. Under this extension, the QMLE remains consistent, while the incidental-parameter bias contains both the OWFE bias component and an additional component induced by the time fixed effects. We derive the corresponding asymptotic distribution, bias correction, and consistent asymptotic covariance matrix estimator.
\end{enumerate}

We examine the finite sample performance of the estimator through Monte Carlo simulations. Furthermore, we apply our model to study an empirical Airbnb listing dataset. We compare the estimated results across different regions and spatial scales to analyze the similarities and differences in the dynamic pricing behavior of Airbnb listings across cities.

The rest of the paper is organized as follows. Section~\ref{section2}
introduces the baseline OWFE model and the concentrated likelihood. Section~\ref{section3}
develops the asymptotic theory, bias correction, and covariance estimation for
the baseline QMLE. Section~\ref{sec:twoway_extension} extends the framework to
TWFE. Section~\ref{section4} reports Monte Carlo simulations, with additional
TWFE results provided in the appendix. Section~\ref{section5} presents the
empirical application, and Section~\ref{section6} concludes. All proofs are
collected in the Supplementary Appendices.


\section*{Notation}
We define $\iota_N$ as the $N$-dimensional column vector of ones, and $I_N$ as the $N$-dimensional identity matrix.
For any matrix $B$, $B_{ij}$ represents the $(i,j)$-th element of the matrix $B$. $B_{i,\cdot}$ and $B_{\cdot,j}$ are the $i$-th row and $j$-th column of $B$, respectively.
Denote $\|B\|_{1}\equiv\sup_{1\leq j\leq N}\sum_{i=1}^N|B_{ij}|$ and $\|B\|_{\infty}\equiv\sup_{1\leq i\leq N}\sum^N_{j=1}|B_{ij}|$ as the one-norm and the infinity-norm for matrix $B$, respectively. 
For a vector \(C\), let
\(\|C\|_1=\sum_{i=1}^N|C_i|\),
\(\|C\|_\infty=\sup_i|C_i|\), and
\(\|C\|_2=(\sum_{i=1}^N C_i^2)^{1/2}\)
denote the \(\ell_1\), infinity, and Euclidean norms of \(C\), respectively.
Let $\mathrm{tr}(B)$ denote the trace of matrix $B$ and $|B|$ denote the determinant of $B$.
Define $\sup_t B_{t}$ as the element-wise supremum of the matrix sequence $\{B_{t}\}$, which means that $(\sup_t B_{t})_{ij}=\sup_t B_{t,ij}$ for $i,j=1,\cdots, N$.
Also, $\mathrm{abs}(B)$ represents the element-wise absolute value of the matrix $B$, i.e., $[\mathrm{abs}(B)]_{ij}=|B_{ij}|$.
For any square matrix $B$, $B^-$ denotes the strictly lower triangular part of $B$, obtained by setting the diagonal and upper triangular entries to zero.
The notation $\mathrm{diag}(b_1,\cdots, b_N)$ represents a diagonal matrix with diagonal entries $b_1,b_2,\cdots,b_N$.
Moreover, $\mathrm{blkdiag}(B_1,\cdots,B_T)$ forms a block-diagonal matrix by the given submatrices.
For brevity, the transposes of symmetrically positioned submatrices in a symmetric matrix are sometimes denoted by $*$.
$\mathrm{Var}$ and $\mathrm{Cov}$ represent the variance and covariance of random variables, respectively.

\section{The Model and Concentrated Likelihood Function}\label{section2}
\subsection{The Model}
Consider a genuinely unbalanced dynamic network panel, where nodes are fully observed during each period (i.e., there are no missing data), and nodes that are not listed do not have any impact on the whole network.
Suppose that we have observations for a network of $N$ nodes over $T$ time periods. 
However, for the $i$-th node, we only have a total of $T_i$ consecutive observations\footnote{
We assume that each individual enters the market only once. This condition can be generalized to allow the same individual to enter and exit multiple times.}. 
In the $t$-th time period, observed nodes may drop out of the network, and new nodes may join the network. We can only observe $N_t$ nodes in period $t$ and let $\mathbb{I}_t = \{i_1, i_2, \cdots, i_{N_t}\}$ denote the set of indices of nodes observed. Their corresponding attributes are $(y_{it},x_{it}')$ for each $i\in\mathbb{I}_t$, where $y_{it}$ is a scalar dependent variable and $x_{it}$ is a $k$-dimensional vector of regressors. The total sample size is $n=\sum_{t=1}^{T}N_t$.


Then we can establish the following model: for $t=1,2,\cdots,T$ and $i\in\mathbb{I}_t$,
\begin{align}\label{model_e}
y_{it}=\rho_0\sum_{j\in\mathbb{I}_t}w_{ij,t}y_{jt}+d_{i,t-1}(\lambda_0\sum_{j\in\mathbb{I}_{t-1}}m_{ij,t}y_{j,t-1}+\nu_0 y_{i,t-1})+d_{it}&(1-d_{i,t-1})\gamma_0+x_{it}'\beta_0+\alpha_{i0}+v_{it},
\end{align}
%
where $w_{ij,t}$ and $m_{ij,t}$ are predefined network weights,
and $d_{it}$ is an indicator that equals $1$ if $y_{it}$ is observed and 0 otherwise. $\gamma_0$ and $\alpha_{i0}$ are unknown listing and individual fixed effects, respectively, where the listing effect only takes effect at the first appearance of each individual, and the effect is identical for each individual. 
The above model assumes that $y_{it}$ is not only affected by its own covariates and the current network structure, but also by the fixed listing effect if node $i$ has just been listed or by its own lagged outcome and the lagged network state if node $i$ was already present. Since we only model the nodes that appear at $t = 0$ or later and cannot observe the data for $t<0$, to simplify the analysis, we ignore the impact of nodes that existed only before $t=0$.

Denote $Y_t=(y_{i_1t},\cdots,y_{i_{N_t}t})'$ as an $N_t$-dimensional column vector, $X_t=(x_{i_1t},\cdots,x_{i_{N_t}t})'$ as an $N_t\times k$ matrix, and $V_t=(v_{i_1t},\cdots,v_{i_{N_t}t})'$ as an $N_t$-dimensional column vector, where $i_1,\cdots,i_{N_t}\in\mathbb{I}_t$.
Define $\alpha_0=(\alpha_{10},\alpha_{20},\cdots,\alpha_{N0})'$. The $N_t\times N$ matrix $D_t$ is obtained by removing the rows corresponding to unobserved nodes during period $t$ from $I_N$.
Then the above model can be written in the following matrix form: for $t=1,2,\cdots,T$,
\begin{align}\label{model_m}
Y_t=\rho_0 W_tY_t+\lambda_0 D_t(I_N-F_t)D_t'M_tY_{t-1}+\nu_0 D_tD_{t-1}'Y_{t-1}+\gamma_0 D_t F_t\iota_N+X_t\beta_0+D_t\alpha_0+V_t.
\end{align}
In the above model, $W_t$ and $M_t$ are $N_t \times N_t$ and $N_t \times N_{t-1}$ network weight matrices, respectively. The elements of $W_t$ and $M_t$ are defined as:  
$$
(W_t)_{k,l} \equiv w_{i_ki_l,t}, \quad \text{for } k,l=1,2,\cdots,N_t \text{ and } i_k,i_l\in\mathbb{I}_t,
$$  
$$
(M_t)_{k,l} \equiv m_{i_kj_l,t}, \quad \text{for } k=1,2,\cdots,N_t\text{, }l=1,2,\cdots,N_{t-1}\text{, }  i_k\in\mathbb{I}_t\text{ and } j_l\in\mathbb{I}_{t-1}.
$$  

The $N$-dimensional diagonal matrix $F_t\equiv D_t'D_t(I_N-D_{t-1}'D_{t-1})=\mathrm{diag}(d_{1t}(1-d_{1,t-1}),\cdots,$ $d_{N,t}(1-d_{N,t-1}))$ is the listing index matrix. $F_t\iota_N$ is sparser than other covariates because, over all $T$ periods, fewer than $N$ individuals can newly join the network.

Let \(S_t\equiv I_{N_t}-\rho_0W_t\), which is assumed to be invertible, and define
\(A_t\equiv D_t[\lambda_0(I_N-F_t)D_t'M_t+\nu_0D_{t-1}']S_{t-1}^{-1}\).
Assume that the following infinite summation is well-defined. Then repeated substitution of model~(2) gives
\begin{align}
Y_t&=\sum_{h=0}^{\infty}S_{t}^{-1}A_t^{(h)}[D_{t-h} (\gamma_0F_{t-h}\iota_N+\alpha_0)+X_{t-h}\beta_0+V_{t-h}]\equiv\mathcal{E}_t+\mathcal{X}_t\beta_0+\mathcal{V}_t,\label{model_sumV}
\end{align}
where $A_t^{(h)}\equiv A_{t}A_{t-1}\cdots A_{t-h+1}$ for $h=1,2,\cdots$ and $A_t^{(0)}\equiv I_{N_t}$, $\mathcal{E}_t\equiv \sum_{h=0}^{\infty}S_{t}^{-1}A_t^{(h)}D_{t-h}(\gamma_0F_{t-h}\iota_N+\alpha_0)$, 
$\mathcal{X}_t\equiv \sum_{h=0}^{\infty}S_{t}^{-1}A_t^{(h)}X_{t-h}$ and $\mathcal{V}_t\equiv \sum_{h=0}^{\infty}S_{t}^{-1}A_t^{(h)}V_{t-h}$. 
The above decomposition can be viewed as an infinite moving average representation of $Y_t$. 
We establish the law of large numbers and the central limit theorem based on this decomposition in the lemmas in Supplementary Appendix A.


\subsection{Quasi-Maximum Likelihood Method}
Define $Z_t\equiv(D_t(I_N-F_t)D_t'M_tY_{t-1},D_tD_{t-1}'Y_{t-1},D_tF_t\iota_N,X_t)$, $\delta_0\equiv(\lambda_0,\nu_0,\gamma_0,\beta_0')'$.
Then model (\ref{model_m}) can be simplified to
\begin{align}\label{model_simp}
Y_t=\rho_0 W_tY_t+ Z_t\delta_0+D_t\alpha_0+V_t.
\end{align}
Stacking the matrices and vectors in model (\ref{model_simp}) for $t=1,2,\cdots,T$, we have 
\begin{align}\label{model_acc}
\mathbb{Y}=\rho_0\mathbb{W}\mathbb{Y}+\mathbb{Z}\delta_0+\mathbb{D}\alpha_0+\mathbb{V},
\end{align}
where $\mathbb{Y}=(Y_1',\cdots,Y_T')'$, $\mathbb{W}=\mathrm{blkdiag}(W_1,\cdots,W_T)$, $\mathbb{Z}=(Z_1',\cdots,Z_T')'$, $\mathbb{D}=(D_1',\cdots,D_T')'$ and $\mathbb{V}=(V_1',\cdots,V_T')'$.
Note that $\mathbb{D'D}=\sum_{t=1}^TD_t'D_t=\sum_{t=1}^T\mathrm{diag}(d_{1t},d_{2t},\cdots,d_{N,t})=\mathrm{diag}(T_1,T_2,\cdots,T_N)\equiv\mathcal{T}$.
Assume $v_{it}$ are i.i.d. random variables with mean $0$ and variance $\sigma^2_0$. Denote the parameters of interest as $\theta=(\rho,\delta',\sigma^2)'$ and let $\zeta=(\rho,\delta')$. Then the log-likelihood function of model (\ref{model_acc}) is
\begin{align}\label{MLE}
L_n(\theta,\alpha)=-\frac{n}{2}\ln2\pi-\frac{n}{2}\ln\sigma^2+\ln|\mathbb{S}(\rho)|-\frac{1}{2\sigma^2}\mathbb{V}'(\zeta,\alpha)\mathbb{V}(\zeta,\alpha),
\end{align}
where $\mathbb{S}(\rho)\equiv I_{n}-\rho\mathbb{W}$ and $\mathbb{V}(\zeta,\alpha)\equiv\mathbb{S(\rho)Y-Z}\delta-\mathbb{D}\alpha$. Given $\zeta$, $L_n(\theta,\alpha)$ is maximized at
$\hat{\alpha}(\zeta)=\mathbb{(D'D)}^{-1} \mathbb{D'}(\mathbb{S}(\rho)\mathbb{Y-Z\delta)}$.
Substituting $\hat{\alpha}$ into (\ref{MLE}), we obtain our 
estimator $\hat{\theta}$ by maximizing the following concentrated log-likelihood function
\begin{equation}
L_n^c(\theta)=-\frac{n}{2}\ln2\pi-\frac{n}{2}\ln\sigma^2+\ln|\mathbb{S}(\rho)|-\frac{1}{2\sigma^2}\widetilde{\mathbb{V}}'(\zeta)\widetilde{\mathbb{V}}(\zeta),\label{CMLE}
\end{equation}
where $\widetilde{\mathbb{V}}(\zeta)\equiv\mathbb{S}(\rho)\mathbb{Y-Z\delta}-\mathbb{D}\hat{\alpha}(\zeta)=\mathbb{Q(S(\rho)Y-Z\delta})$ and $\mathbb{Q}\equiv I_{n}-\mathbb{P}$, $\mathbb{P}\equiv\mathbb{D}(\mathbb{D}'\mathbb{D})^{-1}\mathbb{D}'$.
Moreover, the corresponding expected value function of $L_n^c(\theta)$ is
\begin{align}
Q_n(\theta)\equiv E[L_n^c(\theta)]=-\frac{n}{2}\ln2\pi-\frac{n}{2}\ln\sigma^2+\ln\left|\mathbb{S}(\rho)\right|-\frac{1}{2\sigma^2}E\left[\widetilde{\mathbb{V}}'(\zeta)\widetilde{\mathbb{V}}(\zeta)\right].\label{eq_Q_n}
\end{align}
We derive the asymptotic properties of the estimators based on $L_n^c(\theta)$ and $Q_n(\theta)$. In practice, we further concentrate out $\delta$ and $\sigma^2$ to accelerate computation, that is, we optimize $$L_{n}^{cc}(\rho)\equiv-\frac{n}{2}\ln2\pi-\frac{n}{2}\ln\hat{\sigma}^2+\ln|\mathbb{S}(\rho)|-\frac{n}{2}$$ instead, where
$\hat{\delta}(\rho)
=\left(\mathbb{Z}'\mathbb{Q}\mathbb{Z}
\right)^{-1}\mathbb{Z}'\mathbb{Q}\mathbb{S}(\rho)\mathbb{Y},$ and
$\hat{\sigma}^2(\rho)=\frac{1}{n}\mathbb{Y'S(\rho)}'[\mathbb{Q}-\mathbb{QZ(Z'QZ})^{-1}\mathbb{Z'Q}]\mathbb{S(\rho)Y}$.

\section{Asymptotic properties of QMLE}\label{section3}
\subsection{Consistency of QMLE}
In the proof of consistency, we first establish the consistency of all parameters except $\gamma$, and then prove the consistency of $\gamma$.
Define $\mathbb{Z}^{(-\gamma)}$ as the matrix obtained by removing the third column from $\mathbb{Z}$, and let $\mathbb{Z}^{(\gamma)}$ be the third column of $\mathbb{Z}$.
The following assumptions are needed for consistency.

\begin{assumption}\label{asp_2}
The unbalancedness of the model is nonstochastic, which means that $d_{it}$ and $D_t$ are nonstochastic for all $i$ and $t$. 
There is a lower bound on the number of observations for all individuals, i.e., $T_i\geq c_T T$ for all $i$, where $c_T>0$ is a constant that does not vary with $N$ or $T$. 
Additionally, $N_0<c_NN$, where $N_0$ denotes the number of units observed at $t=0$ and $0<c_N<1$ is a constant that does not vary with $N$ or $T$.
\end{assumption}
\begin{assumption}\label{asp_error}
The error terms $v_{it}$ are i.i.d. across $i$ and $t$ with mean 0, variance $\sigma^2_0$, third moment \(\mu_3\) and fourth moment \(\mu_4\). $E|v_{it}|^{4+2\eta}<\infty$ for some $\eta>0$. 
\end{assumption}
\begin{assumption}\label{asp_4}
$\sup_{\rho\in\Lambda}\sup_{N,T}\sup_t\|\rho W_t\|_{\infty}<1$, where $\Lambda$ is the parameter space of $\rho$. Moreover, $\Lambda$ is compact and $\rho_0$ lies in the interior of $\Lambda$.
\end{assumption}

\begin{assumption}\label{asp_X}
The elements of $x_{it}$ and $\alpha_{i0}$ are nonstochastic and bounded uniformly in $i$, $t$, $N$ and $T$. $\mathbb{Z'QZ}$ is nonsingular. Also, $\lim_{n\rightarrow\infty}\frac{1}{n}E(\mathbb{Z}^{(-\gamma)\prime}\mathbb{QZ}^{(-\gamma)})$ exists and is nonsingular, where $n=\sum_{t=1}^{T}N_t$ is the total sample size.
\end{assumption}

\begin{assumption}\label{asp_UB}
The sequences of matrices $\{W_t\}$, $\{M_t\}$ and $\{S_t^{-1}\}$ are nonstochastic and uniformly bounded in row and column sum norms (for short, UB)
\footnote{We say a $N\times N$ matrix $B$ is UB if $\sup_{N}\max\{\|B\|_1,\|B\|_\infty\}<\infty$. We say a sequence of $N_t\times N_t$ matrices $\{B_{t}\}$ is UB uniformly for $t$ if $\sup_{N,T}\sup_t\max\{\|B_{t}\|_1,\|B_{t}\|_\infty\}<\infty$.}
uniformly for $t$. Moreover,$$\max(\|\sup_t \mathrm{abs}(D_t'S_t^{-1}D_t)\|_\infty,\|\sup_t \mathrm{abs}(D_t'W_tD_t)\|_\infty,\|\sup_t \mathrm{abs}(D_t'M_tD_{t-1})\|_\infty)<\infty.$$
\end{assumption}
\begin{assumption}\label{asp_UBA}
$\sum_{h=1}^{\infty}\mathrm{abs}\left(D_t'A_{t}^{(h)}D_{t-h}\right)$ and $\sum_{h=t}^{\infty}\mathrm{abs}\left(D_h'A_{h}^{(h-t)}D_t\right)$ are UB uniformly for $t$.
\end{assumption}
\begin{assumption}\label{asp_n}
$N$ is a nondecreasing function of $T$, and $N$ will go to infinity as $T$ goes to infinity.
\end{assumption}

Assumption \ref{asp_2} regulates the unbalancedness of the panel, where we assume the unbalancedness to be nonstochastic. We will leave the more complicated case where $d_{it}$ is stochastic for future research. 
A lower bound on $T_i$ is imposed to ensure the consistency of the estimator for $\alpha$, and an upper bound on $N_0$ is imposed to guarantee the consistency of the estimator for $\gamma$, as it requires sufficiently many observed listing effects. If this condition is not satisfied, a similar asymptotic analysis can still be conducted, but the convergence rate of $\hat{\gamma}$ will change.
Assumption \ref{asp_error} provides regularity assumptions for $v_{it}$. 
Similar assumptions are adopted in \cite{https://doi.org/10.1111/j.1468-0262.2004.00558.x} and \cite{YU2008118}. Assumption \ref{asp_4} is a standard assumption for spatial (or network) panel data models. It implies that $S_t(\rho)^{-1}$ exists and
\begin{align}
\sup_{\rho\in\Lambda}\sup_{N,T}\sup_t\|S_t(\rho)^{-1}\|_{\infty}<\infty,\label{eq_asp_S}
\end{align}
which is proved in equation (S.28) in the supplementary materials of \cite{Wu_Jiang_Xu_2024}.
Although the assumption that $\rho_0$ lies in the interior of the parameter space is in fact imposed for simplicity and not required for consistency, we state the conditions jointly for notational convenience.
Assumption \ref{asp_X} rules out multicollinearity, ensuring that $\delta$ is estimable.
We consider fixed design in this paper as in \cite{https://doi.org/10.1111/j.1468-0262.2004.00558.x,LEE2007489}, \cite{YU2008118} and \cite{LEE2014174}, among others. This assumption can be relaxed to allow for stochastic exogenous variables by using the near-epoch dependence theory in \cite{JENISH2012178} or the functional dependence theory in \cite{Wu_Jiang_Xu_2024}. The proof is similar to the proof under fixed design. 
Assumption \ref{asp_UB} is a standard condition in the literature of spatial and network econometrics. See \cite{https://doi.org/10.1111/j.1468-0262.2004.00558.x,LEE2007489}, \cite{YU2008118} and \cite{LEE2014174}. The assumption that $\sup_{N,T}\sup_t\|S_t^{-1}\|_\infty<\infty$ can in fact be ensured by equation (\ref{eq_asp_S}).
Moreover, Assumption \ref{asp_UB} implies that $\{S_{t}(\rho)^{-1}\}$ is UB uniformly in a neighborhood of $\rho_0$, from Lemma A.3 of \cite{https://doi.org/10.1111/j.1468-0262.2004.00558.x}.
Assumption \ref{asp_UBA} restricts the joint effect of network spillovers and temporal dynamics, which is a common assumption in dynamic network panel data, such as Assumption 6 in \cite{YU2008118}, Assumption 7 in \cite{LEE2014174} and Assumption 6 in \cite{lee_qml_2012}. 
Assumptions \ref{asp_UB} and \ref{asp_UBA} are imposed to establish a law of large numbers and a central limit theorem for certain quadratic forms of the network lag spillover term $M_tY_{t-1}$ and the temporal lag term $Y_{t-1}$.
Assumption \ref{asp_n} means that we consider a large-$N$ and large-$T$ setting, in which the QMLE is consistent. For short dynamic network panel data models, other estimation methods, such as GMM, are required.

Denote $\mathbb{S}_0\equiv\mathbb{S}(\rho_0)$, $\mathbb{G}(\rho)\equiv\mathbb{WS}(\rho)^{-1}$ and $\mathbb{G}_0\equiv\mathbb{WS}^{-1}_0$. Define $\mathcal{H}^{(-\gamma)}\equiv\frac{1}{n}(\mathbb{G}_0\mathbb{Z}\delta_0+\mathbb{G}_0\mathbb{D}\alpha_0,\mathbb{Z}^{(-\gamma)})'\mathbb{Q}(\mathbb{G}_0\mathbb{Z}\delta_0+\mathbb{G}_0\mathbb{D}\alpha_0,\mathbb{Z}^{(-\gamma)})$ and  $\sigma_n^2\equiv\allowbreak\frac{\sigma^2_0}{n}tr[(\mathbb{S(\rho)S}_0^{-1})'\mathbb{S(\rho)S}_0^{-1}]$. As in \cite{YU2008118}, the following assumption is needed to establish the global identification condition required for consistency.
\begin{assumption}\label{asp_EH}
At least one of the following two conditions is satisfied: (i) $\lim_{n\rightarrow\infty}E\mathcal{H}^{(-\gamma)}$ exists and is nonsingular.
(ii) $\lim_{n\rightarrow\infty}\left(\frac{1}{n}\ln|\sigma_0^2\mathbb{S}_0^{-1}\mathbb{S}_0^{-1\prime}|-\frac{1}{n}\ln|\sigma_n^2\mathbb{S}(\rho)^{-1}\mathbb{S}(\rho)^{-1\prime}|\right)\neq0$ for $\rho\neq\rho_0$.
\end{assumption}

\begin{theorem}\label{consistency}
Under Assumptions \ref{asp_2}-\ref{asp_EH}, $\hat{\theta}\stackrel{p}{\rightarrow}\theta_0$.
\end{theorem}


\subsection{Asymptotic Distribution of QMLE}\label{section3.2}
Since we have concentrated out the $N$ individual fixed effects from the log likelihood function, the QMLE is subject to the incidental parameters problem, which induces an asymptotic bias in the estimator \citep{Meng21042026}.
We derive the asymptotic distribution of $\hat{\theta}$ by performing a Taylor expansion of $\frac{\partial L_n^c(\hat{\theta})}{\partial\theta}$ around $\theta_0$. The first and second order derivatives of $L_n^c(\theta)$ are derived in Supplementary Appendix  B.
According to (\ref{model_sumV}), denote $Z^\circ_t\equiv\left(D_t(I_N-F_t)D_t'M_t\mathcal{V}_{t-1},D_tD_{t-1}'\mathcal{V}_{t-1},\mathbf{0}_{N_t\times(k+1)}\right)$ and $\mathbb{Z}^\circ\equiv(Z^{\circ\prime}_1,\cdots,Z^{\circ\prime}_T)'$.
We decompose $\mathbb{QZ}$ into $\mathbb{QZ}=\widetilde{\mathbb{Z}}^{*}-\widetilde{\mathbb{Z}}^{\circ}$, where $\widetilde{\mathbb{Z}}^{\circ}\equiv\mathbb{P}\mathbb{Z}^{\circ}$ and $\widetilde{\mathbb{Z}}^{*}=\mathbb{QZ}+\widetilde{\mathbb{Z}}^{\circ}$.
Note that $Z_t^\circ$ is independent of $V_s$ when $t\leq s$, and $\widetilde{\mathbb{Z}}^{*}=\mathbb{QZ}+\widetilde{\mathbb{Z}}^{\circ}=\mathbb{Q(Z-Z}^\circ)+\mathbb{Z}^\circ$, where $\mathbb{Z-Z}^\circ$ is nonstochastic.
Decompose $\widetilde{\mathbb{Z}}^{*}$ into $N_t$-dimensional vectors $\widetilde{Z}_1^*,\cdots,\widetilde{Z}_T^*$ such that $\widetilde{\mathbb{Z}}^{*}=(\widetilde{Z}_1^{*\prime},\cdots,\widetilde{Z}_T^{*\prime})'$. Therefore, $\widetilde{Z}_t^{*}$ is independent of $V_s$ when $t\leq s$. But $\widetilde{\mathbb{Z}}^{\circ}$ is correlated with $\mathbb{V}$, as proved in Lemma 11 in Supplementary Appendix A.

Similarly, we can decompose $\mathbb{QG}_0\mathbb{Z}=\widetilde{\mathbb{G}_0\mathbb{Z}}^{*}-\widetilde{\mathbb{G}_0\mathbb{Z}}^{\circ}$, where
$\widetilde{\mathbb{G}_0\mathbb{Z}}^{\circ}\equiv\mathbb{P}\mathbb{G}_0\mathbb{Z}^{\circ}$ and $\widetilde{\mathbb{G}_0\mathbb{Z}}^{*}=\mathbb{QG}_0\mathbb{Z}+\widetilde{\mathbb{G}_0\mathbb{Z}}^{\circ}$. Further, decompose $\widetilde{\mathbb{G}_0\mathbb{Z}}^{*}$ into $N_t$-dimensional vectors $\widetilde{GZ}_1^*,\cdots,\widetilde{GZ}_T^*$ such that $\widetilde{\mathbb{G}_0\mathbb{Z}}^{*}=(\widetilde{GZ}_1^{*\prime},\cdots,\widetilde{GZ}_T^{*\prime})'$. Similarly, $\widetilde{GZ}_t^{*}$ is independent of $V_s$ when $t\le s$, whereas $\widetilde{\mathbb{G}_0\mathbb{Z}}^{\circ}$ is correlated with $\mathbb{V}$.

Let $\Gamma=\mathrm{blkdiag}(I_3,\sqrt{T},I_{k+1})$ and $\widetilde{\Gamma}=\mathrm{blkdiag}(I_2,\sqrt{T},I_{k})$. Then we can rewrite $\frac{1}{\sqrt{n}}\Gamma\frac{\partial L_n^c(\theta_0)}{\partial\theta}=\frac{1}{\sqrt{n}}\Gamma\frac{\partial L_n^{c*}(\theta_0)}{\partial\theta}-\Delta_n$, where
$$\frac{1}{\sqrt{n}}\Gamma\frac{\partial L_n^{c*}(\theta_0)}{\partial\theta}=\left(\begin{aligned}
&\frac{1}{\sqrt{n}\sigma^2_0}(\mathbb{G}_0\mathbb{D\alpha}_0)'\mathbb{QV}+\frac{1}{\sqrt{n}\sigma_0^2}\mathbb{V}'\mathbb{G}_0\mathbb{V}+\frac{1}{\sqrt{n}\sigma_0^2}(\widetilde{\mathbb{G}_0\mathbb{Z}}^*\delta_0)'\mathbb{V}-\frac{1}{\sqrt{n}}\mathrm{tr}(\mathbb{G}_0)\\
&\frac{1}{\sqrt{n}\sigma_0^2}\widetilde{\Gamma}\widetilde{\mathbb{Z}}^{*\prime}\mathbb{V}\\
&-\frac{n}{2\sqrt{n}\sigma_0^2}+\frac{1}{2\sqrt{n}\sigma_0^4}\mathbb{V'V}
\end{aligned}\right),$$
and
$$\Delta_n=\left(\begin{aligned}
&\frac{1}{\sqrt{n}\sigma_0^2}\mathbb{V}'\mathbb{G}_0'\mathbb{P}\mathbb{V}+\frac{1}{\sqrt{n}\sigma_0^2}(\widetilde{\mathbb{G}_0\mathbb{Z}}^\circ\delta_0)'\mathbb{V}\\
&\frac{1}{\sqrt{n}\sigma_0^2}\widetilde{\Gamma}\widetilde{\mathbb{Z}}^{\circ\prime}\mathbb{V}\\
&\frac{1}{2\sqrt{n}\sigma_0^4}\mathbb{V'}\mathbb{P}\mathbb{V}
\end{aligned}\right).$$
This decomposes $\frac{\partial L_n^c(\theta_0)}{\partial\theta}$ into two components: a mean zero term $\frac{\partial L_n^{c*}(\theta_0)}{\partial\theta}$ and a nonzero mean term $\Delta_n$. $\Delta_n$ can be regarded as the source of asymptotic bias arising from the estimation of fixed effects.
As derived in Supplementary Appendix B.2, $$E\left(\frac{1}{\sqrt{n}}\Gamma\frac{\partial L_n^{c*}(\theta_0)}{\partial\theta}\cdot\frac{1}{\sqrt{n}}\frac{\partial L_n^{c*}(\theta_0)}{\partial\theta'}\Gamma\right)=\Gamma\Sigma_{\theta_0,n}\Gamma+\Gamma\Omega_{\theta_0,n}\Gamma+O\left(\frac{1}{T}\right).$$
Denote $\Sigma_{\theta_0}\equiv\lim_{n\rightarrow\infty}\Gamma\Sigma_{\theta_0,n}\Gamma$, $\Omega_{\theta_0}\equiv\lim_{n\rightarrow\infty}\Gamma\Omega_{\theta_0,n}\Gamma$, then
\begin{align}
\lim_{n\rightarrow\infty}E\left(\frac{1}{\sqrt{n}}\Gamma\frac{\partial L_n^{c*}(\theta_0)}{\partial\theta}\cdot\frac{1}{\sqrt{n}}\frac{\partial L_n^{c*}(\theta_0)}{\partial\theta'}\Gamma\right)=\Sigma_{\theta_0}+\Omega_{\theta_0}.\label{eq_varg}
\end{align}

The asymptotic distribution of $\frac{\partial L_n^{c*}(\theta_0)}{\partial\theta}$ can be derived by applying the central limit theorem for martingale difference sequences, as proved in Lemma 18 in Supplementary Appendix A.
Furthermore, as derived in Supplementary Appendix B.4, we have $\Delta_n=\sqrt{\frac{N}{c_uT}}\varphi_n(\theta_0)+O_p\left(\frac{1}{\sqrt{T}}\right)$, where $c_u\equiv\frac{n}{NT}$, $\varphi_n(\theta_0)=O(1)$ and $\varphi_n(\theta)$ is defined in equation (B.6) in Supplementary Appendix B.4.

The following assumption ensures the nonsingularity of the information matrix $\Sigma_{\theta_0}$.
\begin{assumption}\label{asp_nonsingularity} At least one of the following two conditions is satisfied: (i) $\lim_{n\rightarrow\infty}E\mathcal{H}^{(-\gamma)}$ exists and is nonsingular.
(ii) $\lim_{n\rightarrow\infty}\frac{1}{n}\left(\mathrm{tr}(\mathbb{G}_0^2+\mathbb{G}_0'\mathbb{G}_0)-\frac{2}{n}\left[\mathrm{tr}(\mathbb{G}_0)\right]^2\right)\neq0$.
\end{assumption}

\begin{theorem}\label{Normality}
Under Assumptions \ref{asp_2}-\ref{asp_nonsingularity},
\begin{align}
\sqrt{n}\Gamma^{-1}\left[\hat{\theta}-\left(\theta_0-\frac{\Sigma_{\theta_0,n}^{-1}\varphi_n(\theta_0)}{c_uT}\right)\right]+O_p\left(\max\left\{\frac{1}{\sqrt{T}},\sqrt{\frac{N}{T^3}}\right\}\right)\overset{d}{\rightarrow}N\left(0,\Sigma_{\theta_0}^{-1}(\Sigma_{\theta_0}+\Omega_{\theta_0})\Sigma_{\theta_0}^{-1}\right),\label{eq_Normality}
\end{align}
where $\Sigma_{\theta_0,n}^{-1}\varphi_n(\theta_0)$ is O(1).
\end{theorem}

If error terms $v_{it}$ are normal, $\Omega_{\theta_0}=0$ and the asymptotic variance of equation (\ref{eq_Normality}) becomes $\Sigma_{\theta_0}^{-1}$.
Theorem \ref{Normality} shows that $\hat{\theta}$ exhibits an asymptotic bias $-\Sigma_{\theta_0,n}^{-1}\varphi_n(\theta_0)/(c_uT)$. It also shows that $\hat{\theta}^{(-\gamma)}$ is $\sqrt{n}$-consistent and $\hat{\gamma}$ is $\sqrt{c_uN}$-consistent.
When $N/T\rightarrow0$, $\hat{\theta}$ is consistent and asymptotically centered normal, which means
$\sqrt{n}\Gamma^{-1}\left(\hat{\theta}-\theta_0\right)\overset{d}{\rightarrow}N\left(0,\Sigma_{\theta_0}^{-1}(\Sigma_{\theta_0}+\Omega_{\theta_0})\Sigma_{\theta_0}^{-1}\right)$.


\subsection{Bias Correction}

According to (\ref{eq_Normality}), $\hat{\theta}$ exhibits the bias $-\Sigma_{\theta_0,n}^{-1}\varphi_n(\theta_0)/(c_uT)$. Thus a bias-corrected counterpart of $\hat{\theta}$ takes the form
$$\hat{\theta}^1=\hat{\theta}+\frac{\hat{B}_n}{c_uT},$$
where $\hat{B}_n=\left.\left[\left(-\frac{1}{n}\frac{\partial^2L_n^{c}(\theta)}{\partial\theta\partial\theta'}\right)^{-1}\varphi_n(\theta)\right]\right|_{\theta=\hat{\theta}}$.
\begin{assumption}\label{asp_UBAT}
$\sum_{h=0}^{T-t}D_t'A_t^{(h)}(\theta)D_{t-h}$ and $\frac{\partial}{\partial\theta}\sum_{h=0}^{T-t}D_t'A_t^{(h)}(\theta)D_{t-h}$ are UB uniformly in a neighborhood of $\theta_0$ over $t$.
\end{assumption}
\begin{theorem}\label{bias-corrected_Normality}
Under Assumptions \ref{asp_2}-\ref{asp_UBAT}, if $N/T^3\rightarrow0$,
\begin{align*}
\sqrt{n}\Gamma^{-1}\left(\hat{\theta}^1-\theta_0\right)\overset{d}{\rightarrow}N\left(0,\Sigma_{\theta_0}^{-1}(\Sigma_{\theta_0}+\Omega_{\theta_0})\Sigma_{\theta_0}^{-1}\right).
\end{align*}
\end{theorem}
Hence, the estimator after bias correction is asymptotically normal and centered around $\theta_0$ as long as $T$ grows faster than $N^{\frac{1}{3}}$.

\subsection{Estimation of the Asymptotic Covariance Matrix}

To make the asymptotic distribution in Theorem~\ref{bias-corrected_Normality} operational, we construct a
plug-in estimator of the asymptotic covariance matrix. The population covariance
matrix has the sandwich form
\(
\Sigma_{\theta_0}^{-1}
(\Sigma_{\theta_0}+\Omega_{\theta_0})
\Sigma_{\theta_0}^{-1},
\)
where \(\Sigma_{\theta_0}\) is the limiting information matrix and
\(\Omega_{\theta_0}\) is the additional non-Gaussian correction term. Let
\(\widehat{\Sigma}\) and \(\widehat{\Omega}\) denote the plug-in estimators of
the normalized matrices \(\Gamma\Sigma_{\theta_0,n}\Gamma\) and
\(\Gamma\Omega_{\theta_0,n}\Gamma\), respectively, constructed from the
bias-corrected estimator and the corresponding residual moments. The
explicit formulas are reported in Supplementary Appendix~C.4. Define
\(
\widehat{\Xi}
=
\widehat{\Sigma}^{-1}
(\widehat{\Sigma}+\widehat{\Omega})
\widehat{\Sigma}^{-1}.
\)

\begin{theorem}
Under Assumptions~\ref{asp_2}-\ref{asp_UBAT}, if \(N/T^3\to0\), then
\(
\widehat{\Xi}
\xrightarrow{p}
\Sigma_{\theta_0}^{-1}
(\Sigma_{\theta_0}+\Omega_{\theta_0})
\Sigma_{\theta_0}^{-1}.
\)
Consequently, \(n^{-1}\Gamma\widehat{\Xi}\Gamma\) is a consistent estimator
of the asymptotic covariance matrix of \(\widehat{\theta}^{\,1}\).
\end{theorem}

\section{Extension to TWFE}
\label{sec:twoway_extension}

In some applications, in addition to individual heterogeneity, the dependent variable may also be affected by common period-specific shocks, such as macroeconomic fluctuations, seasonality, or platform-wide demand changes. To accommodate this possibility, we introduce time fixed effects into the baseline genuinely unbalanced model. The individual and time fixed effects are jointly concentrated out from the likelihood, and the asymptotic properties of the resulting TWFE QMLE are then derived. Relative to the OWFE case, the main technical difference is that the incidental-parameter bias must be rederived under the joint projection associated with the two sets of fixed effects.

\subsection{Model Specification}

For $t=1,\ldots,T$ and $i\in \mathbb{I}_t$, consider the following extension of model (\ref{model_e}):
\begin{align}
y_{it}=\rho_0 \sum_{j\in\mathbb{I}_t} w_{ij,t} y_{jt}+d_{i,t-1}
\left(
\lambda_0 \sum_{j\in\mathbb{I}_{t-1}} m_{ij,t} y_{j,t-1}+\nu_0 y_{i,t-1}
\right)+d_{it}(1-d_{i,t-1})\gamma_0\nonumber\\
+x_{it}'\beta_0+\alpha_{i0}+\tau_{t0}+v_{it},\label{model_TWFE}
\end{align}
where $\tau_{t0}$ denotes the time fixed effect in period $t$. As in the baseline model, $\alpha_{i0}$ captures unit-specific heterogeneity, while $\gamma_0$ represents the common listing effect at a unit's first appearance. The new term $\tau_{t0}$ captures shocks common to all units active in period $t$.
To avoid overparameterization, we impose the normalization $\sum_{t=1}^T \tau_{t0}=0.$ Let $\tau_0=(\tau_{20},\ldots,\tau_{T0})'$.
Then, stacking over $t$, the model can be written as
$$\mathbb{Y}=\rho_0 \mathbb{W}\mathbb{Y}+\mathbb{Z}\delta_0+\mathbb{D}\alpha_0+\mathbb{J}\tau_0+\mathbb{V},$$
where $\mathbb{J}\equiv[-\iota_{N_1}\iota_{T-1}';\mathrm{blkdiag}(\iota_{N_2},\iota_{N_3},\cdots,\iota_{N_T})]$, all other objects are defined as in Section \ref{section2} for the baseline model.

\subsection{Quasi-Maximum Likelihood Estimation}

Under the extension with time fixed effects, the corresponding quasi-log-likelihood function is
\begin{align*}
L_n(\theta,\alpha,\tau)
=
&-\frac{n}{2}\ln(2\pi)
-\frac{n}{2}\ln\sigma^2
+\ln|\mathbb{S}(\rho)| \\
&-\frac{1}{2\sigma^2}
\left(
\mathbb{S}(\rho)\mathbb{Y}
-
\mathbb{Z}\delta
-
\mathbb{D}\alpha
-
\mathbb{J}\tau
\right)'
\left(
\mathbb{S}(\rho)\mathbb{Y}
-
\mathbb{Z}\delta
-
\mathbb{D}\alpha
-
\mathbb{J}\tau
\right).
\end{align*}

For given $\zeta$, let
$
(\hat{\alpha}(\zeta)',\hat{\tau}(\zeta)')'
=
\left[
(\mathbb{D},\mathbb{J})'(\mathbb{D},\mathbb{J})
\right]^{-1}
(\mathbb{D},\mathbb{J})'
\left(
\mathbb{S}(\rho)\mathbb{Y}
-
\mathbb{Z}\delta
\right).
$
Accordingly, define
$
\widebar{\mathbb{P}}
\equiv
(\mathbb{D},\mathbb{J})
\left[
(\mathbb{D},\mathbb{J})'(\mathbb{D},\mathbb{J})
\right]^{-1}
(\mathbb{D},\mathbb{J})',
\widebar{\mathbb{Q}}
\equiv
I_n-\widebar{\mathbb{P}}.
$
Substituting out $\alpha$ and $\tau$ yields the concentrated quasi-log-likelihood
\begin{align*}
\widebar{L}_n^c(\rho,\delta,\sigma^2)
=
-\frac{n}{2}\ln(2\pi)
-\frac{n}{2}\ln\sigma^2
+\ln|\mathbb{S}(\rho)|-\frac{1}{2\sigma^2}
\left(
\mathbb{S}(\rho)\mathbb{Y}
-
\mathbb{Z}\delta
\right)'
\widebar{\mathbb{Q}}
\left(
\mathbb{S}(\rho)\mathbb{Y}
-
\mathbb{Z}\delta
\right).
\end{align*}

The concentrated quasi-log-likelihood function $\widebar{L}_n^c(\rho,\delta,\sigma^2)$ serves as the basis for the subsequent asymptotic analysis. Exactly as in the baseline OWFE case, in practical implementation we further concentrate out $\delta$ and $\sigma^2$, so that the optimization is ultimately carried out with respect to $\rho$ alone. Specifically, the profile quasi-log-likelihood for $\rho$ is
$$
\widebar{L}_n^{cc}(\rho)
=
-\frac{n}{2}\ln(2\pi)
-\frac{n}{2}\ln\hat{\sigma}^2(\rho)
+\ln|\mathbb{S}(\rho)|
-\frac{n}{2},
$$
where
$
\hat{\delta}(\rho)
=
\left(
\mathbb{Z}'\widebar{\mathbb{Q}}\mathbb{Z}
\right)^{-1}
\mathbb{Z}'\widebar{\mathbb{Q}}\mathbb{S}(\rho)\mathbb{Y},
$
and $
\hat{\sigma}^2(\rho)
=
\frac{1}{n}
\mathbb{Y}'\mathbb{S}(\rho)'
\left[
\widebar{\mathbb{Q}}
-
\widebar{\mathbb{Q}}\mathbb{Z}
\left(
\mathbb{Z}'\widebar{\mathbb{Q}}\mathbb{Z}
\right)^{-1}
\mathbb{Z}'\widebar{\mathbb{Q}}
\right]
\mathbb{S}(\rho)\mathbb{Y}.
$



\subsection{Asymptotic properties}
The consistency proof proceeds exactly as in the baseline model. To avoid restating conditions that remain unchanged, we continue to maintain Assumptions~\ref{asp_error}, \ref{asp_4}, and \ref{asp_UB}-\ref{asp_n} of the baseline model. Relative to the OWFE case, only Assumptions~\ref{asp_2}, \ref{asp_X}, and \ref{asp_EH} need to be modified under the TWFE specification.

\setcounter{passumption}{0}
\begin{passumption}\label{ass:1prime}
The unbalancedness of the model is nonstochastic, which means that $d_{it}$ and $D_t$ are nonstochastic for all $i$ and $t$. There is a lower bound on the number of observations for all individuals, i.e., $T_i \ge c_T T$ for all $i$, where $c_T > 0$ is a constant that does not vary with $N$ or $T$. Additionally, $N_0 < c_N N$, where $N_0$ denotes the number of units observed at $t=0$ and $0 < c_N < 1$ is a constant that does not vary with $N$ or $T$. There exists a constant $\bar c_N > 0$, not varying with $N$ or $T$, such that $\inf_{1\le t\le T}\frac{N_t}{N}\ge \bar c_N$. Moreover, the matrix \(\mathbb J'\mathbb Q\mathbb J\) is uniformly nonsingular for \(N\) and \(T\).
\end{passumption}

Assumption \ref{ass:1prime} strengthens Assumption \ref{asp_2} in two aspects that are needed for the TWFE specification. The condition \(\inf_{1\le t\le T} \frac{N_t}{N}\ge \bar c_N\) rules out time periods with too few active units, so that each time fixed effect is estimated from a non-negligible cross-sectional sample. The uniform nonsingularity of \(\mathbb{J}'\mathbb{Q}\mathbb{J}\) rules out collinearity between the time fixed effects and the individual fixed effects after projecting out the latter. It ensures that the time fixed effects can be properly concentrated out and that the joint projection matrix is well defined.

The following two assumptions are similar to their OWFE counterparts. We use \(\widebar{\mathbb{Q}}\) and \(\widebar{\mathcal{H}}^{(-\gamma)}\) instead to accommodate the additional component involving \(\mathbb{J}\).

\setcounter{passumption}{3}
\begin{passumption}\label{ass:4prime}
The elements of $x_{it}$, $\alpha_{i0}$ and $\tau_{t0}$ are nonstochastic and bounded uniformly in $i$, $t$, $N$ and $T$. $\mathbb{Z}'\widebar{\mathbb{Q}}\mathbb{Z}$ is nonsingular. Also, $\lim_{n\to\infty}\frac{1}{n}E(\mathbb{Z}^{(-\gamma)\prime}\widebar{\mathbb{Q}}\mathbb{Z}^{(-\gamma)})$ exists and is nonsingular.
\end{passumption}

Define
$\widebar{\mathcal{H}}^{(-\gamma)}
\equiv
\frac{1}{n}
\bigl(
\mathbb{G}_0\mathbb{Z}\delta_0
+
\mathbb{G}_0\mathbb{D}\alpha_0
+
\mathbb{G}_0\mathbb{J}\tau_0,\,
\mathbb{Z}^{(-\gamma)}
\bigr)'
\widebar{\mathbb{Q}}
\bigl(
\mathbb{G}_0\mathbb{Z}\delta_0
+
\mathbb{G}_0\mathbb{D}\alpha_0
+
\mathbb{G}_0\mathbb{J}\tau_0,\,
\mathbb{Z}^{(-\gamma)}
\bigr)$
and $\widebar{\sigma}_n^2
\equiv
\frac{\sigma_0^2}{n}
\mathrm{tr}
\left[
\bigl(\mathbb{S}(\rho)\mathbb{S}_0^{-1}\bigr)'
\mathbb{S}(\rho)\mathbb{S}_0^{-1}
\right]$.

\setcounter{passumption}{7}
\begin{passumption}\label{ass:8prime}
At least one of the following two conditions is satisfied: (i) $\lim_{n\to\infty}E\widebar{\mathcal{H}}^{(-\gamma)}$ exists and is nonsingular. (ii) $\lim_{n\to\infty}
\left(
\frac{1}{n}\ln\bigl|\sigma_0^2\mathbb{S}_0^{-1}\mathbb{S}_0^{-1\prime}\bigr|
-
\frac{1}{n}\ln\bigl|\widebar{\sigma}_n^2\mathbb{S}(\rho)^{-1}\mathbb{S}(\rho)^{-1\prime}\bigr|
\right)
\neq 0
$ for $\rho \neq \rho_0$.
\end{passumption}

\begin{theorem}
Under Assumptions~\ref{ass:1prime}, \ref{asp_error}, \ref{asp_4}, \ref{ass:4prime}, \ref{asp_UB}-\ref{asp_n} and \ref{ass:8prime},
$\hat{\theta}\xrightarrow{p}\theta_0.$
\end{theorem}

The TWFE score decomposition is detailed in Supplementary Appendix~B.5. It is the TWFE counterpart of the decomposition in
Section~\ref{section3.2}, with the OWFE projection \(\mathbb Q\) replaced by the joint
projection \(\widebar{\mathbb Q}\). Specifically, Supplementary
Appendix~B.5 shows that
\[\frac{1}{\sqrt n}\Gamma\frac{\partial\widebar{L}_n^c(\theta_0)}{\partial\theta} = \frac{1}{\sqrt n}\Gamma\frac{\partial\widebar{L}_n^{c*}(\theta_0)}{\partial\theta} -\widebar{\Delta}_n,\] 
where the first term is a mean-zero score component with
limiting covariance
\(\widebar{\Sigma}_{\theta_0}+\widebar{\Omega}_{\theta_0}\). Moreover,
\[ \widebar{\Delta}_n = \sqrt{\frac{N}{c_uT}}\varphi_n(\theta_0) + \sqrt{\frac{T}{c_uN}}\psi_n(\theta_0) + O_p\left(\frac{1}{\sqrt T}\right) + O_p\left(\frac{1}{\sqrt N}\right), \]
where \(c_u=n/(NT)\), \(\varphi_n(\theta_0)=O(1)\) is the OWFE bias
component, and \(\psi_n(\theta_0)=O(1)\) is the additional bias component
generated by the time fixed effects. These results, together with the same
Taylor expansion argument as in the OWFE case, lead to the following asymptotic distribution.


\begin{passumption}\label{asp_nonsingularity'}
At least one of the following two conditions is satisfied:
(i) $\lim_{n\rightarrow\infty}E\widebar{\mathcal{H}}^{(-\gamma)}$ exists and is nonsingular.
(ii) $\lim_{n\rightarrow\infty}\frac{1}{n}\left(\mathrm{tr}(\mathbb{G}_0^2+\mathbb{G}_0'\mathbb{G}_0)-\frac{2}{n}\left[\mathrm{tr}(\mathbb{G}_0)\right]^2\right)\neq0$.
\end{passumption}

\begin{theorem}\label{Normality_twfe}
Under Assumptions~\ref{ass:1prime}, \ref{asp_error}, \ref{asp_4}, \ref{ass:4prime}, \ref{asp_UB}-\ref{asp_n}, and \ref{ass:8prime}-\ref{asp_nonsingularity'},
\begin{align}
\sqrt{n}\Gamma^{-1}
\left[
\hat{\theta}
-
\left(
\theta_0
-
\frac{\widebar{\Sigma}_{\theta_0,n}^{-1}\varphi_n(\theta_0)}{c_uT}
-
\frac{\widebar{\Sigma}_{\theta_0,n}^{-1}\psi_n(\theta_0)}{c_uN}
\right)
\right]
+O_p\left(
\max\left\{
\frac{1}{\sqrt{T}},
\frac{1}{\sqrt{N}},
\sqrt{\frac{N}{T^3}},
\sqrt{\frac{T}{N^3}}
\right\}
\right)\nonumber\\
\overset{d}{\rightarrow}
N\left(
0,
\widebar{\Sigma}_{\theta_0}^{-1}
(\widebar{\Sigma}_{\theta_0}+\widebar{\Omega}_{\theta_0})
\widebar{\Sigma}_{\theta_0}^{-1}
\right).
\label{eq_Normality_twfe}
\end{align}
where $\widebar{\Sigma}_{\theta_0,n}^{-1}\varphi_n(\theta_0)$ and $\widebar{\Sigma}_{\theta_0,n}^{-1}\psi_n(\theta_0)$ are both $O(1)$.
\end{theorem}

If error terms $v_{it}$ are normal, $\widebar{\Omega}_{\theta_0}=0$ and the asymptotic variance of equation (\ref{eq_Normality_twfe}) becomes $\widebar{\Sigma}_{\theta_0}^{-1}$.
Theorem \ref{Normality_twfe} shows that $\hat{\theta}$ exhibits two asymptotic bias components $-\frac{\widebar{\Sigma}_{\theta_0,n}^{-1}\varphi_n(\theta_0)}{c_uT}$ and $-\frac{\widebar{\Sigma}_{\theta_0,n}^{-1}\psi_n(\theta_0)}{c_uN}$.
It also shows that $\hat{\theta}^{(-\gamma)}$ is $\sqrt{n}$-consistent and $\hat{\gamma}$ is $\sqrt{c_uN}$-consistent.

\subsection{Bias Correction}

According to (\ref{eq_Normality_twfe}), $\hat{\theta}$ exhibits the bias
$-\frac{\widebar{\Sigma}_{\theta_0,n}^{-1}\varphi_n(\theta_0)}{c_uT}-\frac{\widebar{\Sigma}_{\theta_0,n}^{-1}\psi_n(\theta_0)}{c_uN}$,
thus a bias-corrected counterpart of $\hat{\theta}$ takes the form
\[
\hat{\theta}^{\,1}
=
\hat{\theta}
+
\frac{\hat{B}_{T,n}}{c_uT}
+
\frac{\hat{B}_{N,n}}{c_uN},
\]
where
\(
\hat{B}_{T,n}
=
\left.
\left[
\left(
-\frac{1}{n}\frac{\partial^2\widebar{L}_n^{c}(\theta)}{\partial\theta\partial\theta'}
\right)^{-1}
\varphi_n(\theta)
\right]
\right|_{\theta=\hat{\theta}}\),
and
\(\hat{B}_{N,n}
=
\left.
\left[
\left(
-\frac{1}{n}\frac{\partial^2\widebar{L}_n^{c}(\theta)}{\partial\theta\partial\theta'}
\right)^{-1}
\psi_n(\theta)
\right]
\right|_{\theta=\hat{\theta}}.
\)

\begin{theorem}\label{bias-corrected_Normality_twfe}
Under Assumptions~\ref{ass:1prime}, \ref{asp_error}, \ref{asp_4}, \ref{ass:4prime}, \ref{asp_UB}-\ref{asp_n}, \ref{ass:8prime}, \ref{asp_nonsingularity'} and \ref{asp_UBAT}, if
$\frac{N}{T^3}\rightarrow0$ and
$\frac{T}{N^3}\rightarrow0$,
then
\begin{align*}
\sqrt{n}\Gamma^{-1}\left(\hat{\theta}^{\,1}-\theta_0\right)
\overset{d}{\rightarrow}
N\left(
0,
\widebar{\Sigma}_{\theta_0}^{-1}
(\widebar{\Sigma}_{\theta_0}+\widebar{\Omega}_{\theta_0})
\widebar{\Sigma}_{\theta_0}^{-1}
\right).
\end{align*}
\end{theorem}
Hence, the estimator after bias correction is asymptotically normal and centered around $\theta_0$ as long as $T$ grows faster than $N^{\frac{1}{3}}$ and $N$ grows faster than $T^{\frac{1}{3}}$.

\subsection{Estimation of the Asymptotic Covariance Matrix}

For feasible inference under the TWFE specification, we use the same plug-in
principle. The TWFE asymptotic covariance matrix has the sandwich form
\(
\widebar{\Sigma}_{\theta_0}^{-1}
(\widebar{\Sigma}_{\theta_0}+\widebar{\Omega}_{\theta_0})
\widebar{\Sigma}_{\theta_0}^{-1}
\),
where \(\widebar{\Sigma}_{\theta_0}\) is the limiting TWFE information matrix
and \(\widebar{\Omega}_{\theta_0}\) is the corresponding non-Gaussian correction
term. Let \(\widehat{\widebar{\Sigma}}\) and
\(\widehat{\widebar{\Omega}}\) denote the plug-in estimators of the normalized
matrices \(\Gamma\widebar{\Sigma}_{\theta_0,n}\Gamma\) and
\(\Gamma\widebar{\Omega}_{\theta_0,n}\Gamma\), respectively, constructed from
the TWFE bias-corrected estimator and the corresponding residual moments. The explicit formulas are reported in Supplementary Appendix~C.8. Define
\(
\widehat{\widebar{\Xi}}
=
\widehat{\widebar{\Sigma}}^{-1}
(\widehat{\widebar{\Sigma}}+\widehat{\widebar{\Omega}})
\widehat{\widebar{\Sigma}}^{-1}.
\)

\begin{theorem}
Under Assumptions~\ref{ass:1prime}, \ref{asp_error}, \ref{asp_4}, \ref{ass:4prime}, \ref{asp_UB}-\ref{asp_n}, \ref{ass:8prime}, \ref{asp_nonsingularity'} and \ref{asp_UBAT},
if \(N/T^3\to0\) and \(T/N^3\to0\), then
\(
\widehat{\widebar{\Xi}}
\xrightarrow{p}
\widebar{\Sigma}_{\theta_0}^{-1}
(\widebar{\Sigma}_{\theta_0}+\widebar{\Omega}_{\theta_0})
\widebar{\Sigma}_{\theta_0}^{-1}.
\)
Consequently, \(n^{-1}\Gamma\widehat{\widebar{\Xi}}\Gamma\) is a consistent
estimator of the asymptotic covariance matrix of \(\widehat{\theta}^{\,1}\).
\end{theorem}

\section{Monte Carlo Simulation}\label{section4}
\subsection{Data-Generating Process}
In our simulation study, we investigate the finite sample performance of the OWFE QMLE, its bias-corrected counterpart, and the feasible inference procedure based on the proposed asymptotic covariance matrix estimator.
In the simulation study, we generate an unbalanced panel dataset of size $N\times(T+21)$, where the first $20$ periods are used only to generate a stable $Y_0$ and the remaining $T+1$ periods are used for estimation (set $t=0$ at the 21st period).
In order to simulate unbalancedness in the panel, we randomly assign to each node a contiguous observation window within the $T+1$ periods. Specifically, we first draw the observation duration for each observed node: $T_i\sim \mathrm{GE}(\frac{p}{T})+2$, where $\mathrm{GE}$ denotes the geometric distribution supported on nonnegative integers and $p$ is the parameter of the geometric distribution. 
Given the draw \(T_i\), if $T_i\geq T+1$, then node $i$ is observed for $T+1$ periods.
Otherwise, we place the contiguous observation window of length $T_i$ uniformly at random within the $(T+1)$-period horizon. 
By tuning the parameter $p$, we can control the unbalancedness proportion of the panel, which is defined as $\mathrm{UP}\equiv1-\frac{n}{NT}$, around a target mean level.


We generate the elements of \(X_t\) and \(\alpha_0\) independently from
standard normal distributions.
We simulate the dependent variable using a 20-period pre-sample burn-in. 
For individuals already observed in period 0, we initialize their outcome at $t=-20$ as $y_{i,-20}\sim N(0,1)$, drawn independently across units.
The subsequent observations ($t=-19,\cdots,0$) are then generated iteratively according to equation (\ref{model_m}).
Given $Y_0$, we can generate $Y_1,\cdots,Y_T$ iteratively according to equation (\ref{model_m}).
For the network weight matrix, we first construct an $N\times N$ adjacency matrix $A$ by rook contiguity, 
and then row-normalize $D_tAD_t'$ and $D_{t}AD_{t-1}'$, respectively to obtain the weight matrices $W_t$ and $M_t$. 

For the supplementary TWFE simulations reported in Supplementary Appendix D.1, we use the same setup as in the OWFE simulations, except that the outcomes are generated according to the TWFE model in \eqref{model_TWFE}. The common time effects are generated from standard normal draws and normalized to satisfy $\sum_{t=1}^T\tau_t=0$. During the burn-in stage, independent standard normal common components are generated analogously.

\begin{figure}[h]
    \centering
    \includegraphics[width=0.85\textwidth]{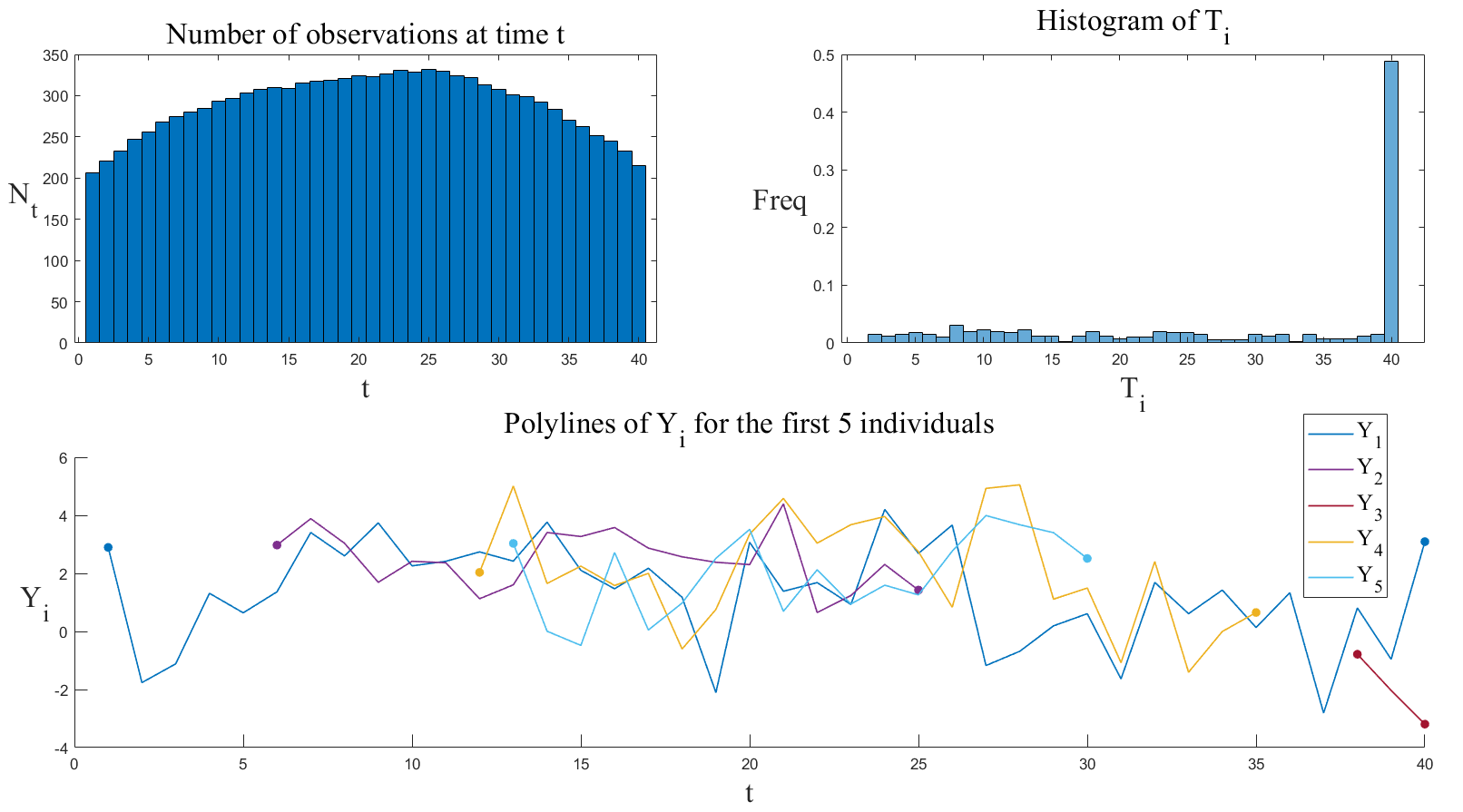}
    \caption{Characteristics of the unbalanced panel we generated}\label{pic_Yt}
\end{figure}
Figure~\ref{pic_Yt} illustrates 
key features of the simulated unbalanced dynamic network panel data with $N=400$, $T=40$ and $\mathrm{UP}=31.7\%$. It reveals that the number of observed nodes peaks during intermediate stages, with sparser observations in initial and terminal phases.
About 50\% of nodes are fully observed over the entire $T+1$ periods, while the remaining nodes have shorter observation windows, whose start times are drawn uniformly conditional on their durations.
\subsection{Simulation Results}

In the main simulation study, we investigate the finite-sample performance of the OWFE QMLE and its bias-corrected counterpart under varying configurations of $N \in \{100, 400, 1600\}$, $T \in \{10, 40\}$, and $\mathrm{UP}\in \{30\%, 60\%\}$. 
For the OWFE simulations, we consider two true parameter settings, $\theta_0=(0.5,0.2,0.1,1,1,1)$ and $\theta_0=(0.1,0.3,0.5,1,1,1)$, where $\theta_0=(\rho_0,\lambda_0,\nu_0,\gamma_0,\beta_0,\sigma^2_0)$. We also consider three error distributions: normal, centered exponential, and Laplace.
Under each setting, the estimation is repeated 1,000 times, and we evaluate the bias, standard deviation (SD), root mean square error (RMSE), and coverage probability (CP) of the estimators.

The OWFE simulation results are presented in Tables \ref{Tab_N130}-\ref{Tab_N160} and Tables D.1--D.6 in Supplementary Appendix D.1. 
Supplementary Appendix D.1 reports additional TWFE simulation results in Tables D.7--D.12 for \(\theta_0=(0.5,0.2,0.1,1,1,1)\) under the same error distributions and UP levels.
From the OWFE simulation tables, we obtain the following findings. 
\begin{enumerate}[(1)]
\item The biases of both the original estimator and the bias-corrected estimator decrease significantly as $T$ increases, but show little change as $N$ increases. Moreover, for fixed $N$ and $T$, a higher UP leads to larger bias. These findings are consistent with the theoretical bias of order $O_p(\frac{1}{c_u T})$.
\item Regardless of whether bias correction is applied, the standard deviations of most estimators are approximately halved when either $N$ or $T$ is quadrupled, consistent with the theoretical convergence rate of $1/\sqrt{n}$. An exception is $\hat{\gamma}$, whose standard deviation remains essentially unchanged as $T$ increases, aligning with the theoretical convergence rate of $1/\sqrt{c_uN}$.
Moreover, the standard deviation increases as the UP rises.
\item The original estimators exhibit substantial bias, especially for $\nu$ and $\sigma^2$, while the bias-corrected estimators significantly reduce the bias. The bias-corrected estimators show notable improvements in both bias and RMSE, and their CPs are also closer to the nominal levels of 95\% and 90\%. 
\item As $N$ increases, there is a downward trend in CPs for both estimators, which is attributable to the asymptotic bias of order $O_p(\sqrt{N/T^3})$.
\end{enumerate}

The supplementary TWFE simulations in Tables D.7--D.12 display a similar pattern. The original TWFE estimator also exhibits noticeable finite-sample bias. After
bias correction, the biases and RMSEs are generally reduced, and the CPs move closer to the nominal levels. These results suggest that the proposed bias correction remains effective after introducing time fixed effects.


To examine the correctness of the asymptotic theory, we plot the kernel density curves of $\frac{\hat{\theta}_i-\theta_{0i}}{\mathrm{SE}_i}$, and $\frac{\hat{\theta}^1_i-\theta_{0i}}{\mathrm{SE}_i}$, where $\hat{\theta}_i$, $\hat{\theta}^1_i$ and $\theta_{0i}$ denote the $i$-th element of $\hat{\theta}$, $\hat{\theta}^1$ and $\theta_{0}$, respectively, and $\mathrm{SE}_i$ denotes the square root of the $i$-th diagonal element of $\frac{1}{n}\Sigma_{\theta_0,n}^{-1}(\Sigma_{\theta_0,n}+\Omega_{\theta_0,n})\Sigma_{\theta_0,n}^{-1}$.
These kernel densities are then compared with the probability density function (PDF) of the standard normal distribution. The results are shown in Figures \ref{Fig_N130}-\ref{Fig_N160} and Figures D.1-D.6 in Supplementary Appendix D.2. As illustrated in the figures, regardless of whether the estimators are bias corrected or not, their distributions become increasingly close to normality as the sample size increases. The bias-corrected estimators exhibit distributions that are closer to a centered normal distribution than the original ones.
\begin{landscape}
\begin{table}[p]
\caption{Results of OWFE estimator with Normal error, UP=30\%, $\theta_0=(0.5,0.2,0.1,1,1,1)$.}
\label{Tab_N130}
\centering
\begin{threeparttable}
{
\resizebox{0.86\linewidth}{!}{
\begin{tabular}{ccrrrrrrccrrrrrr}
\hline
 &  & \multicolumn{1}{c}{$\rho$} & \multicolumn{1}{c}{$\lambda$} & \multicolumn{1}{c}{$\nu$} & \multicolumn{1}{c}{$\gamma$} & \multicolumn{1}{c}{$\beta$} & \multicolumn{1}{c}{$\sigma^2$} &  &  & \multicolumn{1}{c}{$\rho$} & \multicolumn{1}{c}{$\lambda$} & \multicolumn{1}{c}{$\nu$} & \multicolumn{1}{c}{$\gamma$} & \multicolumn{1}{c}{$\beta$} & \multicolumn{1}{c}{$\sigma^2$} \\ \hline
$T=10$ &  & \multicolumn{6}{c}{Before bias correction} & $T=40$ &  & \multicolumn{6}{c}{Before bias correction} \\ \cline{3-8} \cline{11-16} 
$N=100$ & BIAS & -0.0044 & 0.0327 & -0.0753 & -0.0348 & -0.0098 & -0.1594 & $N=100$ & BIAS & 0.0006 & 0.0077 & -0.0191 & -0.0014 & -0.0016 & -0.0399 \\
 & SD & 0.0268 & 0.0367 & 0.0299 & 0.1497 & 0.0427 & 0.0511 &  & SD & 0.0124 & 0.0178 & 0.0146 & 0.1426 & 0.0191 & 0.0275 \\
 & RMSE & 0.0271 & 0.0492 & 0.0810 & 0.1537 & 0.0438 & 0.1674 &  & RMSE & 0.0124 & 0.0194 & 0.0240 & 0.1426 & 0.0192 & 0.0484 \\
 & 95\%CP & 0.9340 & 0.8620 & 0.2810 & 0.9470 & 0.9370 & 0.1530 &  & 95\%CP & 0.9550 & 0.9190 & 0.7070 & 0.9510 & 0.9570 & 0.6860 \\
 & 90\%CP & 0.8640 & 0.7660 & 0.1920 & 0.8900 & 0.8890 & 0.0930 &  & 90\%CP & 0.8910 & 0.8740 & 0.6080 & 0.8970 & 0.9060 & 0.5630 \\ \cline{2-8} \cline{10-16} 
 &  & \multicolumn{6}{c}{After bias correction} &  &  & \multicolumn{6}{c}{After bias correction} \\ \cline{3-8} \cline{11-16} 
 & BIAS & \textbf{-0.0034} & \textbf{0.0061} & \textbf{-0.0159} & \textbf{-0.0064} & \textbf{-0.0005} & \textbf{-0.0397} &  & BIAS & \textbf{0.0002} & \textbf{0.0001} & \textbf{-0.0010} & 0.0019 & \textbf{-0.0007} & \textbf{-0.0053} \\
 & SD & \textbf{0.0267} & 0.0375 & 0.0311 & \textbf{0.1495} & \textbf{0.0426} & 0.0580 &  & SD & 0.0124 & 0.0180 & 0.0148 & \textbf{0.1425} & 0.0191 & 0.0284 \\
 & RMSE & \textbf{0.0269} & \textbf{0.0380} & \textbf{0.0349} & \textbf{0.1497} & \textbf{0.0426} & \textbf{0.0703} &  & RMSE & \textbf{0.0124} & \textbf{0.0180} & \textbf{0.0149} & \textbf{0.1425} & \textbf{0.0191} & \textbf{0.0289} \\
 & 95\%CP & \textbf{0.9360} & \textbf{0.9380} & \textbf{0.9100} & \textit{0.9470} & \textbf{0.9410} & \textbf{0.8770} &  & 95\%CP & \textit{0.9550} & \textbf{0.9490} & \textbf{0.9430} & 0.9530 & \textbf{0.9560} & \textbf{0.9420} \\
 & 90\%CP & \textbf{0.8670} & \textbf{0.9000} & \textbf{0.8380} & \textbf{0.8990} & \textbf{0.8920} & \textbf{0.7780} &  & 90\%CP & 0.8890 & \textbf{0.8950} & \textbf{0.8820} & 0.8960 & \textbf{0.9020} & \textbf{0.8900} \\ \hline
$T=10$ &  & \multicolumn{6}{c}{Before bias correction} & $T=40$ &  & \multicolumn{6}{c}{Before bias correction} \\ \cline{3-8} \cline{11-16} 
$N=400$ & BIAS & -0.0021 & 0.0339 & -0.0748 & -0.0341 & -0.0126 & -0.1525 & $N=400$ & BIAS & 0.0000 & 0.0082 & -0.0194 & -0.0027 & 0.0001 & -0.0364 \\
 & SD & 0.0136 & 0.0189 & 0.0146 & 0.0765 & 0.0207 & 0.0259 &  & SD & 0.0063 & 0.0090 & 0.0070 & 0.0705 & 0.0098 & 0.0142 \\
 & RMSE & 0.0137 & 0.0388 & 0.0762 & 0.0838 & 0.0242 & 0.1546 &  & RMSE & 0.0063 & 0.0122 & 0.0206 & 0.0706 & 0.0098 & 0.0391 \\
 & 95\%CP & 0.9360 & 0.5630 & 0.0010 & 0.9080 & 0.9070 & 0.0000 &  & 95\%CP & 0.9540 & 0.8640 & 0.2120 & 0.9590 & 0.9500 & 0.2440 \\
 & 90\%CP & 0.8730 & 0.4490 & 0.0000 & 0.8450 & 0.8370 & 0.0000 &  & 90\%CP & 0.9080 & 0.7630 & 0.1390 & 0.9020 & 0.8930 & 0.1700 \\ \cline{2-8} \cline{10-16} 
 &  & \multicolumn{6}{c}{After bias correction} &  &  & \multicolumn{6}{c}{After bias correction} \\ \cline{3-8} \cline{11-16} 
 & BIAS & \textbf{-0.0012} & \textbf{0.0064} & \textbf{-0.0152} & \textbf{-0.0065} & \textbf{-0.0033} & \textbf{-0.0319} &  & BIAS & -0.0004 & \textbf{0.0006} & \textbf{-0.0013} & \textbf{0.0004} & 0.0010 & \textbf{-0.0017} \\
 & SD & 0.0136 & 0.0194 & 0.0151 & 0.0768 & \textbf{0.0206} & 0.0294 &  & SD & 0.0063 & 0.0091 & 0.0071 & 0.0707 & \textbf{0.0098} & 0.0146 \\
 & RMSE & \textbf{0.0136} & \textbf{0.0204} & \textbf{0.0214} & \textbf{0.0771} & \textbf{0.0208} & \textbf{0.0434} &  & RMSE & 0.0063 & \textbf{0.0091} & \textbf{0.0072} & 0.0707 & 0.0099 & \textbf{0.0147} \\
 & 95\%CP & 0.9330 & \textbf{0.9350} & \textbf{0.8290} & \textbf{0.9260} & \textbf{0.9540} & \textbf{0.7590} &  & 95\%CP & \textbf{0.9500} & \textbf{0.9460} & \textbf{0.9420} & \textit{0.9590} & 0.9450 & \textbf{0.9310} \\
 & 90\%CP & \textbf{0.8750} & \textbf{0.8670} & \textbf{0.7260} & \textbf{0.8730} & \textbf{0.9000} & \textbf{0.6690} &  & 90\%CP & \textbf{0.9020} & \textbf{0.9040} & \textbf{0.8830} & \textbf{0.9010} & \textbf{0.9010} & \textbf{0.8670} \\ \hline
$T=10$ &  & \multicolumn{6}{c}{Before bias correction} & $T=40$ &  & \multicolumn{6}{c}{Before bias correction} \\ \cline{3-8} \cline{11-16} 
$N=1600$ & BIAS & -0.0020 & 0.0338 & -0.0748 & -0.0343 & -0.0114 & -0.1530 & $N=1600$ & BIAS & 0.0004 & 0.0078 & -0.0192 & -0.0033 & -0.0011 & -0.0371 \\
 & SD & 0.0071 & 0.0090 & 0.0075 & 0.0357 & 0.0102 & 0.0127 &  & SD & 0.0032 & 0.0047 & 0.0034 & 0.0352 & 0.0048 & 0.0067 \\
 & RMSE & 0.0073 & 0.0350 & 0.0751 & 0.0496 & 0.0153 & 0.1535 &  & RMSE & 0.0032 & 0.0090 & 0.0195 & 0.0354 & 0.0049 & 0.0377 \\
 & 95\%CP & 0.9300 & 0.0480 & 0.0000 & 0.8550 & 0.8100 & 0.0000 &  & 95\%CP & 0.9530 & 0.6140 & 0.0010 & 0.9500 & 0.9520 & 0.0010 \\
 & 90\%CP & 0.8630 & 0.0210 & 0.0000 & 0.7690 & 0.6980 & 0.0000 &  & 90\%CP & 0.9020 & 0.4960 & 0.0010 & 0.8970 & 0.8920 & 0.0000 \\ \cline{2-8} \cline{10-16} 
 &  & \multicolumn{6}{c}{After bias correction} &  &  & \multicolumn{6}{c}{After bias correction} \\ \cline{3-8} \cline{11-16} 
 & BIAS & \textbf{-0.0011} & \textbf{0.0062} & \textbf{-0.0152} & \textbf{-0.0067} & \textbf{-0.0022} & \textbf{-0.0325} &  & BIAS & \textbf{0.0000} & \textbf{0.0002} & \textbf{-0.0011} & \textbf{-0.0001} & \textbf{-0.0002} & \textbf{-0.0025} \\
 & SD & 0.0071 & 0.0094 & 0.0078 & 0.0359 & \textbf{0.0102} & 0.0144 &  & SD & 0.0032 & 0.0047 & 0.0034 & 0.0352 & \textbf{0.0048} & 0.0069 \\
 & RMSE & \textbf{0.0072} & \textbf{0.0112} & \textbf{0.0171} & \textbf{0.0365} & \textbf{0.0104} & \textbf{0.0355} &  & RMSE & \textbf{0.0032} & \textbf{0.0047} & \textbf{0.0036} & \textbf{0.0352} & \textbf{0.0048} & \textbf{0.0073} \\
 & 95\%CP & \textbf{0.9310} & \textbf{0.9000} & \textbf{0.4840} & \textbf{0.9530} & \textbf{0.9450} & \textbf{0.3380} &  & 95\%CP & 0.9580 & \textbf{0.9450} & \textbf{0.9450} & 0.9490 & 0.9580 & \textbf{0.9360} \\
 & 90\%CP & \textbf{0.8720} & \textbf{0.8330} & \textbf{0.3690} & \textbf{0.9040} & \textbf{0.9060} & \textbf{0.2280} &  & 90\%CP & \textbf{0.9000} & \textbf{0.8900} & \textbf{0.8830} & 0.8960 & \textbf{0.9070} & \textbf{0.8790} \\ \hline
\end{tabular}}}
\par\smallskip
\begin{minipage}[t]{0.86\linewidth}
\footnotesize
Notes: $\mathrm{UP}=1-\frac{n}{NT}$. In the table, \textbf{bold} numbers indicate that the bias-corrected results have improved, while \textit{italicized} numbers indicate that the bias-corrected results perform equally well as before. 
The 95\%CP and 90\%CP  indicate the frequencies that the true parameter value is contained within the 95\% and 90\% confidence intervals of the estimator, respectively.
\end{minipage}
\end{threeparttable}
\end{table}
\end{landscape}

\begin{landscape}
\begin{table}[p]
\caption{Results of OWFE estimator with Normal error, UP=60\%, $\theta_0=(0.5,0.2,0.1,1,1,1)$.}
\label{Tab_N160}
\centering
\begin{threeparttable}
{
\resizebox{0.86\linewidth}{!}{
\begin{tabular}{ccrrrrrrccrrrrrr}
\hline
 &  & \multicolumn{1}{c}{$\rho$} & \multicolumn{1}{c}{$\lambda$} & \multicolumn{1}{c}{$\nu$} & \multicolumn{1}{c}{$\gamma$} & \multicolumn{1}{c}{$\beta$} & \multicolumn{1}{c}{$\sigma^2$} &  &  & \multicolumn{1}{c}{$\rho$} & \multicolumn{1}{c}{$\lambda$} & \multicolumn{1}{c}{$\nu$} & \multicolumn{1}{c}{$\gamma$} & \multicolumn{1}{c}{$\beta$} & \multicolumn{1}{c}{$\sigma^2$} \\ \hline
$T=10$ &  & \multicolumn{6}{c}{Before bias correction} & $T=40$ &  & \multicolumn{6}{c}{Before bias correction} \\ \cline{3-8} \cline{11-16} 
$N=100$ & BIAS & -0.0093 & 0.0605 & -0.1358 & -0.0929 & -0.0337 & -0.2862 & $N=100$ & BIAS & 0.0004 & 0.0141 & -0.0321 & -0.0147 & -0.0029 & -0.0694 \\
 & SD & 0.0307 & 0.0468 & 0.0440 & 0.1348 & 0.0613 & 0.0613 &  & SD & 0.0141 & 0.0203 & 0.0185 & 0.1098 & 0.0263 & 0.0352 \\
 & RMSE & 0.0321 & 0.0765 & 0.1428 & 0.1638 & 0.0700 & 0.2927 &  & RMSE & 0.0141 & 0.0247 & 0.0370 & 0.1108 & 0.0264 & 0.0778 \\
 & 95\%CP & 0.9100 & 0.7650 & 0.1200 & 0.8810 & 0.9000 & 0.0100 &  & 95\%CP & 0.9470 & 0.9070 & 0.6060 & 0.9570 & 0.9410 & 0.5100 \\
 & 90\%CP & 0.8370 & 0.6700 & 0.0670 & 0.8260 & 0.8310 & 0.0020 &  & 90\%CP & 0.8870 & 0.8320 & 0.4840 & 0.9040 & 0.8990 & 0.3840 \\ \cline{2-8} \cline{10-16} 
 &  & \multicolumn{6}{c}{After bias correction} &  &  & \multicolumn{6}{c}{After bias correction} \\ \cline{3-8} \cline{11-16} 
 & BIAS & \textbf{-0.0062} & \textbf{0.0212} & \textbf{-0.0481} & \textbf{-0.0345} & \textbf{-0.0112} & \textbf{-0.1083} &  & BIAS & \textbf{0.0000} & \textbf{0.0006} & \textbf{-0.0031} & \textbf{-0.0074} & \textbf{-0.0006} & \textbf{-0.0105} \\
 & SD & \textbf{0.0308} & 0.0484 & 0.0458 & \textbf{0.1341} & \textbf{0.0612} & 0.0744 &  & SD & 0.0141 & 0.0206 & 0.0188 & 0.1099 & \textbf{0.0262} & 0.0371 \\
 & RMSE & \textbf{0.0314} & \textbf{0.0528} & \textbf{0.0664} & \textbf{0.1385} & \textbf{0.0623} & \textbf{0.1314} &  & RMSE & 0.0141 & \textbf{0.0206} & \textbf{0.0191} & \textbf{0.1102} & \textbf{0.0262} & \textbf{0.0386} \\
 & 95\%CP & \textbf{0.9200} & \textbf{0.9180} & \textbf{0.7850} & \textbf{0.9380} & \textbf{0.9430} & \textbf{0.6750} &  & 95\%CP & 0.9460 & \textbf{0.9490} & \textbf{0.9500} & 0.9590 & \textbf{0.9460} & \textbf{0.9310} \\
 & 90\%CP & \textbf{0.8490} & \textbf{0.8530} & \textbf{0.6910} & \textbf{0.8710} & \textbf{0.8740} & \textbf{0.5500} &  & 90\%CP & \textbf{0.8880} & \textbf{0.9170} & \textbf{0.8890} & 0.9090 & 0.9030 & \textbf{0.8840} \\ \hline
$T=10$ &  & \multicolumn{6}{c}{Before bias correction} & $T=40$ &  & \multicolumn{6}{c}{Before bias correction} \\ \cline{3-8} \cline{11-16} 
$N=400$ & BIAS & -0.0061 & 0.0583 & -0.1302 & -0.0861 & -0.0316 & -0.2766 & $N=400$ & BIAS & 0.0003 & 0.0145 & -0.0317 & -0.0069 & -0.0026 & -0.0660 \\
 & SD & 0.0153 & 0.0225 & 0.0215 & 0.0645 & 0.0300 & 0.0311 &  & SD & 0.0068 & 0.0101 & 0.0094 & 0.0573 & 0.0129 & 0.0177 \\
 & RMSE & 0.0165 & 0.0625 & 0.1320 & 0.1076 & 0.0435 & 0.2784 &  & RMSE & 0.0068 & 0.0176 & 0.0331 & 0.0577 & 0.0131 & 0.0683 \\
 & 95\%CP & 0.9050 & 0.2960 & 0.0000 & 0.7510 & 0.8230 & 0.0000 &  & 95\%CP & 0.9470 & 0.7290 & 0.0810 & 0.9430 & 0.9450 & 0.0410 \\
 & 90\%CP & 0.8260 & 0.1980 & 0.0000 & 0.6390 & 0.7240 & 0.0000 &  & 90\%CP & 0.8990 & 0.6200 & 0.0390 & 0.8840 & 0.9120 & 0.0220 \\ \cline{2-8} \cline{10-16} 
 &  & \multicolumn{6}{c}{After bias correction} &  &  & \multicolumn{6}{c}{After bias correction} \\ \cline{3-8} \cline{11-16} 
 & BIAS & \textbf{-0.0032} & \textbf{0.0179} & \textbf{-0.0426} & \textbf{-0.0282} & \textbf{-0.0095} & \textbf{-0.0963} &  & BIAS & \textbf{-0.0001} & \textbf{0.0009} & \textbf{-0.0028} & \textbf{0.0004} & \textbf{-0.0003} & \textbf{-0.0071} \\
 & SD & \textbf{0.0152} & 0.0232 & 0.0221 & \textbf{0.0635} & \textbf{0.0299} & 0.0378 &  & SD & 0.0068 & 0.0103 & 0.0096 & 0.0570 & 0.0129 & 0.0187 \\
 & RMSE & \textbf{0.0156} & \textbf{0.0293} & \textbf{0.0479} & \textbf{0.0695} & \textbf{0.0313} & \textbf{0.1035} &  & RMSE & 0.0068 & \textbf{0.0103} & \textbf{0.0100} & \textbf{0.0570} & \textbf{0.0129} & \textbf{0.0200} \\
 & 95\%CP & \textbf{0.9230} & \textbf{0.8910} & \textbf{0.4940} & \textbf{0.9320} & \textbf{0.9360} & \textbf{0.2560} &  & 95\%CP & \textit{0.9470} & \textbf{0.9470} & \textbf{0.9320} & 0.9390 & \textbf{0.9510} & \textbf{0.9250} \\
 & 90\%CP & \textbf{0.8560} & \textbf{0.8090} & \textbf{0.3820} & \textbf{0.8830} & \textbf{0.8850} & \textbf{0.1660} &  & 90\%CP & 0.8940 & \textbf{0.9000} & \textbf{0.8740} & \textbf{0.8980} & \textbf{0.9110} & \textbf{0.8710} \\ \hline
$T=10$ &  & \multicolumn{6}{c}{Before bias correction} & $T=40$ &  & \multicolumn{6}{c}{Before bias correction} \\ \cline{3-8} \cline{11-16} 
$N=1600$ & BIAS & -0.0055 & 0.0591 & -0.1305 & -0.0868 & -0.0329 & -0.2748 & $N=1600$ & BIAS & 0.0003 & 0.0147 & -0.0319 & -0.0083 & -0.0025 & -0.0653 \\
 & SD & 0.0077 & 0.0118 & 0.0108 & 0.0322 & 0.0152 & 0.0155 &  & SD & 0.0035 & 0.0050 & 0.0047 & 0.0268 & 0.0066 & 0.0091 \\
 & RMSE & 0.0095 & 0.0602 & 0.1310 & 0.0926 & 0.0363 & 0.2752 &  & RMSE & 0.0035 & 0.0155 & 0.0323 & 0.0281 & 0.0070 & 0.0659 \\
 & 95\%CP & 0.8560 & 0.0020 & 0.0000 & 0.2400 & 0.4000 & 0.0000 &  & 95\%CP & 0.9500 & 0.1890 & 0.0000 & 0.9500 & 0.9380 & 0.0000 \\
 & 90\%CP & 0.7740 & 0.0010 & 0.0000 & 0.1650 & 0.2880 & 0.0000 &  & 90\%CP & 0.8980 & 0.1120 & 0.0000 & 0.8970 & 0.8770 & 0.0000 \\ \cline{2-8} \cline{10-16} 
 &  & \multicolumn{6}{c}{After bias correction} &  &  & \multicolumn{6}{c}{After bias correction} \\ \cline{3-8} \cline{11-16} 
 & BIAS & \textbf{-0.0028} & \textbf{0.0185} & \textbf{-0.0427} & \textbf{-0.0287} & \textbf{-0.0109} & \textbf{-0.0940} &  & BIAS & \textbf{-0.0002} & \textbf{0.0011} & \textbf{-0.0031} & \textbf{-0.0010} & \textbf{-0.0003} & \textbf{-0.0064} \\
 & SD & \textbf{0.0076} & 0.0122 & 0.0114 & \textbf{0.0320} & \textbf{0.0151} & 0.0187 &  & SD & \textbf{0.0035} & 0.0051 & 0.0048 & 0.0269 & 0.0066 & 0.0096 \\
 & RMSE & \textbf{0.0081} & \textbf{0.0222} & \textbf{0.0442} & \textbf{0.0430} & \textbf{0.0186} & \textbf{0.0959} &  & RMSE & \textbf{0.0035} & \textbf{0.0052} & \textbf{0.0057} & \textbf{0.0269} & \textbf{0.0066} & \textbf{0.0115} \\
 & 95\%CP & \textbf{0.9020} & \textbf{0.6440} & \textbf{0.0230} & \textbf{0.8600} & \textbf{0.8880} & \textit{0.0000} &  & 95\%CP & 0.9480 & \textbf{0.9540} & \textbf{0.8960} & 0.9710 & \textbf{0.9520} & \textbf{0.8850} \\
 & 90\%CP & \textbf{0.8490} & \textbf{0.5270} & \textbf{0.0100} & \textbf{0.7800} & \textbf{0.7980} & \textit{0.0000} &  & 90\%CP & 0.8970 & \textbf{0.9090} & \textbf{0.8180} & 0.9070 & \textbf{0.9080} & \textbf{0.7870} \\ \hline
\end{tabular}
}}
\par\smallskip
\begin{minipage}[t]{0.86\linewidth}
\footnotesize
Notes: $\mathrm{UP}=1-\frac{n}{NT}$. In the table, \textbf{bold} numbers indicate that the bias-corrected results have improved, while \textit{italicized} numbers indicate that the bias-corrected results perform equally well as before. 
The 95\%CP and 90\%CP  indicate the frequencies that the true parameter value is contained within the 95\% and 90\% confidence intervals of the estimator, respectively.
\end{minipage}
\end{threeparttable}
\end{table}
\end{landscape}

\begin{figure}[h]
\centering
\includegraphics[width=\textwidth]{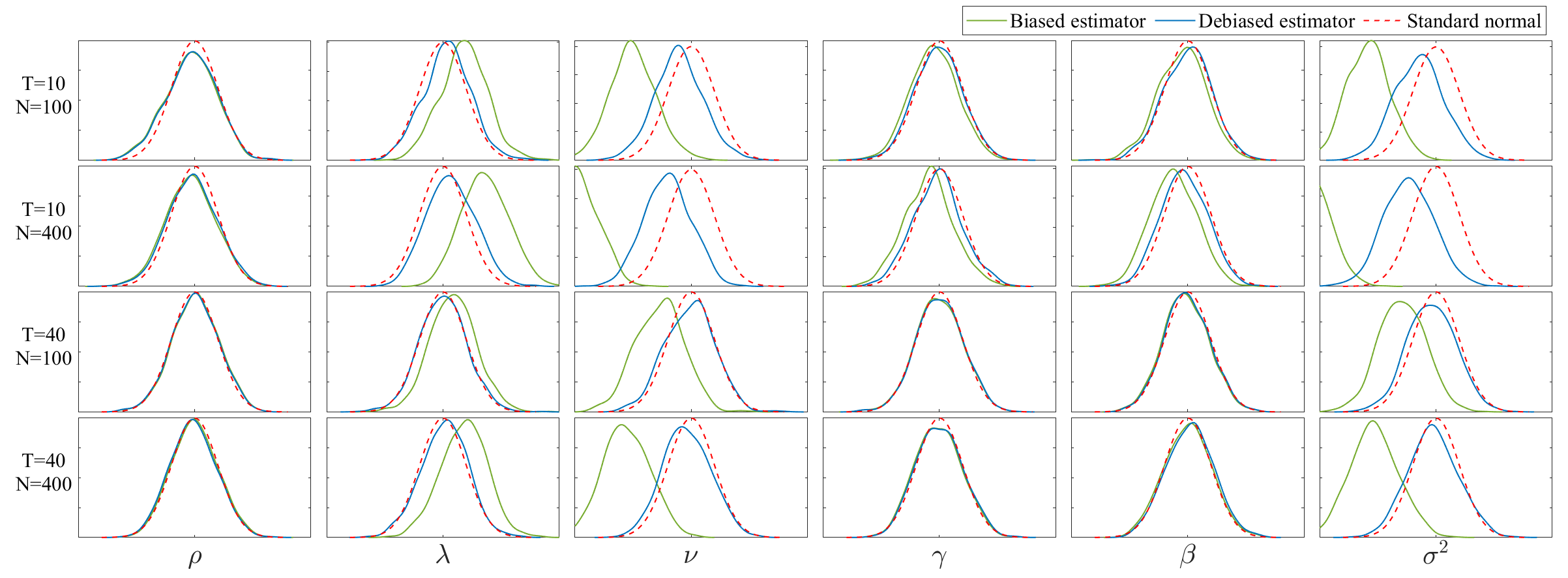}
\caption{PDF of OWFE standardized estimator with normal error, UP=30\%.}
\label{Fig_N130}
\end{figure}

\begin{figure}[h]
\centering
\includegraphics[width=\textwidth]{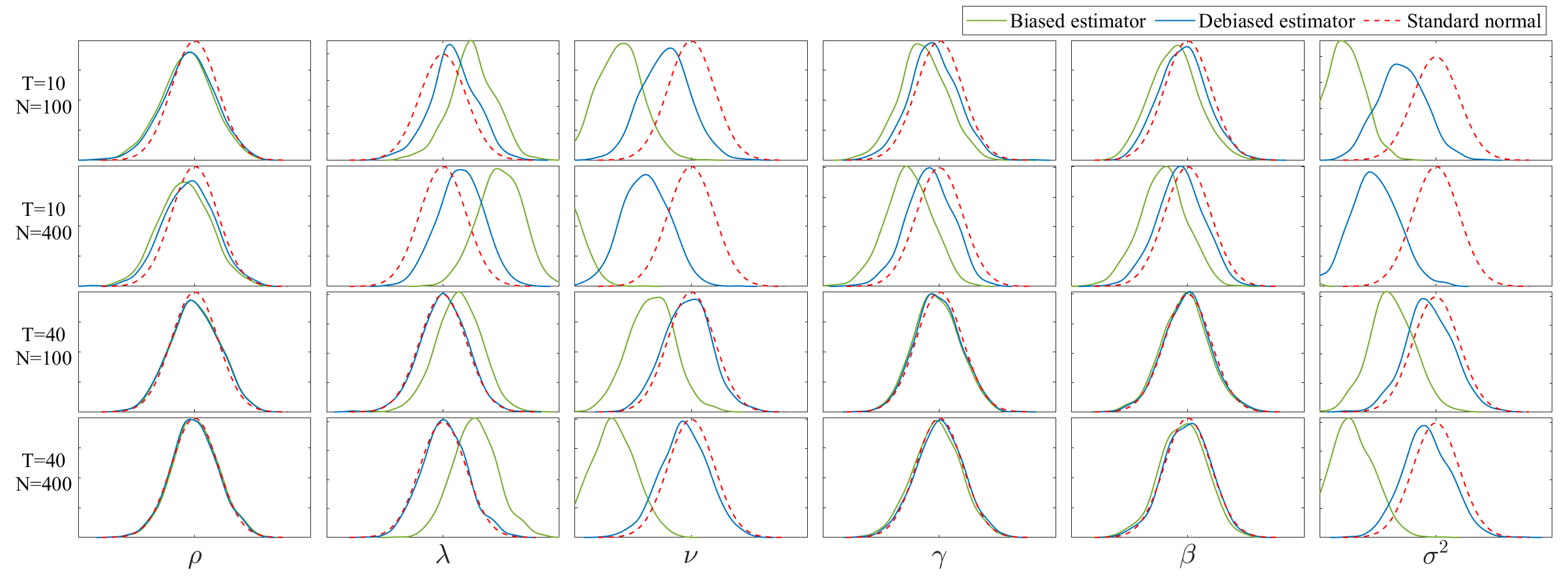}
\caption{PDF of OWFE standardized estimator with normal error, UP=60\%.}
\label{Fig_N160}
\end{figure}

\section{Real Data Analysis}\label{section5}

To assess our model's ability to capture the dynamics of unbalanced spatio-temporal panel data, we take the short-term rental market as the empirical application. In recent years, Airbnb, as a leading global platform for short-term accommodation, has expanded rapidly across cities and regions, and its supply and pricing dynamics have attracted extensive academic attention. 
More importantly, Airbnb data naturally exhibit an unbalanced panel structure: the set of available listings is not identical across time periods, and many listings are not observed throughout the entire sample period. This provides a representative and complex empirical setting for our study.
A large body of literature has already examined the rental pricing of Airbnb properties, such as the studies by \cite{su9091635}, \cite{GYODI2021104319}, and \cite{WANG2023100743}, which apply spatial econometric methods to cross-sectional data of Airbnb listings.
\cite{Juan2021} employ an unbalanced dynamic panel data model to analyze the impact of Airbnb listings on the Spanish tourism markets. Their model does not account for the spatial dependence in the data.
\cite{GUNTER2020104000} use a static spatial panel model to explore the substitution or complementarity relationship between Airbnb and traditional accommodations. However, their analysis is restricted to a balanced subsample of listings that are continuously observed over a long period, which accounts for less than $10\%$ of the total sample. This balanced spatial panel model neglects the spatial spillover effects from short-lived listings and does not account for the temporal dynamics in rental prices. While there have been studies using various models to analyze Airbnb rental prices, none simultaneously account for spatial spillover effects, temporal dynamics (e.g., price lags), and the unbalanced nature of the panel data. This is precisely what our model is designed to achieve.

We collect monthly Airbnb listing data from Inside Airbnb for two regions: New Zealand and New York City\footnote{\url{https://insideairbnb.com}}. The New Zealand raw data contain 31 monthly snapshots from May 13, 2023 to November 7, 2025, and the New York City raw data contain 18 monthly snapshots from June 4, 2024 to November 1, 2025. Each listing includes information such as rental price, geographic coordinates, availability over the next 30 days, and other listing-level characteristics.

To examine how short-term rental market dynamics vary across areas and spatial scales, we focus on six local markets. In New Zealand, the three samples are ward-level local markets: Waitemat\={a} and Gulf Ward, Queenstown-Wakatipu Ward, and Pukeh\={i}nau/Lambton Ward. For readability, we refer to them as Auckland, Queenstown, and Wellington, respectively, in the tables. In New York City, the three samples are the neighborhoods of Williamsburg, Midtown, and Hell's Kitchen. Within each local market, the dependent variable is the logarithm of rental price.

The empirical analysis uses the TWFE extension developed in Section~\ref{sec:twoway_extension}. The covariates in the empirical specification are the listing's own availability over the next 30 days, denoted by avail30d, and a neighborhood-level availability measure, denoted by avgAvail30d1km.

\begin{figure}[h]
    \centering
    \includegraphics[width=\linewidth]{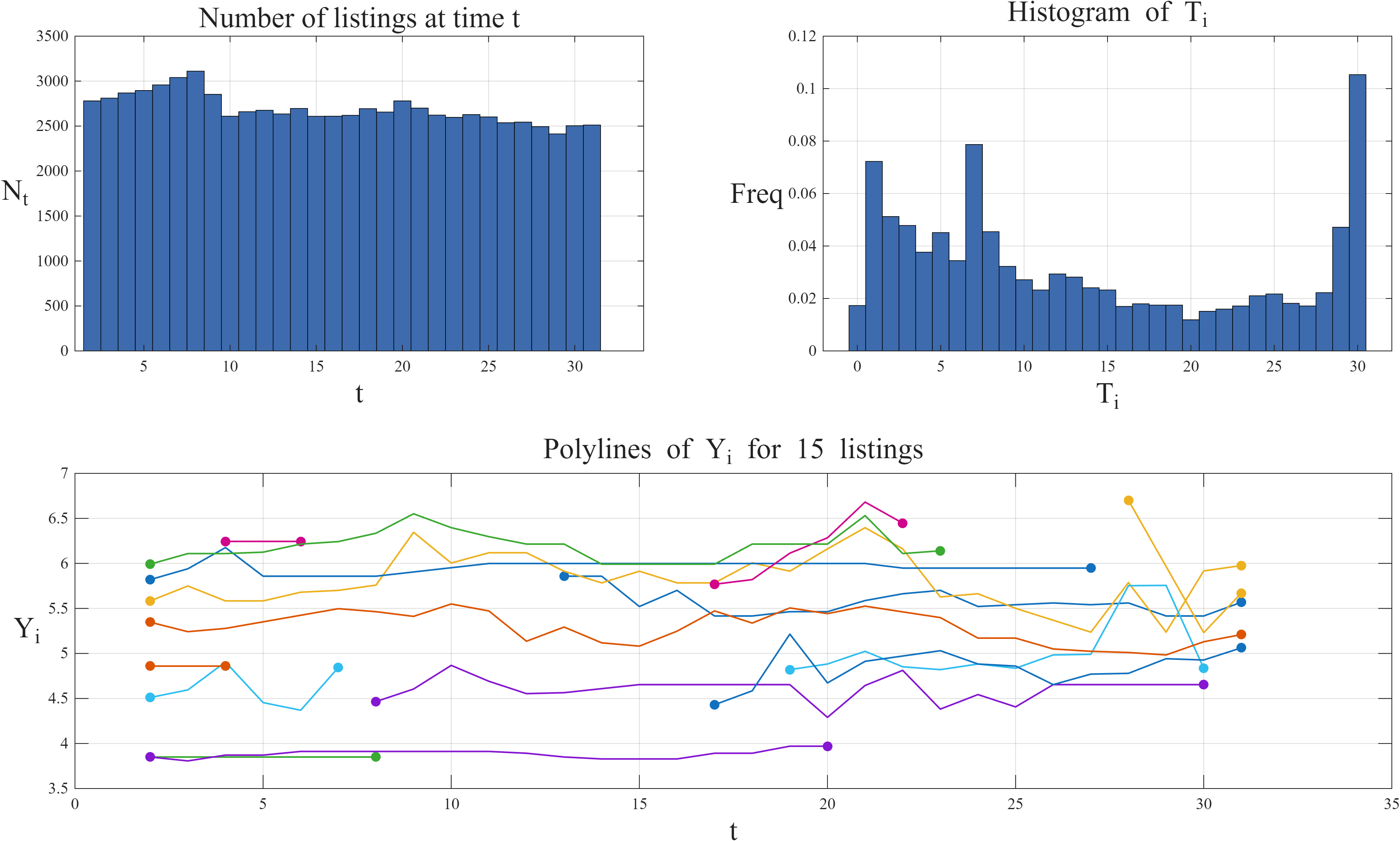}
    \caption{Characteristics of the unbalanced panel of the listing data in Auckland}
    \label{pic_NewZealand_NT}
\end{figure}

The unbalanced panel for Auckland, i.e., Waitemat\={a} and Gulf Ward, is visualized in Figure~\ref{pic_NewZealand_NT}. This sample contains $N=5{,}896$ listings, $T=30$ monthly periods, and $n=80{,}716$ observations, with an unbalancedness proportion of $\mathrm{UP}=54.37\%$. The corresponding sample sizes and unbalancedness proportions for the other local markets are reported in Tables~\ref{Tab_Comv_TWFE} and \ref{Tab_ComV_TWFE}. The descriptive statistics of all six datasets are reported in Supplementary Appendix E, Table E.1.

\begin{figure}[h]
    \centering
    \includegraphics[width=\linewidth]{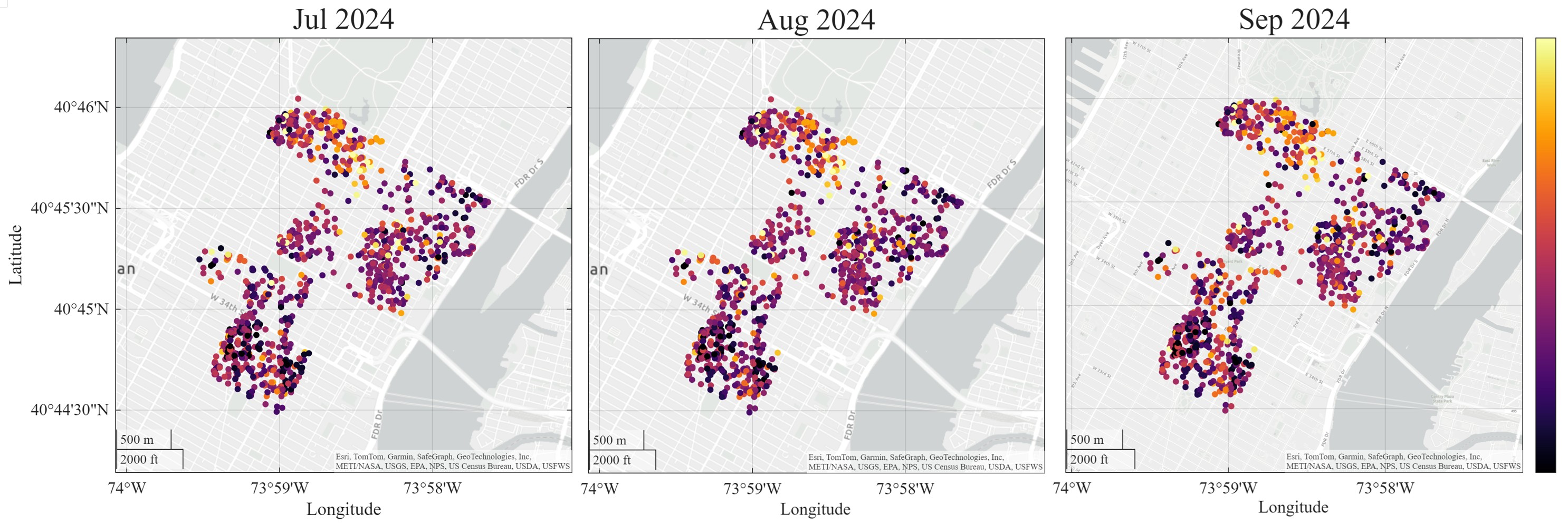}
    \caption{Geographical distribution of log rental prices in Midtown}
    \label{pic_geosc_MT}
\end{figure}

\begin{figure}[h]
    \centering
    \includegraphics[width=\linewidth]{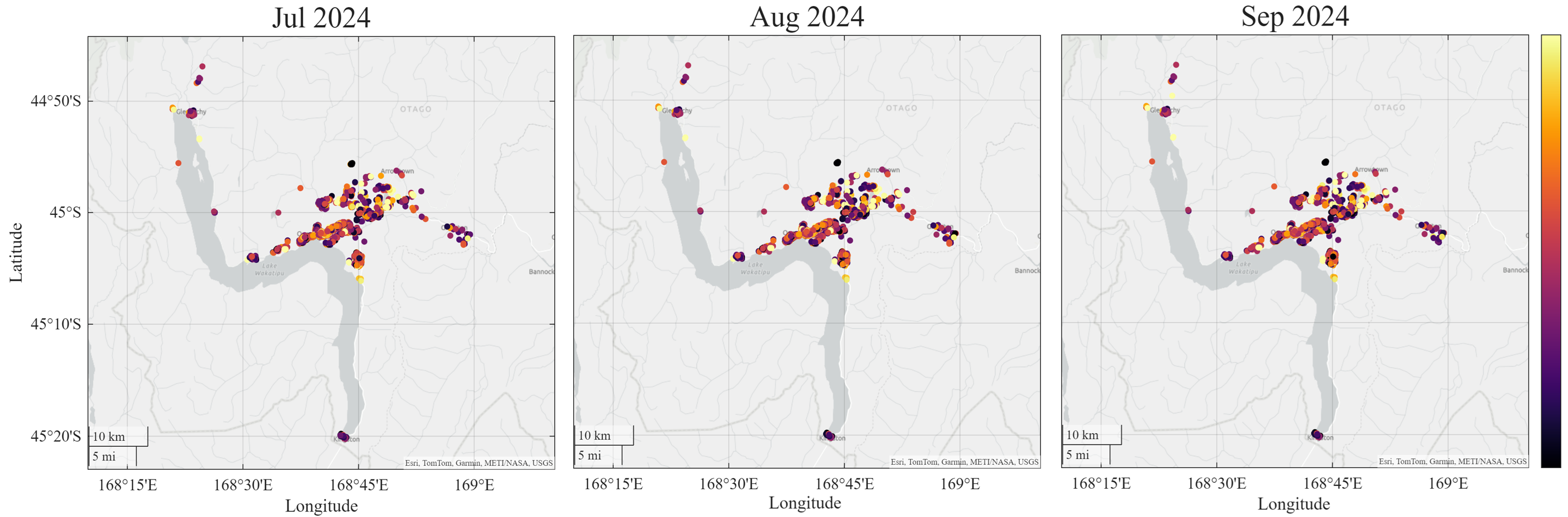}
    \caption{Geographical distribution of log rental prices in Queenstown}
    \label{pic_geosc_QT}
\end{figure}

Figures~\ref{pic_geosc_MT} and~\ref{pic_geosc_QT} display the geographical distribution of log rental prices for listings in Midtown and Queenstown, respectively. The color shading corresponds to the magnitude of log rental price, with darker tones indicating lower values. These figures show that the log rental prices exhibit obvious spatial and temporal correlation.

Based on the data described above, we estimate the TWFE unbalanced dynamic network panel model using the bias-corrected QMLE. For each local market, $W_t$ and $M_t$ are constructed as row-normalized binary distance-band matrices. Specifically, for a given distance threshold $d$, two listings are connected if their geographical distance is no larger than $d$. The contemporaneous matrix $W_t$ is constructed among listings observed in period $t$, with diagonal elements set to zero. The lagged matrix $M_t$ is constructed between listings observed in period $t$ and listings observed in period $t-1$, with links from a listing to itself in the previous period excluded. The resulting binary adjacency matrices are then row-normalized.
Instead of using a common set of distance thresholds for all areas, we choose the thresholds adaptively because the selected local markets differ substantially in spatial scale.
Specifically, the thresholds are selected based on the market-specific median nearest-neighbor distance and then rounded to one decimal place. This procedure yields thresholds of 0.4, 0.7, and 1.1 km for the three New Zealand local markets; 0.4, 0.6, and 1.1 km for Williamsburg; 0.1, 0.2, and 0.4 km for Midtown; and 0.2, 0.3, and 0.5 km for Hell's Kitchen. The main tables report the estimates under the largest distance threshold for each local market, while Supplementary Appendix E reports the estimates under all three thresholds.

Tables~\ref{Tab_Comv_TWFE} and~\ref{Tab_ComV_TWFE} compare the bias-corrected TWFE QMLE results across the selected local markets. The results are based on the same class of TWFE dynamic network model, but the distance thresholds are market-specific because of the adaptive construction described above. We obtain the following findings.

\begin{enumerate}[(1)]
\item The network autoregressive coefficient $\rho$ captures contemporaneous interaction effects among connected listings. It is positive and statistically significant in all six local markets. The estimates are larger in New York City, ranging from 0.4812 to 0.7690, than in New Zealand, where they range from 0.0960 to 0.1612. This difference is consistent with the higher listing density and stronger real-time competitive pressure in New York City. In contrast, listings in the New Zealand local markets are more geographically dispersed, which weakens contemporaneous spatial price transmission.

\item The lagged network autoregressive effect $\lambda$ captures the impact of neighboring prices in the previous period on current rental prices. This effect is significantly negative across all six local markets. Moreover, local markets with a larger contemporaneous network autoregressive coefficient $\rho$ generally exhibit a more negative estimate of $\lambda$. This pattern suggests an intertemporal dampening mechanism in network spillovers, whereby contemporaneous spatial interactions are partially attenuated by subsequent network effects.

\item The temporal lagged effect $\nu$ is positive and statistically significant in all local markets, indicating rental price persistence. The estimates range from 0.2082 to 0.3982 in New Zealand and from 0.3969 to 0.4638 in New York City.
\end{enumerate}

\begin{table}[H]
\centering
\caption{Bias-corrected TWFE QMLE estimates for New Zealand}\label{Tab_Comv_TWFE}
\begin{tabular}{llll}
\hline
\multicolumn{1}{c}{} & \multicolumn{1}{c}{Auckland} & \multicolumn{1}{c}{Queenstown} & \multicolumn{1}{c}{Wellington} \\
\hline
$\rho \quad (W_tY_t)$ & \phantom{-}0.1612(0.0053)*** & \phantom{-}0.0960(0.0055)*** & \phantom{-}0.1591(0.0333)*** \\
$\lambda \quad (M_tY_{t-1})$ & -0.1640(0.0053)*** & -0.0791(0.0052)*** & -0.0735(0.0301)** \\
$\nu \quad (Y_{t-1})$ & \phantom{-}0.3982(0.0031)*** & \phantom{-}0.3006(0.0028)*** & \phantom{-}0.2082(0.0049)*** \\
avail30d & -0.0006(0.0001)*** & -0.0005(0.0001)*** & -0.0003(0.0002)** \\
avgAvail30d1km & -0.0053(0.0004)*** & -0.0058(0.0004)*** & \phantom{-}0.0015(0.0016) \\
Listing effect & \phantom{-}0.2221(0.0059)*** & \phantom{-}0.2335(0.0064)*** & \phantom{-}0.1207(0.0274)*** \\
\cline{2-4}
$N$ & \multicolumn{1}{c}{5896} & \multicolumn{1}{c}{6095} & \multicolumn{1}{c}{1549} \\
$T$ & \multicolumn{1}{c}{30} & \multicolumn{1}{c}{30} & \multicolumn{1}{c}{30} \\
$n$ & \multicolumn{1}{c}{80716} & \multicolumn{1}{c}{95455} & \multicolumn{1}{c}{18723} \\
UP & \multicolumn{1}{c}{54.37\%} & \multicolumn{1}{c}{47.80\%} & \multicolumn{1}{c}{59.71\%} \\
\hline
\end{tabular}
\begin{flushleft}
\footnotesize
Notes: Estimates are presented with their standard errors in parentheses. Statistical significance is indicated by stars: *** for $p<0.01$, ** for $p<0.05$, and * for $p<0.1$.
\end{flushleft}
\end{table}

\begin{table}[H]
\centering
\caption{Bias-corrected TWFE QMLE estimates for New York City}\label{Tab_ComV_TWFE}
\begin{tabular}{llll}
\hline
\multicolumn{1}{c}{} & \multicolumn{1}{c}{Williamsburg} & \multicolumn{1}{c}{Midtown} & \multicolumn{1}{c}{Hell's Kitchen} \\
\hline
$\rho \quad (W_tY_t)$ & \phantom{-}0.7690(0.0310)*** & \phantom{-}0.4812(0.0238)*** & \phantom{-}0.5414(0.0267)*** \\
$\lambda \quad (M_tY_{t-1})$ & -0.5242(0.0397)*** & -0.2740(0.0319)*** & -0.4066(0.0328)*** \\
$\nu \quad (Y_{t-1})$ & \phantom{-}0.4638(0.0064)*** & \phantom{-}0.4597(0.0068)*** & \phantom{-}0.3969(0.0063)*** \\
avail30d & \phantom{-}0.0006(0.0002)*** & -0.0012(0.0003)*** & -0.0002(0.0003) \\
avgAvail30d1km & \phantom{-}0.0035(0.0031) & -0.0020(0.0034) & \phantom{-}0.0013(0.0093) \\
Listing effect & -0.0687(0.0438) & \phantom{-}0.2528(0.0420)*** & -0.0022(0.0380) \\
\cline{2-4}
$N$ & \multicolumn{1}{c}{1566} & \multicolumn{1}{c}{2390} & \multicolumn{1}{c}{1943} \\
$T$ & \multicolumn{1}{c}{17} & \multicolumn{1}{c}{17} & \multicolumn{1}{c}{17} \\
$n$ & \multicolumn{1}{c}{14567} & \multicolumn{1}{c}{23672} & \multicolumn{1}{c}{16862} \\
UP & \multicolumn{1}{c}{45.28\%} & \multicolumn{1}{c}{41.74\%} & \multicolumn{1}{c}{48.95\%} \\
\hline
\end{tabular}
\begin{flushleft}
\footnotesize
Notes: Estimates are presented with their standard errors in parentheses. Statistical significance is indicated by stars: *** for $p<0.01$, ** for $p<0.05$, and * for $p<0.1$.
\end{flushleft}
\end{table}
Furthermore, Tables E.2–E.7 in Supplementary Appendix E report detailed estimation results under different network weight matrix specifications for each area. Within each local market, the estimation results are generally similar across different distance-threshold specifications in terms of signs and statistical significance. A common pattern is that the absolute magnitudes of both the network autoregressive coefficient $\rho$ and the lagged network autoregressive effect $\lambda$ increase as the distance threshold d expands, while the remaining coefficients change only slightly. This pattern is consistent with the role of the distance threshold in defining the network exposure. Larger distance bands include a broader set of local competitors, which strengthens the measured contemporaneous network dependence and the subsequent lagged network adjustment. In contrast, the stability of the non-network coefficients suggests that the estimated effects of listing-level availability, nearby availability, temporal persistence, and listing entry are not driven by the particular distance-band choice.

\section{Conclusion}\label{section6}
In this paper, we study QMLE for genuinely unbalanced dynamic network panel data models as both $N$ and $T$ go to infinity. We first consider a baseline OWFE specification. For this specification, the QMLE is $\sqrt{NT}$-consistent for all parameters except for the listing effect $\hat{\gamma}$, which achieves $\sqrt{N}$-consistency. When $N$ grows proportionally to $T$, the QMLE has an asymptotic bias induced by the incidental parameters. The proposed bias-corrected estimator is asymptotically unbiased and normally distributed as long as $N/T^3 \to 0$. We also construct a consistent estimator of the asymptotic covariance matrix, making feasible inference available for the bias-corrected estimator.

We further extend the framework to a TWFE specification. Under this extension, the QMLE remains consistent, and the asymptotic bias contains an additional component generated by the estimation of time fixed effects. We derive the corresponding bias correction and show that the TWFE bias-corrected estimator is asymptotically unbiased and normally distributed when $N/T^3 \to 0$ and $T/N^3 \to 0$. A consistent estimator of the TWFE asymptotic covariance matrix is also provided.

Several extensions remain for future research. First, GMM estimation for this model is worth studying, because it can accommodate heteroskedastic errors and remains applicable when $T$ is finite. Second, building on the current TWFE extension, individual--time interactive fixed effects provide a natural direction for further development. Third, a more flexible mechanism for the listing effect is another promising extension.



\section*{Funding statement}
This work was supported by the National Natural Science Foundation of China (NSFC) under Grant Nos. 72473118, 72333001, and 71988101.

\section*{Competing interests}
The authors declare none.

\section*{Declaration of Generative AI Use}
During the preparation of this work the authors used ChatGPT in order to enhance the writing quality. After using this tool/service, the authors reviewed and edited the content as needed and take full responsibility for the content of the publication.

\section*{Data Availability Statement}
The data that support the findings of this study are available from Inside Airbnb at \url{https://insideairbnb.com}. These data were derived from publicly available monthly Airbnb listing snapshots.

\phantomsection\label{supplementary-material}
\section*{Supplementary Material}
The supplementary material includes the required lemmas and their proofs, the explicit expressions and properties of the QMLE score function and Hessian matrix, detailed proofs of the theorems, as well as tables and figures for the simulation and empirical results.

\section*{Acknowledgments}
We thank Professor Xiaoyu Meng, Chao Yang, Fei Jin, and Kai Yang for their valuable comments.

\bibliography{bibliography}
\end{document}


\maketitle

\begingroup
\renewcommand{\thefootnote}{}
\footnotetext{\textdagger\ Zhijian Wang and Tuo Liu contributed equally to this work.}
\footnotetext{* Corresponding author: \href{mailto:xuxingbai@xmu.edu.cn}{xuxingbai@xmu.edu.cn}.}
\endgroup
\newsavebox{\mytableaaa}
\savebox{\mytableaaa}{
 \\ \hline
$\rho \quad (W_tY_t)$ & \phantom{-}0.5453(0.0064)*** & \phantom{-}0.5657(0.0078)*** & \phantom{-}0.5466(0.0092)*** & \phantom{-}0.2481(0.0075)*** & \phantom{-}0.3236(0.0113)*** & \phantom{-}0.3944(0.0162)*** \\
$\lambda \quad (M_tY_{t-1})$ & -0.0185(0.0061)*** & -0.0399(0.0075)*** & -0.0577(0.0085)*** & -0.1029(0.0061)*** & -0.1206(0.0081)*** & -0.2116(0.0109)*** \\
$\nu \quad (Y_{t-1})$ & \phantom{-}0.3505(0.0056)*** & \phantom{-}0.3687(0.0058)*** & \phantom{-}0.4161(0.0060)*** & \phantom{-}0.5536(0.0065)*** & \phantom{-}0.5492(0.0066)*** & \phantom{-}0.5686(0.0066)*** \\
avail30d & \phantom{-}0.0005(0.0005) & \phantom{-}0.0008(0.0005) & \phantom{-}0.0011(0.0005)** & \phantom{-}0.0008(0.0006) & \phantom{-}0.0011(0.0006)** & \phantom{-}0.0009(0.0006) \\
avgLogPrice10NN & \phantom{-}0.1181(0.0163)*** & \phantom{-}0.1510(0.0170)*** & \phantom{-}0.1767(0.0178)*** & \phantom{-}0.2493(0.0187)*** & \phantom{-}0.2561(0.0191)*** & \phantom{-}0.2505(0.0195)*** \\
avgAvail30d1km & -0.0295(0.0029)*** & -0.0234(0.0030)*** & -0.0240(0.0032)*** & -0.0635(0.0035)*** & -0.0528(0.0038)*** & -0.0496(0.0041)*** \\
Peak season & \phantom{-}0.0375(0.0085)*** & \phantom{-}0.0372(0.0087)*** & \phantom{-}0.0395(0.0091)*** & \phantom{-}0.0349(0.0097)*** & \phantom{-}0.0329(0.0097)*** & \phantom{-}0.0298(0.0098)*** \\
Listing effect & \phantom{-}0.4780(0.0110)*** & \phantom{-}0.4723(0.0124)*** & \phantom{-}0.5138(0.0139)*** & \phantom{-}0.6577(0.0103)*** & \phantom{-}0.6214(0.0118)*** & \phantom{-}0.5206(0.0151)*** \\ \hline
\end{tabular}}

\newsavebox{\mytableAuckv}
\savebox{\mytableAuckv}{
\begin{tabular}{lccc}
\hline
\multicolumn{1}{c}{} & $d$=0.4km & $d$=0.7km & $d$=1.1km \\
\hline
$\rho \quad (W_tY_t)$ & 0.0350(0.0024)*** & 0.0669(0.0040)*** & 0.0960(0.0055)*** \\
$\lambda \quad (M_tY_{t-1})$ & -0.0298(0.0024)*** & -0.0623(0.0039)*** & -0.0791(0.0052)*** \\
$\nu \quad (Y_{t-1})$ & 0.2984(0.0030)*** & 0.3003(0.0029)*** & 0.3006(0.0028)*** \\
avail30d & -0.0005(0.0001)*** & -0.0005(0.0001)*** & -0.0005(0.0001)*** \\
avgAvail30d1km & -0.0059(0.0004)*** & -0.0059(0.0004)*** & -0.0058(0.0004)*** \\
Listing effect & 0.2822(0.0043)*** & 0.2507(0.0054)*** & 0.2335(0.0064)*** \\
\hline
\end{tabular}}

\newsavebox{\mytableQueenv}
\savebox{\mytableQueenv}{
\begin{tabular}{lccc}
\hline
\multicolumn{1}{c}{} & $d$=0.4km & $d$=0.6km & $d$=1.1km \\
\hline
$\rho \quad (W_tY_t)$ & 0.5475(0.0182)*** & 0.7018(0.0234)*** & 0.7690(0.0310)*** \\
$\lambda \quad (M_tY_{t-1})$ & -0.1517(0.0158)*** & -0.1818(0.0211)*** & -0.5242(0.0397)*** \\
$\nu \quad (Y_{t-1})$ & 0.4306(0.0064)*** & 0.4369(0.0063)*** & 0.4638(0.0064)*** \\
avail30d & 0.0006(0.0002)** & 0.0006(0.0002)** & 0.0006(0.0002)*** \\
avgAvail30d1km & 0.0031(0.0033) & 0.0017(0.0031) & 0.0035(0.0031) \\
Listing effect & 0.3199(0.0176)*** & 0.2925(0.0234)*** & -0.0687(0.0438) \\
\hline
\end{tabular}
}

\newsavebox{\mytableNZv}
\savebox{\mytableNZv}{
\begin{tabular}{lccc}
\hline
\multicolumn{1}{c}{} & $d$=0.2km & $d$=0.3km & $d$=0.5km \\
\hline
$\rho \quad (W_tY_t)$ & 0.3932(0.0129)*** & 0.5211(0.0164)*** & 0.5414(0.0267)*** \\
$\lambda \quad (M_tY_{t-1})$ & -0.2657(0.0153)*** & -0.3238(0.0206)*** & -0.4066(0.0328)*** \\
$\nu \quad (Y_{t-1})$ & 0.3989(0.0065)*** & 0.3948(0.0064)*** & 0.3969(0.0063)*** \\
avail30d & -0.0002(0.0003) & -0.0002(0.0003) & -0.0002(0.0003) \\
avgAvail30d1km & 0.0022(0.0116) & 0.0019(0.0108) & 0.0013(0.0093) \\
Listing effect & 0.1662(0.0179)*** & 0.0926(0.0237)*** & -0.0022(0.0380) \\
\hline
\end{tabular}}

\newsavebox{\mytableNYv}
\savebox{\mytableNYv}{
\begin{tabular}{lllllll}
\hline
\multicolumn{1}{c}{} & \begin{tabular}[c]{c@{}c@{}}KNN matirx\\ ($K$=10)\end{tabular} & \begin{tabular}[c]{c@{}c@{}}KNN matirx\\ ($K$=20)\end{tabular} & \begin{tabular}[c]{c@{}c@{}}KNN matirx\\ ($K$=30)\end{tabular} & \begin{tabular}[c]{c@{}c@{}}distance-band\\ matrix ($d$=0.5km)\end{tabular} & \begin{tabular}[c]{c@{}c@{}}distance-band\\ matrix ($d$=1km)\end{tabular} & \begin{tabular}[c]{c@{}c@{}}distance-band\\ matrix ($d$=2km)\end{tabular} \\ \hline
$\rho \quad (W_tY_t)$ & \phantom{-}0.5329(0.0044)*** & \phantom{-}0.5821(0.0053)*** & \phantom{-}0.5891(0.0062)*** & \phantom{-}0.2046(0.0053)*** & \phantom{-}0.3091(0.0087)*** & \phantom{-}0.4889(0.0151)*** \\
$\lambda \quad (M_tY_{t-1})$ & -0.0189(0.0036)*** & -0.0260(0.0044)*** & -0.0317(0.0050)*** & -0.1015(0.0039)*** & -0.1659(0.0054)*** & -0.2979(0.0062)*** \\
$\nu \quad (Y_{t-1})$ & \phantom{-}0.3310(0.0030)*** & \phantom{-}0.3343(0.0032)*** & \phantom{-}0.3615(0.0034)*** & \phantom{-}0.4961(0.0039)*** & \phantom{-}0.4990(0.0039)*** & \phantom{-}0.5110(0.0039)*** \\
avail30d & \phantom{-}0.0005(0.0002)*** & \phantom{-}0.0005(0.0002)*** & \phantom{-}0.0006(0.0002)*** & \phantom{-}0.0006(0.0002)*** & \phantom{-}0.0006(0.0002)*** & \phantom{-}0.0006(0.0002)*** \\
avgLogPrice10NN & \phantom{-}0.0561(0.0065)*** & \phantom{-}0.0630(0.0068)*** & \phantom{-}0.0800(0.0070)*** & \phantom{-}0.1852(0.0084)*** & \phantom{-}0.1814(0.0083)*** & \phantom{-}0.1669(0.0081)*** \\
avgAvail30d1km & -0.0232(0.0013)*** & -0.0183(0.0013)*** & -0.0163(0.0014)*** & -0.0471(0.0015)*** & -0.0402(0.0017)*** & -0.0309(0.0021)*** \\
Peak season & \phantom{-}0.0218(0.0036)*** & \phantom{-}0.0197(0.0037)*** & \phantom{-}0.0209(0.0038)*** & \phantom{-}0.0306(0.0041)*** & \phantom{-}0.0268(0.0041)*** & \phantom{-}0.0201(0.0041)*** \\
Listing effect & \phantom{-}0.4313(0.0057)*** & \phantom{-}0.4251(0.0065)*** & \phantom{-}0.4544(0.0072)*** & \phantom{-}0.5485(0.0062)*** & \phantom{-}0.4617(0.0071)*** & \phantom{-}0.2969(0.0077)*** \\ \hline
\end{tabular}}

\newsavebox{\mytableDS}
\savebox{\mytableDS}{
\begin{tabular}{ccccccccc}
\hline
 & Sample Size & Mean & Std & Min & Q25 & Median & Q75 & Max \\ \cline{2-9} 
 & \multicolumn{8}{c}{Auckland Dataset} \\ \cline{2-9} 
log rental price & 80716 & 5.421 & 0.8183 & 2.7726 & 4.8363 & 5.3279 & 5.8972 & 11.5129 \\
avail30d & 80716 & 14.0899 & 10.5291 & 0 & 4 & 14 & 24 & 30 \\
avgAvail30d1km & 80716 & 13.9605 & 3.8352 & 0 & 11.127 & 13.8058 & 16.0676 & 30 \\
\cline{2-9} 
 & \multicolumn{8}{c}{Queenstown Dataset} \\ \cline{2-9} 
log rental price & 95455 & 5.9705 & 0.7675 & 2.9957 & 5.4467 & 5.9789 & 6.4425 & 10.6968 \\
avail30d & 95455 & 10.7122 & 9.4054 & 0 & 2 & 9 & 18 & 30 \\
avgAvail30d1km & 95455 & 10.6611 & 3.4676 & 0 & 8.3091 & 10.0317 & 12.7324 & 30 \\
\cline{2-9} 
 & \multicolumn{8}{c}{Wellington Dataset} \\ \cline{2-9} 
log rental price & 18723 & 5.1562 & 0.701 & 3.091 & 4.7707 & 5.2095 & 5.5947 & 8.5172 \\
avail30d & 18723 & 13.5583 & 10.3087 & 0 & 3 & 14 & 23 & 30 \\
avgAvail30d1km & 18723 & 13.782 & 3.9405 & 3.9861 & 10.318 & 14.401 & 17.4 & 22.0946 \\
\cline{2-9} 
\hline
 & Sample Size & Mean & Std & Min & Q25 & Median & Q75 & Max \\ \cline{2-9} 
 & \multicolumn{8}{c}{Williamsburg Dataset} \\ \cline{2-9} 
log rental price & 14567 & 5.1221 & 0.7946 & 2.3026 & 4.6052 & 5.0752 & 5.5759 & 10.5966 \\
avail30d & 14567 & 13.1242 & 12.5974 & 0 & 0 & 9 & 28 & 30 \\
avgAvail30d1km & 14567 & 12.9429 & 2.5332 & 4.7143 & 10.9963 & 12.5299 & 14.7712 & 23 \\
\cline{2-9} 
 & \multicolumn{8}{c}{Midtown Dataset} \\ \cline{2-9} 
log rental price & 23672 & 5.7745 & 0.8484 & 2.0794 & 5.2933 & 5.717 & 6.1431 & 10.8208 \\
avail30d & 23672 & 14.0729 & 12.0784 & 0 & 1 & 13 & 28 & 30 \\
avgAvail30d1km & 23672 & 14.3105 & 2.09 & 8.0087 & 12.7817 & 14.3718 & 15.613 & 20.0807 \\
\cline{2-9} 
 & \multicolumn{8}{c}{Hell's Kitchen Dataset} \\ \cline{2-9} 
log rental price & 16862 & 5.357 & 0.868 & 3.4657 & 4.8283 & 5.2832 & 5.7746 & 10.8235 \\
avail30d & 16862 & 12.8441 & 12.4652 & 0 & 0 & 9 & 28 & 30 \\
avgAvail30d1km & 16862 & 12.5631 & 2.434 & 6.0616 & 10.6333 & 12.6123 & 14.4707 & 18.6227 \\
\hline
\end{tabular}}

\appendix

\renewcommand{\theequation}{\thesection.\arabic{equation}}
\renewcommand{\thetable}{\thesection.\arabic{table}}
\renewcommand{\thefigure}{\thesection.\arabic{figure}}
\counterwithin{table}{section}
\counterwithin{figure}{section}
\counterwithin{equation}{section}
\counterwithin{assumption}{section}
\counterwithin{definition}{section}
\counterwithin{theorem}{section}

\section{Some Lemmas}\label{somelemmas}
\setcounter{assumption}{0}

In order to establish the law of large numbers for the quadratic form of $\mathbb{Z}$ and $\mathbb{V}$,
we need to study the statistical properties of the quadratic form of two random variables formed by the accumulation of infinite historical errors. Define ${P_t}$, ${Q_t}$ as sequences of $N_t\times N_{t-1}$ matrices, and define their cumulative products as $P_{t}^{(h)}\equiv P_{t}P_{t-1}\cdots P_{t-h+1}$ for $h=1,2,\cdots$, $P_{t}^{(0)}=I_{N_t}$ and $Q_{t}^{(h)}\equiv Q_{t}Q_{t-1}\cdots Q_{t-h+1}$ for $h=1,2,\cdots$, $Q_{t}^{(0)}=I_{N_t}$ respectively.
Then denote two different accumulations of infinite historical errors as $$L_t=\sum_{h=0}^\infty P_{th}V_{t-h},\quad R_t=\sum_{h=0}^\infty Q_{th}V_{t-h},$$ 
where $P_{th}\equiv B_{1t}P_{t}^{(h)}$, $Q_{th}\equiv B_{2t}Q_{t}^{(h)}$, and $B_{1t}$ and $B_{2t}$ are $N_t\times N_t$ matrices. Define their respective lagged versions as
$$L_t^\ell=B_{1t}^\ell L_{t-1}, \quad R_t^\ell=B_{2t}^\ell R_{t-1}, $$ 
where $B_{1t}^\ell$ and $B_{2t}^\ell$ are $N_t\times N_{t-1}$ matrices.
Denote their respective stacked versions as $\mathbb{L}\equiv(L_1',L_2',\cdots,$ $L_T')'$, $\mathbb{L}^\ell\equiv(L_1^{\ell\prime},L_2^{\ell\prime},\cdots,L_T^{\ell\prime})'$, and similarly, $\mathbb{R}$ and $\mathbb{R}^\ell$ follow the same form.
Denote the block diagonal matrices $\mathbb{B}_1=\mathrm{blkdiag}(B_{11},$ $B_{12},\cdots,B_{1T})$ and $\mathbb{B}_2=\mathrm{blkdiag}(B_{21},B_{22},\cdots,B_{2T})$. Denote a nonstochastic $n$-dimensional vector as $\mathbb{C}=(C_1',C_2',\cdots,C_T')'$.
The following assumptions will be used in some of the subsequent lemmas:

\begin{assumption}\label{asp_A2}
$\{B_{1t}\}$, $\{B_{2t}\}$, $\{B_{1t}^\ell\}$, $\{B_{2t}^\ell\}$, $\{P_t^{(h)}\}$ and $\{Q_t^{(h)}\}$ are nonstochastic matrix sequences such that $B_{1t}$, $B_{2t}$, $B_{1t}^\ell$, $B_{2t}^\ell$, $\sum_{h=0}^\infty\mathrm{abs}(P_t^{(h)}D_{t-h})$, $\sum_{h=t}^\infty\mathrm{abs}(D_h'P_h^{(h-t)})$,  $\sum_{h=0}^\infty\mathrm{abs}(Q_t^{(h)}D_{t-h})$, $\sum_{h=t}^\infty\mathrm{abs}(D_h'Q_h^{(h-t)})$ are all UB uniformly for $t$. Also, $\|\sup_t \mathrm{abs}(D_t'B_{jt}D_t)\|_\infty<\infty$ and $\|\sup_t \mathrm{abs}(D_t'B_{jt}^\ell D_{t-1})\|_\infty<\infty$, for $j=1,2$.
\end{assumption}

\begin{assumption}\label{asp_C}
$C_t$ is an arbitrary nonstochastic $N_t$-dimensional vector, and elements in $C_t$ are uniformly bounded for $t$ and $N_t$.
\end{assumption}

\begin{lemma}\label{lm_EPQ}
Under Assumption 2, for $t\geq s$, 
$$E(L_tR_s')=\sigma_0^2\left(\sum_{g=0}^\infty P_{t,t-s+g}Q_{sg}'\right),$$
and
$$E(L_t'D_tD_s'R_s)=\sigma_0^2\mathrm{tr}\left(\sum_{g=0}^\infty D_t'P_{t,t-s+g}Q_{sg}'D_s\right).$$ 
\end{lemma}
\begin{proof}
Since $V_t$'s are independent across $t$,
\begin{align*}
E(L_tR_s')&=E\left[\left(\sum_{h=0}^{t-s-1} P_{th}V_{t-h}+\sum_{g=0}^{\infty} P_{t,t-s+g}V_{s-g}\right)\left(\sum_{g=0}^\infty Q_{sg}V_{s-g}\right)'\right]\\
&=E\left[\left(\sum_{g=0}^{\infty} P_{t,t-s+g}V_{s-g}\right)\left(\sum_{g=0}^\infty Q_{sg}V_{s-g}\right)'\right]=\sigma_0^2\left(\sum_{g=0}^\infty P_{t,t-s+g}Q_{sg}'\right).
\end{align*}
Additionally, 
\begin{align*}
E(L_t'D_tD_s'R_s)=&E(R_s'D_sD_t'L_t)=\mathrm{tr}\left[D_t'E(L_tR_s')D_s\right]=\sigma_0^2\mathrm{tr}\left(\sum_{g=0}^\infty D_t'P_{t,t-s+g}Q_{sg}'D_s\right)
\end{align*}
\end{proof}

\begin{lemma}\label{lm_quadratic form}
Under Assumption 2, for any nonstochastic $B_1$ of size $N_t\times N_s$, $B_2$ of size $N_g\times N_h$, we have
\begin{align*}
E\left[(V_{t}^{\prime}B_{1}V_{s})(V_{g}^{\prime}B_{2}V_{h})\right]&=\left\{\begin{array}{ll}(\mu_4-3\sigma_0^4)\sum_{i=1}^{N_t} B_{1,ii}B_{2,ii}&\mathrm{for}\ t=s=g=h\\
+\sigma_0^4(\mathrm{tr}B_{1}\times \mathrm{tr}B_{2}+\mathrm{tr}B_{1}B_{2}+\mathrm{tr}B_{1}B_{2}^{\prime})\\
\sigma_0^4\mathrm{tr}B_{1}\times \mathrm{tr}B_{2}&\mathrm{for}\ t=s\neq g=h\\
\sigma_0^4\mathrm{tr}(B_{1}B_{2}^{\prime})&\mathrm{for}\ t=g\neq s=h\\
\sigma_0^4\mathrm{tr}(B_{1}B_{2})&\mathrm{for}\  t=h\neq s=g\\
0&\rm{otherwise}
\end{array}\right..
\end{align*}
\end{lemma}
\begin{proof}
This lemma is an extension of Lemma 3 in Yu et al. (2008), which assumes that both $B_1$ and $B_2$ are square matrices of the same size. 
Denote the $i$-th element of $V_t$ as $V_{t,i}$. Then
\begin{align*}
E\left[(V_{t}^{\prime}B_{1}V_{s})(V_{g}^{\prime}B_{2}V_{h})\right]&=E\left[\Bigg(\sum_{i=1}^{N_t}\sum_{j=1}^{N_s}V_{t,i}B_{1,ij}V_{s,j}\Bigg)\Bigg(\sum_{k=1}^{N_g}\sum_{l=1}^{N_h}V_{g,k}B_{2,kl}V_{h,l}\Bigg)\right]\\
&=E\left[\sum_{i=1}^{N_t}\sum_{j=1}^{N_s}\sum_{k=1}^{N_{g}}\sum_{l=1}^{N_h}B_{1,ij}B_{2,kl}V_{t,i}V_{s,j}V_{g,k}V_{h,l}\right]
\end{align*}
Since $V_{t,i}$'s are i.i.d across $t$ and $i$, the equation above will not equal to $0$ only when $t=s=g=h$, $t=s\neq g=h$, $t=g\neq s=h$ and $t=h\neq s=g$. For $t=s=g=h$, our conclusion is exactly the same as the corresponding case of Lemma 3 in Yu et al. (2008). Therefore, there is no need for proof. When $t=s\neq g=h$,
\begin{align*}
E\left[(V_{t}^{\prime}B_{1}V_{s})(V_{g}^{\prime}B_{2}V_{h})\right]=&E\left[\sum_{i=1}^{N_t}\sum_{j=1}^{N_t}\sum_{k=1}^{N_g}\sum_{l=1}^{N_g}B_{1,ij}B_{2,kl}V_{t,i}V_{t,j}V_{g,k}V_{g,l}\right]\\
=&E\left[\sum_{i=1}^{N_t}\sum_{k=1}^{N_g}B_{1,ii}B_{2,kk}V_{t,i}^2V_{g,k}^2\right]\\
=&\sigma_0^4\mathrm{tr}B_{1}\times \mathrm{tr}B_{2}.
\end{align*}
When $t=g\neq s=h$,
\begin{align*}
E\left[(V_{t}^{\prime}B_{1}V_{s})(V_{g}^{\prime}B_{2}V_{h})\right]=&E\left[\sum_{i=1}^{N_t}\sum_{j=1}^{N_s}\sum_{k=1}^{N_t}\sum_{l=1}^{N_s}B_{1,ij}B_{2,kl}V_{t,i}V_{s,j}V_{t,k}V_{s,l}\right]\\
=&E\left[\sum_{i=1}^{N_t}\sum_{j=1}^{N_s}B_{1,ij}B_{2,ij}V_{t,i}^2V_{s,j}^2\right]\\
=&\sigma_0^4\mathrm{tr}(B_{1}B_2').
\end{align*}
When $t=h\neq s=g$,
\begin{align*}
E\left[(V_{t}^{\prime}B_{1}V_{s})(V_{g}^{\prime}B_{2}V_{h})\right]=&E\left[\sum_{i=1}^{N_t}\sum_{j=1}^{N_s}\sum_{k=1}^{N_s}\sum_{l=1}^{N_t}B_{1,ij}B_{2,kl}V_{t,i}V_{s,j}V_{s,k}V_{t,l}\right]\\
=&E\left[\sum_{i=1}^{N_t}\sum_{j=1}^{N_s}B_{1,ij}B_{2,ji}V_{t,i}^2V_{s,j}^2\right]\\
=&\sigma_0^4\mathrm{tr}(B_{1}B_2).
\end{align*}
\end{proof}

\begin{lemma}\label{lm_CovPQ}
Under Assumption 2, for $t\geq s$, 
\begin{align*}
&\mathrm{Cov}(L_t'R_t,L_s'R_s)\\
=&\sum_{g=0}^\infty\left[(\mu_4-3\sigma_0^4)\sum_{i=1}^{N_{s-g}}(P_{t,t-s+g}'Q_{t,t-s+g})_{ii}(P_{sg}'Q_{sg})_{ii}\right]+\\
&\sigma_0^4\mathrm{tr}\left[\left(\sum_{g=0}^\infty P_{sg}P_{t,t-s+g}'\right)\left(\sum_{g=0}^\infty Q_{t,t-s+g}Q_{sg}'\right)+\left(\sum_{g=0}^\infty Q_{sg}P_{t,t-s+g}'\right)\left(\sum_{g=0}^\infty Q_{t,t-s+g}P_{sg}'\right)\right].
\end{align*}
\end{lemma}
\begin{proof}
Using Lemma \ref{lm_quadratic form}, we have
\begin{align*}
E(L_{t}^{\prime}R_{t}L_{s}^{\prime}R_{s}) 
=&E\left\lbrack\left(\sum_{h=0}^{\infty}P_{th}V_{t-h}\right)^{\prime}\left(\sum_{h=0}^{\infty}Q_{th}V_{t-h}\right)\left(\sum_{g=0}^{\infty}P_{sg}V_{s-g}\right)^{\prime}\left(\sum_{g=0}^{\infty}Q_{sg}V_{s-g}\right)\right\rbrack\\  
=&E\left\lbrack\left(\sum_{h=0}^{t-s-1}P_{th}V_{t-h}+\sum_{g=0}^{\infty}P_{t,t-s+g}V_{s-g}\right)^{\prime}\left(\sum_{h=0}^{t-s-1}Q_{th}V_{t-h}+\sum_{g=0}^{\infty}Q_{t,t-s+g}V_{s-g}\right)\right.\\  
&\left.\left(\sum_{g=0}^{\infty}P_{sg}V_{s-g}\right)^{\prime}\left(\sum_{g=0}^{\infty}Q_{sg}V_{s-g}\right)\right\rbrack\\  
=&E\left[\left(\sum_{h=0}^{t-s-1}P_{th}V_{t-h}\right)^{\prime}\left(\sum_{h=0}^{t-s-1}Q_{th}V_{t-h}\right)\left(\sum_{g=0}^{\infty}P_{sg}V_{s-g}\right)^{\prime}\left(\sum_{g=0}^{\infty}Q_{sg}V_{s-g}\right)\right]+\\  
&E\left[\left(\sum_{g=0}^{\infty}P_{t,t-s+g}V_{s-g}\right)^{\prime}\left(\sum_{g=0}^{\infty}Q_{t,t-s+g}V_{s-g}\right)\left(\sum_{g=0}^{\infty}P_{sg}V_{s-g}\right)^{\prime}\left(\sum_{g=0}^{\infty}Q_{sg}V_{s-g}\right)\right]\\
\equiv&E_1+E_2,
\end{align*}
where the third equality holds because $V_t$'s are independent across $t$.
Then we have
\begin{align*}
E_1&=E\left\lbrack\left(\sum_{h_1=0}^{t-s-1}P_{th}V_{t-h}\right)^{\prime}\left(\sum_{h_2=0}^{t-s-1}Q_{th}V_{t-h}\right)\left(\sum_{g_1=0}^{\infty}P_{sg}V_{s-g}\right)^{\prime}\left(\sum_{g_2=0}^{\infty}Q_{sg}V_{s-g}\right)\right\rbrack\\
&=E\left\lbrack\left(\sum_{h_1=0}^{t-s-1}\sum_{h_2=0}^{t-s-1}V_{t-h_1}^{\prime}P_{th_1}^{\prime}Q_{th_2}V_{t-h_2}\right)\left(\sum_{g_1=0}^{\infty}\sum_{g_2=0}^{\infty}V_{s-g_1}^{\prime}P_{sg_1}^{\prime}Q_{sg_2}V_{s-g_2}\right)\right\rbrack\\
&=\sigma_0^4\left\lbrack\sum_{h=0}^{t-s-1}\mathrm{tr}\left(P_{th}^{\prime}Q_{th}\right)\right\rbrack\left\lbrack\sum_{g=0}^{\infty}\mathrm{tr}\left(P_{sg}^{\prime}Q_{sg}\right)\right\rbrack,
\end{align*}
where the last equality holds due to the second case in Lemma \ref{lm_quadratic form}. 

\begin{align*}
E_{2}&=E\left\lbrack\left(\sum_{g_1=0}^{\infty}P_{t,t-s+g}V_{s-g}\right)^{\prime}\left(\sum_{g_2=0}^{\infty}Q_{t,t-s+g}V_{s-g}\right)\left(\sum_{g_3=0}^{\infty}P_{sg}V_{s-g}\right)^{\prime}\left(\sum_{g_4=0}^{\infty}Q_{sg}V_{s-g}\right)\right\rbrack\\
&=E\left\lbrack\sum_{g_1=0}^{\infty}\sum_{g_2=0}^{\infty}\sum_{g_3=0}^{\infty}\sum_{g_4=0}^{\infty}\left(V_{s-g_1}^{\prime}P_{t,t-s+g_1}^{\prime}Q_{t,t-s+g_2}V_{s-g_2}\right)\left(V_{s-g_3}^{\prime}P_{sg_3}^{\prime}Q_{sg_4}V_{s-g_4}\right)_{}\right\rbrack\\
&\equiv E_{21}+E_{22}+E_{23}+E_{24},
\end{align*}
where $E_{21}$, $E_{22}$, $E_{23}$ and $E_{24}$ are the four nonzero cases in Lemma \ref{lm_quadratic form}. Specifically, $E_{21}$ corresponds to the case where $g_1 = g_2 = g_3 = g_4$, so
\begin{align*}
E_{21}=&\sum_{g=0}^{\infty}\Bigg[(\mu_4-3\sigma_0^4)\sum_{i=1}^{N_{s-g}}(P_{t,t-s+g}^{\prime}Q_{t,t-s+g})_{ii}(P_{sg}^{\prime}Q_{sg})_{ii}+\sigma_0^4\mathrm{tr}\left(P_{t,t-s+g}^{\prime}Q_{t,t-s+g}\right)\mathrm{tr}\left(P_{sg}^{\prime}Q_{sg}\right)\\
&+\sigma_0^4\mathrm{tr}\left(P_{t,t-s+g}^{\prime}Q_{t,t-s+g}P_{sg}^{\prime}Q_{sg}\right)+\sigma_0^4\mathrm{tr}\left(P_{t,t-s+g}^{\prime}Q_{t,t-s+g}Q_{sg}^{\prime}P_{sg}\right)\Bigg],
\end{align*}
$E_{22}$ corresponds to the case where $g_1 = g_2 \neq g_3 = g_4$,
\begin{align*}
E_{22}=\sigma_0^4\left\lbrack\sum_{g=0}^{\infty}\mathrm{tr}\left(P_{t,t-s+g}^{\prime}Q_{t,t-s+g}\right)\right\rbrack\left\lbrack\sum_{g=0}^{\infty}\mathrm{tr}\left(P_{sg}^{\prime}Q_{sg}\right)\right\rbrack-\sigma_0^4\sum_{g=0}^{\infty}\mathrm{tr}\left(P_{t,t-s+g}^{\prime}Q_{t,t-s+g}\right)\mathrm{tr}\left(P_{sg}^{\prime}Q_{sg}\right).
\end{align*}
$E_{23}$ corresponds to the case where  $g_1 = g_3 \neq g_2 = g_4$,
\begin{align*}
E_{23}&=\sigma_0^4\left\lbrack\sum_{g_1=0}^{\infty}\sum_{g_2=0}^{\infty}\mathrm{tr}\left(P_{t,t-s+g_1}^{\prime}Q_{t,t-s+g_2}Q_{sg_2}^{\prime}P_{sg_1}\right)\right\rbrack-\sigma_0^4\sum_{g=0}^{\infty}\mathrm{tr}\left(P_{t,t-s+g}^{\prime}Q_{t,t-s+g}Q_{sg}^{\prime}P_{sg}\right)\\
&=\sigma_0^4\mathrm{tr}\left\lbrack \left(\sum_{g=0}^{\infty}P_{sg}P_{t,t-s+g}^{\prime}\right)\left(\sum_{g=0}^{\infty}Q_{t,t-s+g}Q_{sg}^{\prime}\right)\right\rbrack-\sigma_0^4\sum_{g=0}^{\infty}\mathrm{tr}\left(P_{t,t-s+g}^{\prime}Q_{t,t-s+g}Q_{sg}^{\prime}P_{sg}\right).
\end{align*}
In addition, $E_{24}$ corresponds to the case where  $g_1 = g_4 \neq g_2 = g_3$,
\begin{align*}
E_{24}&=\sigma_0^4\left[\sum_{g_1=0}^{\infty}\sum_{g_2=0}^{\infty}\mathrm{tr}(P_{t,t-s+g_1}^{\prime}Q_{t,t-s+g_2}P_{sg_2}^{\prime}Q_{sg_1})\right]-\sigma_0^4\sum_{g=0}^{\infty}\mathrm{tr}(P_{t,t-s+g}^{\prime}Q_{t,t-s+g}P_{sg}^{\prime}Q_{sg})\\
&=\sigma_0^4\left[\mathrm{tr}\left(\sum_{g=0}^{\infty}Q_{sg}P_{t,t-s+g}^{\prime}\right)\left(\sum_{g=0}^{\infty}Q_{t,t-s+g}P_{sg}^{\prime}\right)\right]-\sigma_0^4\sum_{g=0}^{\infty}\mathrm{tr}(P_{t,t-s+g}^{\prime}Q_{t,t-s+g}P_{sg}^{\prime}Q_{sg}).
\end{align*}
Therefore, we have
\begin{align*}
&\mathrm{Cov}(L_t'R_t,L_s'R_s)=E(L_t'R_tL_s'R_s)-E(L_t'R_t)E(L_s'R_s)\\
=&E_1+E_{21}+E_{22}+E_{23}+E_{24}-E(L_t'R_t)E(L_s'R_s)\\
=&\sigma_0^4\left\lbrack\sum_{h=0}^{t-s-1}\mathrm{tr}\left(P_{th}^{\prime}Q_{th}\right)\right\rbrack\left\lbrack\sum_{g=0}^{\infty}\mathrm{tr}\left(P_{sg}^{\prime}Q_{sg}\right)\right\rbrack+\sum_{g=0}^{\infty}\left[(\mu_4-3\sigma_0^4)\sum_{i=1}^{N_{s-g}}(P_{t,t-s+g}^{\prime}Q_{t,t-s+g})_{ii}(P_{sg}^{\prime}Q_{sg})_{ii}\right]+\\
&\sigma_0^4\left\lbrack\sum_{g=0}^{\infty}\mathrm{tr}\left(P_{t,t-s+g}^{\prime}Q_{t,t-s+g}\right)\right\rbrack\left\lbrack\sum_{g=0}^{\infty}\mathrm{tr}\left(P_{sg}^{\prime}Q_{sg}\right)\right\rbrack+\sigma_0^4\mathrm{tr}\left\lbrack \left(\sum_{g=0}^{\infty}P_{sg}P_{t,t-s+g}^{\prime}\right)\left(\sum_{g=0}^{\infty}Q_{t,t-s+g}Q_{sg}^{\prime}\right)\right\rbrack+\\
&\sigma_0^4\left[\mathrm{tr}\left(\sum_{g=0}^{\infty}Q_{sg}P_{t,t-s+g}^{\prime}\right)\left(\sum_{g=0}^{\infty}Q_{t,t-s+g}P_{sg}^{\prime}\right)\right]-\sigma_0^4\left\lbrack\sum_{g=0}^{\infty}\mathrm{tr}\left(P_{tg}^{\prime}Q_{tg}\right)\right\rbrack\left\lbrack\sum_{g=0}^{\infty}\mathrm{tr}\left(P_{sg}^{\prime}Q_{sg}\right)\right\rbrack\\
=&\sum_{g=0}^\infty\left[(\mu_4-3\sigma_0^4)\sum_{i=1}^{N_{s-g}}(P_{t,t-s+g}'Q_{t,t-s+g})_{ii}(P_{sg}'Q_{sg})_{ii}\right]+\\
&\sigma_0^4\mathrm{tr}\left[\left(\sum_{g=0}^\infty P_{sg}P_{t,t-s+g}'\right)\left(\sum_{g=0}^\infty Q_{t,t-s+g}Q_{sg}'\right)+\left(\sum_{g=0}^\infty Q_{sg}P_{t,t-s+g}'\right)\left(\sum_{g=0}^\infty Q_{t,t-s+g}P_{sg}'\right)\right],
\end{align*}
where $E(L_t'R_t)$ and $E(L_s'R_s)$ are calculated according to Lemma \ref{lm_EPQ}.
\end{proof}

\begin{lemma}\label{lm33}
Denote $B_t\equiv B_{1t}B_{2t}$ and $B_t^\ell\equiv B_{1t}^\ell B_{2,t-1}$. Under Assumption \ref{asp_A2},
the following facts hold:
\begin{enumerate}[(1)]
\item $\|\sup_t \mathrm{abs}(D_t'B_tD_t)\|_{\infty}<\infty$ and $\|\sup_t \mathrm{abs}(D_t'B_t^\ell D_{t-1})\|_{\infty}<\infty$.
\item $\sum_{h=t}^\infty \mathrm{abs}\left(D_h'B_hP_h^{(h-t)}\right)$ and $\sum_{h=t}^\infty \mathrm{abs}\left(B_t^\ell P_{t-1}^{(h)}D_{t-1-h}\right)$ are UB uniformly for $t$.
\item  $\sum_{h=t}^\infty \mathrm{abs}\left(Q_h^{(h-t)\prime}B_{2h}'B_{1h}P_h^{(h-t)}\right)$ and $\sum_{h=0}^\infty\mathrm{abs}\left(B_{t}^\ell P_{t-1}^{(h)}P_{t-1}^{(h)\prime}B_{t}^{\ell\prime}\right)$ are UB uniformly for $t$.
\item $\sum_{h=0}^\infty \mathrm{abs}(P_t^{(h)})\mathrm{abs}(Q_t^{(h)\prime})$ are UB uniformly for $t$.
\end{enumerate}
\end{lemma}
\begin{proof}
\textbf{Proof of (1):}
According to the definition of $\sup_t \mathrm{abs}(D_t'B_tD_t)$, we have
\begin{align*}
\|\sup_t \mathrm{abs}(D_t'B_tD_t)\|_{\infty}&=\sup_{1\leq i\leq N}\sum_{j=1}^N\sup_{t}\left|(D_t'B_tD_t)_{ij}\right|\\
&\leq\sup_{1\leq i\leq N}\sum_{j=1}^N\sup_{t}\sum_{k=1}^N\left|(D_t'B_{1t}D_t)_{ik}\right|\left|(D_t'B_{2t}D_t)_{kj}\right|\\
&\leq\sup_{1\leq i\leq N}\sum_{j=1}^N\sum_{k=1}^N\left(\sup_{t}\left|(D_t'B_{1t}D_t)_{ik}\right|\right)\left(\sup_{t}\left|(D_t'B_{2t}D_t)_{kj}\right|\right)\\
&\leq\|\sup_t \mathrm{abs}(D_t'B_{1t}D_t)\|_\infty\|\sup_t \mathrm{abs}(D_t'B_{2t}D_t)\|_\infty\\
&<\infty.
\end{align*}
Similarly, we can prove $\|\sup_t \mathrm{abs}(D_t'B_t^\ell D_{t-1})\|_{\infty}<\infty$.

\textbf{Proof of (2):}

Since we assume that $B_{1t}$ and $B_{2t}$,  are UB uniformly for $t$, $B_t$ is UB because $\|B_t\|_p\leq\|B_{1t}\|_p\|B_{2t}\|_p$ holds for both $p=1$ and $p=\infty$.
Moreover, $D_t'B_tD_t$ is also UB uniformly for $t$, because $D_t'B_tD_t$ only adds some zero rows and columns to $B_t$, which means that there exists a constant $c_B>0$, such that
$$\sup_{N,T}\sup_{t\leq T}\max\left(\left\|\sum_{h=t}^\infty \mathrm{abs}\left(D_h'P_h^{(h-t)}\right)\right\|_{1},\left\|\sum_{h=t}^\infty \mathrm{abs}\left(D_h'P_h^{(h-t)}\right)\right\|_{\infty},\left\|D_t'B_tD_t\right\|_{1},\left\|D_t'B_tD_t\right\|_{\infty}\right)\leq c_B.$$
Thus, we have
\begin{align*}
\left\|\sum_{h=t}^\infty \mathrm{abs}\left(D_h'B_hP_h^{(h-t)}\right)\right\|_{1}&\leq\left\|\sum_{h=t}^\infty \mathrm{abs}(D_h'B_hD_h)\mathrm{abs}\left(D_h'P_h^{(h-t)}\right)\right\|_{1}\\
&=\sup_{1\leq j\leq N}\sum_{i=1}^N\sum_{h=t}^\infty\sum_{k=1}^N\left|\left(D_h'B_hD_h\right)_{ik}\right|\left|\left(D_h'P_h^{(h-t)}\right)_{kj}\right|\\
&\leq\sup_{1\leq j\leq N}\sum_{h=t}^\infty\sum_{k=1}^N\left\|D_h'B_hD_h\right\|_1\left|\left(D_h'P_h^{(h-t)}\right)_{kj}\right|\\
&\leq c_{B}\sup_{1\leq j\leq N}\sum_{k=1}^N\sum_{h=t}^\infty\left|\left(D_h'P_h^{(h-t)}\right)_{kj}\right|\\
&=c_B\left\|\sum_{h=t}^\infty \mathrm{abs}\left(D_h'P_h^{(h-t)}\right)\right\|_{1}\\
&<\infty,
\end{align*}
and
\begin{align*}
\left\|\sum_{h=t}^\infty \mathrm{abs}\left(D_h'B_hP_h^{(h-t)}\right)\right\|_{\infty}&\leq\left\|\sum_{h=t}^\infty \mathrm{abs}(D_h'B_hD_h)\mathrm{abs}\left(D_h'P_h^{(h-t)}\right)\right\|_{\infty}\\
&=\sup_{1\leq i\leq N}\sum_{j=1}^{N_t}\sum_{h=t}^\infty\sum_{k=1}^N\left|\left(D_h'B_hD_h\right)_{ik}\right|\left|\left(D_h'P_h^{(h-t)}\right)_{kj}\right|\\
&\leq\sup_{1\leq i\leq N}\sum_{h=t}^\infty\sum_{k=1}^N\sup_{h}\left|\left(D_h'B_hD_h\right)_{ik}\right|\sum_{j=1}^{N_t}\left|\left(D_h'P_h^{(h-t)}\right)_{kj}\right|\\
&\leq\sup_{1\leq i\leq N}\sum_{k=1}^N\sup_{h}\left|\left(D_h'B_hD_h\right)_{ik}\right|\sum_{j=1}^{N_t}\sum_{h=t}^\infty\left|\left(D_h'P_h^{(h-t)}\right)_{kj}\right|\\
&\leq\|\sup_t \mathrm{abs}(D_t'B_{1t}D_t)\|_{\infty}\left\|\sum_{h=t}^\infty \mathrm{abs}\left(D_h'P_h^{(h-t)}\right)\right\|_{\infty}\\
&<\infty.
\end{align*}
Therefore, $\sum_{h=t}^\infty \mathrm{abs}\left(D_h'B_hP_h^{(h-t)}\right)$ is UB uniformly for $t$. Applying the same method, we can prove that $\sum_{h=t}^\infty \mathrm{abs}\left(B_t^\ell P_{t-1}^{(h)}D_{t-1-h}\right)$ is UB uniformly for $t$.

\textbf{Proof of (3):}

According to (2), it is obvious that $\sum_{h=t}^\infty \mathrm{abs}\left(D_h'B_{1h}P_h^{(h-t)}\right)$ and $\sum_{h=t}^\infty \mathrm{abs}\left(D_h'B_{2h}Q_h^{(h-t)}\right)$ are also UB uniformly for $t$. Hence,
\begin{align*}
\left\|\sum_{h=t}^\infty \mathrm{abs}\left(Q_h^{(h-t)\prime}B_{2h}'B_{1h}P_h^{(h-t)}\right)\right\|_p&\leq\left\|\left[\sum_{h=t}^\infty \mathrm{abs}\left(Q_h^{(h-t)\prime}B_{2h}'D_h\right)\right]\left[\sum_{h=t}^\infty \mathrm{abs}\left(D_h'B_{1h}P_h^{(h-t)}\right)\right]\right\|_p\\
&\leq\left\|\sum_{h=t}^\infty \mathrm{abs}\left(Q_h^{(h-t)\prime}B_{2h}'D_h\right)\right\|_p\left\|\sum_{h=t}^\infty \mathrm{abs}\left(D_h'B_{1h}P_h^{(h-t)}\right)\right\|_p\\
&<\infty.
\end{align*}
The above inequality holds for both $p=1$ and $p=\infty$. Thus $\sum_{h=t}^\infty \mathrm{abs}\left(Q_h^{(h-t)\prime}B_{2h}'B_{1h}P_h^{(h-t)}\right)$ is also UB uniformly for $t$. Similarly, we can prove that $\sum_{h=0}^\infty\mathrm{abs}\left(B_{t}^\ell P_{t-1}^{(h)}P_{t-1}^{(h)\prime}B_{t}^{\ell\prime}\right)$ is UB uniformly for $t$.

\textbf{Proof of (4):}

Since $\sum_{h=0}^\infty\mathrm{abs}(P_t^{(h)}D_{t-h})$ and $\sum_{h=0}^\infty\mathrm{abs}(Q_t^{(h)}D_{t-h})$ are also UB uniformly for $t$, we similarly have
\begin{align*}
\left\|\sum_{h=t}^\infty \mathrm{abs}\left(P_t^{(h)}\right)\mathrm{abs}\left(Q_t^{(h)\prime}\right)\right\|_p&\leq\left\|\left[\sum_{h=t}^\infty\mathrm{abs}\left(P_t^{(h)}D_{t-h}\right)\right]\left[\sum_{h=t}^\infty \mathrm{abs}\left(Q_t^{(h)}D_{t-h}\right)'\right]\right\|_p\\
&\leq\left\|\sum_{h=0}^\infty\mathrm{abs}\left(P_t^{(h)}D_{t-h}\right)\right\|_p\left\|\sum_{h=0}^\infty \mathrm{abs}\left(Q_t^{(h)}D_{t-h}\right)'\right\|_p\\
&<\infty.
\end{align*}

\end{proof}

\begin{lemma}\label{lm_OnT}
Under Assumptions 2, 7 and \ref{asp_A2},
$$\mathrm{Var}\left(\sum_{t=1}^TL_t'R_t\right)=\sum_{t=1}^T\sum_{s=1}^T \mathrm{Cov}(L_t'R_t,L_s'R_s)=O(NT)$$
\end{lemma}
\begin{proof}
\begin{align}
\mathrm{Var}\left(\sum_{t=1}^TL_t'R_t\right)=&\sum_{t=1}^T\mathrm{Cov}(L_t'R_t,L_t'R_t)+2\sum_{t=2}^{T}\sum_{s=1}^{t-1}\mathrm{Cov}(L_t'R_t,L_s'R_s)\nonumber\\
\leq&2\sum_{t=1}^{T}\sum_{s=1}^{t}\mathrm{Cov}(L_t'R_t,L_s'R_s)\nonumber\\
=&2\sigma_0^4\sum_{t=1}^T\sum_{s=1}^t\mathrm{tr}\left[\left(\sum_{g=0}^\infty P_{sg}P_{t,t-s+g}'\right)\left(\sum_{g=0}^\infty Q_{t,t-s+g}Q_{sg}'\right)\right]+\nonumber\\
&2\sigma_0^4\sum_{t=1}^T\sum_{s=1}^t\mathrm{tr}\left[\left(\sum_{g=0}^\infty Q_{sg}P_{t,t-s+g}'\right)\left(\sum_{g=0}^\infty Q_{t,t-s+g}P_{sg}'\right)\right]\label{eq_Var}+\\
&2\sum_{g=0}^\infty\left[(\mu_4-3\sigma_0^4)\sum_{t=1}^T\sum_{s=1}^t\sum_{i=1}^{N_{s-g}}(P_{t,t-s+g}'Q_{t,t-s+g})_{ii}(P_{sg}'Q_{sg})_{ii}\right].\nonumber
\end{align}
The last equality holds from Lemma \ref{lm_CovPQ}.
Denote $\mathcal{A}_{t,PP}\equiv\sum_{g=0}^\infty P_t^{(g)} P_t^{(g)\prime}$ and $\mathcal{A}_{t,QQ}\equiv\sum_{g=0}^\infty Q_t^{(g)} Q_t^{(g)\prime}$, and they are UB from Lemma \ref{lm33}. We can derive the order of the first term of Equation (\ref{eq_Var}) as
\begin{align*}
&\left|2\sigma_0^4\sum_{t=1}^T\sum_{s=1}^t\mathrm{tr}\left[\left(\sum_{g=0}^\infty P_{sg}P_{t,t-s+g}'\right)\left(\sum_{g=0}^\infty Q_{t,t-s+g}Q_{sg}'\right)\right]\right|\\
=&\left|2\sigma_0^4\sum_{t=1}^{T}\sum_{s=1}^{t}\mathrm{tr}\left[B_{1s}\left(\sum_{g=0}^{\infty}P_{s}^{(g)}P_{s}^{(g)\prime}\right)P_{t}^{(t-s)\prime}B_{1t}^{\prime}B_{2t}Q_{t}^{(t-s)}\left(\sum_{g=0}^{\infty}Q_{s}^{(g)}Q_{s}^{(g)\prime}\right)B_{2s}^{\prime}\right]\right|\\
=&\left|2\sigma_0^4\sum_{t=1}^{T}\sum_{s=1}^{t}\mathrm{tr}\left[B_{1s}\mathcal{A}_{s,PP}P_{t}^{\left(t-s\right)\prime}B_{1t}^{\prime}B_{2t}Q_{t}^{\left(t-s\right)}\mathcal{A}_{s,QQ}B_{2s}^{\prime}\right]\right|\\
=&\left|2\sigma_0^4\sum_{s=1}^{T}\sum_{t=s}^{T}\mathrm{tr}\left[P_{t}^{\left(t-s\right)\prime}B_{1t}^{\prime}B_{2t}Q_{t}^{\left(t-s\right)}\mathcal{A}_{s,QQ}B_{2s}^{\prime}B_{1s}\mathcal{A}_{s,PP}\right]\right|\\
\leq&2\sigma_0^4\sum_{s=1}^{T}\mathrm{tr}\left[\sum_{t=s}^{T}\mathrm{abs}\left(P_{t}^{\left(t-s\right)\prime}B_{1t}^{\prime}B_{2t}Q_{t}^{\left(t-s\right)}\right)\mathrm{abs}\left(\mathcal{A}_{s,QQ}B_{2s}^{\prime}B_{1s}\mathcal{A}_{s,PP}\right)\right]\\
\leq&2\sigma_0^4\sum_{s=1}^{T}\mathrm{tr}\left[\sum_{t=s}^{\infty}\mathrm{abs}\left(P_{t}^{\left(t-s\right)\prime}B_{1t}^{\prime}B_{2t}Q_{t}^{\left(t-s\right)}\right)\mathrm{abs}\left(\mathcal{A}_{s,QQ}B_{2s}^{\prime}B_{1s}\mathcal{A}_{s,PP}\right)\right]\\
=&O(NT).
\end{align*}
The first equality holds because $P_{th}=B_{1t}P_t^{(h)}$ and $P_t^{(t-s+g)}=P_t^{(t-s)}P_s^{(g)}$.
By (3) in Lemma \ref{lm33}, we know that $\sum_{t=s}^{\infty}\mathrm{abs}\left(P_{t}^{\left(t-s\right)\prime}B_{1t}^{\prime}B_{2t}Q_{t}^{\left(t-s\right)}\right)$ is UB, then the term inside the trace operator of the second-to-last equation is an $N_s \times N_s$ UB matrix, so the last equality holds. Since $P_{th}$ and $Q_{th}$ are arbitrary matrices, under the same assumptions, we can similarly derive the order of the second term in equation (\ref{eq_Var}) as
$$2\sigma_0^4\sum_{t=1}^T\sum_{s=1}^t\mathrm{tr}\left[\left(\sum_{g=0}^\infty Q_{sg}P_{t,t-s+g}'\right)\left(\sum_{g=0}^\infty Q_{t,t-s+g}P_{sg}'\right)\right]=O(NT).$$
Since $\sum_{g=0}^\infty\mathrm{abs}(P_{sg}'Q_{sg})$ is UB by Lemma \ref{lm33}, there exists a finite constant $c_{PQ}$ such that $$\sup_{N}\sup_{s}\sup_{i=1,\cdots,N_s}\sum_{g=0}^\infty\left|(P_{sg}'Q_{sg})_{ii}\right|\leq c_{PQ},$$
therefore we have
\begin{align*}
&\left|\sum_{g=0}^\infty\sum_{t=1}^T\sum_{s=1}^t\sum_{i=1}^{N_{s-g}}(P_{t,t-s+g}'Q_{t,t-s+g})_{ii}(P_{sg}'Q_{sg})_{ii}\right|\\
\leq&c_{PQ}\sum_{t=1}^T\sum_{s=1}^t\sum_{g=0}^\infty\sum_{i=1}^{N_{s-g}}\left|(P_{t,t-s+g}'Q_{t,t-s+g})_{ii}\right|\\
\leq&c_{PQ}\sum_{t=1}^T\sum_{s=1}^t\sum_{g=0}^\infty\mathrm{tr}\left[\mathrm{abs}\left(P_{s}^{(g)\prime}P_{t}^{(t-s)\prime}B_{1t}^{\prime}B_{2t}Q_{t}^{(t-s)}Q_{s}^{(g)}\right)\right]\\
\leq&c_{PQ}\sum_{s=1}^T\sum_{t=s}^T\mathrm{tr}\left[\sum_{g=0}^\infty\mathrm{abs}\left(Q_{s}^{(g)}\right)\mathrm{abs}\left(P_{s}^{(g)\prime}\right)\mathrm{abs}\left(P_{t}^{(t-s)\prime}B_{1t}^{\prime}B_{2t}Q_{t}^{(t-s)}\right)\right]\\
\leq&c_{PQ}\sum_{s=1}^T\mathrm{tr}\left[\sum_{g=0}^\infty\mathrm{abs}\left(Q_{s}^{(g)}\right)\mathrm{abs}\left(P_{s}^{(g)\prime}\right)\sum_{t=s}^\infty\mathrm{abs}\left(P_{t}^{(t-s)\prime}B_{1t}^{\prime}B_{2t}Q_{t}^{(t-s)}\right)\right]\\
=&O(NT).
\end{align*}
The last term in equation (\ref{eq_Var}) is also $O(NT)$, so $\mathrm{Var}\left(\sum_{t=1}^TL_t'R_t\right)=O(NT)$.
\end{proof}

\begin{lemma}\label{lm_ddot}
$\mathbb{L'PR}=\mathbb{L'D(D'D)}^{-1}\mathbb{D'R}=(\sum_{h=0}^\infty\ddot{P}_hV_{T-h})'(\sum_{h=0}^\infty\ddot{Q}_hV_{T-h})$, where 
\begin{align*}
\ddot{P}_h=\begin{cases} 
\mathcal{T}^{-\frac{1}{2}}\sum_{t=T-h}^TD_t'P_{t,t+h-T} & \textrm{for } h<T \\
\mathcal{T}^{-\frac{1}{2}}\sum_{t=1}^TD_t'P_{t,t+h-T} & \text{for } h\geq T
\end{cases},
\end{align*}
$\ddot{Q}_h$ has the same pattern, and $\mathcal{T}\equiv \mathrm{diag}(T_1,T_2,\cdots,T_N)$.
\end{lemma}
\begin{proof}
According to the definitions of $\mathbb{L}$, $\mathbb{R}$ and $\mathbb{D}$, we have
$$\mathbb{D'D}=\sum_{t=1}^TD_t'D_t=\sum_{t=1}^T\mathrm{diag}(d_{1t},d_{2t},\cdots,d_{N,t})=\mathrm{diag}(T_1,T_2,\cdots,T_N)=\mathcal{T}.$$
Then
$$\mathbb{L'D(D'D)}^{-1}\mathbb{D'R}=\left(\mathcal{T}^{-\frac{1}{2}}\sum_{t=1}^TD_t'L_t\right)'\left(\mathcal{T}^{-\frac{1}{2}}\sum_{t=1}^TD_t'R_t\right).$$
Since
\begin{align*}
L_{1}=&P_{10}V_{1}+P_{11}V_{0}+P_{12}V_{-1}+\cdots,\\
L_{2}=P_{20}V_{2}+&P_{21}V_{1}+P_{22}V_{0}+P_{23}V_{-1}+\cdots,\\
&\vdots\\
L_{T}=P_{T0}V_{T}\cdots+&P_{T,T-1}V_{1}+P_{T,T}V_{0}+P_{T,T+1}V_{-1}+\cdots,
\end{align*}
we have
\begin{align*}
\sum_{t=1}^TD_t'L_{t}=&D_T'P_{T0}V_{T}+(D_{T-1}'P_{T-1,0}+D_{T}'P_{T1})V_{T-1}
+\cdots\\
&+(D_1'P_{11}+D_2'P_{22}+\cdots+D_T'P_{TT})V_0+(D_1'P_{12}+D_2'P_{23}+\cdots+D_T'P_{T,T+1})V_{-1}+\cdots.
\end{align*}
Therefore
$$\mathcal{T}^{-\frac{1}{2}}\sum_{t=1}^TD_t'L_t=\mathcal{T}^{-\frac{1}{2}}\sum_{t=1}^TD_t'\sum_{h=0}^\infty P_{th}V_{t-h}=\sum_{h=0}^\infty\ddot{P}_hV_{T-h},$$
where
\begin{align*}
\ddot{P}_h=\begin{cases} 
\mathcal{T}^{-\frac{1}{2}}\sum_{t=T-h}^TD_t'P_{t,t+h-T} & \textrm{for } h<T \\
\mathcal{T}^{-\frac{1}{2}}\sum_{t=1}^TD_t'P_{t,t+h-T} & \text{for } h\geq T
\end{cases}.
\end{align*}
Similarly, $\ddot{Q}_h$ has the same pattern.
\end{proof}

\begin{lemma}\label{lm_PPQ}
Under Assumptions 1, 2, 7 and \ref{asp_A2},
\begin{gather}
\frac{1}{NT}\mathbb{L}'\mathbb{R}-\frac{1}{NT}E\left(\mathbb{L}'\mathbb{R}\right)=O_p\left(\frac{1}{\sqrt{NT}}\right),\label{eq_lr}\\
\frac{1}{NT}\mathbb{L'}\mathbb{PR}-\frac{1}{NT}E\left(\mathbb{L'}\mathbb{PR}\right)=O_p\left(\frac{1}{\sqrt{NT^2}}\right),
\label{eq_lpr}\\
\frac{1}{NT}\mathbb{L'QR}-\frac{1}{NT}E\left(\mathbb{L'QR}\right)=O_p\left(\frac{1}{\sqrt{NT}}\right),\label{eq_lqr}
\end{gather}
where $\frac{1}{NT}E\left(\mathbb{L}'\mathbb{R}\right)=O(1)$ and $\frac{1}{NT}E\left(\mathbb{L'}\mathbb{PR}\right)=O(\frac{1}{T})$.
\end{lemma}
\begin{proof}

\textbf{Proof of equation (\ref{eq_lr})}

According to Lemmas \ref{lm_EPQ} and \ref{lm33}, we have $$\frac{1}{NT}E\left(\mathbb{L}'\mathbb{R}\right)=\frac{1}{NT}E\left(\sum_{t=1}^TL_t'R_t\right)=\frac{1}{NT}\sigma_0^2\sum_{t=1}^T\mathrm{tr}\left(\sum_{h=0}^\infty P_{th}Q_{th}'\right)=O(1)$$
and according to Lemma \ref{lm_OnT}, $\mathrm{Var}(\mathbb{L}'\mathbb{R})=\mathrm{Var}(\sum_{t=1}^TL_t'R_t)=O(NT)$. Using Chebyshev's inequality, we can prove equation (\ref{eq_lr}).

\textbf{Proof of equation (\ref{eq_lpr})}

According to Lemmas \ref{lm_ddot} and \ref{lm_CovPQ}, we have
\begin{align}
\mathrm{Var}\left(\mathbb{L'}\mathbb{PR}\right)=&\mathrm{Cov}\left[\left(\sum_{h=0}^\infty\ddot{P}_hV_{T-h}\right)'\left(\sum_{h=0}^\infty\ddot{Q}_hV_{T-h}\right),\left(\sum_{h=0}^\infty\ddot{P}_hV_{T-h}\right)'\left(\sum_{h=0}^\infty\ddot{Q}_hV_{T-h}\right)\right]\nonumber\\
=&\sigma_0^4\mathrm{tr}\left[\left(\sum_{g=0}^\infty \ddot{Q}_{g}\ddot{P}_{g}'\right)\left(\sum_{g=0}^\infty \ddot{Q}_{g}\ddot{P}_{g}'\right)+\left(\sum_{g=0}^\infty \ddot{P}_{g}\ddot{P}_{g}'\right)\left(\sum_{g=0}^\infty \ddot{Q}_{g}\ddot{Q}_{g}'\right)\right]+\nonumber\\
&\sum_{g=0}^\infty\left[(\mu_4-3\sigma_0^4)\sum_{i=1}^{N_{T-g}}\left(\ddot{P}_{g}'\ddot{Q}_{g}\right)_{ii}\left(\ddot{P}_{g}'\ddot{Q}_{g}\right)_{ii}\right],\label{eq_VLPR}
\end{align}
where $\ddot{P}_g$ and $\ddot{Q}_g$ are defined as in Lemma \ref{lm_ddot}. $\sum_{g=0}^\infty \ddot{Q}_{g}\ddot{P}_{g}'=\sum_{g=0}^{T-1} \ddot{Q}_{g}\ddot{P}_{g}'+\sum_{g=T}^\infty \ddot{Q}_{g}\ddot{P}_{g}'$, $p=1$ and $p=\infty$,
\begin{align*}
\left\|\sum_{g=0}^{T-1} \ddot{Q}_{g}\ddot{P}_{g}'\right\|_p&=\left\|\mathcal{T}^{-\frac{1}{2}}\sum_{g=0}^{T-1} \left(\sum_{t=T-g}^TD_t'Q_{t,t+g-T}\right)\left(\sum_{t=T-g}^TD_t'P_{t,t+g-T}\right)'\mathcal{T}^{-\frac{1}{2}}\right\|_p\\
&\leq\|\mathcal{T}^{-\frac{1}{2}}\|_p^2\sum_{g=0}^{T-1} \left\|\sum_{t=T-g}^TD_t'Q_{t,t+g-T}\right\|_p\left\|\left(\sum_{t=T-g}^TD_t'P_{t,t+g-T}\right)'\right\|_p\\
&\leq\frac{1}{c_T T}\sum_{g=0}^{T-1} \left\|\sum_{t=T-g}^TD_t'B_{2t}Q_t^{(t+g-T)}\right\|_p\left\|\left(\sum_{t=T-g}^TD_t'B_{1t}P_t^{(t+g-T)}\right)'\right\|_p\\
&\leq\frac{1}{c_T T}\sum_{g=0}^{T-1} \left\|\sum_{t=T-g}^\infty \mathrm{abs}\left(D_t'B_{2t}Q_t^{(t+g-T)}\right)\right\|_p\left\|\sum_{t=T-g}^\infty \mathrm{abs}\left(D_t'B_{1t}P_t^{(t+g-T)}\right)'\right\|_p\\
&=O(1),
\end{align*}
where the second inequality holds because $T_i\geq c_T T$ in Assumption 1 and the last equality follows from Lemma \ref{lm33}. Likewise, for $p=1$ and $p=\infty$,
\begin{align*}
\left\|\sum_{g=T}^{\infty} \ddot{Q}_{g}\ddot{P}_{g}'\right\|_p&=\left\|\mathcal{T}^{-\frac{1}{2}}\sum_{g=T}^{\infty} \left(\sum_{t=1}^TD_t'Q_{t,t+g-T}\right)\left(\sum_{t=1}^TD_t'P_{t,t+g-T}\right)'\mathcal{T}^{-\frac{1}{2}}\right\|_p\\
&=\left\|\mathcal{T}^{-\frac{1}{2}}\sum_{g=T}^{\infty} \left(\sum_{t=1}^TD_t'B_{2t}Q_t^{(t)}\right)Q_0^{(g-T)}P_0^{(g-T)\prime}\left(\sum_{t=1}^TD_t'B_{1t}P_t^{(t)}\right)'\mathcal{T}^{-\frac{1}{2}}\right\|_p\\
&\leq\|\mathcal{T}^{-\frac{1}{2}}\|_p^2\left\|\sum_{t=1}^TD_t'B_{2t}Q_t^{(t)}\right\|_p\left\|\sum_{g=T}^{\infty}Q_0^{(g-T)}P_0^{(g-T)\prime}\right\|_p\left\|\left(\sum_{t=1}^TD_t'B_{1t}P_t^{(t)}\right)'\right\|_p\\
&=O\left(\frac{1}{T}\right),
\end{align*}
where the second equality holds because $P_{t,t+g-T}=B_{1t}P_{t}^{(t)}P_{0}^{(g-T)}$ and $Q_{t,t+g-T}=B_{2t}Q_{t}^{(t)}Q_{0}^{(g-T)}$. The last quality holds because of Assumption 1 and Lemma \ref{lm33}. Therefore,
$$\left\|\sum_{g=0}^\infty \ddot{Q}_{g}\ddot{P}_{g}'\right\|_p\leq\left\|\sum_{g=0}^{T-1} \ddot{Q}_{g}\ddot{P}_{g}'\right\|_p+\left\|\sum_{g=T}^\infty \ddot{Q}_{g}\ddot{P}_{g}'\right\|_p=O(1)+O\left(\frac{1}{T}\right)=O(1).$$
Since $P_{th}$ and $Q_{th}$ are arbitrary matrices, under the same assumptions, similarly we have $\left\|\sum_{g=0}^\infty \ddot{P}_{g}\ddot{P}_{g}'\right\|_p=O(1)$ and $\left\|\sum_{g=0}^\infty \ddot{Q}_{g}\ddot{Q}_{g}'\right\|_p=O(1)$ for $p=1$ and $p=\infty$. Then
\begin{align*}
&\left\|\left(\sum_{g=0}^\infty \ddot{Q}_{g}\ddot{P}_{g}'\right)\left(\sum_{g=0}^\infty \ddot{Q}_{g}\ddot{P}_{g}'\right)+\left(\sum_{g=0}^\infty \ddot{P}_{g}\ddot{P}_{g}'\right)\left(\sum_{g=0}^\infty \ddot{Q}_{g}\ddot{Q}_{g}'\right)\right\|_p\\
\leq&\left\|\sum_{g=0}^\infty \ddot{Q}_{g}\ddot{P}_{g}'\right\|_p^2+\left\|\sum_{g=0}^\infty \ddot{P}_{g}\ddot{P}_{g}'\right\|_p\left\|\sum_{g=0}^\infty \ddot{Q}_{g}\ddot{Q}_{g}'\right\|_p\\
=&O(1),
\end{align*}
for $p=1$ and $p=\infty$. So the order of the first term in equation (\ref{eq_VLPR}) can be derived as
$$\sigma_0^4\mathrm{tr}\left[\left(\sum_{g=0}^\infty \ddot{Q}_{g}\ddot{P}_{g}'\right)\left(\sum_{g=0}^\infty \ddot{Q}_{g}\ddot{P}_{g}'\right)+\left(\sum_{g=0}^\infty \ddot{P}_{g}\ddot{P}_{g}'\right)\left(\sum_{g=0}^\infty \ddot{Q}_{g}'\ddot{Q}_{g}'\right)\right]=O(N).$$
Furthermore, 
\begin{align*}
\left|\sum_{g=0}^\infty\sum_{i=1}^{N_{T-g}}\left(\ddot{P}_{g}'\ddot{Q}_{g}\right)_{ii}\left(\ddot{P}_{g}'\ddot{Q}_{g}\right)_{ii}\right|&\leq\sup_{g,i}\left|\left(\ddot{P}_{g}'\ddot{Q}_{g}\right)_{ii}\right|\left(\sum_{i=1}^{N_{T-g}}\left|\sum_{g=0}^\infty\left(D_g'\ddot{P}_{g}'\ddot{Q}_{g}D_g\right)_{ii}\right|\right)\\
&\leq\sup_{g}\left\|D_g'\ddot{P}_{g}'\ddot{Q}_{g}D_g\right\|_1\mathrm{tr}\left[\mathrm{abs}\left(\sum_{g=0}^\infty D_g'\ddot{P}_{g}'\ddot{Q}_{g}D_g\right)\right]\\
&=O(N),
\end{align*}
where the last equality holds because both $\|\sum_{g=0}^\infty D_g'\ddot{P}_{g}'\ddot{Q}_{g}D_g\|_1$ and $\|\sum_{g=0}^\infty D_g'\ddot{P}_{g}'\ddot{Q}_{g}D_g\|_\infty$ are $O(1)$, so the $\sup$ in the expression is $O(1)$, and the trace is $O(N)$.
Thus, the second term in equation \ref{eq_VLPR} is also $O(N)$. Therefore, $\mathrm{Var}(\mathbb{L'}\mathbb{PR})=O(N)$, then
$$\frac{1}{NT}\mathbb{L'}\mathbb{PR}-\frac{1}{NT}E\left(\mathbb{L'}\mathbb{PR}\right)=O_p\left(\frac{1}{\sqrt{NT^2}}\right)$$
according to Chebyshev's inequality. Also, according to Lemma \ref{lm_EPQ},
\begin{align*}
\frac{1}{NT}E\left(\mathbb{L'}\mathbb{PR}\right)=\frac{1}{NT}\sigma_0^2\mathrm{tr}\left(\sum_{h=0}^\infty\ddot{P}_{h}\ddot{Q}_{h}'\right)=O\left(\frac{1}{T}\right).
\end{align*}

\textbf{Proof of equation (\ref{eq_lqr})}

Since $\mathbb{Q}=I_n-\mathbb{P}$, combining equations (\ref{eq_lr}) and (\ref{eq_lpr}), the results follows.

\end{proof}

\begin{lemma}\label{lm_CQL}
Under Assumptions 2, 7, \ref{asp_A2} and \ref{asp_C}, we have

\begin{gather}
\frac{1}{NT}\mathbb{C'L}=O_p\left(\frac{1}{\sqrt{NT}}\right),\label{eq_cl}\\
\frac{1}{NT}\mathbb{C'}\mathbb{Q}\mathbb{L}=O_p\left(\frac{1}{\sqrt{NT}}\right),\label{eq_cql}
\end{gather}
\begin{gather}
\frac{1}{NT}\mathbb{C'V}=O_p\left(\frac{1}{\sqrt{NT}}\right),\label{eq_cv}\\
\frac{1}{NT}\mathbb{C'}\mathbb{QB}_1\mathbb{V}=O_p\left(\frac{1}{\sqrt{NT}}\right),\label{eq_cqv}
\end{gather}
where $E(\mathbb{L})=0$ and $E(\mathbb{V})=0$.
\end{lemma}
\begin{proof}

\textbf{Proof of equations (\ref{eq_cl}) and (\ref{eq_cql})}

Since $E(L_t)=0$ and $C_t$, $B_t$, $D_t$ are nonstochastic, $E(\mathbb{L})=E(\mathbb{C'L})=E(\mathbb{C'}\mathbb{Q}\mathbb{L})=0$. Additionally, 
\begin{align*}
\mathrm{Var}\left(\mathbb{C'L}\right)&=\mathrm{Var}\left(\sum_{t=1}^TC_t'L_t\right)=\sum_{t=1}^T\sum_{s=1}^T\mathrm{Cov}\left(C_t'L_t,C_s'L_s\right)\leq\sum_{t=1}^T\sum_{s=1}^T\left|C_t'E\left(L_tL_s'\right)C_s\right|\\
&\leq2\sum_{t=1}^T\sum_{s=1}^t\left|C_t'E\left(L_tL_s'\right)C_s\right|=2\sigma_0^2\sum_{t=1}^T\sum_{s=1}^t\left|C_t'\left(\sum_{g=0}^\infty P_{t,t-s+g}P_{sg}'\right)C_s\right|,
\end{align*}
where the last equality follows from Lemma \ref{lm_EPQ}. According to Assumption \ref{asp_A2}, $\{B_{1t}\}$ is UB uniformly for $t$ so there exists a constant $c_B>0$ such that
$$\sup_{N,T}\sup_{t\leq T}\max\left(\left\|B_t\right\|_{1},\left\|B_t\right\|_{\infty}\right)\leq c_B.$$
Moreover, by Assumption \ref{asp_C}, elements in $\mathbb{C}$ are uniformly bounded, and the same holds for $\mathbb{QC}$ as $\mathbb{Q}$ is the projection matrix. Thus, there exists a constant $c_C>0$ such that
$$\sup_N\sup_{T}\sup_{1\leq i\leq n}\max\left(\left|(\mathbb{C})_i\right|,\left|(\mathbb{QC})_i\right|\right)\leq c_C.$$
Hence, for any t and $1\leq i\leq N_t$, we have
$$\left|\left(B_{1t}'C_t\right)_i\right|=\left|\sum_{j=1}^{N_t}\left(B_{1t}'\right)_{ij}(C_t)_j\right|\leq c_Bc_C,$$
therefore,
\begin{align*}
\left|C_t'\left(\sum_{g=0}^\infty P_{t,t-s+g}P_{sg}'\right)C_s\right|&=\left|C_t'B_{1t}P_t^{(t-s)}\sum_{g=0}^\infty\left(P_s^{(g)}P_s^{(g)\prime}\right)B_{1s}'C_s\right|
\\
&\leq\sum_{i=1}^{N_t}\sum_{j=1}^{N_s}\left|\left(B_{1t}'C_t\right)_i\left(P_t^{(t-s)}\sum_{g=0}^\infty P_s^{(g)}P_s^{(g)\prime}\right)_{ij}\left(B_{1s}'C_s\right)_j\right|\\
&\leq c_B^2c_C^2\sum_{i=1}^{N_t}\sum_{j=1}^{N_s}\left|\left(P_t^{(t-s)}\sum_{g=0}^\infty P_s^{(g)}P_s^{(g)\prime}\right)_{ij}\right|,
\end{align*}
where the first equation holds by $P_{th}=B_{1t}P_t^{(h)}$ and $P_t^{(t-s+g)}=P_t^{(t-s)}P_s^{(g)}$. So
\begin{align*}
\mathrm{Var}\left(\mathbb{C'L}\right)&\leq2c_B^2c_C^2\sigma_0^2\sum_{t=1}^T\sum_{s=1}^t\sum_{i=1}^{N_t}\sum_{j=1}^{N_s}\left|\left(P_t^{(t-s)}\sum_{g=0}^\infty P_s^{(g)}P_s^{(g)\prime}\right)_{ij}\right|\\
&=2c_B^2c_C^2\sigma_0^2\sum_{s=1}^T\sum_{t=s}^T\sum_{i=1}^{N}\sum_{j=1}^{N_s}\left|\left(D_t'P_t^{(t-s)}\sum_{g=0}^\infty P_s^{(g)}P_s^{(g)\prime}\right)_{ij}\right|\\
&\leq2c_B^2c_C^2\sigma_0^2\sum_{s=1}^T\sum_{i=1}^{N_t}\sum_{j=1}^{N_s}\left|\left[\sum_{t=s}^\infty \mathrm{abs}\left(D_t'P_t^{(t-s)}\right)\sum_{g=0}^\infty \mathrm{abs}\left(P_s^{(g)}P_s^{(g)\prime}\right)\right]_{ij}\right|\\
&=O(NT),
\end{align*}
where the last equality holds because $\sum_{t=s}^\infty \mathrm{abs}\left(D_t'P_t^{(t-s)}\right)\sum_{g=0}^\infty \mathrm{abs}\left(P_s^{(g)}P_s^{(g)\prime}\right)$ is UB by Assumption \ref{asp_A2} and Lemma \ref{lm33}. Then we can obtain equation (\ref{eq_cl}) by Chebyshev's inequality. By the same reasoning, we can obtain the same order for $\mathrm{Var}\left(\mathbb{C'QL}\right)$ by replacing $\mathbb{C}$ with $\mathbb{QC}$ from the previous analysis. Therefore, equation (\ref{eq_cql}) follows.

\textbf{Proof of equations (\ref{eq_cv}) and (\ref{eq_cqv})}

Since $E(V_t)=0$, $E(\mathbb{C'V})=E(\mathbb{C'QB}_1\mathbb{V})=0$. Because the error terms $v_{it}$ are i.i.d. across $i$ and $t$, we have 
\begin{align*}
\mathrm{Var}(\mathbb{C'V})=\mathrm{Var}(\sum_{t=1}^TC_t'V_t)=\sum_{t=1}^T\mathrm{Var}(C_t'V_t)=\sigma_0^2\sum_{t=1}^TC_t'C_t=O(NT),
\end{align*}
where the last equality holds because elements in $C_t$ are uniformly bounded. Then we can obtain equation (\ref{eq_cv}) by Chebyshev's inequality. Since $\{B_{1t}\}$ is UB uniformly for $t$, elements in $\mathbb{B}_1'\mathbb{QC}$ are also uniformly bounded, we therefore have $\mathrm{Var}\left(\mathbb{C'QB}_1\mathbb{V}\right)=O(NT)$ by the same reasoning. Therefore, equation (\ref{eq_cqv}) follows.
\end{proof}

\begin{lemma}\label{lm_Z_gamma}
Under Assumptions 1, 2, 7, \ref{asp_A2}, we have
\begin{align}
\frac{\sqrt{T}}{NT}\mathbb{Z}^{(\gamma)\prime}\mathbb{B}_2'\mathbb{L}=O_p\left(\frac{1}{\sqrt{NT}}\right),\label{eq_ZgBL}\\
\frac{\sqrt{T}}{NT}\mathbb{Z}^{(\gamma)\prime}\mathbb{B}_2'\mathbb{QL}=O_p\left(\frac{1}{\sqrt{NT}}\right),\label{eq_ZgBQL}
\end{align}
and
\begin{align}
\frac{\sqrt{T}}{NT}\mathbb{Z}^{(\gamma)\prime}\mathbb{B}_2'\mathbb{B}_1\mathbb{V}=O_p\left(\frac{1}{\sqrt{NT}}\right),\label{eq_ZgBV}\\
\frac{\sqrt{T}}{NT}\mathbb{Z}^{(\gamma)\prime}\mathbb{B}_2'\mathbb{QB}_1\mathbb{V}=O_p\left(\frac{1}{\sqrt{NT}}\right).\label{eq_ZgBQV}
\end{align}
\end{lemma}
\begin{proof}

\textbf{Proof of equation (\ref{eq_ZgBL}) and (\ref{eq_ZgBQL})}

As $\mathbb{Z}^{(\gamma)}=((D_1F_1\iota_N)',\cdots,(D_TF_T\iota_N)')'$, we have
\begin{align*}
&\mathrm{Var}\left(\mathbb{Z}^{(\gamma)\prime}\mathbb{B}_2'\mathbb{L}\right)=\mathrm{Var}\left[\sum_{t=1}^T(D_tF_t\iota_N)'B_{2t}'L_t\right]\\
\leq&2\sum_{t=1}^T\sum_{s=1}^t\left|\mathrm{Cov}\left[(D_tF_t\iota_N)'B_{2t}'L_t,(D_sF_s\iota_N)'B_{2s}'L_s\right]\right|\\
\leq&2\iota_N'\sum_{t=1}^T\sum_{s=1}^t\mathrm{abs}\left[F_t'D_t'B_{2t}'E\left(L_t L_s^{\prime}\right)B_{2s}D_sF_s\right]\iota_N\\
=&2\sigma_0^2\iota_N'\sum_{s=1}^T\sum_{t=s}^T\mathrm{abs}\left[F_t'D_t'B_{2t}'B_{1t} \left(\sum_{g=0}^\infty P_{t}^{(t-s+g)}P_{s}^{(g)\prime}\right)B_{1s}^{\prime}
B_{2s}D_sF_s\right]\iota_N\\
\leq&2\sigma_0^2\iota_N'\sum_{s=1}^T\sum_{t=s}^T\mathrm{abs}\left(F_t'D_t'B_{2t}'B_{1t}P_{t}^{(t-s)}\right)\mathrm{abs}\left[\left(\sum_{g=0}^\infty P_{s}^{(g)}P_{s}^{(g)\prime}\right)B_{1s}^{\prime}
B_{2s}D_s\right]F_s\iota_N\\
\leq&2\sigma_0^2\sum_{s=1}^T\left\{\iota_N'\left[\sum_{t=s}^\infty\mathrm{abs}\left(D_t'B_{2t}'B_{1t}P_{t}^{(t-s)}\right)\right]\mathrm{abs}\left[\left(\sum_{g=0}^\infty P_{s}^{(g)}P_{s}^{(g)\prime}\right)B_{1s}^{\prime}
B_{2s}D_s\right]\right\}F_s\iota_N\\
=&O(N),
\end{align*}
where the second equality follows from Lemma \ref{lm_EPQ}, and the third inequality is due to $P_{t}^{(t-s+g)}=P_{t}^{(t-s)}P_{s}^{(g)}$. 
The final inequality holds because $F_t$ is a diagonal matrix with diagonal entries equal to either 0 or 1. Furthermore, the last equality holds because $\sum_{t=s}^\infty\mathrm{abs}\left(D_t'B_{2t}'B_{1t}P_{t}^{(t-s)}\right)$ and $\mathrm{abs}\left(\sum_{g=0}^\infty P_{s}^{(g)}P_{s}^{(g)\prime}\right)$ are UB by Lemma \ref{lm33}, then each element of the vector inside the braces is $O(1)$. Therefore, the entire expression is $O(N)$ because $F_s\iota_N$ is a sparse binary vector with fewer than $N$ nonzero entries in total over $s =1,\cdots,T$. Due to $E(\mathbb{Z}^{(\gamma)\prime}\mathbb{B}_2'\mathbb{L})=0$, by Chebyshev's inequality we have $\frac{\sqrt{T}}{NT}\mathbb{Z}^{(\gamma)\prime}\mathbb{B}_2'\mathbb{L}=O_p\left(\frac{1}{\sqrt{NT}}\right)$. Additionally,
\begin{align*}
&\mathrm{Var}(\mathbb{Z}^{(\gamma)\prime}\mathbb{B}_2'\mathbb{PL})\\
=&\mathrm{Var}\left[\left(\sum_{t=1}^TD_t'B_{2t}D_tF_t\iota_N\right)'\mathcal{T}^{-1}\left(\sum_{t=1}^TD_t'B_{1t}\sum_{h=0}^\infty P_{t}^{(h)}V_{t-h}\right)\right]\\
=&\mathrm{Var}\left[\left(\mathcal{T}^{-\frac{1}{2}}\sum_{t=1}^TD_t'B_{2t}D_tF_t\iota_N\right)'\left(\sum_{h=0}^\infty \ddot{P}_{h}V_{T-h}\right)\right]\\
=&\sigma_0^2\left(\mathcal{T}^{-\frac{1}{2}}\sum_{t=1}^TD_t'B_{2t}D_tF_t\iota_N\right)'\left(\sum_{h=0}^\infty\ddot{P}_{h}\ddot{P}_{h}'\right)\left(\mathcal{T}^{-\frac{1}{2}}\sum_{t=1}^TD_t'B_{2t}D_tF_t\iota_N\right)\\
\leq&\frac{\sigma_0^2}{c_TT}\left(\sup_t\mathrm{abs}(D_t'B_{2t}D_t)\sum_{t=1}^TF_t\iota_N\right)'\mathrm{abs}\left(\sum_{h=0}^\infty\ddot{P}_{h}\ddot{P}_{h}'\right)\left(\sup_t\mathrm{abs}(D_t'B_{2t}D_t)\sum_{t=1}^TF_t\iota_N\right)\\
\leq&\sigma_0^2\|\sup_t\mathrm{abs}(D_t'B_{2t}D_t)\|_\infty^2\left[\iota_N'\mathrm{abs}\left(\sum_{h=0}^\infty\ddot{P}_{h}\ddot{P}_{h}'\right)\iota_N\right]\\
=&O\left(\frac{N}{T}\right),
\end{align*}
where $\ddot{P}_h$ is defined in Lemma \ref{lm_ddot}. The first inequality holds because $T_i\geq c_TT$ by Assumption 1. The second inequality above holds due to the sparsity of $F_t\iota_N$, specifically, for each $i$, $(F_{t})_{ii}=d_{it}(1-d_{i,t-1})$ equals 1 at most once over $t = 1, \cdots, T$ (indicating the first appearance of the $i$-th individual) and is zero for other $t$. Therefore, for $i=1,\cdots,N$, we have
$$\left[\sup_t\mathrm{abs}(D_t'B_{2t}D_t)\sum_{t=1}^TF_t\iota_N\right]_i\leq\left[\sup_t\mathrm{abs}(D_t'B_{2t}D_t)\iota_N\right]_i\leq\|\sup_t\mathrm{abs}(D_t'B_{2t}D_t)\|_\infty.$$
In addition, the last equality holds because, similar to the corresponding term in equation (\ref{eq_VLPR}) in the proof of Lemma \ref{lm_PPQ}, $\mathrm{abs}\left(\sum_{h=0}^\infty\ddot{P}_{h}\ddot{P}_{h}'\right)$ is UB, then the value inside the brackets is $O(N)$. Hence, we have $\frac{\sqrt{T}}{NT}\mathbb{Z}^{(\gamma)\prime}\mathbb{B}_2'\mathbb{PL}=O_p\left(\frac{1}{\sqrt{NT^2}}\right)$ by Chebyshev's inequality. Combining with equation (\ref{eq_ZgBL}), we can prove (\ref{eq_ZgBQL}).

\textbf{Proof of equations (\ref{eq_ZgBV}) and (\ref{eq_ZgBQV})}

Since $\mathbb{Z}^{(\gamma)}=\left((D_1F_1\iota_N)',\cdots,(D_TF_T\iota_N)'\right)'$ is a sparse binary vector with at most $N$ nonzero elements and $\mathbb{B}_2'\mathbb{B}_1\mathbb{B}_1'\mathbb{B}_2$ is UB, $\mathrm{Var}(\mathbb{Z}^{(\gamma)\prime}\mathbb{B}_2'\mathbb{B}_1\mathbb{V})=\sigma_0^2\mathbb{Z}^{(\gamma)\prime}\mathbb{B}_2'\mathbb{B}_1\mathbb{B}_1'\mathbb{B}_2\mathbb{Z}^{(\gamma)}=O(N)$. Then $\frac{\sqrt{T}}{NT}\mathbb{Z}^{(\gamma)\prime}\mathbb{B}_2'\mathbb{B}_1\mathbb{V}=O_p\left(\frac{1}{\sqrt{NT}}\right)$ by Chebyshev's inequality. Moreover, since $D_t$ is an $N_t\times N$ matrix obtained by removing the corresponding rows of unobserved nodes during period $t$ from $I_N$, we have
\begin{align*}
\|\mathbb{P}\|_{1}=\|\mathbb{P}\|_{\infty}&=\|\mathrm{abs}\left(\mathbb{D}\mathcal{T}^{-1}\mathbb{D}'\right)\iota_n\|_\infty=\|(D_1',\cdots,D_T')'\mathcal{T}^{-1}(D_1',\cdots,D_T')\iota_n\|_\infty\\
&=\|(D_1',\cdots,D_T')'\mathcal{T}^{-1}(T_1,\cdots,T_N)'\|_\infty=(D_1',\cdots,D_T')'\iota_N=\|\iota_n\|_\infty=1,
\end{align*}
where the third equality holds because all entries of the matrices $\{D_t\}$ and $\mathcal{T}^{-1}$ are all nonnegative. The above means that $\mathbb{P}$ is also UB. Thus,
$\mathrm{Var}(\mathbb{Z}^{(\gamma)\prime}\mathbb{B}_2'\mathbb{P}\mathbb{B}_1\mathbb{V})=\sigma_0^2\mathbb{Z}^{(\gamma)\prime}\mathbb{B}_2'\mathbb{PB}_1\mathbb{B}_1'\mathbb{PB}_2\mathbb{Z}^{(\gamma)}=O(N)$
because $\mathbb{B}_2'\mathbb{PB}_1\mathbb{B}_1'\mathbb{PB}_2$ is UB. By Chebyshev's inequality, we also have $\frac{\sqrt{T}}{NT}\mathbb{Z}^{(\gamma)\prime}\mathbb{B}_2'\mathbb{PB}_1\mathbb{V}=O_p\left(\frac{1}{\sqrt{NT}}\right)$, then equation (\ref{eq_ZgBQV}) follows.

\end{proof}

\begin{lemma}\label{lm_lag}
Lemmas \ref{lm_PPQ}-\ref{lm_Z_gamma} still hold by replacing $\mathbb{L}$ and $\mathbb{R}$ with $\mathbb{L}^\ell$ and $\mathbb{R}^\ell$ respectively.
\end{lemma}
\begin{proof}
By definition, we have $L_t^\ell=\sum_{h=0}^\infty B_{1t}^\ell B_{1,t-1}P_{t-1}^{(h)}V_{t-h-1}$ and $R_t^\ell=\sum_{h=0}^\infty B_{2t}^\ell B_{2,t-1}Q_{t-1}^{(h)}V_{t-h-1}$, where $\|\sup_t\mathrm{abs}(D_t'B_{1t}^\ell B_{1,t-1}D_{t-1})\|_\infty<\infty$ and $\|\sup_t\mathrm{abs}(D_t'B_{2t}^\ell B_{2,t-1}D_{t-1})\|_\infty<\infty$ according to Lemma \ref{lm33}. Moreover, since $v_{it}$'s are i.i.d. with respect to $i$ and $t$, $\mathbb{L}^\ell$ and $\mathbb{R}^\ell$ also satisfy the conditions of Lemmas \ref{lm_PPQ}-\ref{lm_Z_gamma}, leading to the same result.
\end{proof}

\begin{lemma}\label{lm_LQV}
Under Assumptions 1, 2, 7 and \ref{asp_A2},
\begin{gather}
\frac{1}{NT}\mathbb{L}^{\ell\prime}\mathbb{B}_2\mathbb{V}=O_p\left(\frac{1}{\sqrt{NT}}\right),\label{eq_lbv}\\
\frac{1}{NT}\mathbb{L}^{\ell\prime}\mathbb{PB}_2\mathbb{V}-\frac{1}{NT}E\left(\mathbb{L}^{\ell\prime}\mathbb{PB}_2\mathbb{V}\right)=O_p\left(\frac{1}{\sqrt{NT^2}}\right),
\label{eq_lpv}\\
\frac{1}{NT}\mathbb{L}^{\ell\prime}\mathbb{QB}_2\mathbb{V}-\frac{1}{NT}E\left(\mathbb{L}^{\ell\prime}\mathbb{QB}_2\mathbb{V}\right)=O_p\left(\frac{1}{\sqrt{NT}}\right),\label{eq_lqv}
\end{gather}
where 
\begin{align}
E\left(\mathbb{L}^{\ell\prime}\mathbb{B}_2\mathbb{V}\right)=0,\label{eq_ELV}
\end{align} and
\begin{align}
E\left(\mathbb{L}^{\ell\prime}\mathbb{PB}_2\mathbb{V}\right)=\sigma_0^2\mathrm{tr}\left[\mathcal{T}^{-1}\sum_{h=1}^{T-1}\sum_{t=T-h}^{T-1}D_{t+1}'B_{t+1}^\ell P_{t}^{(t+h-T)}B_{2,T-h}'D_{T-h}\right]=O(N).\label{eq_ELPV}
\end{align}
\end{lemma}
\begin{proof}
\textbf{Proof of equation (\ref{eq_lbv})}

For simplicity, let $B_t^\ell\equiv B_{1t}^\ell B_{1,t-1}$.
Since $v_{it}$ are i.i.d across $i$ and $t$,
$$E(\mathbb{L}^{\ell\prime}\mathbb{B}_2\mathbb{V})=E\left(\sum_{t=1}^TL_t^{\ell\prime}B_{2t}V_t\right)=\sum_{t=1}^T\sum_{h=0}^\infty E\left( V_{t-h-1}'P_{t-1}^{(h)\prime}B_{t}^{\ell\prime}B_{2t}V_t\right)=0.$$
Then
\begin{align*}
\mathrm{Var}(\mathbb{L}^{\ell\prime}\mathbb{B}_2\mathbb{V})&=\mathrm{Var}\left(\sum_{t=1}^TL_t^{\ell\prime}B_{2t}V_t\right)=\sum_{t=1}^T\sum_{s=1}^TE\left(L_t^{\ell\prime}B_{2t}V_tL_s^{\ell\prime}B_{2s}V_s\right)\\
&=\sum_{t=1}^T\sum_{s=1}^T\sum_{h=0}^\infty\sum_{g=0}^\infty E\left(V_{t-h-1}'P_{t-1}^{(h)\prime}B_{t}^{\ell\prime}B_{2t}V_tV_{s-g-1}'P_{s-1}^{(g)\prime}B_{s}^{\ell\prime}B_{2s}V_s\right)\\
&=\sum_{t=1}^T\sum_{h=0}^\infty E\left(V_{t-h-1}'P_{t-1}^{(h)\prime}B_{t}^{\ell\prime}B_{2t}V_tV_{t-h-1}'P_{t-1}^{(h)\prime}B_{t}^{\ell\prime}B_{2t}V_t\right)\\
&=\sigma_0^4\sum_{t=1}^T\sum_{h=0}^\infty\mathrm{tr}\left(P_{t-1}^{(h)\prime}B_{t}^{\ell\prime}B_{2t}B_{2t}'B_{t}^{\ell}P_{t-1}^{(h)}\right)\\
&=\sigma_0^4\sum_{t=1}^T\mathrm{tr}\left(B_{t}^{\ell\prime}B_{2t}B_{2t}'B_{t}^{\ell}\sum_{h=0}^\infty P_{t-1}^{(h)}P_{t-1}^{(h)\prime}\right)\\
&=O(NT),
\end{align*}
where the fourth equality holds because the expectation is nonzero only when $t=s$ and $h=g$ according to Lemma \ref{lm_quadratic form}, and the last equality holds because $B_{t}^{\ell\prime}B_{2t}B_{2t}'B_{t}^{\ell}$ and $\sum_{h=0}^\infty P_{t-1}^{(h)}P_{t-1}^{(h)\prime}$ are UB uniformly for $t$ by Assumption \ref{asp_A2} and Lemma \ref{lm33}. Therefore, we can prove equation (\ref{eq_lbv}) by Chebyshev's inequality.

\textbf{Proof of equation (\ref{eq_lpv})}

Similar to Lemma \ref{lm_ddot}, we have
$$\mathbb{L}^{\ell\prime}\mathbb{PB}_2\mathbb{V}=\left(\sum_{t=1}^TD_tL_t^{\ell}\right)'\mathcal{T}^{-1}\left(\sum_{t=1}^TD_t'B_{2t}V_t\right)\equiv\left(\sum_{h=0}^\infty\ddot{P}_hV_{T-h}\right)'\left(\sum_{h=0}^\infty\ddot{Q}_hV_{T-h}\right),$$
where
\begin{align*}
\ddot{P}_h\equiv\begin{cases}
0 & \textrm{for } h=0 \\
\mathcal{T}^{-\frac{1}{2}}\sum_{t=T-h}^{T-1}D_{t+1}'B_{t+1}^\ell P_{t}^{(t+h-T)} & \textrm{for } 0<h<T \\
\mathcal{T}^{-\frac{1}{2}}\sum_{t=0}^{T-1}D_{t+1}'B_{t+1}^\ell P_{t}^{(t+h-T)} & \text{for } h\geq T
\end{cases}\textrm{ and }
\ddot{Q}_h\equiv\begin{cases} 
\mathcal{T}^{-\frac{1}{2}}D_{T-h}'B_{2,T-h} & \textrm{for } h<T \\
0 & \text{for } h\geq T
\end{cases}.
\end{align*}
Therefore, by Lemma \ref{lm_EPQ},
\begin{align*}
&\left|E\left(\mathbb{L}^{\ell\prime}\mathbb{PB}_2\mathbb{V}\right)=\sigma_0^2\mathrm{tr}\left(\sum_{h=0}^\infty\ddot{P}_h\ddot{Q}_h'\right)\right|\\
=&\left|\sigma_0^2\mathrm{tr}\left[\sum_{h=1}^{T-1}\mathcal{T}^{-\frac{1}{2}}\left(\sum_{t=T-h}^{T-1}D_{t+1}'B_{t+1}^\ell P_{t}^{(t+h-T)}\right)B_{2,T-h}'D_{T-h}\mathcal{T}^{-\frac{1}{2}}\right]\right|\\
\leq&\sigma_0^2\mathrm{tr}\left[\mathcal{T}^{-1}\sum_{h=1}^{T-1}\mathrm{abs}\left(\sum_{t=T-h}^{T-1}D_{t+1}'B_{t+1}^\ell P_{t}^{(t+h-T)}\right)\mathrm{abs}\left(B_{2,T-h}'D_{T-h}\right)\right]\\
\leq&\frac{\sigma_0^2}{c_TT}\sum_{h=1}^{T-1}\mathrm{tr}\left[\mathrm{abs}\left(\sum_{t=T-h}^{T-1}D_{t+1}'B_{t+1}^\ell P_{t}^{(t+h-T)}\right)\mathrm{abs}\left(B_{2,T-h}'D_{T-h}\right)\right]\\
=&O(N),
\end{align*}
where the second inequality holds because $T_i>c_TT$ by Assumption 1 and the last equality holds because the matrix under the trace is UB uniformly for $h$ according to Lemma \ref{lm33}.
By Lemma \ref{lm_CovPQ}, we have
\begin{align}
\mathrm{Var}\left(\mathbb{L}^{\ell\prime}\mathbb{PB}_2\mathbb{V}\right)=&\mathrm{Cov}\left[\left(\sum_{h=0}^\infty\ddot{P}_hV_{T-h}\right)'\left(\sum_{h=0}^\infty\ddot{Q}_hV_{T-h}\right),\left(\sum_{h=0}^\infty\ddot{P}_hV_{T-h}\right)'\left(\sum_{h=0}^\infty\ddot{Q}_hV_{T-h}\right)\right]\nonumber\\
=&\sigma_0^4\mathrm{tr}\left[\left(\sum_{g=0}^\infty \ddot{Q}_{g}\ddot{P}_{g}'\right)\left(\sum_{g=0}^\infty \ddot{Q}_{g}\ddot{P}_{g}'\right)+\left(\sum_{g=0}^\infty \ddot{P}_{g}\ddot{P}_{g}'\right)\left(\sum_{g=0}^\infty \ddot{Q}_{g}\ddot{Q}_{g}'\right)\right]+\nonumber\\
&\sum_{g=0}^\infty\left[(\mu_4-3\sigma_0^4)\sum_{i=1}^{N_{T-g}}\left(\ddot{P}_{g}'\ddot{Q}_{g}\right)_{ii}\left(\ddot{P}_{g}'\ddot{Q}_{g}\right)_{ii}\right],\label{eq_Varlpv}
\end{align}
where
\begin{align*}
\left\|\sum_{g=0}^\infty \ddot{Q}_{g}\ddot{P}_{g}'\right\|_p=&\left\|
\sum_{h=1}^{T-1}\mathcal{T}^{-\frac{1}{2}}D_{T-h}'B_{2,T-h}\left(\sum_{t=T-h}^{T-1}D_{t+1}'B_{t+1}^\ell P_{t}^{(t+h-T)}\right)'\mathcal{T}^{-\frac{1}{2}}\right\|_p\\
\leq&\frac{1}{c_TT}\sum_{h=1}^{T-1}\left\|D_{T-h}'B_{2,T-h}\left(\sum_{t=T-h}^\infty D_{t+1}'B_{t+1}^\ell P_{t}^{(t+h-T)}\right)'\right\|_p\\
=&O(1)
\end{align*}
holds for $p=1$ and $p=\infty$.
Since the $\ddot{P}_h$ here can be regarded as a lagged version of that defined in Lemma \ref{lm_ddot}, and $\|\sup_t \mathrm{abs}(B_t^\ell)\|_\infty<\infty$ by Lemma \ref{lm33}, similar to the derivation of the upper bound for the corresponding term in equation (\ref{eq_VLPR}), we have
\begin{align*}
\left\|\sum_{g=0}^\infty \ddot{P}_{g}\ddot{P}_{g}'\right\|_p=O(1)
\end{align*}
for $p=1$ and $p=\infty$. Additionally,
\begin{align*}
\left\|\sum_{g=0}^\infty \ddot{Q}_{g}\ddot{Q}_{g}'\right\|_p&=\left\|\sum_{g=0}^{T-1} \mathcal{T}^{-\frac{1}{2}}D_{T-g}'B_{2,T-g}B_{2,T-g}'D_{T-g}\mathcal{T}^{-\frac{1}{2}}\right\|_p\\
&\leq\frac{1}{c_TT}\sum_{g=0}^{T-1}\left\|D_{T-g}'B_{2,T-g}B_{2,T-g}'D_{T-g}\right\|_p\\
&=\frac{1}{c_TT}\sum_{g=0}^{T-1}\left\|B_{2,T-g}B_{2,T-g}'\right\|_p\\
&=O(1),
\end{align*}
where the first inequality holds by $T_i\geq c_TT$ in Assumption 1, the second one holds because $D_{T-g}'B_{2,T-g}B_{2,T-g}'D_{T-g}$ is merely an augmented version of $B_{2,T-g}B_{2,T-g}'$ with additional all-zero rows and columns, and the last equality holds because $B_{2,T-g}B_{2,T-g}'$ is UB uniformly for $g$. Therefore, the matrix within the trace in the first term of equation (\ref{eq_Varlpv}) is UB, which means that the first term is $O(N)$. For the second term in equation (\ref{eq_Varlpv}), we have
\begin{align*}
\left|\sum_{g=0}^\infty\sum_{i=1}^{N_{T-g}}\left(\ddot{P}_{g}'\ddot{Q}_{g}\right)_{ii}\left(\ddot{P}_{g}'\ddot{Q}_{g}\right)_{ii}\right|&\leq\sup_{g,i}\left|\left(\ddot{P}_{g}'\ddot{Q}_{g}\right)_{ii}\right|\left(\sum_{i=1}^{N_{T-g}}\left|\sum_{g=0}^\infty\left(D_g'\ddot{P}_{g}'\ddot{Q}_{g}D_g\right)_{ii}\right|\right)\\
&\leq\sup_{g}\left\|D_g'\ddot{P}_{g}'\ddot{Q}_{g}D_g\right\|_1\mathrm{tr}\left[\mathrm{abs}\left(\sum_{g=0}^\infty D_g'\ddot{P}_{g}'\ddot{Q}_{g}D_g\right)\right]\\
&=O(N),
\end{align*}
where the last equality holds because $\sum_{g=0}^\infty \ddot{P}_{g}'\ddot{Q}_{g}$ is UB, so $\sup_{g}\left\|D_g'\ddot{P}_{g}'\ddot{Q}_{g}D_g\right\|_1$ is $O(1)$, and the trace above is $O(N)$. So the second term in equation (\ref{eq_Varlpv}) is also $O(N)$, which means that the whole term is $O(N)$. Therefore, we can get equation (\ref{eq_lpv}) by Chebyshev's inequality.

\textbf{Proof of equation (\ref{eq_lqv})}

Since $\mathbb{Q}=I_n-\mathbb{P}$, combining equations (\ref{eq_lbv}) and (\ref{eq_lpv}), the results follows.
\end{proof}


\begin{lemma}\label{lm_VQV}
Under Assumptions
1, 2, 7 and \ref{asp_A2},
\begin{gather}
\frac{1}{NT}\mathbb{V'B}'_1\mathbb{B}_2\mathbb{V}-\frac{1}{NT}E\left(\mathbb{VB}'_1\mathbb{B}_2\mathbb{V}\right)=O_p\left(\frac{1}{\sqrt{NT}}\right),\label{eq_vbv}\\
\frac{1}{NT}\mathbb{V'B}'_1\mathbb{PB}_2\mathbb{V}-\frac{1}{NT}E\left(\mathbb{V'B}'_1\mathbb{PB}_2\mathbb{V}\right)=O_p\left(\frac{1}{\sqrt{NT^2}}\right),
\label{eq_vpv}\\
\frac{1}{NT}\mathbb{V'B}'_1\mathbb{QB}_2\mathbb{V}-\frac{1}{NT}E\left(\mathbb{V'B}'_1\mathbb{QB}_2\mathbb{V}\right)=O_p\left(\frac{1}{\sqrt{NT}}\right),\label{eq_vqv}
\end{gather}
where $\frac{1}{NT}E(\mathbb{V'B}'_1\mathbb{B}_2\mathbb{V})=O(1)$ and $\frac{1}{NT}E(\mathbb{V'B}'_1\mathbb{PB}_2\mathbb{V})=O(\frac{1}{T})$.
\end{lemma}
\begin{proof}

\textbf{Proof of equation (\ref{eq_vbv})}
\begin{align*}
\frac{1}{NT}E\left(\mathbb{V'B}'_1\mathbb{B}_2\mathbb{V}\right)&=\frac{1}{NT}E\left(\sum_{t=1}^T V_t'B_{1t}'B_{2t}V_t\right)=\frac{1}{NT}E\left[\sum_{t=1}^T \mathrm{tr}(B_{2t}V_tV_t'B_{1t}')\right]\\
&=\frac{1}{NT}\sigma^2_0\sum_{t=1}^T \mathrm{tr}(B_{1t}'B_{2t})=O(1),
\end{align*}
where the last equality is because $\{B_{1t}\}$ and $\{B_{2t}\}$ are UB uniform for $t$.
Moreover by Lemma \ref{lm_quadratic form},
\begin{align*}
&\mathrm{Var}\left(\mathbb{V'B}'_1\mathbb{B}_2\mathbb{V}\right)\\
=&\sum_{t=1}^T\sum_{s=1}^T\mathrm{Cov}(V_t'B_{1t}'B_{2t}V_t,V_s'B_{1s}'B_{2s}V_s)\\
=&\sum_{t=1}^T\sum_{s=1}^T \left[E(V_t'B_{1t}'B_{2t}V_tV_s'B_{1s}'B_{2s}V_s)-E(V_t'B_{1t}'B_{2t}V_t)E(V_s'B_{1s}'B_{2s}V_s)\right]\\
=&\sum_{t=1}^T\left[(\mu_4-3\sigma^4_0)\sum_{i=1}^{N_t}(B_{1t}'B_{2t})_{ii}^2+\sigma_0^4\left(\mathrm{tr}(B_{1t}'B_{2t})^2+\mathrm{tr}(B_{1t}'B_{2t}B_{1t}'B_{2t})+\mathrm{tr}(B_{1t}'B_{2t}B_{2t}'B_{1t})\right)\right]+\\
&\sum_{t=1}^T\sum_{s=1}^T\sigma_0^4\mathrm{tr}(B_{1t}'B_{2t})\mathrm{tr}(B_{1s}'B_{2s})-\sum_{t=1}^T\sigma_0^4\mathrm{tr}(B_{1t}'B_{2t})^2-\sigma_0^4\left[\sum_{t=1}^T\mathrm{tr}(B_{1t}'B_{2t})\right]^2\\
=&\sum_{t=1}^T\left[(\mu_4-3\sigma^4_0)\sum_{i=1}^{N_t}(B_{1t}'B_{2t})_{ii}^2+\sigma_0^4\left(\mathrm{tr}(B_{1t}'B_{2t}B_{1t}'B_{2t})+\mathrm{tr}(B_{1t}'B_{2t}B_{2t}'B_{1t})\right)\right]\\
=&O(NT).
\end{align*}
Then we can get equation (\ref{eq_vbv}) by Chebyshev's inequality.

\textbf{Proof of equation (\ref{eq_vpv})}

Similar to Lemma \ref{lm_ddot}, we have
\begin{align*}
\mathbb{V'B}'_1\mathbb{PB}_2\mathbb{V}=\left(\sum_{t=1}^TD_t'B_{1t}V_t\right)'\mathcal{T}^{-1}\left(\sum_{t=1}^TD_t'B_{2t}V_t\right)\equiv\left(\sum_{h=0}^\infty\ddot{P}_hV_{T-h}\right)'\left(\sum_{h=0}^\infty\ddot{Q}_hV_{T-h}\right),
\end{align*}
where
\begin{align*}
\ddot{P}_h\equiv\begin{cases} 
\mathcal{T}^{-\frac{1}{2}}D_{T-h}'B_{1,T-h} & \textrm{for } h<T \\
0 & \text{for } h\geq T
\end{cases}
\end{align*}
and $\ddot{Q}_h$ has the same pattern. Then according to Lemma \ref{lm_EPQ},
\begin{align*}
&\left|\frac{1}{NT}E\left(\mathbb{V'B}'_1\mathbb{PB}_2\mathbb{V}\right)\right|=\frac{1}{NT}\left|E\left[\left(\sum_{h=0}^\infty\ddot{P}_hV_{T-h}\right)'\left(\sum_{h=0}^\infty\ddot{Q}_hV_{T-h}\right)\right]\right|\\
=&\frac{1}{NT}\sigma_0^2\left|\mathrm{tr}\left(\sum_{h=0}^\infty\ddot{P}_{h}\ddot{Q}_{h}'\right)\right|\leq\frac{1}{NT}\sigma_0^2\sum_{t=1}^T\left|\mathrm{tr}\left(\mathcal{T}^{-\frac{1}{2}}D_{t}'B_{1t}
B_{1t}'D_t\mathcal{T}^{-\frac{1}{2}}\right)\right|\\
\leq&\frac{1}{c_TNT^2}\sigma_0^2\sum_{t=1}^T\mathrm{tr}\left(\mathrm{abs}(B_{1t}'B_{2t})\right)=O\left(\frac{1}{T}\right),
\end{align*}
The second equality holds because $V_t$'s are independent across $t$. The second inequality holds due to Assumption 1 with $T_i\geq c_TT$. The last equality holds because $\{B_{1t}\}$ and $\{B_{2t}\}$ are UB uniform for $t$.
Moreover according to Lemma \ref{lm_CovPQ},
\begin{align}
&\mathrm{Var}\left(\mathbb{V'B}'_1\mathbb{PB}_2\mathbb{V}\right)\nonumber\\
=&\mathrm{Cov}\left[\left(\sum_{h=0}^\infty\ddot{P}_hV_{T-h}\right)'\left(\sum_{h=0}^\infty\ddot{Q}_hV_{T-h}\right),\left(\sum_{h=0}^\infty\ddot{P}_hV_{T-h}\right)'\left(\sum_{h=0}^\infty\ddot{Q}_hV_{T-h}\right)\right]\nonumber\\
=&\sigma_0^4\mathrm{tr}\left[\left(\sum_{g=0}^\infty \ddot{Q}_{g}\ddot{P}_{g}'\right)\left(\sum_{g=0}^\infty \ddot{Q}_{g}\ddot{P}_{g}'\right)+\left(\sum_{g=0}^\infty \ddot{P}_{g}\ddot{P}_{g}'\right)\left(\sum_{g=0}^\infty \ddot{Q}_{g}'\ddot{Q}_{g}'\right)\right]+\nonumber\\
&\sum_{g=0}^\infty\left[(\mu_4-3\sigma_0^4)\sum_{i=1}^{N_{T-g}}\left(\ddot{P}_{g}'\ddot{Q}_{g}\right)_{ii}\left(\ddot{P}_{g}'\ddot{Q}_{g}\right)_{ii}\right]\nonumber\\
=&\sigma_0^4\sum_{g=1}^T\sum_{h=1}^T\mathrm{tr}\left[\left(\mathcal{T}^{-\frac{1}{2}}D_{g}'B_{2g}B_{1g}'D_{g}\mathcal{T}^{-\frac{1}{2}}\right)\left( \mathcal{T}^{-\frac{1}{2}}D_{h}'B_{2h}B_{1h}'D_{h}\mathcal{T}^{-\frac{1}{2}}\right)\right]+\nonumber\\
&\sigma_0^4\sum_{g=1}^T\sum_{h=1}^T\mathrm{tr}\left[\left(\mathcal{T}^{-\frac{1}{2}}D_{g}'B_{1g}B_{1g}'D_{g}\mathcal{T}^{-\frac{1}{2}}\right)\left(\mathcal{T}^{-\frac{1}{2}}D_{h}'B_{2h}B_{2h}'D_{h}\mathcal{T}^{-\frac{1}{2}}\right)\right]+\nonumber\\
&\sum_{g=1}^T\left[(\mu_4-3\sigma_0^4)\sum_{i=1}^{N_{g}}\left(B_{1g}'D_g\mathcal{T}^{-1}D_g'B_{2g}\right)_{ii}\left(B_{1g}'D_g\mathcal{T}^{-1}D_g'B_{2g}\right)_{ii}\right].\label{eq_Vvpv}
\end{align}
Since for any $g$ and $h$,
\begin{align*}
&\left|\mathrm{tr}\left[\left(\mathcal{T}^{-\frac{1}{2}}D_{g}'B_{2g}B_{1g}'D_{g}\mathcal{T}^{-\frac{1}{2}}\right)\left(\mathcal{T}^{-\frac{1}{2}}D_{h}'B_{2h}B_{1h}'D_{h}\mathcal{T}^{-\frac{1}{2}}\right)\right]\right|\\
=&\left|\mathrm{tr}\left[D_{h}\mathcal{T}^{-1}D_{g}'B_{2g}B_{1g}'D_{g}\mathcal{T}^{-1}D_{h}'B_{2h}B_{1h}'\right]\right|\\
\leq&\mathrm{tr}\left[\mathrm{abs}\left(D_{h}\mathcal{T}^{-1}D_{g}'\right)\mathrm{abs}\left(B_{2g}B_{1g}'\right)\mathrm{abs}\left(D_{g}\mathcal{T}^{-1}D_{h}'\right)\mathrm{abs}\left(B_{2h}B_{1h}'\right)\right]\\
\leq&\frac{1}{c_T^2T^2}\mathrm{tr}\left[D_{h}D_{g}'\mathrm{abs}\left(B_{2g}B_{1g}'\right)D_{g}D_{h}'\mathrm{abs}\left(B_{2h}B_{1h}'\right)\right]\\
=&O\left(\frac{N}{T^2}\right),
\end{align*}
where the second inequality holds due to Assumption 1 with $T_i\geq c_TT$, and the last equality holds because $\{B_{1t}\}$ and $\{B_{2t}\}$ are UB uniform for $t$. Similarly, we have
$$\mathrm{tr}\left[\left(\mathcal{T}^{-\frac{1}{2}}D_{g}'B_{1g}B_{1g}'D_{g}\mathcal{T}^{-\frac{1}{2}}\right)\left(\mathcal{T}^{-\frac{1}{2}}D_{h}'B_{2h}B_{2h}'D_{h}\mathcal{T}^{-\frac{1}{2}}\right)\right]=O\left(\frac{N}{T^2}\right).$$
Therefore, the first two terms in equation (\ref{eq_Vvpv}) is $O(N)$.
Then, by the same argument, we have
\begin{align*}
\sup_{g,i}\left|\left(B_{1g}'D_g\mathcal{T}^{-1}D_g'B_{2g}\right)_{ii}\right|&\leq\sup_{g,i}\sum_{k=1}^N \mathrm{abs}(B_{1g}'D_g)_{ik}T_k^{-1}\mathrm{abs}(D_g'B_{2g})_{ki}\\
&\leq\frac{1}{c_TT}\sup_{g,i}\sum_{k=1}^N \mathrm{abs}(B_{1g}'D_g)_{ik}\mathrm{abs}(D_g'B_{2g})_{ki}=O\left(\frac{1}{T}\right).   
\end{align*}
Thus,
\begin{align*}
&\left|\sum_{g=1}^T\sum_{i=1}^{N_{g}}\left(B_{1g}'D_g\mathcal{T}^{-1}D_g'B_{2g}\right)_{ii}\left(B_{1g}'D_g\mathcal{T}^{-1}D_g'B_{2g}\right)_{ii}\right|\\
\leq&\sup_{i,g}\left|\left(B_{1g}'D_g\mathcal{T}^{-1}D_g'B_{2g}\right)_{ii}\right|\sum_{g=1}^T\mathrm{tr}\left[\mathrm{abs}\left(B_{1g}'D_g\mathcal{T}^{-1}D_g'B_{2g}\right)\right]\\
=&O\left(\frac{N}{T}\right),
\end{align*}
that is, the last term in equation (\ref{eq_Vvpv}) is $O\left(\frac{N}{T}\right)$. Therefore, $\mathrm{Var}(\mathbb{V'B}'_1\mathbb{PB}_2\mathbb{V})=O(N)+O\left(\frac{N}{T}\right)=O(N).$
Then by Chebyshev's inequality, equation (\ref{eq_vpv}) follows.

\textbf{Proof of equation (\ref{eq_vqv})}

Since $\mathbb{Q}=I_n-\mathbb{P}$, combining equations (\ref{eq_vbv}) and (\ref{eq_vpv}), the results follows.
\end{proof}

\begin{lemma}\label{lm15}
Under Assumptions 1-7 and \ref{asp_A2}, we have
\begin{align}
&\frac{1}{NT}\mathbb{V}'\mathbb{B}_1'\mathbb{Q}\mathbb{B}_2\mathbb{V}-\frac{1}{NT}E[\mathbb{V}'\mathbb{B}_1'\mathbb{Q}\mathbb{B}_2\mathbb{V}]=O_p\left(\frac{1}{\sqrt{NT}}\right),\label{eq_VBV}\\
&\frac{1}{NT}\mathbb{Z}'\mathbb{B}_1'\mathbb{Q}\mathbb{B}_2\mathbb{V}-\frac{1}{NT}E[\mathbb{Z}'\mathbb{B}_1'\mathbb{Q}\mathbb{B}_2\mathbb{V}]=O_p\left(\frac{1}{\sqrt{NT}}\right),\label{eq_ZBV}\\
&\frac{1}{NT}\mathbb{Z}'\mathbb{B}_1'\mathbb{Q}\mathbb{B}_2\mathbb{Z}-\frac{1}{NT}E[\mathbb{Z}'\mathbb{B}_1'\mathbb{Q}\mathbb{B}_2\mathbb{Z}]=O_p\left(\frac{1}{\sqrt{NT}}\right),\label{eq_ZBZ}
\end{align}
where $\frac{1}{NT}E[\mathbb{V}'\mathbb{B}_1'\mathbb{Q}\mathbb{B}_2\mathbb{V}]=O(1)$, $\frac{1}{NT}E[\mathbb{Z}'\mathbb{B}_1'\mathbb{Q}\mathbb{B}_2\mathbb{V}]=O(\frac{1}{T})$ and $\frac{1}{NT}E[\mathbb{Z}'\mathbb{B}_1'\mathbb{Q}\mathbb{B}_2\mathbb{Z}]=O(1)$. 
\end{lemma}
\begin{proof}
\textbf{Proof of equation (\ref{eq_VBV})}

This result is just equation (\ref{eq_vqv}) in Lemma \ref{lm_VQV}.

\textbf{Proof of equation (\ref{eq_ZBV})}

Recall that $Z_t=(D_t(I_N-F_t)D_t'M_tY_{t-1},D_tD_{t-1}'Y_{t-1},D_tF_t\iota_N,X_t)$ and $\mathbb{Z}=(Z_1',Z_2',\cdots,Z_T')'$. Denote $\mathbb{Z}_1$, $\mathbb{Z}_2$ and $\mathbb{Z}^{(\gamma)}$ as the first three columns of $\mathbb{Z}$, and let $\mathbb{Z}_4$ denote the remaining $k$ columns.
Since $$\mathbb{Z}'\mathbb{B}_1'\mathbb{Q}\mathbb{B}_2\mathbb{V}=\left(\begin{array}{c}
\mathbb{Z}_1'\mathbb{B}_1'\mathbb{Q}\mathbb{B}_2\mathbb{V}\\
\mathbb{Z}_2'\mathbb{B}_1'\mathbb{Q}\mathbb{B}_2\mathbb{V}\\
\mathbb{Z}^{(\gamma)\prime}\mathbb{B}_1'\mathbb{Q}\mathbb{B}_2\mathbb{V}\\
\mathbb{Z}_4'\mathbb{B}_1'\mathbb{Q}\mathbb{B}_2\mathbb{V}
\end{array}\right),$$
we only need to derive the quantitative law of large numbers for each element of the above vectors. According to equation (3),
\begin{align*}
Y_t=\sum_{h=0}^{\infty}S_{t}^{-1}A_t^{(h)}[D_{t-h} (\gamma_0F_{t-h}\iota_N+\alpha_0)+X_{t-h}\beta_0+V_{t-h}]=\mathcal{E}_t+\mathcal{X}_t\beta_0+\mathcal{V}_t,
\end{align*}
where $\mathcal{E}_t= \sum_{h=0}^{\infty}S_{t}^{-1}A_t^{(h)}D_{t-h}(\gamma_0F_{t-h}\iota_N+\alpha_0)$, 
$\mathcal{X}_t= \sum_{h=0}^{\infty}S_{t}^{-1}A_t^{(h)}X_{t-h}$ and $\mathcal{V}_t= \sum_{h=0}^{\infty}S_{t}^{-1}A_t^{(h)}V_{t-h}$. Therefore, $\mathbb{B}_1\mathbb{Z}_1$ is the stack version of vector $B_{1t}D_t(I_N-F_t)D_t'M_t(\mathcal{E}_{t-1}+\mathcal{X}_{t-1}\beta_0+\mathcal{V}_{t-1})$, which can be viewed as $\mathbb{C+L}^\ell$, by specifying $C_t=B_{1t}D_t(I_N-F_t)D_t'M_t(\mathcal{E}_{t-1}+\mathcal{X}_{t-1}\beta_0)$ and $L_{t}^\ell=B_{1t}D_t(I_N-F_t)D_t'M_t\mathcal{V}_{t-1}$.
Then we have $$\mathbb{Z}_1'\mathbb{B}_1'\mathbb{Q}\mathbb{B}_2\mathbb{V}\equiv\mathbb{C}'\mathbb{Q}\mathbb{B}_2\mathbb{V}+\mathbb{L}^{\ell\prime}\mathbb{Q}\mathbb{B}_2\mathbb{V},$$ 
where $\frac{1}{NT}\mathbb{C}'\mathbb{Q}\mathbb{B}_2\mathbb{V}=O_p(\frac{1}{\sqrt{NT}})$ by Lemma \ref{lm_CQL}, $\frac{1}{NT}\mathbb{L}^{\ell\prime}\mathbb{QB}_2\mathbb{V}-\frac{1}{NT}E\left(\mathbb{L}^{\ell\prime}\mathbb{QB}_2\mathbb{V}\right)=O_p(\frac{1}{\sqrt{NT}})$, and $\frac{1}{NT}E\left[\mathbb{L}^{\ell\prime}\mathbb{QB}_2\mathbb{V}\right]=O(\frac{1}{T})$ by Lemma \ref{lm_LQV}. Therefore, 
$$\frac{1}{NT}\mathbb{Z}_1'\mathbb{B}_1'\mathbb{Q}\mathbb{B}_2\mathbb{V}-\frac{1}{NT}E\left[\mathbb{Z}_1'\mathbb{B}_1'\mathbb{Q}\mathbb{B}_2\mathbb{V}\right]=O_p\left(\frac{1}{\sqrt{NT}}\right)$$
and $\frac{1}{NT}E\left(\mathbb{Z}_1'\mathbb{B}_1'\mathbb{Q}\mathbb{B}_2\mathbb{V}\right)=O(\frac{1}{T})$.
Similarly, $\mathbb{B}_1\mathbb{Z}_2$ can also be viewed as $\mathbb{C+L}^\ell$ by specifying $C_t=B_{1t}D_tD_{t-1}(\mathcal{E}_{t-1}+\mathcal{X}_{t-1}\beta_0)$ and $L_{t}^\ell=B_{1t}D_tD_{t-1}\mathcal{V}_{t-1}$. Hence,
$$\frac{1}{NT}\mathbb{Z}_2'\mathbb{B}_1'\mathbb{Q}\mathbb{B}_2\mathbb{V}-\frac{1}{NT}E\left[\mathbb{Z}_2'\mathbb{B}_1'\mathbb{Q}\mathbb{B}_2\mathbb{V}\right]=O_p\left(\frac{1}{\sqrt{NT}}\right)$$
and $\frac{1}{NT}E\left(\mathbb{Z}_2'\mathbb{B}_1'\mathbb{Q}\mathbb{B}_2\mathbb{V}\right)=O(\frac{1}{T})$ also hold. Each column of $\mathbb{B}_1\mathbb{Z}^{(\gamma)}$ and $\mathbb{B}_1\mathbb{Z}_4$ can be viewed as $\mathbb{C}$ because they are both vectors whose elements are uniformly bounded.
Then $\mathbb{Z}_3$ and $\mathbb{Z}_4$ satisfy the same conclusion because $\frac{1}{NT}\mathbb{C}'\mathbb{Q}\mathbb{B}_2\mathbb{V}=O_p(\frac{1}{\sqrt{NT}})$ by Lemma \ref{lm_CQL}.
Thus, equation (\ref{eq_ZBV}) holds and $\frac{1}{NT}E\left(\mathbb{Z}'\mathbb{B}_1'\mathbb{Q}\mathbb{B}_2\mathbb{V}\right)=O(\frac{1}{T})$.

\textbf{Proof of equation (\ref{eq_ZBZ})}

We have 
\begin{align*}
\mathbb{Z}'\mathbb{B}_1'\mathbb{Q}\mathbb{B}_2\mathbb{Z}=\left(\begin{array}{cccc}
\mathbb{Z}_1'\mathbb{B}_1'\mathbb{Q}\mathbb{B}_2\mathbb{Z}_1 & \mathbb{Z}_1'\mathbb{B}_1'\mathbb{Q}\mathbb{B}_2\mathbb{Z}_2 & \mathbb{Z}_1'\mathbb{B}_1'\mathbb{Q}\mathbb{B}_2\mathbb{Z}^{(\gamma)} & \mathbb{Z}_1'\mathbb{B}_1'\mathbb{Q}\mathbb{B}_2\mathbb{Z}_4\\
\mathbb{Z}_2'\mathbb{B}_1'\mathbb{Q}\mathbb{B}_2\mathbb{Z}_1 & \mathbb{Z}_2'\mathbb{B}_1'\mathbb{Q}\mathbb{B}_2\mathbb{Z}_2 & \mathbb{Z}_2'\mathbb{B}_1'\mathbb{Q}\mathbb{B}_2\mathbb{Z}^{(\gamma)} & \mathbb{Z}_2'\mathbb{B}_1'\mathbb{Q}\mathbb{B}_2\mathbb{Z}_4\\
\mathbb{Z}^{(\gamma)\prime}\mathbb{B}_1'\mathbb{Q}\mathbb{B}_2\mathbb{Z}_1 & \mathbb{Z}^{(\gamma)\prime}\mathbb{B}_1'\mathbb{Q}\mathbb{B}_2\mathbb{Z}_2 & \mathbb{Z}^{(\gamma)\prime}\mathbb{B}_1'\mathbb{Q}\mathbb{B}_2\mathbb{Z}^{(\gamma)} & \mathbb{Z}^{(\gamma)\prime}\mathbb{B}_1'\mathbb{Q}\mathbb{B}_2\mathbb{Z}_4\\
\mathbb{Z}_4'\mathbb{B}_1'\mathbb{Q}\mathbb{B}_2\mathbb{Z}_1 & \mathbb{Z}_4'\mathbb{B}_1'\mathbb{Q}\mathbb{B}_2\mathbb{Z}_2 & \mathbb{Z}_4'\mathbb{B}_1'\mathbb{Q}\mathbb{B}_2\mathbb{Z}^{(\gamma)} & \mathbb{Z}_4'\mathbb{B}_1'\mathbb{Q}\mathbb{B}_2\mathbb{Z}_4
\end{array}\right).
\end{align*}
Firstly, we can decompose $$\mathbb{Z}_1'\mathbb{B}_1'\mathbb{Q}\mathbb{B}_2\mathbb{Z}_1\equiv(\mathbb{C}_1+\mathbb{L}^\ell)'\mathbb{Q(C}_2+\mathbb{R}^\ell)=\mathbb{C}_1'\mathbb{QC}_2+\mathbb{C}_1'\mathbb{QR}^\ell+\mathbb{C}_2'\mathbb{QL}^\ell+\mathbb{L}^{\ell\prime}\mathbb{QR}^\ell,$$
where $\frac{1}{NT}\mathbb{C'}_1\mathbb{Q}\mathbb{R}^\ell=O_p\left(\frac{1}{\sqrt{NT}}\right)$ and $\frac{1}{NT}\mathbb{C'}_2\mathbb{Q}\mathbb{L}^\ell=O_p\left(\frac{1}{\sqrt{NT}}\right)$ by Lemmas \ref{lm_CQL} and \ref{lm_lag}, and $\frac{1}{NT}\mathbb{L}^{\ell\prime}\mathbb{QR}^\ell-\frac{1}{NT}E\left(\mathbb{L}^{\ell\prime}\mathbb{QR}^\ell\right)=O_p\left(\frac{1}{\sqrt{NT}}\right)$, $\frac{1}{NT}E\left(\mathbb{L}^{\ell\prime}\mathbb{QR}^\ell\right)=O(1)$ by Lemmas \ref{lm_PPQ} and \ref{lm_lag}. We have $\frac{1}{NT}\mathbb{C}_1'\mathbb{QC}_2=O(1)$ because elements in $\mathbb{C}_1$ and $\mathbb{C}_2$ are uniformly bounded and $\mathbb{Q}$ is a projection matrix. Hence,
$$\frac{1}{NT}\mathbb{Z}_1'\mathbb{B}_1'\mathbb{Q}\mathbb{B}_2\mathbb{Z}_1-\frac{1}{NT}E\left[\mathbb{Z}_1'\mathbb{B}_1'\mathbb{Q}\mathbb{B}_2\mathbb{Z}_1\right]=O_p\left(\frac{1}{\sqrt{NT}}\right)$$
and $\frac{1}{NT}E\left(\mathbb{Z}_1'\mathbb{B}_1'\mathbb{Q}\mathbb{B}_2\mathbb{Z}_1\right)=O(1)$.
The same conclusion holds for $\mathbb{Z}_1'\mathbb{B}_1'\mathbb{Q}\mathbb{B}_2\mathbb{Z}_2$, $\mathbb{Z}_2'\mathbb{B}_1'\mathbb{Q}\mathbb{B}_2\mathbb{Z}_1$ and $\mathbb{Z}_2'\mathbb{B}_1'\mathbb{Q}\mathbb{B}_2\mathbb{Z}_2$ as they can undergo similar decomposition.

Secondly, we have $\mathbb{Z}_1'\mathbb{B}_1'\mathbb{Q}\mathbb{B}_2\mathbb{Z}^{(\gamma)}\equiv(\mathbb{C}_1+\mathbb{L}^\ell)'\mathbb{QC}_2=\mathbb{C}_2'\mathbb{QC}_1+\mathbb{C}_2'\mathbb{QL}^\ell$. Similarly, we have
$$\frac{1}{NT}\mathbb{Z}_1'\mathbb{B}_1'\mathbb{Q}\mathbb{B}_2\mathbb{Z}^{(\gamma)}-\frac{1}{NT}E\left[\mathbb{Z}_1'\mathbb{B}_1'\mathbb{Q}\mathbb{B}_2\mathbb{Z}^{(\gamma)}\right]=O_p\left(\frac{1}{\sqrt{NT}}\right)$$
and $\frac{1}{NT}E\left(\mathbb{Z}_1'\mathbb{B}_1'\mathbb{Q}\mathbb{B}_2\mathbb{Z}^{(\gamma)}\right)=O(1)$ by Lemmas \ref{lm_CQL} and \ref{lm_lag}. The same conclusion holds for $\mathbb{Z}^{(\gamma)\prime}\mathbb{B}_1'\mathbb{Q}\mathbb{B}_2\mathbb{Z}_1$, $\mathbb{Z}_1'\mathbb{B}_1'\mathbb{Q}\mathbb{B}_2\mathbb{Z}_4$, $\mathbb{Z}_4'\mathbb{B}_1'\mathbb{Q}\mathbb{B}_2\mathbb{Z}_1$, $\mathbb{Z}_2'\mathbb{B}_1'\mathbb{Q}\mathbb{B}_2\mathbb{Z}^{(\gamma)}$, $\mathbb{Z}^{(\gamma)\prime}\mathbb{B}_1'\mathbb{Q}\mathbb{B}_2\mathbb{Z}_2$, $\mathbb{Z}_2'\mathbb{B}_1'\mathbb{Q}\mathbb{B}_2\mathbb{Z}_4$ and $\mathbb{Z}_4'\mathbb{B}_1'\mathbb{Q}\mathbb{B}_2\mathbb{Z}_2$.

Lastly, 
The scalar $\mathbb{Z}^{(\gamma)\prime}\mathbb{B}_1'\mathbb{Q}\mathbb{B}_2\mathbb{Z}^{(\gamma)}$, as well as each element of the vector $\mathbb{Z}^{(\gamma)\prime}\mathbb{B}_1'\mathbb{Q}\mathbb{B}_2\mathbb{Z}_4$, $\mathbb{Z}_4\mathbb{B}_1'\mathbb{Q}\mathbb{B}_2\mathbb{Z}^{(\gamma)}$, and the matrix $\mathbb{Z}_4'\mathbb{B}_1'\mathbb{Q}\mathbb{B}_2\mathbb{Z}_4$, can be regarded as $\mathbb{C}_1'\mathbb{QC}_2$ and $\frac{1}{NT}\mathbb{C}_1'\mathbb{QC}_2=O(1)$.
Therefore, equation (\ref{eq_ZBZ}) follows and $\frac{1}{NT}E[\mathbb{Z}'\mathbb{B}_1'\mathbb{Q}\mathbb{B}_2\mathbb{Z}]=O(1)$.

\end{proof}

\begin{lemma}\label{lm_ZPZ}
Under Assumptions 1-7 and \ref{asp_A2}, we have
\begin{align}
&\frac{1}{NT}\mathbb{V}'\mathbb{B}_1'\mathbb{P}\mathbb{B}_2\mathbb{V}-\frac{1}{NT}E[\mathbb{V}'\mathbb{B}_1'\mathbb{P}\mathbb{B}_2\mathbb{V}]=O_p\left(\frac{1}{\sqrt{NT^2}}\right),\label{eq_VPV}\\
&\frac{1}{NT}\mathbb{Z}'\mathbb{B}_1'\mathbb{P}\mathbb{B}_2\mathbb{V}-\frac{1}{NT}E[\mathbb{Z}'\mathbb{B}_1'\mathbb{P}\mathbb{B}_2\mathbb{V}]=O_p\left(\frac{1}{\sqrt{NT}}\right),\label{eq_ZPV}\\
&\frac{1}{NT}\mathbb{Z}'\mathbb{B}_1'\mathbb{P}\mathbb{B}_2\mathbb{Z}-\frac{1}{NT}E[\mathbb{Z}'\mathbb{B}_1'\mathbb{P}\mathbb{B}_2\mathbb{Z}]=O_p\left(\frac{1}{\sqrt{NT}}\right),\label{eq_ZPZ}
\end{align}
where $\frac{1}{NT}E[\mathbb{V}'\mathbb{B}_1'\mathbb{P}\mathbb{B}_2\mathbb{V}]=O(\frac{1}{T})$, $\frac{1}{NT}E[\mathbb{Z}'\mathbb{B}_1'\mathbb{P}\mathbb{B}_2\mathbb{V}]=O(\frac{1}{T})$ and $\frac{1}{NT}E[\mathbb{Z}'\mathbb{B}_1'\mathbb{P}\mathbb{B}_2\mathbb{Z}]=O(1)$. 
\end{lemma}
\begin{proof}
\textbf{Proof of equation (\ref{eq_VPV})}

This result is just equation (\ref{eq_vpv}) in Lemma \ref{lm_VQV}.

\textbf{Proof of equation (\ref{eq_ZPV})}

Similar to the proof of Lemma \ref{lm15}, we have
$$\mathbb{Z}'\mathbb{B}_1'\mathbb{P}\mathbb{B}_2\mathbb{V}=\left(\begin{array}{c}
\mathbb{Z}_1'\mathbb{B}_1'\mathbb{P}\mathbb{B}_2\mathbb{V}\\
\mathbb{Z}_2'\mathbb{B}_1'\mathbb{P}\mathbb{B}_2\mathbb{V}\\
\mathbb{Z}^{(\gamma)\prime}\mathbb{B}_1'\mathbb{P}\mathbb{B}_2\mathbb{V}\\
\mathbb{Z}_4'\mathbb{B}_1'\mathbb{P}\mathbb{B}_2\mathbb{V}
\end{array}\right),$$
where $\mathbb{Z}_1'\mathbb{B}_1'\mathbb{P}\mathbb{B}_2\mathbb{V}\equiv\mathbb{C}'\mathbb{V}+\mathbb{L}^{\ell\prime}\mathbb{P}\mathbb{B}_2\mathbb{V}$. We have $\frac{1}{NT}\mathbb{C}'\mathbb{V}=O_p(\frac{1}{\sqrt{NT}})$ by Lemmas \ref{lm_CQL} and \ref{lm_lag}. According to Lemma \ref{lm_LQV}, we have $\frac{1}{NT}\mathbb{L}^{\ell\prime}\mathbb{PB}_2\mathbb{V}-\frac{1}{NT}E\left(\mathbb{L}^{\ell\prime}\mathbb{PB}_2\mathbb{V}\right)=O_p(\frac{1}{\sqrt{NT^2}})$  and $\frac{1}{NT}E\left[\mathbb{L}^{\ell\prime}\mathbb{PB}_2\mathbb{V}\right]=O(\frac{1}{T})$. Therefore, 
$$\frac{1}{NT}\mathbb{Z}_1'\mathbb{B}_1'\mathbb{P}\mathbb{B}_2\mathbb{V}-\frac{1}{NT}E\left[\mathbb{Z}_1'\mathbb{B}_1'\mathbb{P}\mathbb{B}_2\mathbb{V}\right]=O_p\left(\frac{1}{\sqrt{NT}}\right)$$
and $\frac{1}{NT}E\left(\mathbb{Z}_1'\mathbb{B}_1'\mathbb{P}\mathbb{B}_2\mathbb{V}\right)=O(\frac{1}{T})$.
Similarly,
$$\frac{1}{NT}\mathbb{Z}_2'\mathbb{B}_1'\mathbb{P}\mathbb{B}_2\mathbb{V}-\frac{1}{NT}E\left[\mathbb{Z}_2'\mathbb{B}_1'\mathbb{P}\mathbb{B}_2\mathbb{V}\right]=O_p\left(\frac{1}{\sqrt{NT}}\right)$$
holds and $\frac{1}{NT}E\left(\mathbb{Z}_2'\mathbb{B}_1'\mathbb{P}\mathbb{B}_2\mathbb{V}\right)=O(\frac{1}{T})$.

$\mathbb{B}_2'\mathbb{PB}_1\mathbb{Z}^{(\gamma)}$ and each column of $\mathbb{B}_2'\mathbb{PB}_1\mathbb{Z}_4$ can be viewed as $\mathbb{C}$ because elements of all these vectors are uniformly bounded.
Then $\mathbb{Z}_3$ and $\mathbb{Z}_4$ satisfy the same conclusion because $\frac{1}{NT}\mathbb{C}'\mathbb{V}=O_p(\frac{1}{\sqrt{NT}})$ by Lemma \ref{lm_CQL}.
Thus, equation (\ref{eq_ZPV}) holds and $\frac{1}{NT}E\left(\mathbb{Z}'\mathbb{B}_1'\mathbb{P}\mathbb{B}_2\mathbb{V}\right)=O(\frac{1}{T})$.

\textbf{Proof of equation (\ref{eq_ZPZ})}

We have 
\begin{align*}
\mathbb{Z}'\mathbb{B}_1'\mathbb{P}\mathbb{B}_2\mathbb{Z}=\left(\begin{array}{cccc}
\mathbb{Z}_1'\mathbb{B}_1'\mathbb{P}\mathbb{B}_2\mathbb{Z}_1 & \mathbb{Z}_1'\mathbb{B}_1'\mathbb{P}\mathbb{B}_2\mathbb{Z}_2 & \mathbb{Z}_1'\mathbb{B}_1'\mathbb{P}\mathbb{B}_2\mathbb{Z}^{(\gamma)} & \mathbb{Z}_1'\mathbb{B}_1'\mathbb{P}\mathbb{B}_2\mathbb{Z}_4\\
\mathbb{Z}_2'\mathbb{B}_1'\mathbb{P}\mathbb{B}_2\mathbb{Z}_1 & \mathbb{Z}_2'\mathbb{B}_1'\mathbb{P}\mathbb{B}_2\mathbb{Z}_2 & \mathbb{Z}_2'\mathbb{B}_1'\mathbb{P}\mathbb{B}_2\mathbb{Z}^{(\gamma)} & \mathbb{Z}_2'\mathbb{B}_1'\mathbb{P}\mathbb{B}_2\mathbb{Z}_4\\
\mathbb{Z}^{(\gamma)\prime}\mathbb{B}_1'\mathbb{P}\mathbb{B}_2\mathbb{Z}_1 & \mathbb{Z}^{(\gamma)\prime}\mathbb{B}_1'\mathbb{P}\mathbb{B}_2\mathbb{Z}_2 & \mathbb{Z}^{(\gamma)\prime}\mathbb{B}_1'\mathbb{P}\mathbb{B}_2\mathbb{Z}^{(\gamma)} & \mathbb{Z}^{(\gamma)\prime}\mathbb{B}_1'\mathbb{P}\mathbb{B}_2\mathbb{Z}_4\\
\mathbb{Z}_4'\mathbb{B}_1'\mathbb{P}\mathbb{B}_2\mathbb{Z}_1 & \mathbb{Z}_4'\mathbb{B}_1'\mathbb{P}\mathbb{B}_2\mathbb{Z}_2 & \mathbb{Z}_4'\mathbb{B}_1'\mathbb{P}\mathbb{B}_2\mathbb{Z}^{(\gamma)} & \mathbb{Z}_4'\mathbb{B}_1'\mathbb{P}\mathbb{B}_2\mathbb{Z}_4
\end{array}\right).
\end{align*}
Similar to the proof of Lemma \ref{lm15}, we can first decompose $\mathbb{Z}_1'\mathbb{B}_1'\mathbb{P}\mathbb{B}_2\mathbb{Z}_1\equiv(\mathbb{C}_1+\mathbb{L}^\ell)'\mathbb{P(C}_2+\mathbb{R}^\ell)=\mathbb{C}_1'\mathbb{PC}_2+\mathbb{C}_1'\mathbb{PR}^\ell+\mathbb{C}_2'\mathbb{PL}^\ell+\mathbb{L}^{\ell\prime}\mathbb{PR}^\ell$, where $\frac{1}{NT}\mathbb{C'}_1\mathbb{P}\mathbb{R}^\ell=O_p\left(\frac{1}{\sqrt{NT}}\right)$ and $\frac{1}{NT}\mathbb{C'}_2\mathbb{P}\mathbb{L}^\ell=O_p\left(\frac{1}{\sqrt{NT}}\right)$ by Lemmas \ref{lm_CQL} and \ref{lm_lag}, and $\frac{1}{NT}\mathbb{L}^{\ell\prime}\mathbb{PR}^\ell-\frac{1}{NT}E\left(\mathbb{L}^{\ell\prime}\mathbb{PR}^\ell\right)=O_p\left(\frac{1}{\sqrt{NT^2}}\right)$, $\frac{1}{NT}E\left(\mathbb{L}^{\ell\prime}\mathbb{PR}^\ell\right)=O(\frac{1}{T})$ by Lemmas \ref{lm_PPQ} and \ref{lm_lag}. Moreover, 
$\frac{1}{NT}\left|\mathbb{C}_1'\mathbb{PC}_2\right|=O(1)$
because elements in $\mathbb{C}$ are uniformly bounded and $\mathbb{P}$ is a projection matrix. Hence,
$$\frac{1}{NT}\mathbb{Z}_1'\mathbb{B}_1'\mathbb{P}\mathbb{B}_2\mathbb{Z}_1-\frac{1}{NT}E\left[\mathbb{Z}_1'\mathbb{B}_1'\mathbb{P}\mathbb{B}_2\mathbb{Z}_1\right]=O_p\left(\frac{1}{\sqrt{NT}}\right)$$
and $\frac{1}{NT}E\left(\mathbb{Z}_1'\mathbb{B}_1'\mathbb{P}\mathbb{B}_2\mathbb{Z}_1\right)=O(1)$.
The same conclusion holds for $\mathbb{Z}_1'\mathbb{B}_1'\mathbb{P}\mathbb{B}_2\mathbb{Z}_2$, $\mathbb{Z}_2'\mathbb{B}_1'\mathbb{P}\mathbb{B}_2\mathbb{Z}_1$ and $\mathbb{Z}_2'\mathbb{B}_1'\mathbb{P}\mathbb{B}_2\mathbb{Z}_2$ as they can undergo similar decomposition.

Secondly, we have $\mathbb{Z}_1'\mathbb{B}_1'\mathbb{P}\mathbb{B}_2\mathbb{Z}^{(\gamma)}\equiv(\mathbb{C}_1+\mathbb{L}^\ell)'\mathbb{PC}_2=\mathbb{C}_1'\mathbb{PC}_2+\mathbb{C}_2'\mathbb{PL}^\ell$. Similarly, we have
$$\frac{1}{NT}\mathbb{Z}_1'\mathbb{B}_1'\mathbb{P}\mathbb{B}_2\mathbb{Z}^{(\gamma)}-\frac{1}{NT}E\left[\mathbb{Z}_1'\mathbb{B}_1'\mathbb{P}\mathbb{B}_2\mathbb{Z}^{(\gamma)}\right]=O_p\left(\frac{1}{\sqrt{NT}}\right)$$
and $\frac{1}{NT}E\left(\mathbb{Z}_1'\mathbb{B}_1'\mathbb{P}\mathbb{B}_2\mathbb{Z}^{(\gamma)}\right)=O(1)$ by Lemmas \ref{lm_CQL} and \ref{lm_lag}, which also holds for $\mathbb{Z}^{(\gamma)\prime}\mathbb{B}_1'\mathbb{P}\mathbb{B}_2\mathbb{Z}_1$, $\mathbb{Z}_1'\mathbb{B}_1'\mathbb{P}\mathbb{B}_2\mathbb{Z}_4$, $\mathbb{Z}_4'\mathbb{B}_1'\mathbb{P}\mathbb{B}_2\mathbb{Z}_1$, $\mathbb{Z}_2'\mathbb{B}_1'\mathbb{P}\mathbb{B}_2\mathbb{Z}^{(\gamma)}$, $\mathbb{Z}^{(\gamma)\prime}\mathbb{B}_1'\mathbb{P}\mathbb{B}_2\mathbb{Z}_2$, $\mathbb{Z}_2'\mathbb{B}_1'\mathbb{P}\mathbb{B}_2\mathbb{Z}_4$ and $\mathbb{Z}_4'\mathbb{B}_1'\mathbb{P}\mathbb{B}_2\mathbb{Z}_2$.

Lastly, the scalar $\mathbb{Z}^{(\gamma)\prime}\mathbb{B}_1'\mathbb{P}\mathbb{B}_2\mathbb{Z}^{(\gamma)}$, as well as each element of the vector $\mathbb{Z}^{(\gamma)\prime}\mathbb{B}_1'\mathbb{P}\mathbb{B}_2\mathbb{Z}_4$, $\mathbb{Z}_4\mathbb{B}_1'\mathbb{P}\mathbb{B}_2\mathbb{Z}^{(\gamma)}$ and the matrix $\mathbb{Z}_4'\mathbb{B}_1'\mathbb{P}\mathbb{B}_2\mathbb{Z}_4$ is of the form $\mathbb{C}_1'\mathbb{PC}_2$ and $\frac{1}{NT}\mathbb{C}_1'\mathbb{PC}_2=O(1)$.
Therefore, equation (\ref{eq_ZPZ}) follows and $\frac{1}{NT}E[\mathbb{Z}'\mathbb{B}_1'\mathbb{P}\mathbb{B}_2\mathbb{Z}]=O(1)$.
\end{proof}

\begin{lemma}\label{lm_Lambda}
The results in Lemmas \ref{lm15} and \ref{lm_ZPZ} still hold when $\mathbb{Z}$ is replaced by $\mathbb{Z}\widetilde{\Gamma}$, where $\widetilde{\Gamma}=\mathrm{blkdiag}(I_2,\sqrt{T},I_k)$. Furthermore, 
\begin{align*}
&\sqrt{T}\mathbb{Z}^{(-\gamma)\prime}\mathbb{B}_1'\mathbb{QB}_2\mathbb{Z}^{(\gamma)}=O_p\left(N\sqrt{T}\right) & T\mathbb{Z}^{(\gamma)\prime}\mathbb{B}_1'\mathbb{QB}_2\mathbb{Z}^{(\gamma)}=O(NT).
\end{align*}
\end{lemma}

\begin{proof}
For equations (\ref{eq_ZBV}) and (\ref{eq_ZPV}) with $\mathbb{Z}$ replaced by $\mathbb{Z}\widetilde{\Gamma}$, we only need to further show that 
\begin{align*}
&\frac{\sqrt{T}}{NT}\mathbb{Z}^{(\gamma)\prime}\mathbb{B}_1'\mathbb{QB}_2\mathbb{V}=O_p\left(\frac{1}{\sqrt{NT}}\right)
&\text{and\qquad\quad}
&\frac{\sqrt{T}}{NT}\mathbb{Z}^{(\gamma)\prime}\mathbb{B}_1'\mathbb{PB}_2\mathbb{V}=O_p\left(\frac{1}{\sqrt{NT}}\right),
\end{align*}
which have been proved in Lemma \ref{lm_Z_gamma}.

For equations (\ref{eq_ZBZ}) and (\ref{eq_ZPZ}) with $\mathbb{Z}$ replaced by $\mathbb{Z}\widetilde{\Gamma}$, since $\mathbb{Z}_1$ and $\mathbb{Z}_2$ take the same form and $\mathbb{Z}^{(\gamma)}$ and $\mathbb{Z}_4$ are nonstochastic, we need to further show that 
\begin{align*}
&\frac{1}{N}\mathbb{Z}^{(\gamma)\prime}\mathbb{B}_1'\mathbb{QB}_2\mathbb{Z}^{(\gamma)}=O\left(1\right),
&\frac{1}{N}\mathbb{Z}^{(\gamma)\prime}\mathbb{B}_1'\mathbb{PB}_2\mathbb{Z}^{(\gamma)}=O\left(1\right),\\
&\frac{\sqrt{T}}{NT}\mathbb{Z}_4'\mathbb{B}_1'\mathbb{QB}_2\mathbb{Z}^{(\gamma)}=O\left(\frac{1}{\sqrt{T}}\right),
&\frac{\sqrt{T}}{NT}\mathbb{Z}_4'\mathbb{B}_1'\mathbb{PB}_2\mathbb{Z}^{(\gamma)}=O\left(\frac{1}{\sqrt{T}}\right),
\end{align*}
and
\begin{align*}
&\frac{\sqrt{T}}{NT}\mathbb{Z}_1'\mathbb{B}_1'\mathbb{QB}_2\mathbb{Z}^{(\gamma)}-\frac{\sqrt{T}}{NT}E(\mathbb{Z}_1'\mathbb{B}_1'\mathbb{QB}_2\mathbb{Z}^{(\gamma)})=O_p\left(\frac{1}{\sqrt{NT}}\right),\\
&\frac{\sqrt{T}}{NT}\mathbb{Z}_1'\mathbb{B}_1'\mathbb{PB}_2\mathbb{Z}^{(\gamma)}-\frac{\sqrt{T}}{NT}E(\mathbb{Z}_1'\mathbb{B}_1'\mathbb{PB}_2\mathbb{Z}^{(\gamma)})=O_p\left(\frac{1}{\sqrt{NT}}\right).
\end{align*}

Firstly, we can prove $\frac{1}{N}\mathbb{Z}^{(\gamma)\prime}\mathbb{B}_1'\mathbb{QB}_2\mathbb{Z}^{(\gamma)}=O\left(1\right)$ and
$\frac{1}{N}\mathbb{Z}^{(\gamma)\prime}\mathbb{B}_1'\mathbb{PB}_2\mathbb{Z}^{(\gamma)}=O\left(1\right)$ because $\mathbb{B}_1'\mathbb{PB}_2$ and $\mathbb{B}_1'\mathbb{QB}_2$ are UB and $\mathbb{Z}^{(\gamma)}$ is a sparse binary vector with at most $N$ elements.

Secondly, since elements in $\mathbb{Z}_4$ are uniformly bounded, $\|\mathbb{Z}_4\|_{\infty}=O(1)$,
$\|\mathbb{B}_2'\mathbb{QB}_1\mathbb{Z}_4\|_{\infty}\leq\|\mathbb{B}_2'\mathbb{QB}_1\|_{\infty}\|\mathbb{Z}_4\|_{\infty}=O(1)$, which means that elements in $\mathbb{B}_2'\mathbb{QB}_1\mathbb{Z}_4$ are also uniformly bounded. Therefore, $\frac{\sqrt{T}}{NT}\mathbb{Z}_4'\mathbb{B}_1'\mathbb{QB}_2\mathbb{Z}^{(\gamma)}=O\left(\frac{1}{\sqrt{T}}\right)$ by sparsity of $\mathbb{Z}^{(\gamma)}$ and we can similarly prove $\frac{\sqrt{T}}{NT}\mathbb{Z}_4'\mathbb{B}_1'\mathbb{PB}_2\mathbb{Z}^{(\gamma)}=O\left(\frac{1}{\sqrt{T}}\right)$.

Lastly, $\mathbb{B}_1\mathbb{Z}_1$ can be viewed as $\mathbb{L}^\ell+\mathbb{C}$, where $\frac{\sqrt{T}}{NT}\mathbb{C}\mathbb{QB}_2\mathbb{Z}^{(\gamma)}=O\left(\frac{1}{\sqrt{T}}\right)$ and $\frac{\sqrt{T}}{NT}\mathbb{C}\mathbb{PB}_2\mathbb{Z}^{(\gamma)}=O\left(\frac{1}{\sqrt{T}}\right)$. Moreover, we have $\frac{\sqrt{T}}{NT}\mathbb{L}^{\ell\prime}\mathbb{B}_2\mathbb{Z}^{(\gamma)}=O_p\left(\frac{1}{\sqrt{NT}}\right)$ and $\frac{\sqrt{T}}{NT}\mathbb{L}^{\ell\prime}\mathbb{QB}_2\mathbb{Z}^{(\gamma)}=O_p\left(\frac{1}{\sqrt{NT}}\right)$ by Lemma \ref{lm_Z_gamma}. Taken together, we can prove 
\begin{align*}
&\frac{\sqrt{T}}{NT}\mathbb{Z}_1'\mathbb{B}_1'\mathbb{QB}_2\mathbb{Z}^{(\gamma)}-\frac{\sqrt{T}}{NT}E(\mathbb{Z}_1'\mathbb{B}_1'\mathbb{QB}_2\mathbb{Z}^{(\gamma)})=O_p\left(\frac{1}{\sqrt{NT}}\right),\\
&\frac{\sqrt{T}}{NT}\mathbb{Z}_1'\mathbb{B}_1'\mathbb{PB}_2\mathbb{Z}^{(\gamma)}-\frac{\sqrt{T}}{NT}E(\mathbb{Z}_1'\mathbb{B}_1'\mathbb{PB}_2\mathbb{Z}^{(\gamma)})=O_p\left(\frac{1}{\sqrt{NT}}\right),
\end{align*}
and $\frac{\sqrt{T}}{NT}E(\mathbb{Z}_1'\mathbb{B}_1'\mathbb{QB}_2\mathbb{Z}^{(\gamma)})=O\left(\frac{1}{\sqrt{T}}\right)$, $\frac{\sqrt{T}}{NT}E(\mathbb{Z}_1'\mathbb{B}_1'\mathbb{PB}_2\mathbb{Z}^{(\gamma)})=O\left(\frac{1}{\sqrt{T}}\right)$.

The above proof also indicates that $\sqrt{T}\mathbb{Z}^{(-\gamma)\prime}\mathbb{B}_1'\mathbb{QB}_2\mathbb{Z}^{(\gamma)}=O_p\left(N\sqrt{T}\right)$ and $ T\mathbb{Z}^{(\gamma)\prime}\mathbb{B}_1'\mathbb{QB}_2\mathbb{Z}^{(\gamma)}=O(NT)$.
\end{proof}

Based on the definition of $\widetilde{\mathbb{Z}}^*$ in Section 3.2, the following lemma will be used in Appendix \ref{B2} to show that the expectation of the Hessian matrix and the corresponding part of the variance of the score function are asymptotically equivalent up to an $O(1/T)$ remainder.
\begin{lemma}\label{lm_Z*}
Under Assumptions 1-7 and \ref{asp_A2}, we have
\begin{gather}
\frac{1}{NT}E\left(\widetilde{\Gamma}\mathbb{Z'QZ}\widetilde{\Gamma}\right)-\frac{1}{NT}E\left(\widetilde{\Gamma}\widetilde{\mathbb{Z}}^{*\prime}\widetilde{\mathbb{Z}}^{*}\widetilde{\Gamma}\right)=O\left(\frac{1}{T}\right)\label{eq_Z*Z}\\
\frac{1}{NT}E\left(\widetilde{\Gamma}\mathbb{Z'QG}_0\mathbb{Z}\right)-\frac{1}{NT}E\left(\widetilde{\Gamma}\widetilde{\mathbb{Z}}^{*\prime}\widetilde{\mathbb{G}_0\mathbb{Z}}^{*}\right)=O\left(\frac{1}{T}\right)\label{eq_Z*GZ}\\
\frac{1}{NT}E\left(\mathbb{Z'G}_0'\mathbb{QG}_0\mathbb{Z}\right)-\frac{1}{NT}E\left(\widetilde{\mathbb{G}_0\mathbb{Z}}^{*\prime}\widetilde{\mathbb{G}_0\mathbb{Z}}^{*}\right)=O\left(\frac{1}{T}\right)\label{eq_GZ*GZ}
\end{gather}
where $\frac{1}{NT}E\left(\widetilde{\Gamma}\widetilde{\mathbb{Z}}^{*\prime}\widetilde{\mathbb{Z}}^{*}\widetilde{\Gamma}\right)$, $\frac{1}{NT}E\left(\widetilde{\Gamma}\widetilde{\mathbb{Z}}^{*\prime}\widetilde{\mathbb{G}_0\mathbb{Z}}^{*}\right)$ and $\frac{1}{NT}E\left(\widetilde{\mathbb{G}_0\mathbb{Z}}^{*\prime}\widetilde{\mathbb{G}_0\mathbb{Z}}^{*}\right)$ are all $O(1)$.
\end{lemma}
\begin{proof}
Recall that $\widetilde{\mathbb{Z}}^{*}=\mathbb{QZ+PZ}^\circ$,
$$\frac{1}{NT}E\left(\widetilde{\Gamma}\mathbb{Z'QZ}\widetilde{\Gamma}\right)-\frac{1}{NT}E\left(\widetilde{\Gamma}\widetilde{\mathbb{Z}}^{*\prime}\widetilde{\mathbb{Z}}^{*}\widetilde{\Gamma}\right)=\frac{1}{NT}E\left(\widetilde{\Gamma}\mathbb{Z^{\circ\prime}PZ}^\circ\widetilde{\Gamma}\right)=\frac{1}{NT}E\left(\mathbb{Z^{\circ\prime}PZ}^\circ\right),$$
where the last equality holds by $\mathbb{Z}^\circ=(Z^{\circ\prime}_1,\cdots,Z^{\circ\prime}_T)'$ and $Z^\circ_t=\bigl(D_t(I_N-F_t)D_t'M_t\mathcal{V}_{t-1},\allowbreak D_tD_{t-1}'\mathcal{V}_{t-1},\allowbreak\mathbf{0}_{N_t\times(k+1)}\bigr)$. Since $\mathbb{Z}^\circ$ takes the form of $(\mathbb{L}^\ell,\mathbb{R}^\ell,\mathbf{0}_{n\times(k+1)})$, we have $\frac{1}{NT}E\left(\mathbb{Z^{\circ\prime}PZ}^\circ\right)=O\left(\frac{1}{T}\right)$ by Lemma \ref{lm_PPQ}. Additionally, we have $\frac{1}{NT}E\left(\widetilde{\Gamma}\mathbb{Z'QZ}\widetilde{\Gamma}\right)=O(1)$ by Lemmas \ref{lm15} and \ref{lm_Lambda}. Thus, equation (\ref{eq_Z*Z}) holds and $\frac{1}{NT}E\left(\widetilde{\Gamma}\widetilde{\mathbb{Z}}^{*\prime}\widetilde{\mathbb{Z}}^{*}\widetilde{\Gamma}\right)=O(1)$.

Equations (\ref{eq_Z*GZ}) and (\ref{eq_GZ*GZ}) can be proved analogously by the aforementioned method.
\end{proof}

\begin{lemma}\label{lm_EL4}
Under Assumptions 2 and \ref{asp_A2}, $E\left[\left(L_t^\ell\right)_i^4\right]=O(1)$ uniformly for $i,t,N$.
\end{lemma}
\begin{proof}
Denote $e_i$ as an $N_t$ dimensional vector where the $i$-th component equals 1 and all other components are 0. For simplicity, let $B_t^\ell\equiv B_{1t}^\ell B_{1,t-1}$. Thus, $(L_t^\ell)_i=\sum_{h=0}^\infty e_i'B_{t}^\ell P_{t-1}^{(h)}V_{t-1-h}$. Then using Lemma \ref{lm_quadratic form}, we have
\begin{align*}
&E\left[\left(L_t^\ell\right)_i^4\right]\\
=&E\left[\left(\sum_{h_1=0}^\infty e_i'B_{t}^\ell P_{t-1}^{(h_1)}V_{t-1-h_1}\right)'\left(\sum_{h_2=0}^\infty e_i'B_{t}^\ell P_{t-1}^{(h_2)}V_{t-1-h_2}\right)\left(\sum_{h_3=0}^\infty e_i'B_{t}^\ell P_{t-1}^{(h_3)}V_{t-1-h_3}\right)'\right.\\
&\left.\left(\sum_{h_4=0}^\infty e_i'B_{t}^\ell P_{t-1}^{(h_4)}V_{t-1-h_4}\right)\right]\\
=&\sum_{h_1,h_2,h_3,h_4=0}^\infty E\left[\left(V_{t-1-h_1}'P_{t-1}^{(h_1)\prime}B_{t}^{\ell\prime}e_ie_i'B_{t}^\ell P_{t-1}^{(h_2)}V_{t-1-h_2}\right)\left(V_{t-1-h_3}'P_{t-1}^{(h_3)\prime}B_{t}^{\ell\prime}e_ie_i'B_{t}^\ell P_{t-1}^{(h_4)}V_{t-1-h_4}\right)\right]\\
=&\sum_{h=0}^\infty(\mu_4-3\sigma_0^4)\sum_{j=1}^{N_{t-1-h}}\left[\left(P_{t-1}^{(h)\prime}B_{t}^{\ell\prime}e_ie_i'B_{t}^\ell P_{t-1}^{(h)}\right)_{jj}\right]^2+\\
&\sum_{h_1=0}^\infty\sum_{h_2=0}^\infty\sigma_0^4\mathrm{tr}\left(P_{t-1}^{(h_1)\prime}B_{t}^{\ell\prime}e_ie_i'B_{t}^\ell P_{t-1}^{(h_1)}\right)\mathrm{tr}\left(P_{t-1}^{(h_2)\prime}B_{t}^{\ell\prime}e_ie_i'B_{t}^\ell P_{t-1}^{(h_2)}\right)+\\
&\sum_{h_1=0}^\infty\sum_{h_2=0}^\infty\sigma_0^4\mathrm{tr}\left[P_{t-1}^{(h_1)\prime}B_{t}^{\ell\prime}e_ie_i'B_{t}^\ell P_{t-1}^{(h_2)}\left(P_{t-1}^{(h_1)\prime}B_{t}^{\ell\prime}e_ie_i'B_{t}^\ell P_{t-1}^{(h_2)}\right)'\right]+\\
&\sum_{h_1=0}^\infty\sum_{h_2=0}^\infty\sigma_0^4\mathrm{tr}\left[P_{t-1}^{(h_1)\prime}B_{t}^{\ell\prime}e_ie_i'B_{t}^\ell P_{t-1}^{(h_2)}\left(P_{t-1}^{(h_2)\prime}B_{t}^{\ell\prime}e_ie_i'B_{t}^\ell P_{t-1}^{(h_1)}\right)'\right]\\
=&\sum_{h=0}^\infty(\mu_4-3\sigma_0^4)\sum_{j=1}^{N_{t-1-h}}\left[\left(P_{t-1}^{(h)\prime}B_{t}^{\ell\prime}e_ie_i'B_{t}^\ell P_{t-1}^{(h)}\right)_{jj}\right]^2+\\
&\sigma_0^4\sum_{h_1=0}^\infty\sum_{h_2=0}^\infty\left[\left(e_i'B_{t}^\ell P_{t-1}^{(h_1)}P_{t-1}^{(h_1)\prime}B_{t}^{\ell\prime}e_i\right)\left(e_i'B_{t}^\ell P_{t-1}^{(h_2)}P_{t-1}^{(h_2)\prime}B_{t}^{\ell\prime}e_i\right)+\right.\\
&\left.\left(e_i'B_{t}^\ell P_{t-1}^{(h_1)}P_{t-1}^{(h_1)\prime}B_{t}^{\ell\prime}e_ie_i'B_{t}^\ell P_{t-1}^{(h_2)}P_{t-1}^{(h_2)\prime}B_{t}^{\ell\prime}e_i\right)+\left(e_i'B_{t}^\ell P_{t-1}^{(h_2)}P_{t-1}^{(h_1)\prime}B_{t}^{\ell\prime}e_ie_i'B_{t}^\ell P_{t-1}^{(h_2)}P_{t-1}^{(h_1)\prime}B_{t}^{\ell\prime}e_i\right)\right]\\
=&\sum_{h=0}^\infty(\mu_4-3\sigma_0^4)\sum_{j=1}^{N_{t-1-h}}\left[\left(B_{t}^\ell P_{t-1}^{(h)}\right)_{ij}\right]^4+2\sigma_0^4\left[\sum_{h=0}^\infty\left(B_{t}^\ell P_{t-1}^{(h)}P_{t-1}^{(h)\prime}B_{t}^{\ell\prime}\right)_{ii}\right]^2+\\
&\sigma_0^4\sum_{h_1=0}^\infty\sum_{h_2=0}^\infty\left[\left(B_{t}^\ell P_{t-1}^{(h_2)}P_{t-1}^{(h_1)\prime}B_{t}^{\ell\prime}\right)_{ii}\right]^2.
\end{align*}
 
\begin{align*}
\sum_{h=0}^\infty(\mu_4-3\sigma_0^4)\sum_{j=1}^{N_{t-1-h}}\left[\left(B_{t}^\ell P_{t-1}^{(h)}\right)_{ij}\right]^4&=\sum_{h=0}^\infty(\mu_4-3\sigma_0^4)\sum_{j=1}^{N}\left[\left(B_{t}^\ell P_{t-1}^{(h)}D_{t-1-h}\right)_{ij}\right]^4\\
&\leq(\mu_4-3\sigma_0^4)\left[\sum_{j=1}^{N}\sum_{h=0}^\infty\mathrm{abs}\left(B_{t}^\ell P_{t-1}^{(h)}D_{t-1-h}\right)_{ij}\right]^4\\
&=O(1),
\end{align*}
where the last equality holds because matrix $\sum_{h=0}^\infty\mathrm{abs}\left(B_{t}^\ell P_{t-1}^{(h)}D_{t-1-h}\right)$ is UB uniformly for $t$ according to Lemma \ref{lm33}. Moreover,
\begin{align*}
\left[\sum_{h=0}^\infty\left(B_{t}^\ell P_{t-1}^{(h)}P_{t-1}^{(h)\prime}B_{t}^{\ell\prime}\right)_{ii}\right]^2\leq\left\{\left[\sum_{h=0}^\infty\mathrm{abs}\left(B_{t}^\ell P_{t-1}^{(h)}P_{t-1}^{(h)\prime}B_{t}^{\ell\prime}\right)\right]_{ii}\right\}^2=O(1),
\end{align*}
where the equality holds because $\sum_{h=0}^\infty\mathrm{abs}\left(B_{t}^\ell P_{t-1}^{(h)}P_{t-1}^{(h)\prime}B_{t}^{\ell\prime}\right)$ is UB according to Lemma \ref{lm33}. Similarly,
\begin{align*}
&\sum_{h_1=0}^\infty\sum_{h_2=0}^\infty\left[\left(B_{t}^\ell P_{t-1}^{(h_2)}P_{t-1}^{(h_1)\prime}B_{t}^{\ell\prime}\right)_{ii}\right]^2\leq\left\{\left[\sum_{h_1=0}^\infty\sum_{h_2=0}^\infty\mathrm{abs}\left(B_{t}^\ell P_{t-1}^{(h_2)}D_{t-1-h}D_{t-1-h}'P_{t-1}^{(h_1)\prime}B_{t}^{\ell\prime}\right)\right]_{ii}\right\}^2\\
\leq&\left[\left(\sum_{h_2=0}^\infty\mathrm{abs}\left(B_{t}^\ell P_{t-1}^{(h_2)}D_{t-1-h}\right)\right)\left(\sum_{h_1=0}^\infty\mathrm{abs}\left(B_{t}^\ell P_{t-1}^{(h_1)}D_{t-1-h}\right)\right)\right]_{ii}^2=O(1),
\end{align*}
where the equality holds because $\mathrm{abs}\left(B_{t}^\ell P_{t-1}^{(h)}D_{t-1-h}\right)$ is UB uniformly for $t$ according to Lemma \ref{lm33}. Therefore, $E\left[\left(L_t^\ell\right)_i^4\right]=O(1)$.
\end{proof}

%
To establish the central limit theorem for $\Gamma\frac{\partial L_n^{c*}(\theta_0)}{\partial\theta}$, let 
\begin{align}
\mathcal{Q}_n&\equiv\sum_{t=1}^T\left[V_t'B_{3t}V_t+V_t'L_t^\ell+V_t'\widetilde{C}_t-\sigma_0^2\mathrm{tr}(B_{3t})\right]\nonumber\\
&=\sum_{t=1}^T\left[V_t'B_{3t}V_t+V_t'B_{1t}^\ell L_{t-1}+V_t'\left(C_{t}+a\sqrt{T}D_tF_t\iota_N\right)-\sigma_0^2\mathrm{tr}(B_{3t})\right]\equiv\sum_{t=1}^T\sum_{i=1}^{N_t}\mathfrak{q}_{n,ti},\label{eq_clt}
\end{align}
where $B_{3t}$ is an arbitrary symmetric\footnote{The symmetry of $B_3$ can be imposed without loss of generality because $V_t'B_{3t}V_t=V_t'[(B_{3t}+B_{3t}')/2]V_t$.} UB matrix uniformly for $t$. The $N_t$-dimensional vector $\widetilde{C}_t\equiv C_{t}+a\sqrt{T}D_tF_t\iota_N$, where $C_{t}$ is an arbitrary vector satisfying Assumption \ref{asp_C}, and $a$ is a constant.
Furthermore, $B_{1t}^\ell$ is an arbitrary matrix satisfying Assumption \ref{asp_A2}.
$\mathfrak{q}_{n,ti}\equiv2\sum_{j=1}^{i-1}(B_{3t})_{ij}V_{t,j}V_{t,i}+[(L_t^\ell)_i+(\widetilde{C}_t)_i]V_{t,i}+(B_{3t})_{ii}(V_{t,i}^2-\sigma_0^2)$, and $V_{t,i}$ is the $i$-th element in $V_t$. 
Linear combinations of terms satisfying equation (\ref{eq_clt}) can still be written in the same form and do not violate the original assumptions, because 
\begin{align*}
&w^{(1)}\mathcal{Q}_n^{(1)}+w^{(2)}\mathcal{Q}_n^{(2)}\\
=&\sum_{t=1}^T\left[V_t'\left(w^{(1)}B_{3t}^{(1)}+w^{(2)}B_{3t}^{(2)}\right)V_t+V_t'\left(w^{(1)}B_{1t}^{\ell(1)}+w^{(2)}B_{1t}^{\ell(2)}\right) L_{t-1}+\right.\\
&\left.V_t'\left(w^{(1)}C_t^{(1)}+w^{(2)}C_t^{(2)}+(w^{(1)}a^{(1)}+w^{(2)}a^{(2)})\sqrt{T}D_tF_t\iota_N\right)-\sigma_0^2\mathrm{tr}\left(w^{(1)}B_{3t}^{(1)}+w^{(2)}B_{3t}^{(2)}\right)\right].
\end{align*}
Then we have $E(\mathcal{Q}_n)=0$ and the variance is
\begin{align*}
\sigma^2_{\mathcal{Q}_n}=&E\left[\left(\sum_{t=1}^TV_t'B_{3t}V_t+V_t'L_t^\ell+V_t'\widetilde{C}_t-\sigma_0^2\mathrm{tr}(B_{3t})\right)^2\right]\\
=&E\Bigg[\sum_{t=1}^T\sum_{s=1}^T\left(V_t'B_{3t}V_tV_s'B_{3s}V_s+2V_t'B_{3t}V_tV_s'(L_s^\ell+\widetilde{C}_s)-2V_t'B_{3t}V_t\sigma_0^2\mathrm{tr}(B_{3s})+\right.\\
&\left. V_t'(L_t^\ell+\widetilde{C}_t)V_s'(L_s^\ell+\widetilde{C}_s)-2V_t'(L_t^\ell+\widetilde{C}_t)\sigma_0^2\mathrm{tr}(B_{3s})+\sigma_0^4\mathrm{tr}(B_{3t})\mathrm{tr}(B_{3s})\right)\Bigg]\\
=&\sum_{t=1}^T\left[(\mu_4-3\sigma_0^4)\sum_{i=1}^{N_t}(B_{3t})_{ii}^2+2\sigma_0^4\mathrm{tr}(B_{3t}^2)\right]+2\mu_3\sum_{t=1}^T\sum_{i=1}^{N_t}(B_{3t})_{ii}(\widetilde{C}_t)_i+\\
&\sigma_0^4\sum_{t=1}^T\sum_{h=0}^\infty\mathrm{tr}(B_{1t}^\ell P_{t-1,h}P_{t-1,h}'B_{1t}^{\ell\prime})+\sigma_0^2\sum_{t=1}^T\widetilde{C}_t'\widetilde{C}_t\\
=&\sum_{t=1}^T\left[(\mu_4-3\sigma_0^4)\sum_{i=1}^{N_t}(B_{3t})_{ii}^2+2\sigma_0^4\mathrm{tr}(B_{3t}^2)\right]+2\mu_3\sum_{t=1}^T\sum_{i=1}^{N_t}(B_{3t})_{ii}\left[(C_t)_i+a\sqrt{T}(D_tF_t\iota_N)_i\right]+\\
&+\sigma_0^4\sum_{t=1}^T\mathrm{tr}\left[B_{1t}^\ell\sum_{h=0}^\infty\left(P_{t-1,h}P_{t-1,h}'\right)B_{1t}^{\ell\prime}\right]+\sigma_0^2\sum_{t=1}^T\left(C_t+a\sqrt{T}D_tF_t\iota_N\right)'\left(C_t+a\sqrt{T}D_tF_t\iota_N\right)\\
=&O(NT),
\end{align*}
where the last equality holds because $B_{3t}$ is UB uniformly for $t$, elements in $C_t$ are uniformly bounded, $F_t\iota_N$ is a sparse vector with fewer than $N$ nonzero entries in total over $t=1,\cdots,T$, and the matrix inside the trace operation is UB by Lemma \ref{lm33}.

\begin{lemma}\label{lm_CLT}
Under Assumptions 2, 7, \ref{asp_A2} and \ref{asp_C}, if the sequence $\frac{1}{n}\sigma^2_{\mathcal{Q}_n}$ is bounded away from zero, then $\frac{\mathcal{Q}_n}{\sigma_{\mathcal{Q}_n}}\overset{d}{\longrightarrow}N(0,1)$.
\end{lemma}
\begin{proof}
The proof of central limit theorem follows the idea in Yu et al. (2008). Denote the $\sigma$-field
\begin{align*}
\mathcal{F}_{n,ti}=\sigma\left(\left( \bigcup_{s=1}^{\infty}\left\{V_{t-s,j}\mid j=1,2,\cdots,N_{t-s}\right\}\right)\cup\left\{V_{t,j}\mid j=1,2,\cdots,i\right\}\right), \\
\text{ for }t=1,2\cdots,T\text{, and }i=1,2,\cdots,N_t,
\end{align*}
which is generated by the $i$-th node's error observed at time $t$ and all its historical errors, where ``historical" includes both errors from all time periods preceding $t$ and those within period $t$ that are ordered before the $i$-th node. Define $\mathcal{F}_{n,t0}=\mathcal{F}_{n,t-1,N_{t-1}}$.
To merge the two indices $t$ and $i$ into a single index, define a global index $j=\sum_{s=1}^{t-1} N_s+i$, where $j$ ranges from $1$ to $n$. The $\sigma$-field $\mathcal{F}_{n,ti}$ is then simplified as $\mathcal{F}_{n,j}$, and $\mathfrak{q}_{n,ti}$ is simplified as $\mathfrak{q}_{n,j}$.
Thus, $E(\mathfrak{q}_{n,j}|\mathcal{F}_{n,j-1})=0$,
which forms a martingale difference array. To apply the martingale CLT to $\sum_{j=1}^n\mathfrak{q}_{n,j}=\sum_{t=1}^{T}\sum_{i=1}^{N_t}\mathfrak{q}_{n,ti}$, the sufficient conditions are (i) $\frac{1}{\sigma^{2+\eta}_{\mathcal{Q}_n}}\sum_{t=1}^{T}\sum_{i=1}^{N_t}|\mathfrak{q}_{n,ti}|^{2+\eta}\rightarrow0$ and (ii) $\frac{1}{\sigma^{2}_{\mathcal{Q}_n}}\sum_{t=1}^{T}\sum_{i=1}^{N_t}E(\mathfrak{q}_{n,ti}^2|\mathcal{F}_{n,t,i-1})\overset{p}{\rightarrow}1$.

\textbf{Proof of (i)}

For any $p>0$, $q>0$ with $\frac{1}{p}+\frac{1}{q}=1$, we have
\begin{align*}
&|\mathfrak{q}_{n,ti}|\\
\leq&\left(\left|(L_t^\ell)_i\right|+\left|(\widetilde{C}_t)_i\right|\right)|V_{t,i}|+|(B_{3t})_{ii}|^{\frac{1}{p}}|(B_{3t})_{ii}|^{\frac{1}{q}}\left|V_{t,i}^2-\sigma_0^2\right|+\sum_{j=1}^{i-1}|(B_{3t})_{ij}|^{\frac{1}{p}}\left(2|(B_{3t})_{ij}|^{\frac{1}{q}}|V_{t,j}||V_{t,i}|\right)\\
\leq&\left[\left(\left|(L_t^\ell)_i\right|+\left|(\widetilde{C}_t)_i\right|\right)^p+\sum_{j=1}^{i}\left|(B_{3t})_{ij}\right|\right]^{\frac{1}{p}}\left[|V_{t,i}|^{q}+\left|(B_{3t})_{ii}\right|\left|V_{t,i}^2-\sigma_0^2\right|^q+\sum_{j=1}^{i-1}2^q|(B_{3t})_{ij}||V_{t,j}|^q|V_{t,i}|^q\right]^{\frac{1}{q}}
\end{align*}
where the second inequality is derived by applying Hölder's inequality. Thus,
\begin{align*}
E\left[|\mathfrak{q}_{n,ti}|^q\right]\leq&E\left[\left(\left(\left|(L_t^\ell)_i\right|+\left|(\widetilde{C}_t)_i\right|\right)^p+\sum_{j=1}^{i}\left|(B_{3t})_{ij}\right|\right)^{\frac{q}{p}}\right]\times\\
&\left[E\left(|V_{t,i}|^{q}\right)+\left|(B_{3t})_{ii}\right|E\left(\left|V_{t,i}^2-\sigma_0^2\right|^q\right)+2^qE\left(|V_{t,i}|^q\right)\sum_{j=1}^{i-1}|(B_{3t})_{ij}|E\left(|V_{t,j}|^q\right)\right],
\end{align*}
by the independence between $V_t$ and $L_t^\ell$.
Let $q=2+\eta$ with a small $\eta$ defined in Assumption 2, then $p=\frac{2+\eta}{1+\eta}$. There exists $c_1>0$ such that 
$$\sup_{t,i}\left[E|V_{t,i}|^{q}+|(B_{3t})_{ii}|E\left|V_{t,i}^2-\sigma_0^2\right|^q+2^qE|V_{t,i}|^q\sum_{j=1}^{i-1}|(B_{3t})_{ij}|E|V_{t,j}|^q\right]<c_1,$$ due to  $E|v_{it}|^{4+2\eta}<\infty$ by Assumption 2 and $B_{3t}$ is UB uniformly for $t$. Therefore, 
\begin{align*}
&\sum_{t=1}^T\sum_{i=1}^{N_t}E|\mathfrak{q}_{n,ti}|^{2+\eta}\\
\leq&c_1\sum_{t=1}^T\sum_{i=1}^{N_t}E\left[\left(\left(\left|(L_t^\ell)_i\right|+\left|(\widetilde{C}_t)_i\right|\right)^p+\sum_{j=1}^{i}\left|(B_{3t})_{ij}\right|\right)^{\frac{q}{p}}\right]\\
\leq&2^\eta c_1\sum_{t=1}^T\sum_{i=1}^{N_t}E\left[\left(\left|(L_t^\ell)_i\right|+\left|(C_t)_i\right|+\left|a\sqrt{T}(D_tF_t\iota_N)_i\right|\right)^{2+\eta}\right]+\sum_{t=1}^T\sum_{i=1}^{N_t}2^\eta c_1\left(\sum_{j=1}^{i}\left|(B_{3t})_{ij}\right|\right)^{1+\eta}\\
\leq&2^{1+2\eta}c_1\sum_{t=1}^T\sum_{i=1}^{N_t}
\left[E\left(\left|(L_t^\ell)_i\right|^{2+\eta}\right)+\left|(C_t)_i\right|^{2+\eta}+T^{1+\frac{\eta}{2}}\left|a(D_tF_t\iota_N)_i\right|^{2+\eta}+\left(\sum_{j=1}^{n}\left|(B_{3t})_{ij}\right|\right)^{1+\eta}\right]\\
=&O\left(NT^{1+\frac{\eta}{2}}\right),
\end{align*}
where the last two inequalities are derived by applying the $C_r$-inequality.
By Lemma \ref{lm_EL4}, which establishes $E\left(\left|(L_t^\ell)_i\right|^{4}\right) = O(1)$ uniformly for $t$ and $i$, combined with the boundedness of $C_t$, the sparsity of $F_t\iota_N$, and $B_{3t}$ being UB uniformly for $t$, the last equality follows.
Since the sequence $\frac{1}{n}\sigma^2_{\mathcal{Q}_n}$ is bounded away from zero, then $\sigma^{2+\eta}_{\mathcal{Q}_n}=O\left((NT)^{1+\frac{\eta}{2}}\right)$ and it is nonnegligible at this order, we have $\frac{1}{\sigma^{2+\eta}_{\mathcal{Q}_n}}\sum_{j=1}^nE|\mathfrak{q}_{n,j}|^{2+\eta}=O\left(N^{-\frac{\eta}{2}}\right)$, which proves condition (i).

\textbf{Proof of (ii)}

Since
\begin{align*}
E\left(\mathfrak{q}_{n,ti}^2|\mathcal{F}_{n,t,i-1}\right)=&\left[2\sum_{j=1}^{i-1}(B_{3t})_{ij}V_{t,j}+(L_t^\ell)_i+(\widetilde{C}_t)_i\right]^2E(V_{t,i}^2)+(B_{3t})_{ii}^2E\left[(V_{t,i}^2-\sigma_0^2)^2\right]+\\
&2\left[2\sum_{j=1}^{i-1}(B_{3t})_{ij}V_{t,j}+(L_t^\ell)_i+(\widetilde{C}_t)_i\right](B_{3t})_{ii}E\left[V_{t,i}(V_{t,i}^2-\sigma_0^2)\right]\\
=&\sigma_0^2\left[2\sum_{j=1}^{i-1}(B_{3t})_{ij}V_{t,j}+(L_t^\ell)_i+(\widetilde{C}_t)_i\right]^2+(\mu_4-\sigma_0^4)(B_{3t})_{ii}^2+\\
&2\mu_3(B_{3t})_{ii}\left[2\sum_{j=1}^{i-1}(B_{3t})_{ij}V_{t,j}+(L_t^\ell)_i+(\widetilde{C}_t)_i\right].
\end{align*}
Then
\begin{align*}
&\sum_{t=1}^T\sum_{i=1}^{N_t}E\left(\mathfrak{q}_{n,ti}^2|\mathcal{F}_{n,t,i-1}\right)\\
=&\sum_{t=1}^T\sum_{i=1}^{N_t}\sigma_0^2\left[2\sum_{j=1}^{i-1}(B_{3t})_{ij}V_{t,j}+(L_t^\ell)_i+(\widetilde{C}_t)_i\right]^2+\sum_{t=1}^T\sum_{i=1}^{N_t}(\mu_4-\sigma_0^4)(B_{3t})_{ii}^2\\
&+\sum_{t=1}^T\sum_{i=1}^{N_t}2\mu_3(B_{3t})_{ii}\left[2\sum_{j=1}^{i-1}(B_{3t})_{ij}V_{t,j}+(L_t^\ell)_i+(\widetilde{C}_t)_i\right]\\
=&\sum_{t=1}^T\sum_{i=1}^{N_t}4\sigma_0^2\left[\sum_{j=1}^{i-1}(B_{3t})_{ij}V_{t,j}\right]^2+\sum_{t=1}^T\sum_{i=1}^{N_t}\sigma_0^2(L_t^\ell)_i^2+\sum_{t=1}^T\sum_{i=1}^{N_t}\sigma_0^2(\widetilde{C}_t)_i^2\\
&+\sum_{t=1}^T\sum_{i=1}^{N_t}\left[4\sigma_0^2(C_t)_i+4\mu_3(B_{3t})_{ii}\right]\left[\sum_{j=1}^{i-1}(B_{3t})_{ij}V_{t,j}\right]+\sum_{t=1}^T\sum_{i=1}^{N_t}[2\sigma_0^2(C_t)_i+2\mu_3(B_{3t})_{ii}](L_t^\ell)_i\\
&+\sum_{t=1}^T\sum_{i=1}^{N_t}4a\sqrt{T}(D_tF_t\iota_N)_i\left[\sum_{j=1}^{i-1}(B_{3t})_{ij}V_{t,j}\right]+\sum_{t=1}^T\sum_{i=1}^{N_t}2a\sqrt{T}(D_tF_t\iota_N)_i(L_t^\ell)_i\\
&+\sum_{t=1}^T\sum_{i=1}^{N_t}4\sigma_0^2\left[\sum_{j=1}^{i-1}(B_{3t})_{ij}V_{t,j}\right](L_t^\ell)_i+\sum_{t=1}^T(\mu_4-\sigma_0^4)\sum_{i=1}^{N_t}(B_{3t})_{ii}^2+2\mu_3\sum_{t=1}^T\sum_{i=1}^{N_t}(B_{3t})_{ii}(\widetilde{C}_t)_i.
\end{align*}

Denote $\mathbb{B}_{3}\equiv$ $\mathrm{blkdiag}(B_{31},$ $B_{32},\cdots,B_{3T})$, then
$$\sum_{t=1}^T\sum_{i=1}^{N_t}\left[\sum_{j=1}^{i-1}(B_{3t})_{ij}V_{t,j}\right]^2=\sum_{t=1}^T\sum_{i=1}^{N_t}V_t'B_{3t}^{-\prime}B_{3t}^{-}V_t=\mathbb{V}'\mathbb{B}_{3}^{-\prime}\mathbb{B}_{3}^-\mathbb{V},$$ thus by equation (\ref{eq_vbv}) from Lemma \ref{lm_VQV},
$$\frac{1}{NT}\mathbb{V}'\mathbb{B}_{3}^{-\prime}\mathbb{B}_{3}^-\mathbb{V}-\frac{1}{NT}E(\mathbb{V}'\mathbb{B}_{3}^{-\prime}\mathbb{B}_{3}^-\mathbb{V})=O_p\left(\frac{1}{\sqrt{NT}}\right),$$
where $E(\mathbb{V}'\mathbb{B}_{3}^{-\prime}\mathbb{B}_{3}^-\mathbb{V})=\sigma_0^2\sum_{t=1}^T\mathrm{tr}(B_{3t}^{-\prime}B_{3t}^{-})$.
Additionally, $\sum_{t=1}^T\sum_{i=1}^{N_t}\sigma_0^2(L_t^\ell)_i^2=\sigma_0^2\sum_{t=1}^TL_t^{\ell\prime}L_t^\ell=\sigma_0^2\mathbb{L}^{\ell\prime}\mathbb{L}^\ell$, according to equation (\ref{eq_lr}) and Lemma \ref{lm_lag},
$$\frac{1}{NT}\mathbb{L}^{\ell\prime}\mathbb{L}^\ell-\frac{1}{NT}E(\mathbb{L}^{\ell\prime}\mathbb{L}^\ell)=O_P\left(\frac{1}{\sqrt{NT}}\right),$$
where $E(\mathbb{L}^{\ell\prime}\mathbb{L}^\ell)=\sigma_0^2\sum_{t=1}^T\mathrm{tr}(\sum_{h=0}^\infty B_{1t}^\ell P_{t-1,h}P_{t-1,h}'B_{1t}^{\ell\prime})$.
Denote $C_{2t}$ as an $N_t$ dimensional vector such that elements in $C_{2t}$ are uniformly bounded and $(C_{2t})_i=4\sigma_0^2(C_t)_i+4\mu_3(B_{3t})_{ii}$, then $$\sum_{t=1}^T\sum_{i=1}^{N_t}\left[4\sigma_0^2(C_t)_i+4\mu_3(B_{3t})_{ii}\right]\left[\sum_{j=1}^{i-1}(B_{3t})_{ij}V_{t,j}\right]=\sum_{t=1}^TC_{2t}'B_{3t}^-V_t=O_p\left(\sqrt{NT}\right)$$ by equation (\ref{eq_cv}) in Lemma \ref{lm_CQL}.
Similarly, 
$$\sum_{t=1}^T\sum_{i=1}^{N_t}[2\sigma_0^2(C_t)_i+2\mu_3(B_{3t})_{ii}](L_t^\ell)_i=\sum_{t=1}^TC_{2t}'L_t^\ell=O_p\left(\sqrt{NT}\right)$$
by equation (\ref{eq_cl}) from Lemma \ref{lm_CQL}. Moreover 
$$\sum_{t=1}^T\sum_{i=1}^{N_t}a\sqrt{T}(D_tF_t\iota_N)_i\left[\sum_{j=1}^{i-1}(B_{3t})_{ij}V_{t,j}\right]=\sum_{t=1}^Ta\sqrt{T}(D_tF_t\iota_N)'B_{3t}^-V_t=a\sqrt{T}\mathbb{Z}^{(\gamma)\prime}\mathbb{B}_3^-\mathbb{V}$$ and $$\sum_{t=1}^T\sum_{i=1}^{N_t}a\sqrt{T}(D_tF_t\iota_N)_i(L_t^\ell)_i=a\sqrt{T}\mathbb{Z}^{(\gamma)\prime}\mathbb{L}^\ell,$$ which are both $O_p(\sqrt{NT})$ by equations (\ref{eq_ZgBV}) and (\ref{eq_ZgBL}) by Lemma \ref{lm_Z_gamma}.
Furthermore, we have $$\sum_{t=1}^T\sum_{i=1}^{N_t}\left[\sum_{j=1}^{i-1}(B_{3t})_{ij}V_{t,j}\right](L_t^\ell)_i=\sum_{t=1}^TL_t^{\ell\prime} B_{3t}^-V_t=\mathbb{L}^{\ell\prime}\mathbb{B}_3^-\mathbb{V}=O_p\left(\sqrt{NT}\right)$$ by equation (\ref{eq_lbv}) in Lemma \ref{lm_LQV}.
Therefore,
\begin{align*}
&\sum_{t=1}^T\sum_{i=1}^{N_t}E\left(\mathfrak{q}_{n,ti}^2|\mathcal{F}_{n,t,i-1}\right)\\
=&\sum_{t=1}^T4\sigma_0^4\mathrm{tr}\left(B_{3t}^{-\prime}B_{3t}^{-}\right)+\sigma_0^4\sum_{t=1}^T\mathrm{tr}\left(\sum_{h=0}^\infty B_{1t}^\ell P_{t-1,h}P_{t-1,h}'B_{1t}^{\ell\prime}\right)+\sum_{t=1}^T\sum_{i=1}^{N_t}\sigma_0^2(\widetilde{C}_t)_i^2\\
&+\sum_{t=1}^T(\mu_4-\sigma_0^4)\sum_{i=1}^{N_t}(B_{3t})_{ii}^2+2\mu_3\sum_{t=1}^T\sum_{i=1}^{N_t}(B_{3t})_{ii}(\widetilde{C}_t)_i+O_p(\sqrt{NT}).
\end{align*}
Since $B_{3t}$ is symmetric, which implies that $\mathrm{tr}(B_{3t}^2)=2tr(B_{3t}^{-\prime}B_{3t}^-)+\sum_{i=1}^{N_t}(B_{3t})_{ii}^2$. So
$$\frac{1}{NT}E\left(\mathfrak{q}_{n,ti}^2|\mathcal{F}_{n,t,i-1}\right)=\frac{1}{NT}\sigma^2_{\mathcal{Q}_n}+O_p\left(\frac{1}{\sqrt{NT}}\right),$$
which proves condition (ii).
\end{proof}

\subsection*{Additional lemmas for the TWFE case}
Recall that $\mathbb{L}^\ell$ and $\mathbb{R}^\ell$ are defined above. We can write $\mathbb{L}^\ell = \mathbb{B}_1^\ell \mathcal{P}_\infty^\ell \mathbb{V}_\infty$ and $\mathbb{R}^\ell = \mathbb{B}_2^\ell \mathcal{Q}_\infty^\ell \mathbb{V}_\infty$, where $\mathbb{B}_1^\ell = \mathrm{blkdiag}(B_{11}^\ell,\ldots,B_{1T}^\ell)$, $\mathbb{B}_2^\ell = \mathrm{blkdiag}(B_{21}^\ell,\ldots,B_{2T}^\ell)$ are $n\times(\sum_{t=0}^{T-1}N_t)$ matrices,
$\mathbb{V}_{\infty}\equiv
(V_T',V_{T-1}',\ldots,V_1'$, $V_0',V_{-1}',\ldots)'$ is infinity dimensional vector,
$$\mathcal{P}_\infty^\ell = \begin{pmatrix}
\mathbf{0}_{N_0 \times N_T} & \mathbf{0}_{N_0 \times N_{T-1}} & \dots & \mathbf{0}_{N_0 \times N_1} & B_{10}P_0^{(0)} & B_{10}P_0^{(1)} &  \dots \\
\mathbf{0}_{N_1 \times N_T} & \mathbf{0}_{N_1 \times N_{T-1}} & \dots & B_{11}P_1^{(0)} & B_{11}P_1^{(1)} & B_{11}P_1^{(2)} &  \dots \\
\vdots & \vdots & \iddots & \vdots & \vdots & \vdots &  \ddots \\
\mathbf{0}_{N_{T-1} \times N_T} & B_{1,T-1}P_{T-1}^{(0)} & \dots & B_{1,T-1}P_{T-1}^{(T-2)} & B_{1,T-1}P_{T-1}^{(T-1)} & B_{1,T-1}P_{T-1}^{(T)} &  \dots
\end{pmatrix},$$ is $(\sum_{t=0}^{T-1}N_t)\times\infty$ matrix, and $\mathcal{Q}_{\infty}^\ell$ is defined analogously.
\begin{lemma}\label{lem:Pinfty-UB}
Under Assumption~\ref{asp_A2}, both $\mathcal{P}_{\infty}^\ell$ and $\mathcal{Q}_{\infty}^\ell$ are UB. Consequently,
$\mathbb{B}_1^\ell \mathcal{P}_\infty^\ell$ and $\mathbb{B}_2^\ell \mathcal{Q}_\infty^\ell$ are also UB.
\end{lemma}

\begin{proof}
We only prove the claim for $\mathcal{P}_{\infty}^\ell$, since the proof for $\mathcal{Q}_{\infty}^\ell$ is identical.
By Assumption~\ref{asp_A2} and Lemma \ref{lm33},
\[
\sup_{N,T}\|\mathcal{P}_{\infty}^\ell\|_{\infty}
\leq
\sup_{N,T}\sup_{0\le t\le T-1}
\left\|
\sum_{h=0}^{\infty}\mathrm{abs}\!\left(B_{1t}P_t^{(h)}D_{t-h}\right)
\right\|_{\infty}
<\infty,
\]
and
\[
\sup_{N,T}\|\mathcal{P}_{\infty}^\ell\|_{1}
\leq
\sup_{N,T}\sup_{t\le T-1}
\left\|
\sum_{h=t}^{\infty}\mathrm{abs}\!\left(D_h'B_{1h}P_h^{(h-t)}\right)
\right\|_{1}
<\infty.
\]
Hence $\mathcal{P}_{\infty}^\ell$ is UB. Since $\mathbb{B}_1^\ell$ is UB by Assumption~\ref{asp_A2},
$\mathbb{B}_1^\ell\mathcal{P}_{\infty}^\ell$ is also UB. The proof for
$\mathbb{B}_2^\ell\mathcal{Q}_{\infty}^\ell$ is the same.
\end{proof}
\begin{lemma}\label{lem:infinite-qf-homo}
Suppose Assumption~2 holds. Let $\Pi_n$ be an $n\times n$ nonstochastic matrix that is UB, and let $\mathbb{A}_{1,\infty}$ and $\mathbb{A}_{2,\infty}$ be two nonstochastic $n\times\infty$ matrices that are UB. Define $\mathbb{M}_n\equiv \mathbb{A}_{1,\infty}'\Pi_n\mathbb{A}_{2,\infty}$. Then
\begin{align}
E\!\left[
(\mathbb{A}_{1,\infty}\mathbb{V}_{\infty})'
\Pi_n
(\mathbb{A}_{2,\infty}\mathbb{V}_{\infty})
\right]
=
\sigma_0^2\,\mathrm{tr}(\mathbb{M}_n),\label{eq_Einfty}    
\end{align}
and
\[
\mathrm{Var}\!\left[
(\mathbb{A}_{1,\infty}\mathbb{V}_{\infty})'
\Pi_n
(\mathbb{A}_{2,\infty}\mathbb{V}_{\infty})
\right]
=
(\mu_4-3\sigma_0^4)\sum_{j=1}^{\infty}
(\mathbb{M}_n)_{jj}^2
+
\sigma_0^4\,\mathrm{tr}(\mathbb{M}_n\mathbb{M}_n')
+
\sigma_0^4\,\mathrm{tr}(\mathbb{M}_n^2).
\]
\end{lemma}

\begin{proof}
We have
\[
(\mathbb{A}_{1,\infty}\mathbb{V}_{\infty})'
\Pi_n
(\mathbb{A}_{2,\infty}\mathbb{V}_{\infty})
=
\mathbb{V}_{\infty}'\mathbb{M}_n\mathbb{V}_{\infty}
=
\sum_{j=1}^{\infty}\sum_{k=1}^{\infty}
(\mathbb{M}_n)_{jk}
(\mathbb{V}_{\infty})_j
(\mathbb{V}_{\infty})_k .
\]

Since $\Pi_n$, $\mathbb{A}_{1,\infty}$ and $\mathbb{A}_{2,\infty}$ are UB, the $n\times n$ matrix $\Pi_n\mathbb{A}_{2,\infty}\mathbb{A}_{1,\infty}'$ is UB, and hence
$\mathrm{tr}\!\left(
\Pi_n\mathbb{A}_{2,\infty}\mathbb{A}_{1,\infty}'
\right)=O(n)$.
Similarly, the $n\times n$ matrices
$\Pi_n\mathbb{A}_{2,\infty}\mathbb{A}_{2,\infty}'\Pi_n'\mathbb{A}_{1,\infty}\mathbb{A}_{1,\infty}'$
and
$\Pi_n\mathbb{A}_{2,\infty}\mathbb{A}_{1,\infty}'
\Pi_n\mathbb{A}_{2,\infty}\mathbb{A}_{1,\infty}'$
are also UB, and thus their traces are $O(n)$.
Therefore, we have
\[
\sum_{j=1}^{\infty}\sum_{k=1}^{\infty}
\bigl|(\mathbb{M}_n)_{jk}\bigr|^2
=
\mathrm{tr}(\mathbb{M}_n\mathbb{M}_n')
=
\mathrm{tr}\!\left(
\Pi_n\mathbb{A}_{2,\infty}\mathbb{A}_{2,\infty}'\Pi_n'
\mathbb{A}_{1,\infty}\mathbb{A}_{1,\infty}'
\right)<\infty.
\]
where the cyclic permutation of factors is legitimate because the matrix $\Pi_n\mathbb{A}_{2,\infty}\mathbb{A}_{2,\infty}'\Pi_n'\mathbb{A}_{1,\infty}\mathbb{A}_{1,\infty}'
$ is
$n\times n$ and hence finite-dimensional.
Additionally,
$\sum_{j=1}^{\infty}
\bigl|(\mathbb{M}_n)_{jj}\bigr|^2<\infty,$ and
\[
\sum_{j=1}^{\infty}\sum_{k=1}^{\infty}
\bigl|(\mathbb{M}_n)_{jk}(\mathbb{M}_n)_{kj}\bigr|
\le
\frac12
\sum_{j=1}^{\infty}\sum_{k=1}^{\infty}
\left[
(\mathbb{M}_n)_{jk}^2+(\mathbb{M}_n)_{kj}^2
\right]
=
\sum_{j=1}^{\infty}\sum_{k=1}^{\infty}
(\mathbb{M}_n)_{jk}^2
=
\mathrm{tr}(\mathbb{M}_n\mathbb{M}_n')
<\infty.
\]

Hence the relevant series are absolutely summable. In particular, Tonelli's theorem is used in the
following two steps. First, for the expectation,
\[
E\!\left[
\sum_{j=1}^{\infty}\sum_{k=1}^{\infty}
(\mathbb{M}_n)_{jk}
(\mathbb{V}_{\infty})_j
(\mathbb{V}_{\infty})_k
\right]
=
\sum_{j=1}^{\infty}\sum_{k=1}^{\infty}
(\mathbb{M}_n)_{jk}
E\!\left[
(\mathbb{V}_{\infty})_j
(\mathbb{V}_{\infty})_k
\right].
\]
Since the elements of $\mathbb{V}_{\infty}$ are i.i.d. with mean zero and variance $\sigma_0^2$,
only the terms with $j=k$ survive, so
\[
E\!\left[
(\mathbb{A}_{1,\infty}\mathbb{V}_{\infty})'
\Pi_n
(\mathbb{A}_{2,\infty}\mathbb{V}_{\infty})
\right]
=
\sigma_0^2
\sum_{j=1}^{\infty}
(\mathbb{M}_n)_{jj}
=
\sigma_0^2\,\mathrm{tr}(\mathbb{M}_n).
\]

Second, for the second moment, Tonelli's theorem justifies the expansion
\[
E\!\left[
\left(
\sum_{j=1}^{\infty}\sum_{k=1}^{\infty}
(\mathbb{M}_n)_{jk}
(\mathbb{V}_{\infty})_j
(\mathbb{V}_{\infty})_k
\right)^2
\right]
=
\sum_{j,k,l,m\ge1}
(\mathbb{M}_n)_{jk}
(\mathbb{M}_n)_{l m}
E\!\left[
(\mathbb{V}_{\infty})_j
(\mathbb{V}_{\infty})_k
(\mathbb{V}_{\infty})_l
(\mathbb{V}_{\infty})_m
\right].
\]
The same case-by-case classification as in the proof of Lemma~\ref{lm_quadratic form} applies. The only nonzero
contributions are
\[
j=k=l=m,\qquad
j=k\neq l=m,\qquad
j=l\neq k=m,\qquad
j=m\neq k=l.
\]
Therefore,
\[
\mathrm{Var}\!\left[
(\mathbb{A}_{1,\infty}\mathbb{V}_{\infty})'
\Pi_n
(\mathbb{A}_{2,\infty}\mathbb{V}_{\infty})
\right]
=
(\mu_4-3\sigma_0^4)
\sum_{j=1}^{\infty}
(\mathbb{M}_n)_{jj}^2
+
\sigma_0^4
\sum_{j=1}^{\infty}\sum_{k=1}^{\infty}
(\mathbb{M}_n)_{jk}^2
+
\sigma_0^4
\sum_{j=1}^{\infty}\sum_{k=1}^{\infty}
(\mathbb{M}_n)_{jk}(\mathbb{M}_n)_{kj}.
\]
Finally,
\[
\sum_{j=1}^{\infty}\sum_{k=1}^{\infty}
(\mathbb{M}_n)_{jk}^2
=
\mathrm{tr}(\mathbb{M}_n\mathbb{M}_n'),
\qquad
\sum_{j=1}^{\infty}\sum_{k=1}^{\infty}
(\mathbb{M}_n)_{jk}(\mathbb{M}_n)_{kj}
=
\mathrm{tr}(\mathbb{M}_n^2),
\]
which gives the desired result.
\end{proof}

\begin{lemma}\label{lem:twfe-master-qf}
Suppose Assumptions~1$'$, 2, 7 and \ref{asp_A2} hold. Let \(\Pi_n\) be either \(\widebar{\mathbb{P}}\) or
\(\widebar{\mathbb{Q}}\), and let \(\mathbb{A}_{1,\infty}\) and \(\mathbb{A}_{2,\infty}\) be two nonstochastic
\(n\times\infty\) matrices that are UB. Then
\[
E\Bigl[(\mathbb{A}_{1,\infty}\mathbb{V}_{\infty})'\Pi_n(\mathbb{A}_{2,\infty}\mathbb{V}_{\infty})\Bigr]=
O\bigl(\mathrm{rank}(\Pi_n)\bigr).
\]
Moreover,
\[
\mathrm{Var}\Bigl[(\mathbb{A}_{1,\infty}\mathbb{V}_{\infty})'\Pi_n(\mathbb{A}_{2,\infty}\mathbb{V}_{\infty})\Bigr]
=O\bigl(\mathrm{rank}(\Pi_n)\bigr).
\]
Consequently,
\[
\frac{1}{NT}
(\mathbb{A}_{1,\infty}\mathbb{V}_{\infty})'
\widebar{\mathbb{Q}}
(\mathbb{A}_{2,\infty}\mathbb{V}_{\infty})
-
\frac{1}{NT}E\Bigl[
(\mathbb{A}_{1,\infty}\mathbb{V}_{\infty})'
\widebar{\mathbb{Q}}
(\mathbb{A}_{2,\infty}\mathbb{V}_{\infty})
\Bigr]
=
O_p\!\left(\frac{1}{\sqrt{NT}}\right),
\]
\[
\frac{1}{NT}
(\mathbb{A}_{1,\infty}\mathbb{V}_{\infty})'
\widebar{\mathbb{P}}
(\mathbb{A}_{2,\infty}\mathbb{V}_{\infty})
-\frac{1}{NT}
E\Bigl[
(\mathbb{A}_{1,\infty}\mathbb{V}_{\infty})'
\widebar{\mathbb{P}}
(\mathbb{A}_{2,\infty}\mathbb{V}_{\infty})
\Bigr]
=O_p\left(\frac{1}{T\sqrt{N}}+\frac{1}{N\sqrt{T}}\right),
\]
where
\(
\frac{1}{NT}
E\Bigl[
(\mathbb{A}_{1,\infty}\mathbb{V}_{\infty})'
\widebar{\mathbb{Q}}
(\mathbb{A}_{2,\infty}\mathbb{V}_{\infty})
\Bigr]
=
O(1)
\)
and
\(
\frac{1}{NT}
E\Bigl[
(\mathbb{A}_{1,\infty}\mathbb{V}_{\infty})'
\widebar{\mathbb{P}}
(\mathbb{A}_{2,\infty}\mathbb{V}_{\infty})
\Bigr]
=
O\!\left(\frac{1}{T}\right)
+
O\!\left(\frac{1}{N}\right).
\)
\end{lemma}

\begin{proof}
By Lemma~B.3 of Meng and Yang (2026b), both $\widebar{\mathbb{P}}$ and $\widebar{\mathbb{Q}}$ are UB.
Hence \(\Pi_n\) is UB in either case.
The expectation formula follows directly from Lemma~\ref{lem:infinite-qf-homo}:
\[
E\Bigl[(\mathbb{A}_{1,\infty}\mathbb{V}_{\infty})'\Pi_n(\mathbb{A}_{2,\infty}\mathbb{V}_{\infty})\Bigr]
=
\sigma_0^2\,\mathrm{tr}\Bigl(\Pi_n\mathbb{A}_{2,\infty}\mathbb{A}_{1,\infty}'\Bigr).
\]
Moreover,
\begin{align}
\bigl|\mathrm{tr}\bigl(\Pi_n\mathbb{A}_{2,\infty}\mathbb{A}_{1,\infty}'\bigr)\bigr|
&\le
\mathrm{rank}\bigl(\Pi_n\mathbb{A}_{2,\infty}\mathbb{A}_{1,\infty}'\bigr)\,
\bigl\|\Pi_n\mathbb{A}_{2,\infty}\mathbb{A}_{1,\infty}'\bigr\|_2
\nonumber\\
&\le
\mathrm{rank}(\Pi_n)\,
\Bigl\|
\Pi_n\mathbb{A}_{2,\infty}\mathbb{A}_{1,\infty}'
\Bigr\|_1^{1/2}
\Bigl\|
\Pi_n\mathbb{A}_{2,\infty}\mathbb{A}_{1,\infty}'
\Bigr\|_\infty^{1/2}
=
O\bigl(\mathrm{rank}(\Pi_n)\bigr),\label{eq_rank}
\end{align}
Therefore,
\[
E\Bigl[(\mathbb{A}_{1,\infty}\mathbb{V}_{\infty})'\Pi_n(\mathbb{A}_{2,\infty}\mathbb{V}_{\infty})\Bigr]
=
O\bigl(\mathrm{rank}(\Pi_n)\bigr).
\]

For the variance, let
$\mathbb{M}_n\equiv \mathbb{A}_{1,\infty}'\Pi_n\mathbb{A}_{2,\infty}.$
By Lemma~\ref{lem:infinite-qf-homo},
\begin{align*}
\mathrm{Var}\Bigl[(\mathbb{A}_{1,\infty}\mathbb{V}_{\infty})'\Pi_n(\mathbb{A}_{2,\infty}\mathbb{V}_{\infty})\Bigr]
&=
(\mu_4-3\sigma_0^4)\sum_{j=1}^{\infty}(\mathbb{M}_n)_{jj}^2
+\sigma_0^4\,\mathrm{tr}(\mathbb{M}_n\mathbb{M}_n')
+\sigma_0^4\,\mathrm{tr}(\mathbb{M}_n^2).
\end{align*}
As in the proof of Lemma~\ref{lem:infinite-qf-homo},
$$\mathrm{tr}(\mathbb{M}_n\mathbb{M}_n')
=
\mathrm{tr}\!\left(
\Pi_n\mathbb{A}_{2,\infty}\mathbb{A}_{2,\infty}'\Pi_n'
\mathbb{A}_{1,\infty}\mathbb{A}_{1,\infty}'
\right),\qquad\mathrm{tr}(\mathbb{M}_n^2)
=
\mathrm{tr}\!\left(
\Pi_n\mathbb{A}_{2,\infty}\mathbb{A}_{1,\infty}'
\Pi_n\mathbb{A}_{2,\infty}\mathbb{A}_{1,\infty}'
\right).$$ Moreover, $\sum_{j=1}^{\infty}(\mathbb{M}_n)_{jj}^2
\le
\mathrm{tr}(\mathbb{M}_n\mathbb{M}_n').$
Hence, by the same argument as for the equation (\ref{eq_rank}),
\[
\bigl|\mathrm{tr}(\mathbb{M}_n\mathbb{M}_n')\bigr|
=
O\bigl(\mathrm{rank}(\Pi_n)\bigr),
\qquad
\bigl|\mathrm{tr}(\mathbb{M}_n^2)\bigr|
=
O\bigl(\mathrm{rank}(\Pi_n)\bigr).
\]
Therefore,
\[
\mathrm{Var}\Bigl[(\mathbb{A}_{1,\infty}\mathbb{V}_{\infty})'\Pi_n(\mathbb{A}_{2,\infty}\mathbb{V}_{\infty})\Bigr]
=O\bigl(\mathrm{rank}(\Pi_n)\bigr).
\]
Since $\widebar{\mathbb{P}}$ is the orthogonal projection matrix onto the column space of the
two-way fixed effects design matrix $(\mathbb{D},\mathbb{J})$, we have
\(
\mathrm{rank}(\widebar{\mathbb{P}})
=
N+T-1,
\)
where the subtraction of one degree of freedom is due to the normalization on the time fixed effects. Moreover,
\(
\mathrm{rank}(\widebar{\mathbb{Q}})=n-N-T+1=O(NT),
\)
so the stated expectation and stochastic orders follow. The stochastic orders follow from Chebyshev's inequality.
\end{proof}

\begin{lemma}\label{lem:twfe-LV-lln}
Under the Assumptions~1$'$, 2, 7 and \ref{asp_A2}, the following results hold:
\begin{align}
\frac{1}{NT}\mathbb{L}^{\ell\prime}\widebar{\mathbb{Q}}\mathbb{R}^{\ell}-\frac{1}{NT}E\bigl(\mathbb{L}^{\ell\prime}\widebar{\mathbb{Q}}\mathbb{R}^{\ell}\bigr)
&=O_p\left(\frac{1}{\sqrt{NT}}\right), \label{eq:twfe-LV-1} \\
\frac{1}{NT}\mathbb{L}^{\ell\prime}\widebar{\mathbb{P}}\mathbb{R}^{\ell}-\frac{1}{NT}E\bigl(\mathbb{L}^{\ell\prime}\widebar{\mathbb{P}}\mathbb{R}^{\ell}\bigr)
&=O_p\left(\frac{1}{T\sqrt{N}}+\frac{1}{N\sqrt{T}}\right), \label{eq:twfe-LV-2} \\
\frac{1}{NT}\mathbb{L}^{\ell\prime}\widebar{\mathbb{Q}}\mathbb{B}_2\mathbb{V}-\frac{1}{NT}E\bigl(\mathbb{L}^{\ell\prime}\widebar{\mathbb{Q}}\mathbb{B}_2\mathbb{V}\bigr)
&=O_p\left(\frac{1}{\sqrt{NT}}\right), \label{eq:twfe-LV-3} \\
\frac{1}{NT}\mathbb{L}^{\ell\prime}\widebar{\mathbb{P}}\mathbb{B}_2\mathbb{V}-\frac{1}{NT}E\bigl(\mathbb{L}^{\ell\prime}\widebar{\mathbb{P}}\mathbb{B}_2\mathbb{V}\bigr)
&=O_p\left(\frac{1}{T\sqrt{N}}+\frac{1}{N\sqrt{T}}\right). \label{eq:twfe-LV-4}\\
\frac{1}{NT}
\mathbb{L}^{\ell\prime}
\mathbb{Q}\mathbb{J}
(\mathbb{J}'\mathbb{Q}\mathbb{J})^{-1}
\mathbb{J}'\mathbb{Q}
\mathbb{R}^{\ell}
-
\frac{1}{NT}
E
\left(
\mathbb{L}^{\ell\prime}
\mathbb{Q}\mathbb{J}
(\mathbb{J}'\mathbb{Q}\mathbb{J})^{-1}
\mathbb{J}'\mathbb{Q}
\mathbb{R}^{\ell}
\right)
&=
O_p\left(\frac{1}{N\sqrt{T}}\right), \label{eq:twfe-LV-5} \\
\frac{1}{NT}
\mathbb{L}^{\ell\prime}
\mathbb{Q}\mathbb{J}
(\mathbb{J}'\mathbb{Q}\mathbb{J})^{-1}
\mathbb{J}'\mathbb{Q}
\mathbb{B}_2\mathbb{V}
-
\frac{1}{NT}
E
\left(
\mathbb{L}^{\ell\prime}
\mathbb{Q}\mathbb{J}
(\mathbb{J}'\mathbb{Q}\mathbb{J})^{-1}
\mathbb{J}'\mathbb{Q}
\mathbb{B}_2\mathbb{V}
\right)
&=
O_p\left(\frac{1}{N\sqrt{T}}\right). \label{eq:twfe-LV-6}
\end{align}

The expectations satisfy:
\begin{align}
\frac{1}{NT}E\bigl(\mathbb{L}^{\ell\prime}\widebar{\mathbb{Q}}\mathbb{R}^{\ell}\bigr)&=O(1), &
\frac{1}{NT}E\bigl(\mathbb{L}^{\ell\prime}\widebar{\mathbb{P}}\mathbb{R}^{\ell}\bigr)&=O\left(\frac{1}{T}\right)+O\left(\frac{1}{N}\right), \label{eq:twfe-LV-7} \\
\frac{1}{NT}E\bigl(\mathbb{L}^{\ell\prime}\widebar{\mathbb{Q}}\mathbb{B}_2\mathbb{V}\bigr)&=O\left(\frac{1}{T}\right)+O\left(\frac{1}{N}\right). \label{eq:twfe-LV-8b}
\end{align}
\begin{align}
\frac{1}{NT}
E
\left(
\mathbb{L}^{\ell\prime}
\mathbb{Q}\mathbb{J}
(\mathbb{J}'\mathbb{Q}\mathbb{J})^{-1}
\mathbb{J}'\mathbb{Q}
\mathbb{R}^{\ell}
\right)
=O\left(\frac{1}{N}\right),\qquad\frac{1}{NT}
E
\left(
\mathbb{L}^{\ell\prime}
\mathbb{Q}\mathbb{J}
(\mathbb{J}'\mathbb{Q}\mathbb{J})^{-1}
\mathbb{J}'\mathbb{Q}
\mathbb{B}_2\mathbb{V}
\right)
=O\left(\frac{1}{N}\right). \label{eq:twfe-LV-10}
\end{align}
Specifically,
\begin{align}
\frac{1}{NT}E\bigl(\mathbb{L}^{\ell\prime}\widebar{\mathbb{P}}\mathbb{B}_2\mathbb{V}\bigr) &= \frac{\sigma_0^2}{NT}\mathrm{tr}\left[\mathcal{T}^{-1}\sum_{h=1}^{T-1}\sum_{t=T-h}^{T-1}D_{t+1}'B_{t+1}^\ell P_{t}^{(t+h-T)}B_{2,T-h}'D_{T-h}\right] \notag\\
&\quad + \frac{\sigma_0^2}{NT}\mathrm{tr}\left( \mathbb{Q}\mathbb{J}(\mathbb{J}'\mathbb{Q}\mathbb{J})^{-1}\mathbb{J}'\mathbb{Q} \mathcal{B}_{2\infty} \mathcal{P}_{\infty}^{\ell\prime} \mathbb{B}_1^{\ell\prime}\right) = O\left(\frac{1}{T}\right)+O\left(\frac{1}{N}\right), \label{eq:twfe-LV-8a}
\end{align}
where $B_{t+1}^\ell \equiv B_{1,t+1}^\ell B_{1,t}$.
\end{lemma}

\begin{proof}
We have $\mathbb{L}^\ell = \mathbb{B}_1^\ell \mathcal{P}_\infty^\ell \mathbb{V}_\infty$ and $\mathbb{R}^\ell = \mathbb{B}_2^\ell \mathcal{Q}_\infty^\ell \mathbb{V}_\infty$.
Moreover, we can write $\mathbb{B}_2 \mathbb{V} = \mathcal{B}_{2\infty}\mathbb{V}_\infty$, where $\mathcal{B}_{2\infty}$ is an $n \times \infty$ matrix defined as
$$\mathcal{B}_{2\infty} = \begin{pmatrix}
\mathbf{0}_{N_1 \times N_T} & \dots & \mathbf{0}_{N_1 \times N_2} & B_{21} & \mathbf{0}_{N_1 \times N_0} & \dots \\
\mathbf{0}_{N_2 \times N_T} & \dots & B_{22} & \mathbf{0}_{N_2 \times N_1} & \mathbf{0}_{N_2 \times N_0} & \dots \\
\vdots & \iddots & \vdots & \vdots & \vdots & \ddots \\
B_{2T} & \dots & \mathbf{0}_{N_T \times N_2} & \mathbf{0}_{N_T \times N_1} & \mathbf{0}_{N_T \times N_0} & \dots
\end{pmatrix}.$$
Since the matrices $\mathbb{B}_1^\ell \mathcal{P}_\infty^\ell$, $\mathbb{B}_2^\ell \mathcal{Q}_\infty^\ell$, and $\mathcal{B}_{2\infty}$ are all UB according to Assumption~\ref{asp_A2} and Lemma \ref{lem:Pinfty-UB}, Lemma~\ref{lem:twfe-master-qf} can be directly applied to establish \eqref{eq:twfe-LV-1}--\eqref{eq:twfe-LV-4}, and \eqref{eq:twfe-LV-7}--\eqref{eq:twfe-LV-8b}.

To derive the explicit expression \eqref{eq:twfe-LV-8a}, we notice that $\widebar{\mathbb{P}}$ is the projection matrix onto the column space of $[\mathbb{D}, \mathbb{J}]$. We can naturally decompose this into the projection onto $\mathbb{D}$ plus the projection onto the part of $\mathbb{J}$ that is orthogonal to $\mathbb{D}$. Therefore, we have
\(
\widebar{\mathbb{P}} = \mathbb{P} + \mathbb{Q}\mathbb{J}(\mathbb{J}'\mathbb{Q}\mathbb{J})^{-1}\mathbb{J}'\mathbb{Q}.
\)
Then the expectation splits into two terms:
\[
E\bigl(\mathbb{L}^{\ell\prime}\widebar{\mathbb{P}}\mathbb{B}_2\mathbb{V}\bigr) = E\bigl(\mathbb{L}^{\ell\prime}\mathbb{P}\mathbb{B}_2\mathbb{V}\bigr) + E\bigl(\mathbb{L}^{\ell\prime}\mathbb{Q}\mathbb{J}(\mathbb{J}'\mathbb{Q}\mathbb{J})^{-1}\mathbb{J}'\mathbb{Q}\mathbb{B}_2\mathbb{V}\bigr).
\]
By Lemma~\ref{lm_LQV}, the first term yields the double-sum trace formula exactly as in \eqref{eq_ELPV}. For the second term, applying equation \eqref{eq_Einfty}, we have:
\begin{align*}
E\bigl(\mathbb{L}^{\ell\prime}\mathbb{Q}\mathbb{J}(\mathbb{J}'\mathbb{Q}\mathbb{J})^{-1}\mathbb{J}'\mathbb{Q}\mathbb{B}_2\mathbb{V}\bigr)
&= E\bigl( \mathbb{V}_{\infty}^{\prime} \mathcal{P}_{\infty}^{\ell\prime} \mathbb{B}_1^{\ell\prime} \mathbb{Q}\mathbb{J}(\mathbb{J}'\mathbb{Q}\mathbb{J})^{-1}\mathbb{J}'\mathbb{Q} \mathcal{B}_{2\infty} \mathbb{V}_{\infty} \bigr)\\
&= \sigma_0^2 \, \mathrm{tr}\bigl( \mathbb{Q}\mathbb{J}(\mathbb{J}'\mathbb{Q}\mathbb{J})^{-1}\mathbb{J}'\mathbb{Q} \mathcal{B}_{2\infty} \mathcal{P}_{\infty}^{\ell\prime} \mathbb{B}_1^{\ell\prime} \bigr).
\end{align*}
Combining these two parts directly yields \eqref{eq:twfe-LV-8a}.

The same argument also applies to the residualized time-effect projection
\(
\mathbb{Q}\mathbb{J}
(\mathbb{J}'\mathbb{Q}\mathbb{J})^{-1}
\mathbb{J}'\mathbb{Q}
\). Since
\(\mathrm{rank}(\mathbb{Q}\mathbb{J}
(\mathbb{J}'\mathbb{Q}\mathbb{J})^{-1}
\mathbb{J}'\mathbb{Q})=T-1\), Lemma~\ref{lem:twfe-master-qf} gives \eqref{eq:twfe-LV-5}--\eqref{eq:twfe-LV-6}, and their expectations satisfy \eqref{eq:twfe-LV-10}.
\end{proof}

\begin{lemma}\label{lem:twfe-ZV-ZZ}
Under Assumptions~1$'$, 2, 3, 4$'$, 5--7 and \ref{asp_A2}, we have
\begin{align}
\frac{1}{NT}\mathbb{Z}'\mathbb{B}_1'\widebar{\mathbb{Q}}\mathbb{B}_2\mathbb{V}-\frac{1}{NT}E\bigl(\mathbb{Z}'\mathbb{B}_1'\widebar{\mathbb{Q}}\mathbb{B}_2\mathbb{V}\bigr)
&=O_p\left(\frac{1}{\sqrt{NT}}\right),
\label{eq:twfe-ZV-ZZ-1}
\\
\frac{1}{NT}\mathbb{Z}'\mathbb{B}_1'\widebar{\mathbb{P}}\mathbb{B}_2\mathbb{V}-\frac{1}{NT}E\bigl(\mathbb{Z}'\mathbb{B}_1'\widebar{\mathbb{P}}\mathbb{B}_2\mathbb{V}\bigr)
&=O_p\left(\frac{1}{\sqrt{NT}}\right),
\label{eq:twfe-ZV-ZZ-2}
\\
\frac{1}{NT}\mathbb{Z}'\mathbb{B}_1'\widebar{\mathbb{Q}}\mathbb{B}_2\mathbb{Z}-\frac{1}{NT}E\bigl(\mathbb{Z}'\mathbb{B}_1'\widebar{\mathbb{Q}}\mathbb{B}_2\mathbb{Z}\bigr)
&=O_p\left(\frac{1}{\sqrt{NT}}\right),
\label{eq:twfe-ZV-ZZ-3}
\\
\frac{1}{NT}\mathbb{Z}'\mathbb{B}_1'\widebar{\mathbb{P}}\mathbb{B}_2\mathbb{Z}-\frac{1}{NT}E\bigl(\mathbb{Z}'\mathbb{B}_1'\widebar{\mathbb{P}}\mathbb{B}_2\mathbb{Z}\bigr)
&=O_p\left(\frac{1}{\sqrt{NT}}\right).
\label{eq:twfe-ZV-ZZ-4}
\end{align}
Moreover,
\begin{align}
\frac{1}{NT}E\bigl(\mathbb{Z}'\mathbb{B}_1'\widebar{\mathbb{Q}}\mathbb{B}_2\mathbb{V}\bigr)
&=O\left(\frac{1}{T}\right)+O\left(\frac{1}{N}\right),
\qquad
\frac{1}{NT}E\bigl(\mathbb{Z}'\mathbb{B}_1'\widebar{\mathbb{P}}\mathbb{B}_2\mathbb{V}\bigr)
=O\left(\frac{1}{T}\right)+O\left(\frac{1}{N}\right),
\label{eq:twfe-ZV-ZZ-5}
\end{align}
\begin{align}
\frac{1}{NT}E\bigl(\mathbb{Z}'\mathbb{B}_1'\widebar{\mathbb{Q}}\mathbb{B}_2\mathbb{Z}\bigr)
&=O(1),
\qquad
\frac{1}{NT}E\bigl(\mathbb{Z}'\mathbb{B}_1'\widebar{\mathbb{P}}\mathbb{B}_2\mathbb{Z}\bigr)
=O(1).
\label{eq:twfe-ZV-ZZ-6}
\end{align}
The results in equations \eqref{eq:twfe-ZV-ZZ-1}--\eqref{eq:twfe-ZV-ZZ-4} remain valid when \(\mathbb{Z}\) is replaced by
\(\mathbb{Z}\widetilde{\Gamma}\), where \(\widetilde{\Gamma}=\mathrm{blkdiag}(I_2,\sqrt{T},I_k)\).
Furthermore,
\begin{align}
\sqrt{T}\,\mathbb{Z}^{(-\gamma)'}\mathbb{B}_1'\widebar{\mathbb{Q}}\mathbb{B}_2\mathbb{Z}^{(\gamma)}=O_p(N\sqrt{T}),\qquad
T\,\mathbb{Z}^{(\gamma)'}\mathbb{B}_1'\widebar{\mathbb{Q}}\mathbb{B}_2\mathbb{Z}^{(\gamma)}=O(NT).\label{eq:twfe-ZV-ZZ-7}
\end{align}
\end{lemma}

\begin{proof}
Write
\(
\mathbb{Z}=(\mathbb{Z}_1,\mathbb{Z}_2,\mathbb{Z}^{(\gamma)},\mathbb{Z}_4),
\)
where \(\mathbb{Z}^{(\gamma)}\) is the listing regressor and \(\mathbb{Z}_4\) collects the remaining \(k\) columns.
As in the one-way case, \(\mathbb{B}_1\mathbb{Z}_1,\mathbb{B}_1\mathbb{Z}_2,
\mathbb{B}_2\mathbb{Z}_1\), and \(\mathbb{B}_2\mathbb{Z}_2\) are all of the form
\(\mathbb{C}+\mathbb{L}^{\ell}\) or \(\mathbb{C}+\mathbb{R}^{\ell}\), while
\(\mathbb{Z}^{(\gamma)}\) and the columns of \(\mathbb{Z}_4\) are of the form \(\mathbb{C}\).

We first consider the \(\mathbb{Z}\)-\(\mathbb{V}\) terms. Let
\(\Pi_n\in\{\widebar{\mathbb{Q}},\widebar{\mathbb{P}}\}\). Every element of
\(\mathbb{Z}'\mathbb{B}_1'\Pi_n\mathbb{B}_2\mathbb{V}\) can be reduced to terms of the form
\[
\mathbb{C}'\Pi_n\mathbb{B}_2\mathbb{V},
\qquad
\mathbb{L}^{\ell\prime}\Pi_n\mathbb{B}_2\mathbb{V}.
\]
Since Lemma~B.3 of Meng and Yang (2026) implies that
\(\widebar{\mathbb{P}}\) and \(\widebar{\mathbb{Q}}\) are UB, elements in \(\mathbb{B}_2'\Pi_n\mathbb{C}\)
are still uniformly bounded. Therefore, by equation~\eqref{eq_cv} in
Lemma~\ref{lm_CQL},
\[
\frac{1}{NT}\mathbb{C}'\Pi_n\mathbb{B}_2\mathbb{V}
=O_p\left(\frac{1}{\sqrt{NT}}\right).
\]
For the second component, Lemma~\ref{lem:twfe-LV-lln} gives
\[
\frac{1}{NT}\mathbb{L}^{\ell\prime}\Pi_n\mathbb{B}_2\mathbb{V}
-
\frac{1}{NT}E\bigl(\mathbb{L}^{\ell\prime}\Pi_n\mathbb{B}_2\mathbb{V}\bigr)
=O_p\left(\frac{1}{\sqrt{NT}}\right),
\]
with
\[
\frac{1}{NT}E\bigl(\mathbb{L}^{\ell\prime}\Pi_n\mathbb{B}_2\mathbb{V}\bigr)
=O\left(\frac{1}{T}\right)+O\left(\frac{1}{N}\right).
\]
Combining these bounds proves equations \eqref{eq:twfe-ZV-ZZ-1}, \eqref{eq:twfe-ZV-ZZ-2}, and
\eqref{eq:twfe-ZV-ZZ-5}.

Next consider the \(\mathbb{Z}\)-\(\mathbb{Z}\) terms. Expanding block by block, every element of
\(\mathbb{Z}'\mathbb{B}_1'\Pi_n\mathbb{B}_2\mathbb{Z}\) is a sum of terms of the following types:
\[
\mathbb{C}_1'\Pi_n\mathbb{C}_2,
\qquad
\mathbb{C}'\Pi_n\mathbb{L}^{\ell},
\qquad
\mathbb{L}^{\ell\prime}\Pi_n\mathbb{R}^{\ell}.
\]
The first type is deterministic, and
\(
\mathbb{C}_1'\Pi_n\mathbb{C}_2=O(NT),
\)
because \(\Pi_n\) is UB and \(\mathbb{C}_1,\mathbb{C}_2\) have uniformly bounded elements.

For the second type, \(\Pi_n\mathbb{C}\) is deterministic with uniformly bounded elements. 
Applying equation (\ref{eq_cl}), together with Lemma~\ref{lm_lag}, gives
\[
\frac{1}{NT}\mathbb{C}'\Pi_n\mathbb{L}^{\ell}
=O_p\left(\frac{1}{\sqrt{NT}}\right).
\]
The expectation is zero. 

The third type is controlled by equations
\eqref{eq:twfe-LV-1}--\eqref{eq:twfe-LV-2} and the expectation bounds in \eqref{eq:twfe-LV-7}. Therefore, equations \eqref{eq:twfe-ZV-ZZ-3} and \eqref{eq:twfe-ZV-ZZ-4} hold, and their expectations satisfy equation
\eqref{eq:twfe-ZV-ZZ-6}.

It remains to verify the statements involving \(\widetilde{\Gamma}\). The matrix \(\widetilde{\Gamma}\)
only multiplies the listing column by \(\sqrt{T}\). Since
\(\Pi_n\in\{\widebar{\mathbb{Q}},\widebar{\mathbb{P}}\}\) is still UB, the same sparse-vector calculations as in
Lemma~\ref{lm_Lambda} give
\[
\frac{1}{N}\mathbb{Z}^{(\gamma)'}\mathbb{B}_1'\Pi_n\mathbb{B}_2\mathbb{Z}^{(\gamma)}=O(1),
\qquad
\frac{\sqrt{T}}{NT}\mathbb{C}'\Pi_n\mathbb{B}_2\mathbb{Z}^{(\gamma)}
=O\left(\frac{1}{\sqrt{T}}\right),
\]
and
\[
\frac{\sqrt{T}}{NT}\mathbb{L}^{\ell\prime}\Pi_n\mathbb{B}_2\mathbb{Z}^{(\gamma)}
=O_p\left(\frac{1}{\sqrt{NT}}\right),
\qquad
\frac{\sqrt{T}}{NT}\mathbb{Z}^{(\gamma)'}\mathbb{B}_1'\Pi_n\mathbb{B}_2\mathbb{V}
=O_p\left(\frac{1}{\sqrt{NT}}\right).
\]
Together with the bounds already established for the non-listing columns, these equations imply that 
\eqref{eq:twfe-ZV-ZZ-1}--\eqref{eq:twfe-ZV-ZZ-4} remain valid after replacing \(\mathbb{Z}\) by
\(\mathbb{Z}\widetilde{\Gamma}\). They also imply
\[
\sqrt{T}\,\mathbb{Z}^{(-\gamma)'}\mathbb{B}_1'\widebar{\mathbb{Q}}\mathbb{B}_2\mathbb{Z}^{(\gamma)}
=O_p\left(N\sqrt{T}\right),
\qquad
T\,\mathbb{Z}^{(\gamma)'}\mathbb{B}_1'\widebar{\mathbb{Q}}\mathbb{B}_2\mathbb{Z}^{(\gamma)}
=O(NT),
\]
which is equation \eqref{eq:twfe-ZV-ZZ-7}.
\end{proof}

\begin{lemma}\label{lem:twfe-bridge}
Under Assumptions~1$'$, 2, 3, 4$'$, 5--7 and \ref{asp_A2}, we have
\begin{align}
\frac{1}{NT}E\bigl(\widetilde{\Gamma}\mathbb{Z}'\widebar{\mathbb{Q}}\mathbb{Z}\widetilde{\Gamma}\bigr)
-
\frac{1}{NT}E\bigl(\widetilde{\Gamma}\widebar{\widetilde{\mathbb{Z}}}^{*\prime}\widebar{\widetilde{\mathbb{Z}}}^{*}\widetilde{\Gamma}\bigr)
&=O\left(\frac{1}{T}\right)+O\left(\frac{1}{N}\right),
\label{eq:twfe-bridge-1}
\\
\frac{1}{NT}E\bigl(\widetilde{\Gamma}\mathbb{Z}'\widebar{\mathbb{Q}}\mathbb{G}_0\mathbb{Z}\bigr)
-
\frac{1}{NT}E\bigl(\widetilde{\Gamma}\widebar{\widetilde{\mathbb{Z}}}^{*\prime}\widebar{\widetilde{\mathbb{G}_0\mathbb{Z}}}^{*}\bigr)
&=O\left(\frac{1}{T}\right)+O\left(\frac{1}{N}\right),
\label{eq:twfe-bridge-2}
\\
\frac{1}{NT}E\bigl(\mathbb{Z}'\mathbb{G}_0'\widebar{\mathbb{Q}}\mathbb{G}_0\mathbb{Z}\bigr)
-
\frac{1}{NT}E\bigl(\widebar{\widetilde{\mathbb{G}_0\mathbb{Z}}}^{*\prime}\widebar{\widetilde{\mathbb{G}_0\mathbb{Z}}}^{*}\bigr)
&=O\left(\frac{1}{T}\right)+O\left(\frac{1}{N}\right),
\label{eq:twfe-bridge-3}
\end{align}
where
\(
\frac{1}{NT}E\bigl(\widetilde{\Gamma}\widebar{\widetilde{\mathbb{Z}}}^{*\prime}\widebar{\widetilde{\mathbb{Z}}}^{*}\widetilde{\Gamma}\bigr)=O(1),\) \(
\frac{1}{NT}E\bigl(\widetilde{\Gamma}\widebar{\widetilde{\mathbb{Z}}}^{*\prime}\widebar{\widetilde{\mathbb{G}_0\mathbb{Z}}}^{*}\bigr)=O(1),\) and
\(
\frac{1}{NT}E\bigl(\widebar{\widetilde{\mathbb{G}_0\mathbb{Z}}}^{*\prime}\widebar{\widetilde{\mathbb{G}_0\mathbb{Z}}}^{*}\bigr)=O(1).
\)
\end{lemma}
\begin{proof}
Recall that
\(
\widebar{\widetilde{\mathbb{Z}}}^{\circ}=\widebar{\mathbb{P}}\mathbb{Z}^{\circ}\), \(\widebar{\widetilde{\mathbb{Z}}}^{*}=\widebar{\mathbb{Q}}\mathbb{Z}+\widebar{\widetilde{\mathbb{Z}}}^{\circ},
\)
and similarly
\(
\widebar{\widetilde{\mathbb{G}_0\mathbb{Z}}}^{\circ}=\widebar{\mathbb{P}}\mathbb{G}_0\mathbb{Z}^{\circ}\), \(\widebar{\widetilde{\mathbb{G}_0\mathbb{Z}}}^{*}=\widebar{\mathbb{Q}}\mathbb{G}_0\mathbb{Z}+\widebar{\widetilde{\mathbb{G}_0\mathbb{Z}}}^{\circ}.
\)
Therefore,
\[
\frac{1}{NT}E\bigl(\widetilde{\Gamma}\mathbb{Z}'\widebar{\mathbb{Q}}\mathbb{Z}\widetilde{\Gamma}\bigr)
-
\frac{1}{NT}E\bigl(\widetilde{\Gamma}\widebar{\widetilde{\mathbb{Z}}}^{*\prime}\widebar{\widetilde{\mathbb{Z}}}^{*}\widetilde{\Gamma}\bigr)
=
\frac{1}{NT}E\bigl(\widetilde{\Gamma}\mathbb{Z}^{\circ\prime}\widebar{\mathbb{P}}\mathbb{Z}^{\circ}\widetilde{\Gamma}\bigr),
\]
where \(\mathbb{Z}^{\circ}\) has the same structure as in the one-way case, namely
\(
\mathbb{Z}^{\circ}=\bigl(\mathbb{L}^{\ell},\mathbb{R}^{\ell},0_{n\times (k+1)}\bigr),
\)
up to relabeling of the coefficient matrices. Hence, Lemma~\ref{lem:twfe-LV-lln} implies that the right-hand
side is of order \(O(\frac{1}{T})+O(\frac{1}{N})\), which proves \eqref{eq:twfe-bridge-1}. Since $\frac{1}{NT}E\bigl(\widetilde{\Gamma}\mathbb{Z}'\widebar{\mathbb{Q}}\mathbb{Z}\widetilde{\Gamma}\bigr)$ is $O(1)$ by Lemma~\ref{lem:twfe-ZV-ZZ}, therefore $\frac{1}{NT}E\bigl(\widetilde{\Gamma}\widebar{\widetilde{\mathbb{Z}}}^{*\prime}\widebar{\widetilde{\mathbb{Z}}}^{*}\widetilde{\Gamma}\bigr)=O(1)$.

Equations \eqref{eq:twfe-bridge-2} and \eqref{eq:twfe-bridge-3} are proved in the same way after replacing
\(\mathbb{Z}^{\circ}\) by \(\mathbb{G}_0\mathbb{Z}^{\circ}\). Under Assumptions~3 and 5, \(\mathbb{G}_0=\mathbb{W}\mathbb{S}_0^{-1}\)
is UB, so Lemmas~\ref{lem:twfe-LV-lln} and
\ref{lem:twfe-ZV-ZZ} continue to apply.
\end{proof}

\section{QMLE}
\subsection{The first and second order conditions of QMLE}\label{Foc}
For the concentrated likelihood function (7), the first order derivatives are
$$\frac{\partial L_n^{c}(\theta)}{\partial\theta}=\left(\begin{aligned}
\frac{\partial L_n^{c}(\theta)}{\partial \rho}\\
\frac{\partial L_n^{c}(\theta)}{\partial \delta}\\
\frac{\partial L_n^{c}(\theta)}{\partial \sigma^2}
\end{aligned}\right)=\left(\begin{aligned}
&\frac{1}{\sigma^2}\mathbb{(WY)'}\mathbb{\widetilde{V}}(\zeta)-\mathrm{tr}\left(\mathbb{S(\rho)}^{-1}\mathbb{W}\right)\\
&\frac{1}{\sigma^2}\mathbb{Z}'\mathbb{\widetilde{V}}(\zeta)\\
&-\frac{n}{2\sigma^2}+\frac{1}{2\sigma^4}\mathbb{\widetilde{V}}(\zeta)'\mathbb{\widetilde{V}}(\zeta)
\end{aligned}\right),$$
where we denote $\mathbb{\widetilde{V}}(\zeta)=\mathbb{Q(S(\rho)Y-Z}\delta)$ for simplicity. The second order derivatives are
{
\renewcommand{\arraystretch}{2}
\begin{align*}
\frac{\partial^2L_n^{c}(\theta)}{\partial\theta\partial\theta'}=
-\begin{pmatrix}
\displaystyle\frac{1}{\sigma^2}\mathbb{(WY)'}\mathbb{QWY}+\mathrm{tr}\left(\mathbb{S(\rho)}^{-1}\mathbb{W}\mathbb{S(\rho)}^{-1}\mathbb{W}\right) & \displaystyle* & \displaystyle*\\
\displaystyle\frac{1}{\sigma^2}\mathbb{Z}'\mathbb{QWY} & \displaystyle\frac{1}{\sigma^2}\mathbb{Z'QZ} & \displaystyle*\\
\displaystyle\frac{1}{\sigma^4}\mathbb{\widetilde{V}}(\zeta)'\mathbb{QWY} & \displaystyle\frac{1}{\sigma^4}\mathbb{\widetilde{V}}(\zeta)'\mathbb{QZ} & \displaystyle-\frac{n}{2\sigma^4}+\frac{1}{\sigma^6}\mathbb{\widetilde{V}}(\zeta)'\mathbb{\widetilde{V}}(\zeta)
\end{pmatrix}.    
\end{align*} 
At $\theta_0$, we have
$$\frac{\partial L_n^{c}(\theta_0)}{\partial\theta}=\left(\begin{aligned}
&\frac{1}{\sigma_0^2}\mathbb{(WY)'QV}-\mathrm{tr}(\mathbb{G}_0)\\
&\frac{1}{\sigma_0^2}\mathbb{Z'QV}\\
&-\frac{n}{2\sigma_0^2}+\frac{1}{2\sigma_0^4}\mathbb{V'QV}
\end{aligned}\right)=\left(\begin{aligned}
&\frac{1}{\sigma_0^2}\mathbb{(G}_0\mathbb{D\alpha}_0+\mathbb{G}_0\mathbb{V})'\mathbb{QV}+\frac{1}{\sigma_0^2}(\mathbb{G}_0\mathbb{Z\delta}_0)'\mathbb{QV}-\mathrm{tr}(\mathbb{G}_0)\\
&\frac{1}{\sigma_0^2}\mathbb{Z'QV}\\
&-\frac{n}{2\sigma_0^2}+\frac{1}{2\sigma_0^4}\mathbb{V'QV}
\end{aligned}\right),$$
and the information matrix is $\Sigma_{\theta_0,n}\equiv E\left(-\frac{1}{n}\frac{\partial^2L_n^{c}(\theta_0)}{\partial\theta\partial\theta'}\right)=\Sigma_{\theta_0,n}^{(1)}-\Sigma_{\theta_0,n}^{(2)}$, where $\Sigma_{\theta_0,n}^{(1)}=$
\begin{align*}
\begin{pmatrix}
\displaystyle\frac{1}{n\sigma_0^2}E\left[\mathbb{(G}_0\mathbb{Z}\delta_0+\mathbb{G}_0\mathbb{D}\alpha_0)'\mathbb{Q(G}_0\mathbb{Z}\delta_0+\mathbb{G}_0\mathbb{D}\alpha_0)\right]+\frac{1}{n}\mathrm{tr}\left(\mathbb{G}_0^2+\mathbb{G}_0'\mathbb{G}_0\right) & \displaystyle* & \displaystyle*\\
\displaystyle\frac{1}{n\sigma_0^2}E\left[\mathbb{Z}'\mathbb{Q(G}_0\mathbb{Z}\delta_0+\mathbb{G}_0\mathbb{D}\alpha_0)\right] & \displaystyle\frac{1}{n\sigma_0^2}E\left(\mathbb{Z'QZ}\right) & \displaystyle*\\
\displaystyle\frac{1}{n\sigma_0^2}\mathrm{tr}\left(\mathbb{G}_0\right) & \displaystyle\mathbf{0}_{1\times(k+3)} & \displaystyle\frac{1}{2\sigma_0^4}
\end{pmatrix},
\end{align*}
$\Sigma_{\theta_0,n}^{(2)}=$
\begin{align*}
\begin{pmatrix}
\displaystyle\frac{2}{n\sigma_0^2}E\left[\mathbb{(G}_0\mathbb{Z}\delta_0)'\mathbb{PG}_0\mathbb{V}\right]+\frac{1}{n}\mathrm{tr}\left(\mathbb{G}_0'\mathbb{P}\mathbb{G}_0\right) & \displaystyle* & \displaystyle*\\
\displaystyle\frac{1}{n\sigma_0^2}E\left[\mathbb{Z'}\mathbb{PG}_0\mathbb{V}\right] & \displaystyle\mathbf{0}_{(k+3)\times(k+3)} & \displaystyle*\\
\displaystyle\frac{1}{n\sigma_0^4}E\left(\mathbb{V}'\mathbb{PG}_0\mathbb{Z}\delta_0\right)+\frac{1}{n\sigma_0^2}\mathrm{tr}\left(\mathbb{P}\mathbb{G}_0\right)& \displaystyle\frac{1}{n\sigma_0^4}E\left(\mathbb{V}'\mathbb{P}\mathbb{Z}\right) & \displaystyle\frac{N}{n\sigma_0^4}
\end{pmatrix}.
\end{align*}

By Lemma \ref{lm_ZPZ}, all expectation terms in $\Sigma_{\theta_0,n}^{(2)}$ are $O(\frac{1}{T})$.
Since $\mathbb{G}_0=\mathrm{blkdiag}(W_1S_1^{-1},\cdots,W_TS_T^{-1})$, 
\begin{align*}
\frac{1}{n}\left|\mathrm{tr}\left(\mathbb{P}\mathbb{G}_0\right)\right|&=\frac{1}{n}\left|\mathrm{tr}\left(\mathbb{D}\mathcal{T}^{-1}\mathbb{D}'\mathbb{G}_0\right)\right|=\frac{1}{n}\left|\mathrm{tr}\left(\mathcal{T}^{-1}\mathbb{D}'\mathbb{G}_0\mathbb{D}\right)\right|\leq\frac{1}{c_TnT}\left|\mathrm{tr}\left(\mathbb{D}'\mathbb{G}_0\mathbb{D}\right)\right|\\
&\leq\frac{1}{c_TnT}\sum_{t=1}^T\left|\mathrm{tr}\left(D_t'W_tS_t^{-1}D_t\right)\right|=O\left(\frac{1}{T}\right),
\end{align*}
where the first inequality holds because $T_i>c_TT$ by Assumption 1 and the last equality holds because $D_t'W_tS_t^{-1}D_t$ is an N-dimensional UB matrix. Similarly, we can prove $\frac{1}{n}\mathrm{tr}\left(\mathbb{G}_0'\mathbb{P}\mathbb{G}_0\right)=O\left(\frac{1}{T}\right)$. Thus, $\Sigma_{\theta_0,n}^{(2)}=O(\frac{1}{T})$, which means that $\Sigma_{\theta_0,n}=\Sigma_{\theta_0,n}^{(1)}+O(\frac{1}{T})$.
We also have $\Gamma\Sigma_{\theta_0,n}\Gamma=\Gamma\Sigma_{\theta_0,n}^{(1)}\Gamma+O(\frac{1}{T})$ by Lemmas \ref{lm15}-\ref{lm_Lambda}.

\subsection{The variance of the gradient}\label{B2}
Denote 
$$E\left(\frac{1}{\sqrt{n}}\frac{\partial L_n^{c*}(\theta_0)}{\partial\theta}\cdot\frac{1}{\sqrt{n}}\frac{\partial L_n^{c*}(\theta_0)}{\partial\theta'}\right)\equiv\left(\begin{array}{ccc}
E_{\rho\rho}&*&*\\E_{\delta\rho}& E_{\delta\delta'}&*\\
E_{\sigma^2\rho}& E_{\sigma^2\delta'}&E_{\sigma^2\sigma^2}
\end{array}\right),$$
where
\begin{align*}
E_{\rho\rho}=&\frac{1}{n}E\left[\left(\frac{1}{\sigma^2_0}(\mathbb{G}_0\mathbb{D\alpha}_0)'\mathbb{QV}+\frac{1}{\sigma_0^2}\mathbb{V}'\mathbb{G}_0'\mathbb{V}+\frac{1}{\sigma_0^2}(\widetilde{\mathbb{G}_0\mathbb{Z}}^*\delta_0)'\mathbb{V}-\mathrm{tr}(\mathbb{G}_0)\right)^2\right]\\
=&\frac{1}{n}E\left[\frac{1}{\sigma_0^4}(\mathbb{G}_0\mathbb{D\alpha}_0)'\mathbb{QV}\mathbb{V}'\mathbb{Q}\mathbb{G}_0\mathbb{D\alpha}_0+\frac{1}{\sigma_0^4}\mathbb{V}'\mathbb{G}_0'\mathbb{V}\mathbb{V}'\mathbb{G}_0'\mathbb{V}+\frac{1}{\sigma_0^4}(\widetilde{\mathbb{G}_0\mathbb{Z}}^*\delta_0)'\mathbb{V}\mathbb{V}'\widetilde{\mathbb{G}_0\mathbb{Z}}^*\delta_0+\mathrm{tr}(\mathbb{G}_0)^2+\right.\\
&\frac{2}{\sigma_0^4}(\mathbb{G}_0\mathbb{D\alpha}_0)'\mathbb{QV}\mathbb{V}'\mathbb{G}_0'\mathbb{V}+\frac{2}{\sigma_0^4}(\mathbb{G}_0\mathbb{D\alpha}_0)'\mathbb{QV}\mathbb{V}'\widetilde{\mathbb{G}_0\mathbb{Z}}^*\delta_0-\frac{2}{\sigma_0^2}(\mathbb{G}_0\mathbb{D\alpha}_0)'\mathbb{QV}\mathrm{tr}(\mathbb{G}_0)+\\
&\left.\frac{2}{\sigma_0^4}\mathbb{V}'\mathbb{G}_0'\mathbb{V}\mathbb{V}'\widetilde{\mathbb{G}_0\mathbb{Z}}^*\delta_0-\frac{2}{\sigma_0^2}\mathbb{V}'\mathbb{G}_0'\mathbb{V}\mathrm{tr}(\mathbb{G}_0)
-\frac{2}{\sigma_0^2}(\widetilde{\mathbb{G}_0\mathbb{Z}}^*\delta_0)'\mathbb{V}\mathrm{tr}(\mathbb{G}_0)\right]\\
=&\frac{1}{n\sigma_0^2}E\left[\left(\widetilde{\mathbb{G}_0\mathbb{Z}}^*\delta_0+\mathbb{Q}\mathbb{G}_0\mathbb{D\alpha}_0\right)'\left(\widetilde{\mathbb{G}_0\mathbb{Z}}^*\delta_0+\mathbb{Q}\mathbb{G}_0\mathbb{D\alpha}_0\right)\right]+\frac{\mu_4-3\sigma_0^4}{n\sigma_0^4}\sum_{i=1}^n\mathbb{G}_{0,ii}^2\\
&+\frac{1}{n}\mathrm{tr}(\mathbb{G}_0^2+\mathbb{G}_0\mathbb{G}_0')+\frac{2\mu_3}{n\sigma_0^4}\sum_{i=1}^n(\mathbb{Q}\mathbb{G}_0\mathbb{D\alpha}_0)_i\mathbb{G}_{0,ii}+\frac{2\mu_3}{n\sigma_0^4}\sum_{i=1}^nE(\widetilde{\mathbb{G}_0\mathbb{Z}}^*\delta_0)_i\mathbb{G}_{0,ii},
\end{align*}
\begin{align*}
E_{\delta\rho}=&\frac{1}{n}E\left[\frac{1}{\sigma^2_0}\widetilde{\mathbb{Z}}^{*\prime}\mathbb{V}\left(\frac{1}{\sigma^2_0}(\mathbb{G}_0\mathbb{D\alpha}_0)'\mathbb{QV}+\frac{1}{\sigma_0^2}\mathbb{V}'\mathbb{G}_0'\mathbb{V}+\frac{1}{\sigma_0^2}(\widetilde{\mathbb{G}_0\mathbb{Z}}^*\delta_0)'\mathbb{V}-\mathrm{tr}(\mathbb{G}_0)\right)\right]\\
=&\frac{1}{n}E\left[\frac{1}{\sigma^4_0}\widetilde{\mathbb{Z}}^{*\prime}\mathbb{V}(\mathbb{G}_0\mathbb{D\alpha}_0)'\mathbb{QV}+\frac{1}{\sigma_0^4}\widetilde{\mathbb{Z}}^{*\prime}\mathbb{V}\mathbb{V}'\mathbb{G}_0'\mathbb{V}+\frac{1}{\sigma_0^4}\widetilde{\mathbb{Z}}^{*\prime}\mathbb{V}(\widetilde{\mathbb{G}_0\mathbb{Z}}^*\delta_0)'\mathbb{V}-\frac{1}{\sigma_0^2}\widetilde{\mathbb{Z}}^{*\prime}\mathbb{V}\mathrm{tr}(\mathbb{G}_0)\right]\\
=&\frac{1}{n\sigma_0^2}E\left(\widetilde{\mathbb{Z}}^{*\prime}\mathbb{QG}_0\mathbb{D\alpha}_0\right)+\frac{\mu_3}{n\sigma_0^4}\sum_{i=1}^nE(\widetilde{\mathbb{Z}}^{*\prime})_{\cdot,i}\mathbb{G}_{0,ii}+\frac{1}{n\sigma_0^2}E\left(\widetilde{\mathbb{Z}}^{*\prime}\widetilde{\mathbb{G}_0\mathbb{Z}}^*\delta_0\right),
\end{align*}
\begin{align*}
E_{\sigma^2\rho}=&\frac{1}{n}E\left[\left(-\frac{n}{2\sigma_0^2}+\frac{1}{2\sigma_0^4}\mathbb{V'V}\right)\left(\frac{1}{\sigma^2_0}(\mathbb{G}_0\mathbb{D\alpha}_0)'\mathbb{QV}+\frac{1}{\sigma_0^2}\mathbb{V}'\mathbb{G}_0'\mathbb{V}+\frac{1}{\sigma_0^2}(\widetilde{\mathbb{G}_0\mathbb{Z}}^*\delta_0)'\mathbb{V}-\mathrm{tr}(\mathbb{G}_0)\right)\right]\\
=&\frac{1}{n}E\left[-\frac{n}{2\sigma_0^4}(\mathbb{G}_0\mathbb{D\alpha}_0)'\mathbb{QV}-\frac{n}{2\sigma_0^4}\mathbb{V}'\mathbb{G}_0'\mathbb{V}-\frac{n}{2\sigma_0^4}(\widetilde{\mathbb{G}_0\mathbb{Z}}^*\delta_0)'\mathbb{V}+\frac{n}{2\sigma_0^2}\mathrm{tr}(\mathbb{G}_0)+\right.\\
&\left.\frac{1}{2\sigma_0^6}\mathbb{V}'\mathbb{V}(\mathbb{G}_0\mathbb{D\alpha}_0)'\mathbb{QV}
+\frac{1}{2\sigma_0^6}\mathbb{V}'\mathbb{V}\mathbb{V}'\mathbb{G}_0'\mathbb{V}+\frac{1}{2\sigma_0^6}\mathbb{V}'\mathbb{V}(\widetilde{\mathbb{G}_0\mathbb{Z}}^*\delta_0)'\mathbb{V}-\frac{1}{2\sigma_0^4}\mathbb{V}'\mathbb{V}\mathrm{tr}(\mathbb{G}_0)\right]\\
=&\frac{\mu_4-\sigma_0^4}{2n\sigma_0^6}\mathrm{tr}(\mathbb{G}_0)+\frac{\mu_3}{2n\sigma_0^6}\sum_{i=1}^n(\mathbb{QG}_0\mathbb{D\alpha}_0)_i+\frac{\mu_3}{2n\sigma_0^6}\sum_{i=1}^nE(\widetilde{\mathbb{G}_0\mathbb{Z}}^*\delta_0)_i,
\end{align*}
\begin{align*}
E_{\delta\delta'}=&\frac{1}{n}E\left(\frac{1}{\sigma^4_0}\widetilde{\mathbb{Z}}^{*\prime}\mathbb{VV'}\widetilde{\mathbb{Z}}^{*}\right)=
\frac{1}{n}E\left(\frac{1}{\sigma^4_0}\sum_{t=1}^T\sum_{s=1}^T\widetilde{Z}^{*\prime}_tV_tV'_s\widetilde{Z}^{*}_s\right)\\
=&\frac{1}{n}E\left(\frac{1}{\sigma^4_0}\sum_{t=1}^T\widetilde{Z}^{*\prime}_tV_tV'_t\widetilde{Z}^{*}_t\right)=
\frac{1}{n\sigma^2_0}E\left(\widetilde{\mathbb{Z}}^{*\prime}\widetilde{\mathbb{Z}}^{*}\right),
\end{align*}
where the third equality holds because the fact that $\widetilde{Z}^{*}_t$ is independent with $V_s$ when $t\leq s$.
\begin{align*}
E_{\sigma^2\delta'}=&\frac{1}{n}E\left[\left(-\frac{n}{2\sigma_0^2}+\frac{1}{2\sigma_0^4}\mathbb{V'V}\right)\mathbb{V'}\widetilde{\mathbb{Z}}^{*}\right]=\frac{\mu_3}{2n\sigma_0^6}\sum_{i=1}^nE(\widetilde{\mathbb{Z}}^*)_{i,\cdot},
\end{align*}
\begin{align*}
E_{\sigma^2\sigma^2}=&\frac{1}{n}E\left[\left(-\frac{n}{2\sigma_0^2}+\frac{1}{2\sigma_0^4}\mathbb{V'V}\right)^2\right]=\frac{\mu_4-\sigma_0^4}{4\sigma_0^8}.
\end{align*}
Therefore, 
{
\renewcommand{\arraystretch}{2}
\begin{align*}
&E\left(\frac{1}{\sqrt{n}}\Gamma\frac{\partial L_n^{c*}(\theta_0)}{\partial\theta}\cdot\frac{1}{\sqrt{n}}\frac{\partial L_n^{c*}(\theta_0)}{\partial\theta'}\Gamma\right)=\\
&\begin{pmatrix}
\left\{\begin{gathered}
\frac{1}{n\sigma_0^2}E\left[\left(\widetilde{\mathbb{G}_0\mathbb{Z}}^*\delta_0+\mathbb{Q}\mathbb{G}_0\mathbb{D\alpha}_0\right)'\left(\widetilde{\mathbb{G}_0\mathbb{Z}}^*\delta_0+\mathbb{Q}\mathbb{G}_0\mathbb{D\alpha}_0\right)\right]\\
+\frac{1}{n}\mathrm{tr}(\mathbb{G}_0^2+\mathbb{G}_0\mathbb{G}_0')
\end{gathered}\right\}& \displaystyle* & \displaystyle*\\
\displaystyle\frac{1}{n\sigma_0^2}E\left[\widetilde{\Gamma}\widetilde{\mathbb{Z}}^{*\prime}\left(\widetilde{\mathbb{G}_0\mathbb{Z}}^*\delta_0+\mathbb{QG}_0\mathbb{D\alpha}_0\right)\right] & \displaystyle\frac{1}{n\sigma^2_0}\widetilde{\Gamma}E\left(\widetilde{\mathbb{Z}}^{*\prime}\widetilde{\mathbb{Z}}^{*}\right)\widetilde{\Gamma} & \displaystyle*\\
\displaystyle\frac{1}{n\sigma_0^2}\mathrm{tr}(\mathbb{G}_0) & \displaystyle\mathbf{0}_{1\times(k+3)} & \displaystyle\frac{1}{2\sigma_0^4}
\end{pmatrix}+\\
&\begin{pmatrix}
\displaystyle\frac{\mu_4-3\sigma_0^4}{n\sigma_0^4}\sum_{i=1}^n\mathbb{G}_{0,ii}^2+\frac{2\mu_3}{n\sigma_0^4}\sum_{i=1}^n\left[E(\widetilde{\mathbb{G}_0\mathbb{Z}}^*\delta_0)+\mathbb{Q}\mathbb{G}_0\mathbb{D\alpha}_0\right]_i\mathbb{G}_{0,ii}& \displaystyle* & \displaystyle*\\
\displaystyle\frac{\mu_3}{n\sigma_0^4}\sum_{i=1}^nE(\widetilde{\Gamma}\widetilde{\mathbb{Z}}^{*\prime})_{\cdot,i}\mathbb{G}_{0,ii} & \displaystyle\mathbf{0}_{(k+3)\times(k+3)} & \displaystyle*\\
\displaystyle\frac{\mu_4-3\sigma_0^4}{2n\sigma_0^6}\mathrm{tr}(\mathbb{G}_0)+\frac{\mu_3}{2n\sigma_0^6}\sum_{i=1}^n\left[E(\widetilde{\mathbb{G}_0\mathbb{Z}}^*\delta_0)+\mathbb{Q}\mathbb{G}_0\mathbb{D\alpha}_0\right]_i & \displaystyle\frac{\mu_3}{2n\sigma_0^6}\sum_{i=1}^nE(\widetilde{\mathbb{Z}}^*\widetilde{\Gamma})_{i,\cdot} & \displaystyle\frac{\mu_4-3\sigma_0^4}{4\sigma_0^8}
\end{pmatrix}.
\end{align*}
}
The first matrix above is equal to $\Gamma\Sigma_{\theta_0,n}\Gamma+O\left(\frac{1}{T}\right)$ by Lemma \ref{lm_Z*}. Denote the second matrix as $\Gamma\Omega_{\theta_0,n}\Gamma$, 
Since $\widetilde{\mathbb{G}_0\mathbb{Z}}^*=\mathbb{QG}_0\mathbb{Z}+\widetilde{\mathbb{G}_0\mathbb{Z}}^\circ$ and $\widetilde{\mathbb{Z}}^*=\mathbb{Q}\mathbb{Z}+\widetilde{\mathbb{Z}}^\circ$, where $E(\widetilde{\mathbb{G}_0\mathbb{Z}}^\circ)=E(\widetilde{\mathbb{Z}}^\circ)=0$, it follows that
$\sum_{i=1}^nE(\widetilde{\mathbb{G}_0\mathbb{Z}}^*)_{i,\cdot}=\sum_{i=1}^nE(\widetilde{\mathbb{Z}}^*\widetilde{\Gamma})_{i,\cdot}=0$ and $\sum_{i=1}^n\left(\mathbb{Q}\mathbb{G}_0\mathbb{D\alpha}_0\right)_i=0$ because $\iota_n'\mathbb{Q}=0$. Therefore, we have
\begin{align*}
\Gamma\Omega_{\theta_0,n}\Gamma=\begin{pmatrix}
\displaystyle\frac{\mu_4-3\sigma_0^4}{n\sigma_0^4}\sum_{i=1}^n\mathbb{G}_{0,ii}^2+\frac{2\mu_3}{n\sigma_0^4}\sum_{i=1}^n\left[E(\widetilde{\mathbb{G}_0\mathbb{Z}}^*\delta_0)+\mathbb{Q}\mathbb{G}_0\mathbb{D\alpha}_0\right]_i\mathbb{G}_{0,ii}& \displaystyle* & \displaystyle*\\
\displaystyle\frac{\mu_3}{n\sigma_0^4}\sum_{i=1}^nE(\widetilde{\Gamma}\widetilde{\mathbb{Z}}^*)_{\cdot,i}\mathbb{G}_{0,ii} & \displaystyle\mathbf{0}_{(k+3)\times(k+3)} & \displaystyle*\\
\displaystyle\frac{\mu_4-3\sigma_0^4}{2n\sigma_0^6}\mathrm{tr}(\mathbb{G}_0) & \displaystyle\mathbf{0}_{1\times(k+3)} & \displaystyle\frac{\mu_4-3\sigma_0^4}{4\sigma_0^8}
\end{pmatrix}.
\end{align*}
When $v_{it}$ are normally distributed, $\Gamma\Omega_{\theta_0,n}\Gamma=0$ because $\mu_3=0$ and $\mu_4-3\sigma^4=0$ for normal distribution. 
Hence, 
$$E\left(\frac{1}{\sqrt{n}}\Gamma\frac{\partial L_n^{c*}(\theta_0)}{\partial\theta}\cdot\frac{1}{\sqrt{n}}\frac{\partial L_n^{c*}(\theta_0)}{\partial\theta'}\Gamma\right)=\Gamma\Sigma_{\theta_0,n}\Gamma+\Gamma\Omega_{\theta_0,n}\Gamma+O\left(\frac{1}{T}\right).$$

\subsection{Some properties of the Hessian matrix}\label{B.3}
For Hessian matrix $\frac{1}{n}\Gamma\frac{\partial^2 L_n^c(\theta)}{\partial\theta\partial\theta'}\Gamma$, we have the following results:
\begin{enumerate}[(1).]
\item Hessian matrix is Lipschitz continuous at $\theta_0$, which means
\begin{align}
-\frac{1}{n}\Gamma\frac{\partial^2 L_n^c(\theta)}{\partial\theta\partial\theta'}\Gamma-\left(-\frac{1}{n}\Gamma\frac{\partial^2 L_n^c(\theta_0)}{\partial\theta\partial\theta'}\Gamma\right)=\|\Gamma^{-1}(\theta-\theta_0)\|_2\cdot O_p(1),\label{eq_H_conti}
\end{align}
\item Hessian matrix is consistent at $\theta_0$, which means
\begin{align}
-\frac{1}{n}\Gamma\frac{\partial^2 L_n^c(\theta_0)}{\partial\theta\partial\theta'}\Gamma-\Gamma\Sigma_{\theta_0,n}\Gamma=O_p\left(\frac{1}{\sqrt{NT}}\right),\label{eq_H_consi}
\end{align}
\item The information matrix $\Sigma_{\theta_0}\equiv\lim_{n\rightarrow\infty}\Gamma\Sigma_{\theta_0,n}\Gamma$ is nonsingular.
\end{enumerate}

\begin{proof}
\textbf{Proof of (1)}

Since 
\begin{align*}
&-\frac{1}{n}\Gamma\frac{\partial^2 L_n^c(\theta)}{\partial\theta\partial\theta'}\Gamma-\left(-\frac{1}{n}\Gamma\frac{\partial^2 L_n^c(\theta_0)}{\partial\theta\partial\theta'}\Gamma\right)=-\frac{1}{n}\times\\
&\begin{pmatrix}
\displaystyle\left(\frac{1}{\sigma^2}-\frac{1}{\sigma^2_0}\right)\mathbb{(WY)'}\mathbb{QWY}+\mathrm{tr}\left[\mathbb{G(\rho)}^2-\mathbb{G}_0^2\right] & \displaystyle* & \displaystyle*\\
\displaystyle\left(\frac{1}{\sigma^2}-\frac{1}{\sigma^2_0}\right)\widetilde{\Gamma}\mathbb{Z}'\mathbb{QWY} & \displaystyle\left(\frac{1}{\sigma^2}-\frac{1}{\sigma^2_0}\right)\widetilde{\Gamma}\mathbb{Z'QZ}\widetilde{\Gamma} & \displaystyle*\\
\displaystyle\left[\frac{1}{\sigma^4}\mathbb{\widetilde{V}}(\zeta)-\frac{1}{\sigma_0^4}\mathbb{V}\right]'\mathbb{QWY} & \displaystyle\left[\frac{1}{\sigma^4}\mathbb{\widetilde{V}}(\zeta)-\frac{1}{\sigma_0^4}\mathbb{V}\right]'\mathbb{QZ}\widetilde{\Gamma} & \left[\begin{gathered}
\displaystyle-\frac{n}{2\sigma^4_0}+\frac{1}{\sigma^6}\mathbb{\widetilde{V}}(\zeta)'\mathbb{\widetilde{V}}(\zeta)\\
+\frac{n}{2\sigma^4}
-\frac{1}{\sigma^6_0}\mathbb{V'QV}
\end{gathered}\right]
\end{pmatrix}
\end{align*}
where $(\frac{1}{n\sigma^2}-\frac{1}{n\sigma^2_0})\mathbb{(WY)'}\mathbb{QWY}=\frac{1}{n\bar{\sigma}^4}\mathbb{(WY)'}\mathbb{QWY}(\sigma_0^2-\sigma^2)$, $\bar{\sigma}^2$ lies between $\sigma^2$ and $\sigma_0^2$. Since $\mathbb{WY}=\mathbb{G}_0(\mathbb{Z}\delta_0+\mathbb{D}\alpha_0+\mathbb{V})$, the term $\mathbb{(WY)'}\mathbb{QWY}$ can be decomposed as a sum of quadratic forms involving $\mathbb{Z}$ and $\mathbb{V}$, as shown in equation (\ref{eq_WYQWY}). According to Lemma \ref{lm15}, it follows that $\frac{1}{n\bar{\sigma}^4}\mathbb{(WY)'}\mathbb{QWY}=O_p(1)$. Thus, $\|(\frac{1}{n\sigma^2}-\frac{1}{n\sigma^2_0})\mathbb{(WY)'}\mathbb{QWY}\|_2\leq \|(\sigma^2_0-\sigma^2)\|_2\cdot O_p(1)$.
Similarly, we can prove $\|(\frac{1}{n\sigma^2}-\frac{1}{n\sigma^2_0})\widetilde{\Gamma}\mathbb{Z}'\mathbb{QWY}\|_2\leq\|(\sigma^2_0-\sigma^2)\|_2\cdot O_p(1)$ and $\|(\frac{1}{n\sigma^2}-\frac{1}{n\sigma^2_0})\widetilde{\Gamma}\mathbb{Z'QZ}\widetilde{\Gamma}\|_2\leq\|(\sigma^2_0-\sigma^2)\|_2\cdot O_p(1)$ by Lemmas \ref{lm15} and \ref{lm_Lambda}.
Additionally, $\frac{1}{n}\mathrm{tr}[\mathbb{G(\rho)}^2-\mathbb{G}_0^2]=\frac{2}{n}\mathrm{tr}[\mathbb{G(\bar{\rho})}^3](\rho-\rho_0)$, where $\bar{\rho}$ lies between $\rho$ and $\rho_0$. Note that
\begin{align}
\sup_{N,T}\|\mathbb{G}(\bar{\rho})\|_\infty&=\sup_{N,T}\|\mathbb{WS}^{-1}(\bar{\rho})\|_\infty\leq\sup_{N,T}\|\mathbb{W}\|_{\infty}\sup_{N,T}\|\mathbb{S}^{-1}(\Bar{\rho})\|_{\infty}\nonumber\\
&=\left(\sup_{N,T}\sup_t\|W_t\|_{\infty}\right)\left(\sup_{N,T}\sup_t\|S_t(\Bar{\rho})^{-1}\|_{\infty}\right)<\infty,\label{eq_Grho}
\end{align}
where the second equality holds because $\mathbb{W}$ and $\mathbb{S}^{-1}(\Bar{\rho})$ are block diagonal matrix and the last inequality holds by equation (9) and Assumption 5.
Thus, $\sup_{N,T}\|\mathbb{G}(\bar{\rho})^3\|_\infty\leq\sup_{N,T}\|\mathbb{G}(\bar{\rho})\|_\infty^3\leq\infty$, which implies that $\frac{2}{n}\mathrm{tr}[\mathbb{G(\bar{\rho})}^3]=O(1)$. 
So $\|\frac{1}{n}\mathrm{tr}[\mathbb{G(\rho)}^2-\mathbb{G}_0^2]\|_2\leq\|\rho-\rho_0\|_2\cdot O(1)$.
Moreover,
\begin{align*}
&\frac{1}{n}\left[\frac{1}{\sigma^4}\mathbb{\widetilde{V}}(\zeta)-\frac{1}{\sigma_0^4}\mathbb{V}\right]'\mathbb{QWY}\\
=&\frac{1}{n}\left[\frac{1}{\sigma^4}(\rho_0-\rho)\mathbb{WY}+\frac{1}{\sigma^4}\mathbb{Z}(\delta_0-\delta)+\left(\frac{1}{\sigma^4}-\frac{1}{\sigma_0^4}\right)\mathbb{V}\right]'\mathbb{QWY}\\
=&(\rho_0-\rho)\frac{1}{n\sigma^4}(\mathbb{WY})'\mathbb{QWY}+[\widetilde{\Gamma}^{-1}(\delta_0-\delta)]'\frac{1}{n\sigma^4}\widetilde{\Gamma}\mathbb{Z'QWY}-(\sigma^2-\sigma^2_0)\frac{2}{n\bar{\sigma}^6}\mathbb{V'QWY}.
\end{align*}
Since all the coefficients of $\Gamma^{-1}(\theta-\theta_0)$ in the above expression are $O_p(1)$ by Lemmas \ref{lm15} and \ref{lm_Lambda}, we therefore have $\|\frac{1}{n}[\frac{1}{\sigma^4}\mathbb{\widetilde{V}}(\zeta)-\frac{1}{\sigma_0^4}\mathbb{V}]'\mathbb{QWY}\|_2\leq\|\Gamma^{-1}(\theta-\theta_0)\|_2\cdot O_p(1)$ by Cauchy-Schwarz inequality.
Therefore, we can similarly prove that the remaining elements of the above matrix are $\|\Gamma^{-1}(\theta-\theta_0)\|_2\cdot O_p(1)$.


\textbf{Proof of (2)}

We have
\begin{align*}
&-\frac{1}{n}\Gamma\frac{\partial^2 L_n^c(\theta_0)}{\partial\theta\partial\theta'}\Gamma-\Gamma\Sigma_{\theta_0,n}\Gamma=-\frac{1}{n}\times\\
&\begin{pmatrix}
\displaystyle\frac{1}{\sigma^2_0}\left[\mathbb{(WY)'}\mathbb{QWY}-E\left(\mathbb{(WY)'}\mathbb{QWY}\right)\right] & \displaystyle* & \displaystyle*\\
\displaystyle\frac{1}{\sigma^2_0}\left[\widetilde{\Gamma}\mathbb{Z}'\mathbb{QWY}-E\left(\widetilde{\Gamma}\mathbb{Z}'\mathbb{QWY}\right)\right] & \displaystyle\frac{1}{\sigma^2_0}\left[\widetilde{\Gamma}\mathbb{Z'QZ}\widetilde{\Gamma}-E\left(\widetilde{\Gamma}\mathbb{Z'QZ}\widetilde{\Gamma}\right)\right] & \displaystyle*\\
\displaystyle\frac{1}{\sigma_0^4}\left[\mathbb{V}'\mathbb{QWY}-E\left(\mathbb{V}'\mathbb{QWY}\right)\right] & \displaystyle\frac{1}{\sigma_0^4}\left[\mathbb{V}'\mathbb{QZ}\widetilde{\Gamma}-E\left(\mathbb{V}'\mathbb{QZ}\widetilde{\Gamma}\right)\right] & \displaystyle\frac{1}{\sigma^6_0}\left[\mathbb{V'QV}-E\left(\mathbb{V'QV}\right)\right]
\end{pmatrix}.
\end{align*}
Since $\mathbb{WY}=\mathbb{G}_0(\mathbb{Z}\delta_0+\mathbb{D}\alpha_0+\mathbb{V})$, which is a combination of $\mathbb{Z}$ and $\mathbb{V}$, it follows that elements in the matrix above are $O_p(\frac{1}{\sqrt{NT}})$ by Lemmas \ref{lm15} and \ref{lm_Lambda}.

\textbf{Proof of (3)}

The information matrix is nonsingular if and only if for any vector $\boldsymbol{a}$, $\Sigma_{\theta_0}\boldsymbol{a} = 0$ implies $\boldsymbol{a}=0$. Let $\boldsymbol{a}=(a_1,\boldsymbol{a}_2',a_3)'$, where $a_1$ and $a_3$ are scalars and $\boldsymbol{a}_2$ is $(k+3)$-dimensional vector. As
\begin{align*}
\Sigma_{\theta_0}=&\lim_{n\rightarrow\infty}\Gamma\Sigma_{\theta_0,n}\Gamma=\lim_{n\rightarrow\infty}\Gamma\Sigma_{\theta_0,n}^{(1)}\Gamma\\
\equiv&\frac{1}{\sigma_0^2}\times
\begin{pmatrix}
\displaystyle\mathcal{H}_{\rho\rho}
+\lim_{n\rightarrow\infty}\frac{\sigma_0^2}{n}\mathrm{tr}\left(\mathbb{G}_0^2+\mathbb{G}_0'\mathbb{G}_0\right) & \displaystyle* & \displaystyle*\\
\displaystyle\mathcal{H}_{\rho\delta} & \displaystyle\mathcal{H}_{\delta\delta} & \displaystyle*\\
\displaystyle\lim_{n\rightarrow\infty}\frac{1}{n}\mathrm{tr}\left(\mathbb{G}_0\right) & \displaystyle\mathbf{0}_{1\times(k+3)} & \displaystyle\frac{1}{2\sigma_0^2}
\end{pmatrix},
\end{align*}
where the second equality holds because $\Gamma\Sigma_{\theta_0,n}\Gamma=\Gamma\left(\Sigma_{\theta_0,n}^{(1)}-\Sigma_{\theta_0,n}^{(2)}\right)\Gamma=\Gamma\Sigma_{\theta_0,n}^{(1)}\Gamma+O(\frac{1}{T})$, as derived in \ref{Foc}. Additionally, $\mathcal{H}_{\rho\rho}\equiv\lim_{n\rightarrow\infty}\frac{1}{n}E\left[\mathbb{(G}_0\mathbb{Z}\delta_0+\mathbb{G}_0\mathbb{D}\alpha_0)'\mathbb{Q(G}_0\mathbb{Z}\delta_0+\mathbb{G}_0\mathbb{D}\alpha_0)\right]$, $\mathcal{H}_{\rho\delta}\equiv\lim_{n\rightarrow\infty}\frac{1}{n}E[\widetilde{\Gamma}\mathbb{Z}'\mathbb{Q(G}_0\mathbb{Z}\delta_0+\mathbb{G}_0\mathbb{D}\alpha_0)]$ and $\mathcal{H}_{\delta\delta}\equiv\lim_{n\rightarrow\infty}\frac{1}{n}E(\widetilde{\Gamma}\mathbb{Z'QZ}\widetilde{\Gamma})$. 
$\Sigma_{\theta_0}\boldsymbol{a}=0$ means
$$\begin{cases}
\displaystyle\frac{1}{\sigma_0^2}\left[\mathcal{H}_{\rho\rho}
+\lim_{n\rightarrow\infty}\frac{\sigma_0^2}{n}\mathrm{tr}\left(\mathbb{G}_0^2+\mathbb{G}_0'\mathbb{G}_0\right)\right]a_1+\frac{1}{\sigma_0^2}\mathcal{H}_{\rho\delta}'\boldsymbol{a}_2+\frac{1}{\sigma_0^2}\lim_{n\rightarrow\infty}\frac{1}{n}\mathrm{tr}\left(\mathbb{G}_0\right)a_3=0\\
\displaystyle\frac{1}{\sigma_0^2}\mathcal{H}_{\rho\delta}a_1+\frac{1}{\sigma_0^2}\mathcal{H}_{\delta\delta}\boldsymbol{a}_2=0\\
\displaystyle\frac{1}{\sigma_0^2}\lim_{n\rightarrow\infty}\frac{1}{n}\mathrm{tr}\left(\mathbb{G}_0\right)a_1+\frac{1}{2\sigma_0^4}a_3=0
\end{cases}.$$
According to the last two equations, we have $a_3=-2\sigma_0^2\lim_{n\rightarrow\infty}\frac{1}{n}\mathrm{tr}(\mathbb{G}_0)a_1$ and $\boldsymbol{a}_2=-\mathcal{H}_{\delta\delta}^{-1}\mathcal{H}_{\rho\delta}a_1$. Substituting them into the first equation gives
\begin{align}
\left[\mathcal{H}_{\rho\rho}-\mathcal{H}_{\rho\delta}'\mathcal{H}_{\delta\delta}^{-1}\mathcal{H}_{\rho\delta}
+\sigma_0^2\lim_{n\rightarrow\infty}\frac{1}{n}\left(\mathrm{tr}(\mathbb{G}_0^2+\mathbb{G}_0'\mathbb{G}_0)-\frac{2}{n}\left(\mathrm{tr}(\mathbb{G}_0)\right)^2\right)\right]a_1=0.\label{eq_a1}
\end{align}
Note that
\begin{align}
&\left(\begin{matrix}
\mathcal{H}_{\rho\rho} &\mathcal{H}_{\rho\delta}'\\
\mathcal{H}_{\rho\delta} & \mathcal{H}_{\delta\delta}
\end{matrix}\right)\nonumber\\
=&\lim_{n\rightarrow\infty}\frac{1}{n}E\left[\mathbb{(G}_0\mathbb{Z}\delta_0+\mathbb{G}_0\mathbb{D}\alpha_0,\widetilde{\Gamma}\mathbb{Z})'\mathbb{Q(G}_0\mathbb{Z}\delta_0+\mathbb{G}_0\mathbb{D}\alpha_0,\widetilde{\Gamma}\mathbb{Z})\right]\nonumber\\
=&\lim_{n\rightarrow\infty}\frac{1}{n}\nonumber\\
&E\begin{pmatrix}
\displaystyle\mathbb{(G}_0\mathbb{Z}\delta_0+\mathbb{G}_0\mathbb{D}\alpha_0)'\mathbb{Q(G}_0\mathbb{Z}\delta_0+\mathbb{G}_0\mathbb{D}\alpha_0) & \displaystyle* & \displaystyle* & \displaystyle* & \displaystyle*\\
\displaystyle\mathbb{Z}_1'\mathbb{Q(G}_0\mathbb{Z}\delta_0+\mathbb{G}_0\mathbb{D}\alpha_0) & \displaystyle\mathbb{Z}_1'\mathbb{QZ}_1 & \displaystyle* & \displaystyle* & \displaystyle*\\
\displaystyle\mathbb{Z}_2'\mathbb{Q(G}_0\mathbb{Z}\delta_0+\mathbb{G}_0\mathbb{D}\alpha_0) & \displaystyle\mathbb{Z}_2'\mathbb{QZ}_1 & \displaystyle\mathbb{Z}_2'\mathbb{QZ}_2 & \displaystyle* & \displaystyle*\\
\displaystyle\sqrt{T}\mathbb{Z}^{(\gamma)\prime}\mathbb{Q(G}_0\mathbb{Z}\delta_0+\mathbb{G}_0\mathbb{D}\alpha_0) & \displaystyle\sqrt{T}\mathbb{Z}^{(\gamma)\prime}\mathbb{QZ}_1 & \displaystyle\sqrt{T}\mathbb{Z}^{(\gamma)\prime}\mathbb{QZ}_2 & \displaystyle T\mathbb{Z}^{(\gamma)\prime}\mathbb{QZ}^{(\gamma)} & \displaystyle*\\
\displaystyle\mathbb{Z}_4'\mathbb{Q(G}_0\mathbb{Z}\delta_0+\mathbb{G}_0\mathbb{D}\alpha_0) & \displaystyle\mathbb{Z}_4'\mathbb{QZ}_1 & \displaystyle\mathbb{Z}_4'\mathbb{QZ}_2 & \displaystyle\sqrt{T}\mathbb{Z}_4'\mathbb{QZ}^{(\gamma)} & \displaystyle\mathbb{Z}_4'\mathbb{QZ}_4\\
\end{pmatrix}\label{eq_Matrix},
\end{align}
where $\mathbb{Z}_1$, $\mathbb{Z}_2$ and $\mathbb{Z}^{(\gamma)}$ are the first three columns of $\mathbb{Z}$, and $\mathbb{Z}_4$ denotes the remaining $k$ columns of $\mathbb{Z}$. Then $\mathcal{H}_{\rho\rho}-\mathcal{H}_{\rho\delta}'\mathcal{H}_{\delta\delta}^{-1}\mathcal{H}_{\rho\delta}\geq0$ because $\mathbb{(G}_0\mathbb{Z}\delta_0+\mathbb{G}_0\mathbb{D}\alpha_0,\widetilde{\Gamma}\mathbb{Z})'\mathbb{Q(G}_0\mathbb{Z}\delta_0+\mathbb{G}_0\mathbb{D}\alpha_0,\widetilde{\Gamma}\mathbb{Z})$ is positive semi-definite.
Moreover, note that
$$\mathrm{tr}(\mathbb{G}_0^2+\mathbb{G}_0'\mathbb{G}_0)-\frac{2}{n}\mathrm{tr}(\mathbb{G}_0)^2=\frac{1}{2}\mathrm{tr}\left[\left(\mathbb{G}_0+\mathbb{G}_0'-\frac{2}{n}\mathrm{tr}(\mathbb{G}_0)I_n\right)\left(\mathbb{G}_0+\mathbb{G}_0'-\frac{2}{n}\mathrm{tr}(\mathbb{G}_0)I_n\right)'\right]\geq0.$$

Under the above conditions, we will use Assumption 9 (i) and Assumption 9 (ii) separately to show that either $\mathcal{H}_{\rho\rho}-\mathcal{H}_{\rho\delta}'\mathcal{H}_{\delta\delta}^{-1}\mathcal{H}_{\rho\delta}$ or $\lim_{n\rightarrow\infty}\frac{1}{n}\left(\mathrm{tr}(\mathbb{G}_0^2+\mathbb{G}_0'\mathbb{G}_0)-\frac{2}{n}(\mathrm{tr}(\mathbb{G}_0))^2\right)$ is greater than 0, which then implies from equation (\ref{eq_a1}) that $a_1=0$.
\begin{enumerate}
\item (Using Assumption 9 (i)) Because $\mathbb{Z}^{(\gamma)}$ is sparse binary vector with at most $N$ nonzero elements, elements in $E(\mathbb{Z})$ and $\alpha_0$ are uniformly bounded, and $\mathbb{G}_0$ is UB, we have
$\sqrt{T}E[\mathbb{Z}^{(\gamma)\prime}\mathbb{Q(G}_0\mathbb{Z}\delta_0+\mathbb{G}_0\mathbb{D}\alpha_0)]=O(N\sqrt{T})$, $\sqrt{T}E[\mathbb{Z}^{(\gamma)\prime}\mathbb{QZ}_1]=O(N\sqrt{T})$, $\sqrt{T}E[\mathbb{Z}^{(\gamma)\prime}\mathbb{QZ}_2]=O(N\sqrt{T})$, $\sqrt{T}\mathbb{Z}_4'\mathbb{QZ}^{(\gamma)}=O(N\sqrt{T})$ and $T\mathbb{Z}^{(\gamma)\prime}\mathbb{QZ}^{(\gamma)}=O(NT)$. Moreover,
\begin{align*}
\frac{T}{n}\mathbb{Z}^{(\gamma)\prime}\mathbb{QZ}^{(\gamma)}&=\frac{T}{n}\sum_{t=1}^T\iota_N'F_t'D_t'D_tF_t\iota_N-\frac{T}{n}\left(\sum_{t=1}^TD_t'D_tF_t\iota_N\right)'\mathcal{T}^{-1}\left(\sum_{t=1}^TD_t'D_tF_t\iota_N\right)\\
&=\frac{T}{n}\sum_{t=1}^T\iota_N'F_t\iota_N-\frac{T}{n}\mathrm{tr}\left[\mathcal{T}^{-1}\left(\sum_{t=1}^TF_t\iota_N\right)\left(\sum_{t=1}^TF_t\iota_N\right)'\right]\\
&\geq\frac{T(1-c_N)N}{n}-\frac{T}{n}\sum_{i=1}^N\frac{1}{T_i}\\
&\geq \frac{1-c_N}{c_u}-\frac{1}{c_uc_TT},
\end{align*}
where the second equality holds because $D_t'D_t=\mathrm{diag}(d_{1t},\cdots,d_{Nt})$ and $F_t=\mathrm{diag}(d_{1t}(1-d_{1,t-1}),\cdots,$ $d_{Nt}(1-d_{N,t-1}))$. The first inequality follows from Assumption 1, which states that the listing effect appears at least $(1-c_N)N$ times and at most $N$ times over the entire panel. Thus
$\sum_{t=1}^T \iota_N' F_t \iota_N \ge (1-c_N)N$,
and elements in vector $\sum_{t=1}^T F_t \iota_N$ is bounded above by 1. The second inequality holds by $T_i\geq c_TT$ in Assumption 1. Therefore, $\lim_{n\rightarrow\infty}\frac{T}{n}\mathbb{Z}^{(\gamma)\prime}\mathbb{QZ}^{(\gamma)}$ exists and bounded away from zero.

Then according to Assumption 9 (i), the determinant of the matrix (\ref{eq_Matrix}) is, $$\lim_{n\rightarrow\infty}\frac{1}{n}T\mathbb{Z}^{(\gamma)\prime}\mathbb{QZ}^{(\gamma)}\left|\frac{1}{n}E\left[\mathcal{H}^{(-\gamma)}\right]\right|>0.$$
Then the matrix (\ref{eq_Matrix}) is positive definite, and therefore $\mathcal{H}_{\rho\rho}-\mathcal{H}_{\rho\delta}'\mathcal{H}_{\delta\delta}^{-1}\mathcal{H}_{\rho\delta}>0$.
\item (Using Assumption 9 (ii)) It is trivial.
\end{enumerate}

Once we have $a_1=0$, we then have $\boldsymbol{a}_2=\mathbf{0}_{(k+3)\times1}$ and $a_3=0$, which means that the information matrix is nonsingular.
\end{proof}

\subsection{Asymptotic Representation of 
$\Delta_n$}\label{B.4}
Given $$\Delta_n=\left(\begin{aligned}
&\frac{1}{\sqrt{n}\sigma_0^2}\mathbb{V}'\mathbb{G}_0'\mathbb{P}\mathbb{V}+\frac{1}{\sqrt{n}\sigma_0^2}(\widetilde{\mathbb{G}_0\mathbb{Z}}^\circ\delta_0)'\mathbb{V}\\
&\frac{1}{\sqrt{n}\sigma_0^2}\widetilde{\Gamma}\widetilde{\mathbb{Z}}^{\circ\prime}\mathbb{V}\\
&\frac{1}{2\sqrt{n}\sigma_0^4}\mathbb{V'}\mathbb{P}\mathbb{V}
\end{aligned}\right),$$
where 
\begin{align*}
\frac{1}{\sqrt{n}\sigma_0^2}\mathbb{V}'\mathbb{G}_0'\mathbb{P}\mathbb{V}&=\frac{1}{\sqrt{n}}\mathrm{tr}(\mathbb{G}_0'\mathbb{P})+O_p\left(\frac{1}{\sqrt{T}}\right)=\sqrt{\frac{N}{c_uT}}\frac{1}{N}\mathrm{tr}(\mathbb{G}_0'\mathbb{P})+O_p\left(\frac{1}{\sqrt{T}}\right),\\
\frac{1}{2\sqrt{n}\sigma_0^4}\mathbb{V}'\mathbb{P}\mathbb{V}&=\frac{1}{2\sqrt{n}\sigma_0^2}\mathrm{tr}(\mathbb{P})+O_p\left(\frac{1}{\sqrt{T}}\right)=\sqrt{\frac{N}{c_uT}}\frac{1}{2\sigma_0^2}+O_p\left(\frac{1}{\sqrt{T}}\right)
\end{align*}
by equation (\ref{eq_vpv}) in Lemma \ref{lm_VQV}. Additionally, we have
\begin{align*}
\frac{1}{\sqrt{n}\sigma_0^2}(\widetilde{\mathbb{G}_0\mathbb{Z}}^\circ\delta_0)'\mathbb{V}&=\sqrt{\frac{N}{c_uT}}\frac{1}{N}E\left[(\widetilde{\mathbb{G}_0\mathbb{Z}}^\circ\delta_0)'\mathbb{V}\right]+O_p\left(\frac{1}{\sqrt{T}}\right),\\
\frac{1}{\sqrt{n}\sigma_0^2}\left(\widetilde{\Gamma}\widetilde{\mathbb{Z}}^{\circ\prime}\mathbb{V}\right)&=\sqrt{\frac{N}{c_uT}}\frac{1}{N}E\left(\widetilde{\Gamma}\widetilde{\mathbb{Z}}^{\circ\prime}\mathbb{V}\right)+O_p\left(\frac{1}{\sqrt{T}}\right)\\
\end{align*}
by equation (\ref{eq_lpv}) in Lemma \ref{lm_LQV}. 
Since $\widetilde{\mathbb{Z}}^{\circ}=\mathbb{PZ}^{\circ}$, and $Z_t^\circ=(D_t(I_N-F_t)D_t'M_t\mathcal{V}_t,D_tD_{t-1}'\mathcal{V}_t,\mathbf{0}_{N_t\times(k+1)})$, thus
\begin{align*}
E[\widetilde{\Gamma}\widetilde{\mathbb{Z}}^{\circ\prime}\mathbb{V}]=\left(\begin{aligned}
&\frac{1}{N}\mathrm{tr}\left(
\mathcal{T}^{-1}\sum_{h=1}^{T-1}\sum_{t=T-h}^{T-1}D_{t+1}'D_{t+1}(I_N-F_{t+1})D_{t+1}'M_{t+1}S_t(\rho)^{-1}A_{t}^{(t+h-T)}(\theta)D_{T-h}\right)\\
&\frac{1}{N}\mathrm{tr}\left(
\mathcal{T}^{-1}\sum_{h=1}^{T-1}\sum_{t=T-h}^{T-1}D_{t+1}'D_{t+1}D_{t}'S_t(\rho)^{-1}A_{t}^{(t+h-T)}(\theta)D_{T-h}\right)\\
&\mathbf{0}_{(k+1)\times1}
\end{aligned}\right)
\end{align*}
by equations (\ref{eq_ELV}) and (\ref{eq_ELPV}) in Lemma \ref{lm_LQV}. Similarly, we have 
\begin{align*}
E\left[(\widetilde{\mathbb{G}_0\mathbb{Z}}^\circ\delta_0)'\mathbb{V}\right]
=&\mathrm{tr}\left(
\mathcal{T}^{-1}\sum_{h=1}^{T-1}\sum_{t=T-h}^{T-1} \lambda D_{t+1}'G_{t+1}D_{t+1}(I_N-F_{t+1})D_{t+1}'M_{t+1}S_t^{-1}A_{t}^{(t+h-T)}D_{T-h}\right)\\
&+\mathrm{tr}\left(\mathcal{T}^{-1}\sum_{h=1}^{T-1}\sum_{t=T-h}^{T-1}\nu D_{t+1}'G_{t+1}D_{t+1}D_{t}'S_t^{-1}A_{t}^{(t+h-T)}D_{T-h}\right),
\end{align*}
where $G_t\equiv W_tS_t^{-1}$. Thus, $\Delta_n=\sqrt{\frac{N}{c_uT}}\varphi_n(\theta_0)+O_p\left(\frac{1}{\sqrt{T}}\right)$, $\varphi_n(\theta_0)=O(1)$ and
\begin{align}
&\varphi_n(\theta)=\nonumber\\
&\left(\begin{aligned}
&\left[\begin{aligned}
&\frac{1}{N}\mathrm{tr}\left(
\mathcal{T}^{-1}\sum_{h=1}^{T-1}\sum_{t=T-h}^{T-1} \lambda D_{t+1}'G_{t+1}(\rho)D_{t+1}(I_N-F_{t+1})D_{t+1}'M_{t+1}S_t(\rho)^{-1}A_{t}^{(t+h-T)}(\theta)D_{T-h}\right)\\
&+\frac{1}{N}\mathrm{tr}\left(\mathcal{T}^{-1}\sum_{h=1}^{T-1}\sum_{t=T-h}^{T-1}\nu D_{t+1}'G_{t+1}(\rho)D_{t+1}D_{t}'S_t(\rho)^{-1}A_{t}^{(t+h-T)}(\theta)D_{T-h}\right)+\frac{1}{N}\mathrm{tr}(\mathbb{G}(\rho)'\mathbb{P})    
\end{aligned}\right]\\
&\frac{1}{N}\mathrm{tr}\left(
\mathcal{T}^{-1}\sum_{h=1}^{T-1}\sum_{t=T-h}^{T-1}D_{t+1}'D_{t+1}(I_N-F_{t+1})D_{t+1}'M_{t+1}S_t(\rho)^{-1}A_{t}^{(t+h-T)}(\theta)D_{T-h}\right)\\
&\frac{1}{N}\mathrm{tr}\left(
\mathcal{T}^{-1}\sum_{h=1}^{T-1}\sum_{t=T-h}^{T-1}D_{t+1}'D_{t+1}D_{t}'S_t(\rho)^{-1}A_{t}^{(t+h-T)}(\theta)D_{T-h}\right)\\
&\mathbf{0}_{(k+1)\times1}\\
&\frac{1}{2\sigma^2}
\end{aligned}\right),\label{eq_phi_n}
\end{align}
where $G_t(\rho)=W_tS_t(\rho)^{-1}$, $A_t^{(h)}(\theta)=A_t(\theta)A_{t-1}(\theta)\cdots A_{t-h+1}(\theta)$ for $h=1,2,\cdots$, $A_t^{(0)}(\theta)=I_{N_t}$, and $A_t(\theta)=D_t[\lambda_0 (I_N-F_t)D_t'M_t+\nu_0 D_{t-1}']S_{t-1}(\rho)^{-1}$.

\subsection{Score, variance, and Hessian formulas for the TWFE QMLE}\label{secB5}

This subsection provides the TWFE counterparts of the formulas in Sections \ref{Foc}--\ref{B.4}.
Recall that \(\widebar{\mathbb{P}}\) is the projection matrix onto the space spanned by the individual and time effects, and \(\widebar{\mathbb{Q}}=I_n-\widebar{\mathbb{P}}\). For \(\zeta=(\rho,\delta')'\), denote
\[
\widebar{\mathbb{V}}(\zeta)
=
\widebar{\mathbb{Q}}
\left(
\mathbb{S}(\rho)\mathbb{Y}-\mathbb{Z}\delta
\right).
\]
Then the TWFE concentrated log-likelihood function is
\[
\widebar{L}_n^c(\theta)
=
-\frac{n}{2}\log(2\pi)
-\frac{n}{2}\log\sigma^2
+\log|\mathbb{S}(\rho)|
-\frac{1}{2\sigma^2}
\widebar{\mathbb{V}}(\zeta)'\widebar{\mathbb{V}}(\zeta).
\]

\textbf{The first and second order conditions}

For simplicity, denote \(\mathbb{G}(\rho)=\mathbb{W}\mathbb{S}^{-1}(\rho)\). The first order derivatives are
\[
\frac{\partial\widebar{L}_n^c(\theta)}{\partial\theta}
=
\left(
\begin{array}{c}
\displaystyle
\frac{1}{\sigma^2}
(\mathbb{W}\mathbb{Y})'
\widebar{\mathbb{V}}(\zeta)
-
\operatorname{tr}\{\mathbb{G}(\rho)\}
\\
\displaystyle
\frac{1}{\sigma^2}
\mathbb{Z}'
\widebar{\mathbb{V}}(\zeta)
\\
\displaystyle
-\frac{n}{2\sigma^2}
+
\frac{1}{2\sigma^4}
\widebar{\mathbb{V}}(\zeta)'\widebar{\mathbb{V}}(\zeta)
\end{array}
\right).
\]
The second order derivatives are
\[
\frac{\partial^2\widebar{L}_n^c(\theta)}
{\partial\theta\partial\theta'}
=
-
\left(
\begin{array}{ccc}
\displaystyle
\frac{1}{\sigma^2}
(\mathbb{W}\mathbb{Y})'
\widebar{\mathbb{Q}}
\mathbb{W}\mathbb{Y}
+
\operatorname{tr}\{\mathbb{G}^2(\rho)\}
&
*
&
*
\\
\displaystyle
\frac{1}{\sigma^2}
\mathbb{Z}'
\widebar{\mathbb{Q}}
\mathbb{W}\mathbb{Y}
&
\displaystyle
\frac{1}{\sigma^2}
\mathbb{Z}'
\widebar{\mathbb{Q}}
\mathbb{Z}
&
*
\\
\displaystyle
\frac{1}{\sigma^4}
\widebar{\mathbb{V}}(\zeta)'
\widebar{\mathbb{Q}}
\mathbb{W}\mathbb{Y}
&
\displaystyle
\frac{1}{\sigma^4}
\widebar{\mathbb{V}}(\zeta)'
\widebar{\mathbb{Q}}
\mathbb{Z}
&
\displaystyle
-\frac{n}{2\sigma^4}
+
\frac{1}{\sigma^6}
\widebar{\mathbb{V}}(\zeta)'\widebar{\mathbb{V}}(\zeta)
\end{array}
\right).
\]

Moreover, the information matrix is
\(
\widebar{\Sigma}_{\theta_0,n}
\equiv
E\left(
-\frac{1}{n}
\frac{\partial^2\widebar{L}_n^c(\theta_0)}
{\partial\theta\partial\theta'}
\right)
=
\widebar{\Sigma}_{\theta_0,n}^{(1)}
-
\widebar{\Sigma}_{\theta_0,n}^{(2)},
\)
where \(\widebar{\Sigma}_{\theta_0,n}^{(1)}=\)
\[
\left(
\begin{array}{ccc}
\displaystyle
\left[\begin{aligned}
\frac{1}{n\sigma_0^2}
E\left[
\left\{
\mathbb{G}_0\mathbb{Z}\delta_0
+
\mathbb{G}_0\mathbb{D}\alpha_0
+
\mathbb{G}_0\mathbb{J}\tau_0
\right\}'
\widebar{\mathbb{Q}}
\left\{
\mathbb{G}_0\mathbb{Z}\delta_0
+
\mathbb{G}_0\mathbb{D}\alpha_0
+
\mathbb{G}_0\mathbb{J}\tau_0
\right\}
\right]
\\
+
\frac{1}{n}
\operatorname{tr}
\left(
\mathbb{G}_0^2+\mathbb{G}_0'\mathbb{G}_0
\right)
\end{aligned}
\right]
&
*
&
*
\\
\displaystyle
\frac{1}{n\sigma_0^2}
E\left[
\mathbb{Z}'
\widebar{\mathbb{Q}}
\left\{
\mathbb{G}_0\mathbb{Z}\delta_0
+
\mathbb{G}_0\mathbb{D}\alpha_0
+
\mathbb{G}_0\mathbb{J}\tau_0
\right\}
\right]
&
\displaystyle
\frac{1}{n\sigma_0^2}
E\left(
\mathbb{Z}'\widebar{\mathbb{Q}}\mathbb{Z}
\right)
&
*
\\
\displaystyle
\frac{1}{n\sigma_0^2}
\operatorname{tr}\left(\mathbb{G}_0\right)
&
\displaystyle
\mathbf{0}_{1\times(k+3)}
&
\displaystyle
\frac{1}{2\sigma_0^4}
\end{array}
\right),
\]
and \(\widebar{\Sigma}_{\theta_0,n}^{(2)}=\)
\[
\left(
\begin{array}{ccc}
\displaystyle
\frac{2}{n\sigma_0^2}
E\left[
\left(
\mathbb{G}_0\mathbb{Z}\delta_0
\right)'
\widebar{\mathbb{P}}\mathbb{G}_0\mathbb{V}
\right]
+
\frac{1}{n}
\operatorname{tr}
\left(
\mathbb{G}_0'\widebar{\mathbb{P}}\mathbb{G}_0
\right)
&
*
&
*
\\
\displaystyle
\frac{1}{n\sigma_0^2}
E\left[
\mathbb{Z}'
\widebar{\mathbb{P}}\mathbb{G}_0\mathbb{V}
\right]
&
\displaystyle
\mathbf{0}_{(k+3)\times(k+3)}
&
*
\\
\displaystyle
\frac{1}{n\sigma_0^4}
E\left(
\mathbb{V}'\widebar{\mathbb{P}}\mathbb{G}_0\mathbb{Z}\delta_0
\right)
+
\frac{1}{n\sigma_0^2}
\operatorname{tr}
\left(
\widebar{\mathbb{P}}\mathbb{G}_0
\right)
&
\displaystyle
\frac{1}{n\sigma_0^4}
E\left(
\mathbb{V}'\widebar{\mathbb{P}}\mathbb{Z}
\right)
&
\displaystyle
\frac{N+T-1}{n\sigma_0^4}
\end{array}
\right).
\]
By Lemma~\ref{lem:twfe-ZV-ZZ}, all expectation terms in
\(\widebar{\Sigma}_{\theta_0,n}^{(2)}\) are
\(
O\left(\frac{1}{T}\right)
+
O\left(\frac{1}{N}\right).
\)
It remains to control the trace terms. Since \(\widebar{\mathbb{P}}\) is an orthogonal
projection matrix with rank \(N+T-1\), and \(\mathbb{G}_0\), \(\mathbb{G}_0'\mathbb{G}_0\) are UB under Assumptions~3 and 5, we have
\[
\begin{aligned}
\frac{1}{n}
\left|
\operatorname{tr}
\left(
\widebar{\mathbb{P}}\mathbb{G}_0
\right)
\right|
\leq
\frac{1}{n}
\operatorname{rank}\left(\widebar{\mathbb{P}}\right)
\left\|\mathbb{G}_0\right\|_2=
O\left(\frac{N+T}{NT}\right)
=
O\left(\frac{1}{T}\right)
+
O\left(\frac{1}{N}\right).
\end{aligned}
\]
Similarly,
\[
\frac{1}{n}
\left|
\operatorname{tr}
\left(
\mathbb{G}_0'\widebar{\mathbb{P}}\mathbb{G}_0
\right)
\right|
\leq
\frac{1}{n}
\operatorname{rank}\left(\widebar{\mathbb{P}}\right)
\left\|\mathbb{G}_0\right\|_2^2
=
O\left(\frac{1}{T}\right)
+
O\left(\frac{1}{N}\right).
\]
Thus,
\(
\widebar{\Sigma}_{\theta_0,n}^{(2)}
=
O\left(\frac{1}{T}\right)
+
O\left(\frac{1}{N}\right),
\)
which means that
\(
\widebar{\Sigma}_{\theta_0,n}
=
\widebar{\Sigma}_{\theta_0,n}^{(1)}
+
O\left(\frac{1}{T}\right)
+
O\left(\frac{1}{N}\right).
\)
We also have
\(
\Gamma\widebar{\Sigma}_{\theta_0,n}\Gamma
=
\Gamma\widebar{\Sigma}_{\theta_0,n}^{(1)}\Gamma
+
O\left(\frac{1}{T}\right)
+
O\left(\frac{1}{N}\right)
\)
by Lemma \ref{lem:twfe-ZV-ZZ}.

\textbf{TWFE score decomposition}

Let \(\widebar{\mathbb P}=I_n-\widebar{\mathbb Q}\). Define
\(\widebar{\widetilde{\mathbb{Z}}}^{\circ}
=\widebar{\mathbb{P}}\mathbb{Z}^{\circ}\),
\(\widebar{\widetilde{\mathbb{Z}}}^{*}
=\widebar{\mathbb{Q}}\mathbb{Z}
+\widebar{\widetilde{\mathbb{Z}}}^{\circ}\),
\(\widebar{\widetilde{\mathbb{G}_0\mathbb{Z}}}^{\circ}
=\widebar{\mathbb{P}}\mathbb{G}_0\mathbb{Z}^{\circ}\), and
\(\widebar{\widetilde{\mathbb{G}_0\mathbb{Z}}}^{*}
=\widebar{\mathbb{Q}}\mathbb{G}_0\mathbb{Z}
+\widebar{\widetilde{\mathbb{G}_0\mathbb{Z}}}^{\circ}\).
At \(\theta_0\), the scaled TWFE score can be decomposed as
\[
\frac{1}{\sqrt{n}}
\Gamma
\frac{\partial\widebar{L}_n^c(\theta_0)}{\partial\theta}
=
\frac{1}{\sqrt{n}}
\Gamma
\frac{\partial\widebar{L}_n^{c*}(\theta_0)}{\partial\theta}
-
\widebar{\Delta}_n,
\]
where
\[
\frac{1}{\sqrt{n}}
\Gamma
\frac{\partial\widebar{L}_n^{c*}(\theta_0)}{\partial\theta}
=
\left(
\begin{array}{c}
\displaystyle
\frac{1}{\sqrt{n}\sigma_0^2}
\left[
\left\{
\widebar{\mathbb{Q}}\mathbb{G}_0
(\mathbb{D}\alpha_0+\mathbb{J}\tau_0)
+
\widebar{\widetilde{\mathbb{G}_0\mathbb{Z}}}^{*}\delta_0
\right\}'\mathbb{V}
+
\mathbb{V}'\mathbb{G}_0\mathbb{V}
\right]
-
\frac{1}{\sqrt{n}}\operatorname{tr}(\mathbb{G}_0)
\\[1.5ex]
\displaystyle
\frac{1}{\sqrt{n}\sigma_0^2}
\widetilde{\Gamma}'
\widebar{\widetilde{\mathbb{Z}}}^{*\prime}\mathbb{V}
\\[1.5ex]
\displaystyle
-\frac{\sqrt{n}}{2\sigma_0^2}
+
\frac{1}{2\sqrt{n}\sigma_0^4}
\mathbb{V}'\mathbb{V}
\end{array}
\right),
\]
and
\[
\widebar{\Delta}_n
=
\left(
\begin{array}{c}
\displaystyle
\frac{1}{\sqrt{n}\sigma_0^2}
\mathbb{V}'\mathbb{G}_0'
\widebar{\mathbb{P}}\mathbb{V}
+
\frac{1}{\sqrt{n}\sigma_0^2}
\left(
\widebar{\widetilde{\mathbb{G}_0\mathbb{Z}}}^{\circ}\delta_0
\right)'\mathbb{V}
\\[1.5ex]
\displaystyle
\frac{1}{\sqrt{n}\sigma_0^2}
\widetilde{\Gamma}'
\widebar{\widetilde{\mathbb{Z}}}^{\circ\prime}\mathbb{V}
\\[1.5ex]
\displaystyle
\frac{1}{2\sqrt{n}\sigma_0^4}
\mathbb{V}'\widebar{\mathbb{P}}\mathbb{V}
\end{array}
\right).
\]
The first component is a mean-zero score component, while
\(\widebar{\Delta}_n\) collects the nonzero mean terms generated by
concentrating out the individual and time fixed effects.

\textbf{Variance of the mean-zero score}

The mean-zero score can be written as
\[
\frac{1}{\sqrt{n}}
\Gamma
\frac{\partial\widebar{L}_n^{c*}(\theta_0)}{\partial\theta}
=
\left(
\begin{array}{c}
\displaystyle
\frac{1}{\sqrt{n}\sigma_0^2}
\left[
\widebar{\mathbb{Q}}\mathbb{G}_0
\left(
\mathbb{D}\alpha_0+\mathbb{J}\tau_0
\right)
+
\widebar{\widetilde{\mathbb{G}_0\mathbb{Z}}}^{*}\delta_0
\right]'
\mathbb{V}
+
\frac{1}{\sqrt{n}\sigma_0^2}
\mathbb{V}'\mathbb{G}_0\mathbb{V}
-
\frac{1}{\sqrt{n}}\operatorname{tr}(\mathbb{G}_0)
\\
\displaystyle
\frac{1}{\sqrt{n}\sigma_0^2}
\widetilde{\Gamma}
\widebar{\widetilde{\mathbb{Z}}}^{*\prime}\mathbb{V}
\\
\displaystyle
-\frac{\sqrt{n}}{2\sigma_0^2}
+
\frac{1}{2\sqrt{n}\sigma_0^4}
\mathbb{V}'\mathbb{V}
\end{array}
\right).
\]
By the same calculation as in Section~B.2 and Lemma \ref{lem:twfe-bridge}, with \(\mathbb{Q}\) replaced by \(\widebar{\mathbb{Q}}\) and \(\mathbb{D}\alpha_0\) replaced by \(\mathbb{D}\alpha_0+\mathbb{J}\tau_0\), we have
\begin{align}
&E
\left[
\frac{1}{\sqrt{n}}
\Gamma
\frac{\partial\widebar{L}_n^{c*}(\theta_0)}{\partial\theta}
\frac{\partial\widebar{L}_n^{c*}(\theta_0)}{\partial\theta'}
\Gamma
\frac{1}{\sqrt{n}}
\right]\nonumber
\\
=&
\left(
\begin{array}{ccc}
\widebar{\Sigma}_{\rho\rho,n}
&
\widebar{\Sigma}_{\rho\delta,n}
&
\widebar{\Sigma}_{\rho\sigma,n}
\\
*
&
\widebar{\Sigma}_{\delta\delta,n}
&
0
\\
*
&
*
&
\displaystyle
\frac{1}{2\sigma_0^4}
\end{array}
\right)+
\left(
\begin{array}{ccc}
\widebar{\Omega}_{\rho\rho,n}
&
\widebar{\Omega}_{\rho\delta,n}
&
\widebar{\Omega}_{\rho\sigma,n}
\\
*
&
\mathbf{0}_{(k+3)\times(k+3)}
&
\widebar{\Omega}_{\delta\sigma,n}
\\
*
&
*
&
\displaystyle
\frac{\mu_4-3\sigma_0^4}{4\sigma_0^8}
\end{array}
\right)+
O\left(\frac{1}{T}\right)
+
O\left(\frac{1}{N}\right),\label{eq_TWFE_VAR}
\end{align}
where
\[
\begin{aligned}
\widebar{\Sigma}_{\rho\rho,n}
=&
\frac{1}{n\sigma_0^2}
E
\left[
\left\{
\widebar{\mathbb{Q}}\mathbb{G}_0
\left(
\mathbb{D}\alpha_0+\mathbb{J}\tau_0
\right)
+
\widebar{\widetilde{\mathbb{G}_0\mathbb{Z}}}^{*}\delta_0
\right\}'
\left\{
\widebar{\mathbb{Q}}\mathbb{G}_0
\left(
\mathbb{D}\alpha_0+\mathbb{J}\tau_0
\right)
+
\widebar{\widetilde{\mathbb{G}_0\mathbb{Z}}}^{*}\delta_0
\right\}
\right]
\\
&+
\frac{1}{n}
\operatorname{tr}
\left(
\mathbb{G}_0^2+\mathbb{G}_0'\mathbb{G}_0
\right),
\\
\widebar{\Sigma}_{\rho\delta,n}
=&
\frac{1}{n\sigma_0^2}
E
\left[
\left\{
\widebar{\mathbb{Q}}\mathbb{G}_0
\left(
\mathbb{D}\alpha_0+\mathbb{J}\tau_0
\right)
+
\widebar{\widetilde{\mathbb{G}_0\mathbb{Z}}}^{*}\delta_0
\right\}'
\widebar{\widetilde{\mathbb{Z}}}^{*}
\right]
\widetilde{\Gamma},
\\
\widebar{\Sigma}_{\delta\delta,n}
=&
\frac{1}{n\sigma_0^2}
\widetilde{\Gamma}
E
\left(
\widebar{\widetilde{\mathbb{Z}}}^{*\prime}
\widebar{\widetilde{\mathbb{Z}}}^{*}
\right)
\widetilde{\Gamma},
\\
\widebar{\Sigma}_{\sigma\rho,n}
=&
\frac{1}{n\sigma_0^2}
\operatorname{tr}(\mathbb{G}_0),
\\
\widebar{\Omega}_{\rho\rho,n}
=&
\frac{\mu_4-3\sigma_0^4}{n\sigma_0^4}
\sum_{i=1}^{n}
(\mathbb{G}_0)_{ii}^2
+
\frac{2\mu_3}{n\sigma_0^4}
\sum_{i=1}^{n}
E
\left[
\left\{
\widebar{\mathbb{Q}}\mathbb{G}_0
\left(
\mathbb{D}\alpha_0+\mathbb{J}\tau_0
\right)
+
\widebar{\widetilde{\mathbb{G}_0\mathbb{Z}}}^{*}\delta_0
\right\}_i
\right]
(\mathbb{G}_0)_{ii},
\\
\widebar{\Omega}_{\rho\delta,n}
=&
\frac{\mu_3}{n\sigma_0^4}
\sum_{i=1}^{n}
(\mathbb{G}_0)_{ii}
E
\left(
\widebar{\widetilde{\mathbb{Z}}}^{*}\widetilde{\Gamma}
\right)_{i,\cdot},
\\
\widebar{\Omega}_{\rho\sigma,n}
=&
\frac{\mu_4-3\sigma_0^4}{2n\sigma_0^6}
\operatorname{tr}(\mathbb{G}_0)
+
\frac{\mu_3}{2n\sigma_0^6}
\sum_{i=1}^{n}
E
\left[
\left\{
\widebar{\mathbb{Q}}\mathbb{G}_0
\left(
\mathbb{D}\alpha_0+\mathbb{J}\tau_0
\right)
+
\widebar{\widetilde{\mathbb{G}_0\mathbb{Z}}}^{*}\delta_0
\right\}_i
\right],
\\
\widebar{\Omega}_{\delta\sigma,n}
=&
\frac{\mu_3}{2n\sigma_0^6}
\sum_{i=1}^{n}
E
\left[
\left(
\widebar{\widetilde{\mathbb{Z}}}^{*}\widetilde{\Gamma}
\right)_{i,\cdot}^{\prime}
\right].
\end{aligned}
\]
Since
\(
\widebar{\widetilde{\mathbb{Z}}}^{*}
=
\widebar{\mathbb{Q}}\mathbb{Z}
+
\widebar{\widetilde{\mathbb{Z}}}^{\circ}\), and
\(
\widebar{\widetilde{\mathbb{G}_0\mathbb{Z}}}^{*}
=
\widebar{\mathbb{Q}}\mathbb{G}_0\mathbb{Z}
+
\widebar{\widetilde{\mathbb{G}_0\mathbb{Z}}}^{\circ}\),
where
\(
E\!\left(\widebar{\widetilde{\mathbb{Z}}}^{\circ}\right)=0\) and
\(E\!\left(\widebar{\widetilde{\mathbb{G}_0\mathbb{Z}}}^{\circ}\right)=0,
\)
and since \(\iota_n'\widebar{\mathbb{Q}}=0\), we have
\[
\begin{aligned}
\sum_{i=1}^{n}
E
\left[
\left\{
\widebar{\mathbb{Q}}\mathbb{G}_0
\left(
\mathbb{D}\alpha_0+\mathbb{J}\tau_0
\right)
+
\widebar{\widetilde{\mathbb{G}_0\mathbb{Z}}}^{*}\delta_0
\right\}_i
\right]
=
\iota_n'
\widebar{\mathbb{Q}}\mathbb{G}_0
\left(
\mathbb{D}\alpha_0+\mathbb{J}\tau_0
\right)
+
\iota_n'
\widebar{\mathbb{Q}}\mathbb{G}_0E(\mathbb{Z})\delta_0
=0,
\end{aligned}
\]
and
\[
\sum_{i=1}^{n}
E
\left[
\left(
\widebar{\widetilde{\mathbb{Z}}}^{*}\widetilde{\Gamma}
\right)_{i,\cdot}^{\prime}
\right]
=
\left\{
\iota_n'
\widebar{\mathbb{Q}}E(\mathbb{Z})\widetilde{\Gamma}
\right\}^{\prime}
=
\mathbf{0}_{(k+3)\times1}.
\]
Therefore,
\[
\widebar{\Omega}_{\rho\sigma,n}
=
\frac{\mu_4-3\sigma_0^4}{2n\sigma_0^6}
\operatorname{tr}(\mathbb{G}_0),
\qquad
\widebar{\Omega}_{\delta\sigma,n}
=
\mathbf{0}_{(k+3)\times1}.
\]

If \(v_{it}\)'s are normally distributed, then \(\mu_3=0\) and \(\mu_4-3\sigma_0^4=0\). Hence
\(
\Gamma\widebar{\Omega}_{\theta_0,n}\Gamma=0.
\)

\textbf{Some properties of the Hessian matrix}

For the Hessian matrix \(
\frac{1}{n}
\Gamma
\frac{\partial^2\widebar{L}_n^c(\theta)}
{\partial\theta\partial\theta'}
\Gamma,
\)
we have the following results.

(1). The Hessian matrix is Lipschitz continuous at \(\theta_0\), that is,
\begin{align}
-\frac{1}{n}
\Gamma
\frac{\partial^2\widebar{L}_n^c(\theta)}
{\partial\theta\partial\theta'}
\Gamma
-
\left(
-\frac{1}{n}
\Gamma
\frac{\partial^2\widebar{L}_n^c(\theta_0)}
{\partial\theta\partial\theta'}
\Gamma
\right)
=
\left\|
\Gamma^{-1}(\theta-\theta_0)
\right\|_2O_p(1).
\label{B.7}
\end{align}

(2). The Hessian matrix is consistent at \(\theta_0\), that is,
\begin{align}
-\frac{1}{n}
\Gamma
\frac{\partial^2\widebar{L}_n^c(\theta_0)}
{\partial\theta\partial\theta'}
\Gamma
-
\Gamma\widebar{\Sigma}_{\theta_0,n}\Gamma
=
O_p\left(\frac{1}{\sqrt{NT}}\right).
\label{B.8}
\end{align}

(3). The information matrix
\(
\widebar{\Sigma}_{\theta_0}
=
\lim_{n\to\infty}
\Gamma\widebar{\Sigma}_{\theta_0,n}\Gamma
\)
is nonsingular.

\begin{proof}
For (1), the proof is the same as that of the Lipschitz continuity of the Hessian matrix in
Section~B.3. The only difference is that \(\mathbb{Q}\) is replaced by
\(\widebar{\mathbb{Q}}\). Since \(\widebar{\mathbb{Q}}\) is also an orthogonal projection matrix,
the TWFE linear--quadratic form bounds in Lemma~\ref{lem:twfe-ZV-ZZ} give the same orders.
Therefore,
\[
-\frac{1}{n}
\Gamma
\frac{\partial^2\widebar{L}_n^c(\theta)}
{\partial\theta\partial\theta'}
\Gamma
-
\left(
-\frac{1}{n}
\Gamma
\frac{\partial^2\widebar{L}_n^c(\theta_0)}
{\partial\theta\partial\theta'}
\Gamma
\right)
=
\left\|
\Gamma^{-1}(\theta-\theta_0)
\right\|_2O_p(1).
\]

For (2), at \(\theta_0\), we have
\[
-\frac{1}{n}
\Gamma
\frac{\partial^2\widebar{L}_n^c(\theta_0)}
{\partial\theta\partial\theta'}
\Gamma
=
\left(
\begin{array}{ccc}
\displaystyle
\frac{1}{n\sigma_0^2}
(\mathbb{W}\mathbb{Y})'
\widebar{\mathbb{Q}}
\mathbb{W}\mathbb{Y}
+
\frac{1}{n}
\operatorname{tr}(\mathbb{G}_0^2)
&
\displaystyle
\frac{1}{n\sigma_0^2}
(\mathbb{W}\mathbb{Y})'
\widebar{\mathbb{Q}}
\mathbb{Z}\widetilde{\Gamma}
&
\displaystyle
\frac{1}{n\sigma_0^4}
(\mathbb{W}\mathbb{Y})'
\widebar{\mathbb{Q}}
\mathbb{V}
\\
*
&
\displaystyle
\frac{1}{n\sigma_0^2}
\widetilde{\Gamma}
\mathbb{Z}'
\widebar{\mathbb{Q}}
\mathbb{Z}
\widetilde{\Gamma}
&
\displaystyle
\frac{1}{n\sigma_0^4}
\widetilde{\Gamma}
\mathbb{Z}'
\widebar{\mathbb{Q}}
\mathbb{V}
\\
*
&
*
&
\displaystyle
-\frac{1}{2\sigma_0^4}
+
\frac{1}{n\sigma_0^6}
\mathbb{V}'
\widebar{\mathbb{Q}}
\mathbb{V}
\end{array}
\right).
\]
Moreover,
\(
\mathbb{W}\mathbb{Y}
=
\mathbb{G}_0\mathbb{Z}\delta_0
+
\mathbb{G}_0
\left(
\mathbb{D}\alpha_0+\mathbb{J}\tau_0
\right)
+
\mathbb{G}_0\mathbb{V}.
\)
Hence, after substituting this expression into the above Hessian matrix, its stochastic blocks are finite linear combinations of TWFE linear--quadratic forms involving \(\mathbb{Z}\) and \(\mathbb{V}\), together with nonstochastic terms. By Lemma~\ref{lem:twfe-ZV-ZZ}, we have
\[
-\frac{1}{n}
\Gamma
\frac{\partial^2\widebar{L}_n^c(\theta_0)}
{\partial\theta\partial\theta'}
\Gamma
-
\Gamma\widebar{\Sigma}_{\theta_0,n}\Gamma
=
O_p\left(\frac{1}{\sqrt{NT}}\right).
\]
This proves (2).

For (3), the proof is the same as the proof of nonsingularity in Section~B.3. Specifically, after
replacing the OWFE information matrix by
\(\Gamma\widebar{\Sigma}_{\theta_0,n}\Gamma\), the Schur complement argument gives the
same two alternatives as in Assumption~9$'$. Thus, under Assumption~9$'\),
\(
\widebar{\Sigma}_{\theta_0}\)
is nonsingular. This completes the proof of (1)--(3).
\end{proof}

\textbf{TWFE bias representation}

Using the decomposition of \(\widebar{\Delta}_n\) above and the identity
\(
\widebar{\mathbb{P}}
=
\mathbb{P}
+
\mathbb{Q}\mathbb{J}
(\mathbb{J}'\mathbb{Q}\mathbb{J})^{-1}
\mathbb{J}'\mathbb{Q}
\),
we have
\[
\widebar{\widetilde{\mathbb{Z}}}^{\circ}
=
\widetilde{\mathbb{Z}}^{\circ}
+
\mathbb{Q}\mathbb{J}
(\mathbb{J}'\mathbb{Q}\mathbb{J})^{-1}
\mathbb{J}'\mathbb{Q}
\mathbb{Z}^{\circ},
\]
and
\[
\widebar{\widetilde{\mathbb{G}_0\mathbb{Z}}}^{\circ}
=
\widetilde{\mathbb{G}_0\mathbb{Z}}^{\circ}
+
\mathbb{Q}\mathbb{J}
(\mathbb{J}'\mathbb{Q}\mathbb{J})^{-1}
\mathbb{J}'\mathbb{Q}
\mathbb{G}_0\mathbb{Z}^{\circ}.
\]
Therefore, \(\widebar{\Delta}_n\) can be decomposed into the OWFE component
generated by \(\mathbb{P}\) and the time-effect component generated by
\(\mathbb{Q}\mathbb{J}(\mathbb{J}'\mathbb{Q}\mathbb{J})^{-1}\mathbb{J}'\mathbb{Q}\).
The leading deterministic part of the OWFE component is summarized by
\(\varphi_n(\theta)\), whose explicit expression is given in
equation~(\ref{eq_phi_n}).


For the time-effect component, let \(K_t\) be the \(N_t\times(T-1)\) matrix such that
\((K_1', K_2',\cdots,K_T')'=\mathbb{Q}\mathbb{J}\).
Then according to equations (\ref{eq:twfe-LV-5}) and (\ref{eq:twfe-LV-6}), the additional time-effect bias component is
\begin{align}
\psi_n(\theta)
\equiv
\left(
\begin{array}{c}
\displaystyle
\begin{aligned}
&
\frac{1}{T}
\sum_{t=1}^{T}
\operatorname{tr}
\left[
(\mathbb{J}'\mathbb{Q}\mathbb{J})^{-1}
K_t'
G_t(\rho)
K_t
\right]
+
\frac{1}{T}
\sum_{t=1}^{T-1}
\sum_{h=T-t}^{T-1}
\operatorname{tr}
\left[
(\mathbb{J}'\mathbb{Q}\mathbb{J})^{-1}
K_{t+1}'
\mathcal{M}_{\rho,t,h}(\theta)
K_{T-h}
\right]    
\end{aligned}
\\
\displaystyle
\frac{1}{T}
\sum_{t=1}^{T-1}
\sum_{h=T-t}^{T-1}
\operatorname{tr}
\left[
(\mathbb{J}'\mathbb{Q}\mathbb{J})^{-1}
K_{t+1}'
\mathcal{M}_{\lambda,t,h}(\theta)
K_{T-h}
\right]
\\
\displaystyle
\frac{1}{T}
\sum_{t=1}^{T-1}
\sum_{h=T-t}^{T-1}
\operatorname{tr}
\left[
(\mathbb{J}'\mathbb{Q}\mathbb{J})^{-1}
K_{t+1}'
\mathcal{M}_{\nu,t,h}(\theta)
K_{T-h}
\right]
\\
\displaystyle
\mathbf{0}_{k+1}
\\
\displaystyle
\frac{T-1}{2T\sigma^2}
\end{array}
\right),
\label{B.9}
\end{align}
where
\[
\begin{aligned}
\mathcal{M}_{\lambda,t,h}(\theta)
=&
D_{t+1}(I_N-F_{t+1})D_{t+1}'M_{t+1}S_t^{-1}(\rho)
A_t^{(t+h-T)}(\theta),
\\
\mathcal{M}_{\nu,t,h}(\theta)
=&
D_{t+1}D_t'S_t^{-1}(\rho)
A_t^{(t+h-T)}(\theta),
\\
\mathcal{M}_{\rho,t,h}(\theta)
=&
G_{t+1}(\rho)
\left[
\lambda\mathcal{M}_{\lambda,t,h}(\theta)
+
\nu\mathcal{M}_{\nu,t,h}(\theta)
\right].
\end{aligned}
\]
We have \(\psi_n(\theta_0)=O(1)\) by equation (\ref{eq:twfe-LV-10}).
Thus, by \eqref{eq:twfe-LV-4} and \eqref{eq:twfe-LV-8a},
\begin{align}
\widebar{\Delta}_n
=
\sqrt{\frac{N}{c_uT}}
\varphi_n(\theta_0)
+
\sqrt{\frac{T}{c_uN}}
\psi_n(\theta_0)
+
O_p\left(\frac{1}{\sqrt{T}}\right)
+
O_p\left(\frac{1}{\sqrt{N}}\right),
\label{B.10}
\end{align}
where \(c_u=n/(NT)\), \(\varphi_n(\theta_0)=O(1)\), and
\(\psi_n(\theta_0)=O(1)\).

Furthermore, by the nonsingularity of
\(\Gamma\widebar{\Sigma}_{\theta_0,n}\Gamma\) established above and the same argument as in
Section~\ref{B.4},
\(
\widebar{\Sigma}_{\theta_0,n}^{-1}\varphi_n(\theta_0)=O_p(1),\) and
\(\widebar{\Sigma}_{\theta_0,n}^{-1}\psi_n(\theta_0)=O_p(1).
\)

\section{Proof of Theorems}

\subsection{Proof of Theorem 1}
Since only $O(N)$ observations contain information about $\gamma$, this parameter exhibits a slower convergence rate compared to others. 
Denote $\theta^{(-\gamma)}$ as the vector obtained by removing $\gamma$ from $\theta$, with $\hat{\theta}^{(-\gamma)}$ and $\theta^{(-\gamma)}_0$ defined analogously. We will first prove that $\hat{\theta}^{(-\gamma)}\overset{p}{\rightarrow}\theta^{(-\gamma)}_0$, then the consistency of $\hat{\gamma}$ follows directly from its explicit expression in terms of $\hat{\theta}^{(-\gamma)}$.

Recall that
$$Q_n(\theta)= E[L_n^c(\theta)]=-\frac{n}{2}\ln2\pi-\frac{n}{2}\ln\sigma^2+\ln\left|\mathbb{S}(\rho)\right|-\frac{1}{2\sigma^2}E\left[\widetilde{\mathbb{V}}'(\zeta)\widetilde{\mathbb{V}}(\zeta)\right].$$
The consistency of $\hat{\theta}^{(-\gamma)}$ comes from uniform convergence and identifiable uniqueness, we first show the uniform convergence of $\frac{1}{n}(L_n^c(\theta)-Q_n(\theta))$, and then show the identifiable uniqueness by proving the equicontinuity of $Q_n(\theta)$ and the global identified condition. Our proof follows the idea in Yu et al. (2008).

\textbf{Proof of uniform convergence}

The uniform convergence is specifically defined as $\sup_{\theta\in\Theta}|\frac{1}{n}\left(L_n^c(\theta)-Q_n(\theta)\right)|\overset{p}{\to}0.$
Denote the true value of each parameter as $\theta_0=(\rho_0,\delta_0',\sigma_0^2)$, $\zeta_0=(\rho_0,\delta_0')$. As

$$\frac{1}{n}\left(L_n^c(\theta)-Q_n(\theta)\right)=\frac{1}{2\sigma^2}\left(\frac{1}{n}\widetilde{\mathbb{V}}(\zeta)'\widetilde{\mathbb{V}}(\zeta)-\frac{1}{n}E\widetilde{\mathbb{V}}(\zeta)'\widetilde{\mathbb{V}}(\zeta)\right),$$
we will first show $\sup_{\theta\in\Theta}\frac{1}{n}\left|\widetilde{\mathbb{V}}(\zeta)'\widetilde{\mathbb{V}}(\zeta)-E\left[\widetilde{\mathbb{V}}(\zeta)'\widetilde{\mathbb{V}}(\zeta)\right]\right|\stackrel{p}{\rightarrow}0$.
Since $\widetilde{\mathbb{V}}(\zeta)=\mathbb{Q(Y-\rho WY-Z\delta})$, substituting $\mathbb{Y}$ for $\mathbb{Y}=\rho_0\mathbb{W}\mathbb{Y}+\mathbb{Z}\delta_0+\mathbb{D}\alpha_0+\mathbb{V}$, we have $\widetilde{\mathbb{V}}(\zeta)=\mathbb{Q[V-(}\rho-\rho_0)\mathbb{WY-Z}(\delta-\delta_0)]$. Therefore
\begin{align}
\widetilde{\mathbb{V}}(\zeta)'\widetilde{\mathbb{V}}(\zeta)=&\mathbb{V'QV}+(\rho-\rho_0)^2\mathbb{(WY)'QWY}+(\delta-\delta_0)'\mathbb{Z'QZ}(\delta-\delta_0)-\nonumber\\
&2(\rho-\rho_0)\mathbb{V'QWY}-2\mathbb{V'QZ}(\delta-\delta_0)+2(\rho-\rho_0)\mathbb{(WY)'QZ}(\delta-\delta_0).\label{eq_V'V}
\end{align}
Since $\mathbb{WY}=\mathbb{G}_0(\mathbb{Z}\delta_0+\mathbb{D}\alpha_0+\mathbb{V})$, $\mathbb{V'QWY}=\mathbb{V'Q}\mathbb{G}_0\mathbb{Z}\delta_0+\mathbb{V'Q}\mathbb{G}_0\mathbb{D}\alpha_0+\mathbb{V'Q}\mathbb{G}_0\mathbb{V}$, $\mathbb{(WY)'QZ}=(\mathbb{G}_0\mathbb{Z}\delta_0)'\mathbb{QZ}+(\mathbb{G}_0\mathbb{D}\alpha_0)'\mathbb{QZ}+(\mathbb{G}_0\mathbb{V)'QZ}$, and
\begin{align}
\mathbb{(WY)'QWY}=&(\mathbb{G}_0\mathbb{Z}\delta_0)'\mathbb{QG}_0\mathbb{Z}\delta_0+(\mathbb{G}_0\mathbb{D}\alpha_0)'\mathbb{QG}_0\mathbb{D}\alpha_0+(\mathbb{G}_0\mathbb{V})'\mathbb{QG}_0\mathbb{V}+\nonumber\\
&2(\mathbb{G}_0\mathbb{Z}\delta_0)'\mathbb{QG}_0\mathbb{D}\alpha_0+2(\mathbb{G}_0\mathbb{Z}\delta_0)'\mathbb{QG}_0\mathbb{V}+2(\mathbb{G}_0\mathbb{D}\alpha_0)'\mathbb{QG}_0\mathbb{V}.\label{eq_WYQWY}
\end{align}
Since $\mathbb{G}_0=\mathbb{WS}^{-1}_0=\mathrm{blkdiag}(W_1S_1^{-1},W_2S_2^{-1},\cdots,W_TS_T^{-1})$, where $W_tS_t^{-1}$ is UB uniformly for $t$ and $\|\sup_t \mathrm{abs}(W_tS_t^{-1})\|_\infty<\infty$ is ensured by Assumption 5 and Lemma \ref{lm33}, it follows from Lemma \ref{lm15} that
\begin{align*}
&\frac{1}{n}\mathbb{V'QV}-\frac{1}{n}E[\mathbb{V'QV}]\stackrel{p}{\rightarrow}0,
&\frac{1}{n}\mathbb{V'Q}\mathbb{G}_0\mathbb{V}-\frac{1}{n}E[\mathbb{V'Q}\mathbb{G}_0\mathbb{V}]\stackrel{p}{\rightarrow}0,\\
&\frac{1}{n}(\mathbb{G}_0\mathbb{V})'\mathbb{QG}_0\mathbb{V}-\frac{1}{n}E[(\mathbb{G}_0\mathbb{V})'\mathbb{QG}_0\mathbb{V}]\stackrel{p}{\rightarrow}0,
&\frac{1}{n}\mathbb{Z'QZ}-\frac{1}{n}E[\mathbb{Z'QZ}]\stackrel{p}{\rightarrow}0,\\
&\frac{1}{n}(\mathbb{G}_0\mathbb{Z}\delta_0)'\mathbb{QZ}-\frac{1}{n}E[(\mathbb{G}_0\mathbb{Z}\delta_0)'\mathbb{QZ}]\stackrel{p}{\rightarrow}0,
&\frac{1}{n}(\mathbb{G}_0\mathbb{Z}\delta_0)'\mathbb{QG}_0\mathbb{Z}\delta_0-\frac{1}{n}E[(\mathbb{G}_0\mathbb{Z}\delta_0)'\mathbb{QG}_0\mathbb{Z}\delta_0]\stackrel{p}{\rightarrow}0,\\
&\frac{1}{n}\mathbb{V'QZ}-\frac{1}{n}E[\mathbb{V'QZ}]\stackrel{p}{\rightarrow}0,
&\frac{1}{n}\mathbb{V'Q}\mathbb{G}_0\mathbb{Z}\delta_0-\frac{1}{n}E[\mathbb{V'Q}\mathbb{G}_0\mathbb{Z}\delta_0]\stackrel{p}{\rightarrow}0,\\
&\frac{1}{n}(\mathbb{G}_0\mathbb{V)'QZ}-\frac{1}{n}E[(\mathbb{G}_0\mathbb{V)'QZ}]\stackrel{p}{\rightarrow}0,
&\frac{1}{n}(\mathbb{G}_0\mathbb{Z}\delta_0)'\mathbb{QG}_0\mathbb{V}-\frac{1}{n}E[(\mathbb{G}_0\mathbb{Z}\delta_0)'\mathbb{QG}_0\mathbb{V}]\stackrel{p}{\rightarrow}0.
\end{align*}
Additionally, according to Lemma \ref{lm_CQL}, 
\begin{align*}
&\frac{1}{n}\mathbb{V'Q}\mathbb{G}_0\mathbb{D}\alpha_0\stackrel{p}{\rightarrow}0,
&\frac{1}{n}(\mathbb{G}_0\mathbb{D}\alpha_0)'\mathbb{QG}_0\mathbb{V}\stackrel{p}{\rightarrow}0,
\\
&\frac{1}{n}(\mathbb{G}_0\mathbb{D}\alpha_0)'\mathbb{QZ}-\frac{1}{n}E[(\mathbb{G}_0\mathbb{D}\alpha_0)'\mathbb{QZ}]\stackrel{p}{\rightarrow}0,
&\frac{1}{n}(\mathbb{G}_0\mathbb{Z}\delta_0)'\mathbb{QG}_0\mathbb{D}\alpha_0-\frac{1}{n}E[(\mathbb{G}_0\mathbb{Z}\delta_0)'\mathbb{QG}_0\mathbb{D}\alpha_0]\stackrel{p}{\rightarrow}0.
\end{align*}
Therefore, all polynomial coefficients in $\frac{1}{n}\widetilde{\mathbb{V}}(\zeta)'\widetilde{\mathbb{V}}(\zeta)$ will converge to their expected values. 
Since $\rho$ and $\delta$ are bounded in $\Theta$, it follows that $\sup_{\theta\in\Theta}\frac{1}{n}|\widetilde{\mathbb{V}}(\zeta)'\widetilde{\mathbb{V}}(\zeta)-E[\widetilde{\mathbb{V}}(\zeta)'\widetilde{\mathbb{V}}(\zeta)]|\stackrel{p}{\rightarrow}0$. Since $\sigma^2$ is bounded away from zero, we therefore have $\sup_{\theta\in\Theta}\frac{1}{n}|L_n^c(\theta)-Q_n(\theta)|\stackrel{p}{\rightarrow}0$.

\textbf{Proof of equicontinuity}

Next we prove that $\frac{1}{n}Q_n(\theta)$ is uniformly equicontinuous for $\theta\in\Theta$, which is specifically defined as for any $\epsilon>0$, there exists $\eta>0$ and $n_0\in\mathbb{N}$, such that whenever $d(\theta,\bar{\theta})<\eta$ and $n>n_0$, we have 
$$\left|\frac{1}{n}\left(Q_n(\theta)-Q_n(\bar{\theta})\right)\right|<\epsilon.$$
Since 
$$\frac{1}{n}Q_n(\theta)=-\frac{1}{2}\ln2\pi-\frac{1}{2}\ln\sigma^2+\frac{1}{n}\ln|\mathbb{S}(\rho)|-\frac{1}{2n\sigma^2}E\left[\widetilde{\mathbb{V}}'(\zeta)\widetilde{\mathbb{V}}(\zeta)\right],$$
where $E[\widetilde{\mathbb{V}}'(\zeta)\widetilde{\mathbb{V}}(\zeta)]=\mathbb{(\mathbb{S(\rho)}Y-Z\delta)'Q(\mathbb{S(\rho)}Y-Z\delta})$, substitute $\mathbb{Y}$ for $\mathbb{Y=S}_0^{-1}(\mathbb{Z}\delta_0+\mathbb{D}\alpha_0+\mathbb{V})$, then
\begin{align}
E[\widetilde{\mathbb{V}}(\zeta)'\widetilde{\mathbb{V}}(\zeta)]=&E\left[\left(\mathbb{S(\rho)S}_0^{-1}(\mathbb{Z}\delta_0+\mathbb{D}\alpha_0+\mathbb{V})-\mathbb{Z\delta}\right)'\mathbb{Q}\left(\mathbb{S(\rho)S}_0^{-1}(\mathbb{Z}\delta_0+\mathbb{D}\alpha_0+\mathbb{V})-\mathbb{Z\delta}\right)\right]\nonumber\\
=&E\left[\left(\mathbb{S(\rho)S}_0^{-1}(\mathbb{Z}\delta_0+\mathbb{D}\alpha_0)-\mathbb{Z\delta}\right)'\mathbb{Q}\left(\mathbb{S(\rho)S}_0^{-1}(\mathbb{Z}\delta_0+\mathbb{D}\alpha_0)-\mathbb{Z\delta}\right)\right]\nonumber\\
&+E\left[\left(\mathbb{S(\rho)S}_0^{-1}\mathbb{V}\right)'\mathbb{QS(\rho)S}_0^{-1}\mathbb{V}\right]
+2E\left[\left(\mathbb{S(\rho)S}_0^{-1}(\mathbb{Z}\delta_0+\mathbb{D}\alpha_0)-\mathbb{Z\delta}\right)'\mathbb{QS(\rho)S}_0^{-1}\mathbb{V}\right],\label{eq_EVTV}
\end{align}
where $E\left[\left(\mathbb{S(\rho)S}_0^{-1}\mathbb{V}\right)'\mathbb{QS(\rho)S}_0^{-1}\mathbb{V}\right]=\sigma^2_0\mathrm{tr}\left[\left(\mathbb{S(\rho)S}_0^{-1}\right)'\mathbb{QS(\rho)S}_0^{-1}\right]$,
and $$E\left[\left(\mathbb{S(\rho)S}_0^{-1}(\mathbb{Z}\delta_0+\mathbb{D}\alpha_0)-\mathbb{Z\delta}\right)'\mathbb{QS(\rho)S}_0^{-1}\mathbb{V}\right]=O(N)$$
by Lemma \ref{lm15}.

Denote $\sigma^2_n\equiv\frac{\sigma^2_0}{n}\mathrm{tr}[(\mathbb{S(\rho)S}_0^{-1})'\mathbb{S(\rho)S}_0^{-1}]$. Since
\begin{align*}
\frac{\sigma^2_0}{n}\mathrm{tr}\left[\left(\mathbb{S(\rho)S}_0^{-1}\right)'\mathbb{PS(\rho)S}_0^{-1}\right]&=\frac{\sigma^2_0}{n}\mathrm{tr}\left[\mathbb{D}\mathcal{T}^{-1}\mathbb{D'S(\rho)S}_0^{-1}\left(\mathbb{S(\rho)S}_0^{-1}\right)'\right]\\
&\leq\frac{\sigma^2_0}{c_TnT}\mathrm{tr}\left[\mathrm{abs}\left(\mathbb{D'S(\rho)S}_0^{-1}\left(\mathbb{S(\rho)S}_0^{-1}\right)'\mathbb{D}\right)\right]=O\left(\frac{1}{T}\right),
\end{align*}
where the inequality holds because $T_i>c_TT$ by Assumption 1 and the matrix inside the trace operation is UB. Thus, $\frac{\sigma^2_0}{n}\mathrm{tr}\left[\left(\mathbb{S(\rho)S}_0^{-1}\right)'\mathbb{QS(\rho)S}_0^{-1}\right]=\sigma^2_n+O\left(\frac{1}{T}\right)$. Further substitute $\mathbb{S(\rho)S}_0^{-1}$ for $\mathbb{S(\rho)S}_0^{-1}=I_n+(\rho_0-\rho)\mathbb{G}_0$, then we have
\begin{align}
\frac{1}{n}E[\widetilde{\mathbb{V}}(\zeta)'\widetilde{\mathbb{V}}(\zeta)]=&\frac{1}{n}E\left\{\left[(\rho_0-\rho)\mathbb{G}_0(\mathbb{Z}\delta_0+\mathbb{D}\alpha_0)+\mathbb{Z}(\delta_0-\delta)\right]'\mathbb{Q}\left[(\rho_0-\rho)\mathbb{G}_0(\mathbb{Z}\delta_0+\mathbb{D}\alpha_0)+\mathbb{Z}(\delta_0-\delta)\right]\right\}\nonumber\\
&+\sigma^2_n+o(1),\label{EVV}
\end{align}
where the first term is equal to $(\rho-\rho_0,\delta'-\delta_0')E[\mathcal{H}](\rho-\rho_0,\delta'-\delta_0')'$, and $\mathcal{H}\equiv\frac{1}{n}(\mathbb{G}_0\mathbb{Z}\delta_0+\mathbb{G}_0\mathbb{D}\alpha_0,\mathbb{Z})'\mathbb{Q}(\mathbb{G}_0\mathbb{Z}\delta_0+\mathbb{G}_0\mathbb{D}\alpha_0,\mathbb{Z})$.
Therefore, the uniform equicontinuity of $\frac{1}{n}Q_n(\theta)$ is sufficient to show the following is uniformly equicontinuous: (i) $-\frac{1}{2}\ln\sigma^2$; (ii) $\frac{1}{n}\ln|\mathbb{S(\rho)}|$;
(iii) $(\rho-\rho_0,\delta'-\delta_0')E[\mathcal{H}](\rho-\rho_0,\delta'-\delta_0')'$; (iv) $\sigma^2_n$.

Since $\sigma^2$ is bounded away from zero, (i) is obviously uniformly equicontinuous.
For (ii), $\frac{1}{n}(\ln|\mathbb{S(\rho}_1)|-\ln|\mathbb{S(\rho}_2)|)=(\rho_1-\rho_2)\frac{1}{n}\mathrm{tr}(\mathbb{WS}^{-1}(\Bar{\rho}))$, where $\Bar{\rho}$ lies between $\rho_1$ and $\rho_2$. According to equation (\ref{eq_Grho}), 
$\sup_{N,T}\|\mathbb{W}\mathbb{S}^{-1}(\Bar{\rho})\|_{\infty}<\infty$, so $\frac{1}{n}\mathrm{tr}(\mathbb{WS}^{-1}(\Bar{\rho}))$ is bounded, which means that (ii) is uniformly equicontinuous.
(iii) is uniformly equicontinuous because the parameter space $(\rho,\delta)$ is bounded and $E[\mathcal{H}]$ is $O(1)$. 
For (iv), 
\begin{align*}
&\frac{\sigma^2_0}{n}\{\mathrm{tr}[(\mathbb{S(\rho}_1)\mathbb{S}_0^{-1})'\mathbb{QS(\rho}_1)\mathbb{S}_0^{-1}]-\mathrm{tr}[(\mathbb{S(\rho}_2)\mathbb{S}_0^{-1})'\mathbb{QS(\rho}_2)\mathbb{S}_0^{-1}]\}\\
=&\frac{\sigma^2_0}{n}\{(\rho_0-\rho_1)^2\mathrm{tr}(\mathbb{G}_0'\mathbb{QG}_0)+2(\rho_0-\rho_1)\mathrm{tr}(\mathbb{G}_0\mathbb{Q})-(\rho_0-\rho_2)^2\mathrm{tr}(\mathbb{G}_0'\mathbb{QG}_0)-2(\rho_0-\rho_2)\mathrm{tr}(\mathbb{G}_0\mathbb{Q})\}\\
=&2(\rho_2-\rho_1)\sigma^2_0\frac{\mathrm{tr}(\mathbb{G}_0\mathbb{Q})}{n}+(2\rho_0-\rho_1-\rho_2)(\rho_2-\rho_1)\sigma^2_0\frac{\mathrm{tr}(\mathbb{G}_0'\mathbb{QG}_0)}{n}.
\end{align*}
Since $\mathbb{G}_0$ and $\mathbb{Q}$ are UB, similarly (iv) is uniformly equicontinuous.

\textbf{Proof of global identification condition}

Then we prove that $\theta_0$, except $\gamma_0$, is identified, which is specifically defined as $\forall\theta^{(-\gamma)}\neq\theta^{(-\gamma)}_0$,
$$\limsup_{n\rightarrow\infty}\frac{1}{n}[Q_{n}(\theta)-Q_{n}(\theta_{0})]<0.$$ 
Note that
\begin{align}
\frac{1}{n}\left(Q_n(\theta)-Q_n(\theta_0)\right)=-\frac{1}{2}\left(\ln\sigma^2-\ln\sigma^2_0\right)+\frac{1}{n}\left(\ln|\mathbb{S}(\rho)|-\ln|\mathbb{S}_0|\right)-\frac{1}{2n\sigma^2}E\left[\widetilde{\mathbb{V}}(\zeta)'\widetilde{\mathbb{V}}(\zeta)\right]+\frac{c_{u}T-1}{2c_{u}T},\label{Q-Q0}
\end{align}
where $c_u\equiv\frac{n}{NT}$. According to equation (\ref{EVV}), we can decompose equation (\ref{Q-Q0}) into
\begin{align*}
\frac{1}{n}\left(Q_n(\theta)-Q_n(\theta_0)\right)=L_1(\rho,\sigma^2)-\frac{1}{2\sigma^2}L_2(\rho,\delta)+o(1),
\end{align*}
where $L_1(\rho,\sigma^2)\equiv-\frac{1}{2}\left(\ln\sigma^2-\ln\sigma^2_0\right)+\frac{1}{n}\left(\ln|\mathbb{S}(\rho)|-\ln|\mathbb{S}_0|\right)-\frac{1}{2\sigma^2}(\sigma^2_n-\sigma^2)$ and $L_2(\rho,\delta)\equiv\frac{1}{n}E\{[(\rho_0-\rho)\mathbb{G}_0(\mathbb{Z}\delta_0+\mathbb{D}\alpha_0)+\mathbb{Z}(\delta_0-\delta)]'\mathbb{Q}[(\rho_0-\rho)\mathbb{G}_0(\mathbb{Z}\delta_0+\mathbb{D}\alpha_0)+\mathbb{Z}(\delta_0-\delta)]\}$.


For a spatial autoregressive model 
$\mathbb{Y}=\rho_0\mathbb{WY+V}$ with $v_{it}\sim \mathrm{N}(0,\sigma_0^2)$, the information inequality of this model ensures that
$$-\frac{1}{2}\left(\ln\sigma^2-\ln\sigma^2_0\right)+\frac{1}{n}\left(\ln|\mathbb{S}(\rho)|-\ln|\mathbb{S}_0|\right)-\frac{1}{2\sigma^2}(\sigma^2_n-\sigma^2)\leq0.$$
Note that left-hand side of the above equation is precisely $L_1(\rho,\sigma^2)$.

Denote $\delta^{(-\gamma)}$ as the vector obtained by removing $\gamma$ from $\delta$, with $\hat{\delta}^{(-\gamma)}$ and $\delta^{(-\gamma)}_0$ defined analogously. By decomposing $\mathbb{Z}\delta$ into $\mathbb{Z}^{(-\gamma)}\delta^{(-\gamma)}+ \mathbb{Z}^{(\gamma)}\gamma$, $L_2(\rho,\delta)$ can be viewed as a quadratic function of $\rho$, $\delta^{(-\gamma)}$ and $\gamma$, i.e.,
\begin{align*}
L_2(\rho,\delta^{(-\gamma)},\gamma)=&\left(\rho-\rho_0,\delta^{(-\gamma)\prime}-\delta^{(-\gamma)\prime}_0\right)E\left[\mathcal{H}^{(-\gamma)}\right]\left(\rho-\rho_0,\delta^{(-\gamma)\prime}-\delta^{(-\gamma)\prime}_0\right)'+\\
&+\frac{2}{n}\left[(\rho_0-\rho)\mathbb{G}_0(E\mathbb{Z}\delta_0+\mathbb{D}\alpha_0)+E\mathbb{Z}^{(-\gamma)}(\delta^{(-\gamma)}-\delta^{(-\gamma)}_0)\right]'\mathbb{QZ}^{(\gamma)}\gamma+\frac{1}{n}\gamma'\mathbb{Z}^{(\gamma)\prime}\mathbb{QZ}^{(\gamma)}\gamma,
\end{align*}
where $\mathcal{H}^{(-\gamma)}=\frac{1}{n}'(\mathbb{G}_0\mathbb{Z}\delta_0+\mathbb{G}_0\mathbb{D}\alpha_0,\mathbb{Z}^{(-\gamma)})'\mathbb{Q}(\mathbb{G}_0\mathbb{Z}\delta_0+\mathbb{G}_0\mathbb{D}\alpha_0,\mathbb{Z}^{(-\gamma)})$, and $\mathbb{Z}^{(\gamma)}=\left((D_1F_1\iota_N)',\cdots,(D_TF_T\iota_N)'\right)'$ is a sparse binary vector with at most $N$ nonzero elements, as we can only observe at most $N$ listed nodes. Therefore, $$\frac{2}{n}\left[(\rho_0-\rho)\mathbb{G}_0(E\mathbb{Z}\delta_0+\mathbb{D}\alpha_0)+E\mathbb{Z}^{(-\gamma)}(\delta^{(-\gamma)}-\delta^{(-\gamma)}_0)\right]'\mathbb{QZ}^{(\gamma)}\gamma+\frac{1}{n}\gamma'\mathbb{Z}^{(\gamma)\prime}\mathbb{QZ}^{(\gamma)}\gamma=o(1),$$
because elements in $E\mathbb{Z}$ and $E\mathbb{Z}^{(-\gamma)}$ are uniformly bounded and $\mathbb{G}_0$ is UB. Thus
$$\lim_{n\rightarrow\infty} L_2(\rho,\delta^{(-\gamma)},\gamma)=\left(\rho-\rho_0,\delta^{(-\gamma)\prime}-\delta^{(-\gamma)\prime}_0\right)E\left[\mathcal{H}^{(-\gamma)}\right]\left(\rho-\rho_0,\delta^{(-\gamma)\prime}-\delta^{(-\gamma)\prime}_0\right)'\geq0,$$
because $\mathcal{H}^{(-\gamma)}$ is a quadratic form.
Below we will show that $(\rho,\delta^{(-\gamma)})$ is globally identified by using either Assumption 8 (i) or Assumption 8 (ii).
\begin{enumerate}
\item (Using Assumption 8 (i))
According to Assumption 8 (i), we have $\lim_{n\rightarrow\infty} L_2(\rho,\delta^{(-\gamma)},\gamma)>0$ whenever $(\rho,\delta^{(-\gamma)})\neq(\rho_0,\delta^{(-\gamma)}_0)$. Combined with $L_1(\rho,\sigma^2)\leq0$, it implies that $(\rho,\delta^{(-\gamma)})$ is globally identified.
\item (Using Assumption 8 (ii)) Since $\sigma^2_n$ is uniquely maximizer of $L_1(\rho,\sigma^2)$ over $\sigma^2$, the information inequality of the SAR model imples that $L_1(\rho,\sigma^2)\leq L_1(\rho,\sigma^2_n)\leq0$. Moreover, $L_1(\rho,\sigma^2_n)=0$ if and only if the pdf $f(\mathbb{Y}|\rho,\sigma^2_n)= f(\mathbb{Y}|\rho_0,\sigma^2_0)$, where
$$f(\mathbb{Y}|\rho,\sigma^2_n)=\frac{1}{(2\pi)^\frac{n}{2}\left|\sigma_n^2\mathbb{S}(\rho)^{-1}\mathbb{S}(\rho)^{-1\prime}\right|^{\frac{1}{2}}}\exp\left[-\frac{1}{2\sigma_n^2}\mathbb{Y'S}(\rho)'\mathbb{S}(\rho)\mathbb{Y}\right].$$
If Assumption 8 (ii) holds, $f(\mathbb{Y}|\rho,\sigma^2_n)\neq f(\mathbb{Y}|\rho_0,\sigma^2_0)$ for $\rho\neq\rho_0$ as $n\rightarrow\infty$. Then $\limsup_{n\rightarrow\infty}L_1(\rho,\sigma^2)<0$ whenever $\rho\neq\rho_0$. Combined with $\lim_{n\rightarrow\infty} L_2(\rho,\delta^{(-\gamma)},\gamma)\geq0$, $\rho_0$ is globally identified. Then $\delta^{(-\gamma)}$ will also be identified because, $$\lim_{n\rightarrow\infty} L_2(\rho_0,\delta^{(-\gamma)},\gamma)=\left(\delta^{(-\gamma)}-\delta^{(-\gamma)}_0\right)'\left[\lim_{n\rightarrow\infty}\frac{1}{n}E\left(\mathbb{Z}^{(-\gamma)\prime}\mathbb{QZ}^{(-\gamma)}\right)\right]\left(\delta^{(-\gamma)}-\delta^{(-\gamma)}_0\right)>0$$
whenever $\delta^{(-\gamma)}\neq\delta^{(-\gamma)}_0$ by Assumption 4.
\end{enumerate}

Once $(\rho,\delta^{(-\gamma)})$ is globally identified, $\sigma^2$ is also globally identified because $\limsup_{n\rightarrow\infty}L_1(\rho_0,\sigma^2)=\frac{1}{2}(\ln\frac{\sigma_0^2}{\sigma^2}-\frac{\sigma_0^2}{\sigma^2}+1)\leq0$ and the equality holds if and only if $\sigma^2=\sigma^2_0$, because $\ln x-x+1\leq0$ and the equality holds only when $x=1$.
Therefore, $\limsup_{n\rightarrow\infty}\frac{1}{n}\left(Q_n(\theta)-Q_n(\theta_0)\right)<0$ for any $\theta^{(-\gamma)}\neq\theta^{(-\gamma)}_0$, which means $\theta^{(-\gamma)}$ is globally identified.


Since $\Theta$ is compact and $Q_n(\theta)$ is uniformly equicontinuous, combined with uniformly convergence and global identification of $\theta^{(-\gamma)}$, we can prove the consistency of $\hat{\theta}^{(-\gamma)}$. Given $\hat{\theta}^{(-\gamma)}$, $L_n^c(\theta)$ is maximized at 
\begin{align*}
\hat{\gamma}&=\left(\mathbb{Z}^{(\gamma)\prime}\mathbb{QZ}^{(\gamma)}\right)^{-1}\mathbb{Z}^{(\gamma)\prime}\mathbb{Q\left(S(\hat{\rho})Y-Z^{(-\gamma)}\hat{\delta}^{(-\gamma)}\right)}\\
&=\left(\mathbb{Z}^{(\gamma)\prime}\mathbb{QZ}^{(\gamma)}\right)^{-1}\mathbb{Z}^{(\gamma)\prime}\mathbb{Q}\left(\mathbb{S(\hat{\rho})S}_0^{-1}\left(\mathbb{Z^{(-\gamma)}}\delta_0^{(-\gamma)}+\mathbb{Z}^{(\gamma)}\gamma_0+\mathbb{D}\alpha_0+\mathbb{V}\right)-\mathbb{Z^{(-\gamma)}\hat{\delta}^{(-\gamma)}}\right).
\end{align*}
Since $\hat{\theta}^{(-\gamma)}\overset{p}{\rightarrow}\theta_0^{(-\gamma)}$, 
$
\hat{\gamma}=\gamma_0+(\mathbb{Z}^{(\gamma)\prime}\mathbb{QZ}^{(\gamma)})^{-1}\mathbb{Z}^{(\gamma)\prime}\mathbb{QV}+o_p(1)$.
Additionally,
$(\mathbb{Z}^{(\gamma)\prime}\mathbb{QZ}^{(\gamma)})^{-1}\mathbb{Z}^{(\gamma)\prime}\mathbb{QV}=o_p(1)$ because $E[(\mathbb{Z}^{(\gamma)\prime}\mathbb{QZ}^{(\gamma)})^{-1}\mathbb{Z}^{(\gamma)\prime}\mathbb{QV}]=0$ and
\begin{align*}
\mathrm{Var}\left[\left(\mathbb{Z}^{(\gamma)\prime}\mathbb{QZ}^{(\gamma)}\right)^{-1}\mathbb{Z}^{(\gamma)\prime}\mathbb{QV}\right]&=\sigma_0^2\left(\mathbb{Z}^{(\gamma)\prime}\mathbb{QZ}^{(\gamma)}\right)^{-1}=\sigma_0^2\left(\mathbb{Z}^{(\gamma)\prime}\mathbb{Z}^{(\gamma)}-\mathbb{Z^{(\gamma)\prime}D}\mathcal{T}^{-1}\mathbb{D'Z}^{(\gamma)}\right)^{-1}\\
&\leq\sigma_0^2\left[(1-c_N)N-\frac{N}{c_TT}\right]^{-1}=O\left(\frac{1}{N}\right),
\end{align*}
where the inequality holds because $\mathbb{Z}^{(\gamma)}$ is a binary vector with at least $(1-c_N)N$ nonzero elements and $T_i\geq c_TT$ by Assumption 1.
Therefore, $\hat{\gamma}\overset{p}{\rightarrow}\gamma_0$, so $\hat{\theta}\overset{p}{\rightarrow}\theta_0$.


\subsection{Proof of Theorem 2}
According to the Taylor expansion
$\frac{\partial L_n^{c}(\hat{\theta})}{\partial\theta}=\frac{\partial L_n^{c}(\theta_0)}{\partial\theta}+\frac{\partial^2L_n^{c}(\bar{\theta})}{\partial\theta\partial\theta'}(\hat{\theta}-\theta_0)$, where $\frac{\partial L_n^{c}(\hat{\theta})}{\partial\theta}=0$ and $\bar{\theta}$ lies between $\hat{\theta}$ and $\theta_0$, we further have
\begin{align}
\sqrt{n}\Gamma^{-1}(\hat{\theta}-\theta_0)=\left(-\frac{1}{n}\Gamma\frac{\partial^2L_n^{c}(\bar{\theta})}{\partial\theta\partial\theta'}\Gamma\right)^{-1}\left(\frac{1}{\sqrt{n}}\Gamma\frac{\partial L_n^{c*}(\theta_0)}{\partial\theta}-\Delta_n\right),\label{eq_Taylor}
\end{align}
where $\Delta_n=\sqrt{\frac{N}{c_uT}}\varphi_n(\theta_0)+O_p\left(\frac{1}{\sqrt{T}}\right)$, and
$$\frac{1}{\sqrt{n}}\Gamma\frac{\partial L_n^{c*}(\theta_0)}{\partial\theta}=\left(\begin{aligned}
&\frac{1}{\sqrt{n}\sigma^2_0}(\mathbb{G}_0\mathbb{D\alpha}_0)'\mathbb{QV}+\frac{1}{\sqrt{n}\sigma_0^2}\mathbb{V}'\mathbb{G}_0'\mathbb{V}+\frac{1}{\sqrt{n}\sigma_0^2}(\widetilde{\mathbb{G}_0\mathbb{Z}}^*\delta_0)'\mathbb{V}-\frac{1}{\sqrt{n}}\mathrm{tr}(\mathbb{G}_0)\\
&\frac{1}{\sqrt{n}\sigma_0^2}\widetilde{\Gamma}\widetilde{\mathbb{Z}}^{*\prime}\mathbb{V}\\
&-\frac{n}{2\sqrt{n}\sigma_0^2}+\frac{1}{2\sqrt{n}\sigma_0^4}\mathbb{V'V}
\end{aligned}\right).$$ 
Since $\widetilde{\mathbb{Z}}^*=\mathbb{QZ}+\widetilde{\mathbb{Z}}^\circ=\mathbb{Z-P(Z-Z}^\circ)$ and $\widetilde{\mathbb{G}_0\mathbb{Z}}^{*}=\mathbb{QG}_0\mathbb{Z}+\widetilde{\mathbb{G}_0\mathbb{Z}}^{\circ}=\mathbb{G}_0\mathbb{Z}-\mathbb{PG}_0(\mathbb{Z-Z}^\circ)$, where
$$Z_t-Z_t^\circ=\left(D_t(I_N-F_t)D_t'M_t(\mathcal{X}_{t-1}\beta_0+\mathcal{E}_{t-1}),D_tD_{t-1}'(\mathcal{X}_{t-1}\beta_0+\mathcal{E}_{t-1}),D_tF_t\iota_N,X_t\right)$$
can be viewed as a constant matrix, elements in $\frac{1}{\sqrt{n}}\frac{\partial L_n^{c*}(\theta_0)}{\partial\theta}$ follow the form of equation (\ref{eq_clt}). Moreover, since linear combinations of different $\mathcal{Q}_{n}$ still preserve the form of equation (\ref{eq_clt}), and by Lemma \ref{lm_CLT} and Cramér-Wold Theorem, we have 
$$\frac{1}{\sqrt{n}}\Gamma\frac{\partial L_n^{c*}(\theta_0)}{\partial\theta}\overset{d}{\rightarrow}N(0,\Sigma_{\theta_0}+\Omega_{\theta_0}),$$
where $\Sigma_{\theta_0}+\Omega_{\theta_0}$ is the asymptotic variance of $\frac{1}{\sqrt{n}}\Gamma\frac{\partial L_n^{c*}(\theta_0)}{\partial\theta}$ according to equation (10). Moreover,
\begin{align*}
-\frac{1}{n}\Gamma\frac{\partial^2L_n^{c}(\bar{\theta})}{\partial\theta\partial\theta'}\Gamma
=\left[-\frac{1}{n}\Gamma\frac{\partial^2L_n^{c}(\bar{\theta})}{\partial\theta\partial\theta'}\Gamma-\left(-\frac{1}{n}\Gamma\frac{\partial^2L_n^{c}(\theta_0)}{\partial\theta\partial\theta'}\Gamma\right)\right]+\left(-\frac{1}{n}\Gamma\frac{\partial^2L_n^{c}(\theta_0)}{\partial\theta\partial\theta'}\Gamma-\Gamma\Sigma_{\theta_0,n}\Gamma\right)+\Gamma\Sigma_{\theta_0,n}\Gamma,
\end{align*}
where the first term is $\|\Gamma^{-1}(\bar{\theta}-\theta_0)\|_2\cdot O_p(1)$ and the second term is $O_p(\frac{1}{\sqrt{NT}})$ by equations (\ref{eq_H_conti}) and (\ref{eq_H_consi}), respectively. By the consistency of $\hat{\theta}$, $-\frac{1}{n}\Gamma\frac{\partial^2L_n^{c}(\bar{\theta})}{\partial\theta\partial\theta'}\Gamma=\Gamma\Sigma_{\theta_0,n}\Gamma+o_p(1)$. Moreover, $\Gamma\Sigma_{\theta_0,n}\Gamma$ is nonsingular, which has been proved in Appendix \ref{B.3}. So
$\sqrt{n}\Gamma^{-1}(\hat{\theta}-\theta_0)=O_p(1)\cdot[O_p(1)+O(\sqrt{N/T})]$ according to equation (\ref{eq_Taylor}), which means that
\begin{align}
\Gamma^{-1}(\hat{\theta}-\theta_0)=O_p\left(\max\left\{\frac{1}{\sqrt{NT}},\frac{1}{T}\right\}\right).\label{eq_dtheta}
\end{align}
Hence,
\begin{align*}
\sqrt{n}\Gamma^{-1}(\hat{\theta}-\theta_0)&=\left[\Gamma\Sigma_{\theta_0,n}\Gamma+O_p\left(\max\left\{\frac{1}{\sqrt{NT}},\frac{1}{T}\right\}\right)\right]^{-1}\left(\frac{1}{\sqrt{n}}\Gamma\frac{\partial L_n^{c*}(\theta_0)}{\partial\theta}-\Delta_n\right)\\
&=\left[(\Gamma\Sigma_{\theta_0,n}\Gamma)^{-1}+O_p\left(\max\left\{\frac{1}{\sqrt{NT}},\frac{1}{T}\right\}\right)\right]\left(\frac{1}{\sqrt{n}}\Gamma\frac{\partial L_n^{c*}(\theta_0)}{\partial\theta}-\Delta_n\right),
\end{align*}
where the second equality holds by the fact that for any matrix sequences $A_n$ and $B_n$ such that $\lim_{n\rightarrow\infty}A_n$ exists and nonsingular,  and $B_n$ is $O_p(n^{-\eta})$ for some $\eta>0$, we have
\begin{align}
(A_n+B_n)^{-1}=(I+A_n^{-1}B_n)^{-1}A_n^{-1}=A_n^{-1}-A_n^{-1}B_nA_n^{-1}+A_n^{-1}B_nA_n^{-1}B_nA_n^{-1}-\cdots=A_n^{-1}+O_p(n^{-\eta}).\label{eq_inv_A+B}
\end{align}
Therefore, we have
\begin{align*}
\sqrt{n}\Gamma^{-1}(\hat{\theta}-\theta_0)+\left[(\Gamma\Sigma_{\theta_0,n}\Gamma)^{-1}+O_p\left(\max\left\{\frac{1}{\sqrt{NT}},\frac{1}{T}\right\}\right)\right]\Delta_n
=\left[(\Gamma\Sigma_{\theta_0,n}\Gamma)^{-1}+o_p(1)\right]\frac{1}{\sqrt{n}}\Gamma\frac{\partial L_n^{c*}(\theta_0)}{\partial\theta}.
\end{align*}
As $\lim_{n\rightarrow\infty}\Gamma\Sigma_{\theta_0,n}\Gamma=\Sigma_{\theta_0}$, and $\Delta_n=\sqrt{\frac{N}{c_uT}}\varphi_n(\theta_0)+O_p\left(\frac{1}{\sqrt{T}}\right)$, combined with the limiting distribution of $\frac{1}{\sqrt{n}}\Gamma\frac{\partial L_n^{c*}(\theta_0)}{\partial\theta}$, we finally have
$$\sqrt{n}\Gamma^{-1}(\hat{\theta}-\theta_0)+\sqrt{\frac{N}{c_uT}}(\Gamma\Sigma_{\theta_0,n}\Gamma)^{-1}\varphi_n(\theta_0)+O_p\left(\max\left\{\frac{1}{\sqrt{T}},\sqrt{\frac{N}{T^3}}\right\}\right)\overset{d}{\rightarrow}N\left(0,\Sigma_{\theta_0}^{-1}(\Sigma_{\theta_0}+\Omega_{\theta_0})\Sigma_{\theta_0}^{-1}\right),$$
which means that
$$\sqrt{n}\Gamma^{-1}\left[\hat{\theta}-\left(\theta_0-\frac{\Sigma_{\theta_0,n}^{-1}\varphi_n(\theta_0)}{c_uT}\right)\right]+O_p\left(\max\left\{\frac{1}{\sqrt{T}},\sqrt{\frac{N}{T^3}}\right\}\right)\overset{d}{\rightarrow}N\left(0,\Sigma_{\theta_0}^{-1}(\Sigma_{\theta_0}+\Omega_{\theta_0})\Sigma_{\theta_0}^{-1}\right),$$
where one of the multipliers $\Gamma$ can be omitted because the $\gamma$-component of the vector $\varphi_n(\theta_0)$ is zero.
\subsection{Proof of Theorem 3}
Since $\hat{\theta}^1=\hat{\theta}+\frac{\hat{B}_n}{c_uT}$, where $\hat{B}_n=\left.\left[\left(-\frac{1}{n}\frac{\partial^2L_n^{c}(\theta)}{\partial\theta\partial\theta'}\right)^{-1}\varphi_n(\theta)\right]\right|_{\theta=\hat{\theta}}$. According to Theorem 2, if $\frac{N}{T^3}\rightarrow0$, we need to prove
\begin{align}
\sqrt{\frac{N}{c_uT}}\Gamma^{-1}\left[\left(-\frac{1}{n}\frac{\partial^2L_n^{c}(\hat{\theta})}{\partial\theta\partial\theta'}\right)^{-1}\varphi_n(\hat{\theta})-\Sigma_{\theta_0,n}^{-1}\varphi_n(\theta_0)\right]\overset{p}{\rightarrow}0.\label{eq_B_consi}
\end{align}
Based on equations (\ref{eq_H_conti}) and (\ref{eq_H_consi}), we have
\begin{align*}
-\Gamma\frac{1}{n}\frac{\partial^2L_n^{c}(\hat{\theta})}{\partial\theta\partial\theta'}\Gamma-\Gamma\Sigma_{\theta_0,n}\Gamma&=-\Gamma\frac{1}{n}\frac{\partial^2L_n^{c}(\hat{\theta})}{\partial\theta\partial\theta'}\Gamma-\left(-\Gamma\frac{1}{n}\frac{\partial^2L_n^{c}(\theta_0)}{\partial\theta\partial\theta'}\Gamma\right)+\left(-\Gamma\frac{1}{n}\frac{\partial^2L_n^{c}(\theta_0)}{\partial\theta\partial\theta'}\Gamma\right)-\Gamma\Sigma_{\theta_0,n}\Gamma\\
&=\|\Gamma^{-1}(\hat{\theta}-\theta_0)\|_2\cdot O_p(1)+O_p\left(\frac{1}{\sqrt{NT}}\right)\\
&=O_p\left(\max\left\{\frac{1}{\sqrt{NT}},\frac{1}{T}\right\}\right),
\end{align*}
where the last equality holds by equation (\ref{eq_dtheta}). Then using equation (\ref{eq_inv_A+B}), we have
$$\left(-\Gamma\frac{1}{n}\frac{\partial^2L_n^{c}(\hat{\theta})}{\partial\theta\partial\theta'}\Gamma\right)^{-1}=(\Gamma\Sigma_{\theta_0,n}\Gamma)^{-1}+O_p\left(\max\left\{\frac{1}{\sqrt{NT}},\frac{1}{T}\right\}\right).$$
Recall that $\Gamma^{-1}\varphi_n(\theta)=\varphi_n(\theta)$, thus
\begin{align*}
&\sqrt{\frac{N}{c_uT}}\Gamma^{-1}\left[\left(-\frac{1}{n}\frac{\partial^2L_n^{c}(\hat{\theta})}{\partial\theta\partial\theta'}\right)^{-1}\varphi_n(\hat{\theta})-\Sigma_{\theta_0,n}^{-1}\varphi_n(\theta_0)\right]\\
=&\sqrt{\frac{N}{c_uT}}\left[\left(-\frac{1}{n}\Gamma\frac{\partial^2L_n^{c}(\hat{\theta})}{\partial\theta\partial\theta'}\Gamma\right)^{-1}\varphi_n(\hat{\theta})-(\Gamma\Sigma_{\theta_0,n}\Gamma)^{-1}\varphi_n(\theta_0)\right]\\
=&\sqrt{\frac{N}{c_uT}}(\Gamma\Sigma_{\theta_0,n}\Gamma)^{-1}\left[\varphi_n(\hat{\theta})-\varphi_n(\theta_0)\right]+\sqrt{\frac{N}{c_uT}}\varphi_n(\hat{\theta})\times O_p\left(\max\left\{\frac{1}{\sqrt{NT}},\frac{1}{T}\right\}\right)\\
=&O_p\left(\sqrt{\frac{N}{T}}\right)\cdot\left[\frac{\partial\varphi_n(\bar{\theta})}{\partial\theta'}\Gamma\right]\left[\Gamma^{-1}(\hat{\theta}-\theta_0)\right]+\varphi_n(\hat{\theta})\times O_p\left(\max\left\{\frac{1}{T},\sqrt{\frac{N}{T^3}}\right\}\right),
\end{align*}
where $\bar{\theta}$ lies between $\hat{\theta}$ and $\theta_0$. Since $\Gamma^{-1}(\hat{\theta}-\theta_0)=O_p\left(\max\left\{\frac{1}{\sqrt{NT}},\frac{1}{T}\right\}\right)$, in order to prove equation (\ref{eq_B_consi}), it suffices to show that $\frac{\partial\varphi_n(\bar{\theta})}{\partial\theta'}\Gamma$ is $O_p(1)$.
By exchanging the order of summation in equation (\ref{eq_phi_n}) and adjusting the summation indices, we obtain
\begin{align*}
&\varphi_n(\theta)=\\
&\left(\begin{aligned}
&\left[\begin{aligned}
&\frac{1}{N}\mathrm{tr}\left(
\mathcal{T}^{-1}\sum_{t=1}^{T-1} \lambda D_{t+1}'G_{t+1}(\rho)D_{t+1}(I_N-F_{t+1})D_{t+1}'M_{t+1}S_t(\rho)^{-1}\sum_{h=0}^{T-t}A_{t}^{(h)}(\theta)D_{t-h}\right)\\
&+\frac{1}{N}\mathrm{tr}\left(\mathcal{T}^{-1}\sum_{t=1}^{T-1}\nu D_{t+1}'G_{t+1}(\rho)D_{t+1}D_{t}'S_t(\rho)^{-1}\sum_{h=0}^{T-t}A_{t}^{(h)}(\theta)D_{t-h}\right)+\frac{1}{N}\mathrm{tr}(\mathbb{G}(\rho)'\mathbb{P})
\end{aligned}\right]\\
&\frac{1}{N}\mathrm{tr}\left(
\mathcal{T}^{-1}\sum_{t=T-h}^{T-1}D_{t+1}'D_{t+1}(I_N-F_{t+1})D_{t+1}'M_{t+1}S_t(\rho)^{-1}\sum_{h=0}^{T-t}A_{t}^{(h)}(\theta)D_{t-h}\right)\\
&\frac{1}{N}\mathrm{tr}\left(
\mathcal{T}^{-1}\sum_{t=T-h}^{T-1}D_{t+1}'D_{t+1}D_{t}'S_t(\rho)^{-1}\sum_{h=0}^{T-t}A_{t}^{(h)}(\theta)D_{t-h}\right)\\
&\mathbf{0}_{(k+1)\times1}\\
&\frac{1}{2\sigma^2}
\end{aligned}\right).
\end{align*}
Firstly, we will prove that elements in $\frac{\partial\varphi_n(\bar{\theta})}{\partial\rho}$ are $O_p(1)$. As
\begin{align*}
&\Bigl|\frac{\partial}{\partial\rho}\frac{1}{N}\mathrm{tr}\Bigl(
\mathcal{T}^{-1}\sum_{t=1}^{T-1} \lambda D_{t+1}'G_{t+1}(\rho)D_{t+1}
\\&\qquad(I_N-F_{t+1})D_{t+1}'M_{t+1}S_t(\rho)^{-1}\sum_{h=0}^{T-t}A_{t}^{(h)}(\theta)D_{t-h}\Bigr)\Bigr|\\
\leq&\frac{1}{N}\Bigl|\mathrm{tr}\Bigl[
\mathcal{T}^{-1}\sum_{t=1}^{T-1}\lambda D_{t+1}'W_{t+1}\Bigl(\frac{\partial}{\partial\rho}S_{t+1}^{-1}(\rho)\Bigr)D_{t+1}
\\&\qquad(I_N-F_{t+1})D_{t+1}'M_{t+1}S_t(\rho)^{-1}\sum_{h=0}^{T-t}A_{t}^{(h)}(\theta)D_{t-h}\Bigr]\Bigr|\\
&+\frac{1}{N}\Bigl|\mathrm{tr}\Bigl[
\mathcal{T}^{-1}\sum_{t=1}^{T-1} \lambda D_{t+1}'G_{t+1}(\rho)D_{t+1}
\\&\qquad(I_N-F_{t+1})D_{t+1}'M_{t+1}\Bigl(\frac{\partial}{\partial\rho}S_{t}^{-1}(\rho)\Bigr)\sum_{h=0}^{T-t}A_{t}^{(h)}(\theta)D_{t-h}\Bigr]\Bigr|\\
&+\frac{1}{N}\Bigl|\mathrm{tr}\Bigl[
\mathcal{T}^{-1}\sum_{t=1}^{T-1} \lambda D_{t+1}'G_{t+1}(\rho)D_{t+1}
\\&\qquad(I_N-F_{t+1})D_{t+1}'M_{t+1}S_{t}^{-1}(\rho)\Bigl(\frac{\partial}{\partial\rho}\sum_{h=0}^{T-t}A_{t}^{(h)}(\theta)D_{t-h}\Bigr)\Bigr]\Bigr|
\\
\leq&\frac{1}{c_TNT}\sum_{t=1}^{T-1}\mathrm{tr}\Bigl\{
\mathrm{abs}\Bigl[\lambda D_{t+1}'W_{t+1}\Bigl(\frac{\partial}{\partial\rho}S_{t+1}^{-1}(\rho)\Bigr)D_{t+1}
\\&\qquad(I_N-F_{t+1})D_{t+1}'M_{t+1}S_t(\rho)^{-1}\Bigr]
\\&\qquad\mathrm{abs}\Bigl[\sum_{h=0}^{T-t}A_{t}^{(h)}(\theta)D_{t-h}\Bigr]\Bigr\}\\
&+\frac{1}{c_TNT}\sum_{t=1}^{T-1} \mathrm{tr}\Bigl\{\mathrm{abs}\Bigl[
\lambda D_{t+1}'G_{t+1}(\rho)D_{t+1}
\\&\qquad(I_N-F_{t+1})D_{t+1}'M_{t+1}\Bigl(\frac{\partial}{\partial\rho}S_{t}^{-1}(\rho)\Bigr)\Bigr]
\\&\qquad\mathrm{abs}\Bigl[\sum_{h=0}^{T-t}A_{t}^{(h)}(\theta)D_{t-h}\Bigr]\Bigr\}\\
&+\frac{1}{c_TNT}\sum_{t=1}^{T-1} \mathrm{tr}\Bigl\{\mathrm{abs}\Bigl[
\lambda D_{t+1}'G_{t+1}(\rho)D_{t+1}
\\&\qquad(I_N-F_{t+1})D_{t+1}'M_{t+1}S_{t}^{-1}(\rho)\Bigr]
\\&\qquad\mathrm{abs}\Bigl[\frac{\partial}{\partial\rho}\sum_{h=0}^{T-t}A_{t}^{(h)}(\theta)D_{t-h}\Bigr]\Bigr\}
\end{align*}
where the last inequality holds because $T_i\geq c_TT$ by Assumption 1. The last equality holds because $W_t$ and $M_t$ are UB uniformly for $t$ by Assumption 5. Additionally, $S_t(\rho)^{-1}$, $G_t(\rho)$, $\frac{\partial}{\partial\rho}S^{-1}_t(\rho)=S^{-1}_t(\rho)W_tS^{-1}_t(\rho)$, $\sum_{h=0}^{T-t}A_{t}^{(h)}(\theta)D_{t-h}$ and $\frac{\partial}{\partial\rho}\sum_{h=0}^{T-t}A_{t}^{(h)}(\theta)D_{t-h}$ are all UB uniformly in a neighborhood of $\rho_0$ for $t$ by Assumptions 5 and 10. Therefore, all the matrices inside the trace operations are UB uniformly for $t$, which means that the derivative above is $O(1)$ in a neighborhood of $\rho_0$.
Since $\hat{\theta}\overset{p}{\rightarrow}\theta_0$, and $\bar{\theta}$ lies between $\theta_0$ and $\hat{\theta}$, $\bar{\theta}$ lies in a neighborhood of $\theta_0$ with probability approaching one. Thus, the derivative above is $O_p(1)$ when evaluated at $\bar{\theta}$.
Using the same approach, we can similarly show that all elements in $\frac{\partial\varphi_n(\bar{\theta})}{\partial\rho}$, as well as $\frac{\partial\varphi_n(\bar{\theta})}{\partial\lambda}$ and $\frac{\partial\varphi_n(\bar{\theta})}{\partial\nu}$ are $O_p(1)$. Since $\sqrt{T}\frac{\partial\varphi_n(\bar{\theta})}{\partial\gamma}=\mathbf{0}_{(k+5)\times1}$, $\frac{\partial\varphi_n(\bar{\theta})}{\partial\beta'}=\mathbf{0}_{(k+5)\times k}$ and $\frac{\partial\varphi_n(\bar{\theta})}{\partial\sigma^2}=(\mathbf{0}_{1\times(k+4)},-\frac{1}{2\bar{\sigma}^4})'$, it follows that elements in $\frac{\partial\varphi_n(\bar{\theta})}{\partial\theta'}\Gamma$ are $O_p(1)$, which completes the proof.

\subsection{Proof of Theorem 4}\label{sec_T4}
Let \(\widehat{\theta}^{\,1}=(\widehat{\rho}^{\,1},
\widehat{\delta}^{\,1\prime},\widehat{\sigma}^{2,1})'\) be the
bias-corrected estimator. Define
\(
\widehat{\alpha}^{\,1}
=
(\mathbb D'\mathbb D)^{-1}\mathbb D'
\{\mathbb S(\widehat\rho^{\,1})\mathbb Y-\mathbb Z\widehat\delta^{\,1}\}\), and
\(\widehat{\mathbb V}
=
\mathbb Q\{\mathbb S(\widehat\rho^{\,1})\mathbb Y-\mathbb Z\widehat\delta^{\,1}\}
\).
Let \(\check\sigma^2=n^{-1}\widehat{\mathbb V}'\widehat{\mathbb V}\),
\(\check\mu_3=n^{-1}\sum_{r=1}^n\widehat{\mathbb V}_r^3\),
\(\check\mu_4=n^{-1}\sum_{r=1}^n\widehat{\mathbb V}_r^4\), and
\(\check c=(\check\mu_4-3\check\sigma^4)/\check\sigma^4\), where
\(\check\sigma^4=(\check\sigma^2)^2\) and
\(\check\sigma^6=(\check\sigma^2)^3\). Let
\(\widehat{\mathbb G}=\mathbb W\mathbb S(\widehat\rho^{\,1})^{-1}\),
\(\widehat g=(\widehat{\mathbb G}_{11},\ldots,\widehat{\mathbb G}_{nn})'\). Define
\[
\widehat{\Sigma}
=
\begin{pmatrix}
\displaystyle
\frac{1}{n\check\sigma^2}
\{\widehat{\mathbb G}(\mathbb Z\widehat\delta^{\,1}
+\mathbb D\widehat\alpha^{\,1})\}'
\mathbb Q
\{\widehat{\mathbb G}(\mathbb Z\widehat\delta^{\,1}
+\mathbb D\widehat\alpha^{\,1})\}
+
\frac{1}{n}\operatorname{tr}\{
\widehat{\mathbb G}(\widehat{\mathbb G}+\widehat{\mathbb G}')
\}
& * & *
\\[2.2ex]
\displaystyle
\frac{1}{n\check\sigma^2}
\widetilde{\Gamma}'\mathbb Z'\mathbb Q
\widehat{\mathbb G}(\mathbb Z\widehat\delta^{\,1}
+\mathbb D\widehat\alpha^{\,1})
&
\displaystyle
\frac{1}{n\check\sigma^2}
\widetilde{\Gamma}'\mathbb Z'\mathbb Q\mathbb Z\widetilde{\Gamma}
& *
\\[2.2ex]
\displaystyle
\frac{1}{n\check\sigma^2}\operatorname{tr}(\widehat{\mathbb G})
&
\mathbf{0}_{1\times(k+3)}
&
\displaystyle
\frac{1}{2\check\sigma^4}
\end{pmatrix}.
\]
Next, define
\[
\widehat{\Omega}
=
\begin{pmatrix}
\displaystyle
\frac{\check c}{n}\widehat g'\widehat g
+
\frac{2\check\mu_3}{n\check\sigma^4}
\left[
\widehat{\mathbb G}\mathbb Z\widehat\delta^{\,1}
+
\mathbb Q\widehat{\mathbb G}\mathbb D\widehat\alpha^{\,1}
\right]'
\widehat g
& * & *
\\[2.2ex]
\displaystyle
\frac{\check\mu_3}{n\check\sigma^4}
\widetilde{\Gamma}'\mathbb Z'\widehat g
&
\mathbf{0}_{(k+3)\times(k+3)}
& *
\\[2.2ex]
\displaystyle
\frac{\check c}{2n\check\sigma^2}
\operatorname{tr}(\widehat{\mathbb G})
&
\displaystyle
\mathbf{0}_{1\times(k+3)}
&
\displaystyle
\frac{\check c}{4\check\sigma^4}
\end{pmatrix}.
\]
Set
\[
\widehat{\Xi}
=
\widehat{\Sigma}^{-1}
(\widehat{\Sigma}+\widehat{\Omega})
\widehat{\Sigma}^{-1}.
\]
\begin{proof}
We first show the consistency of the residual moment estimators. Since
\[
\widehat{\mathbb{V}}
-
\mathbb{Q}\mathbb{V}
=
\mathbb{Q}
\left[
-(\widehat{\rho}^{\,1}-\rho_0)\mathbb{W}\mathbb{Y}
-
\mathbb{Z}(\widehat{\delta}^{\,1}-\delta_0)
\right],
\]
we need to control the average moments of the right-hand side. For any $n\times n$ UB matrix \(B\) and any vector \(a\), H\"older's inequality implies that, for \(q=2,3,4\),
\begin{align}
\sum_{r=1}^n |(Ba)_r|^q
&=
\sum_{r=1}^n
\left|
\sum_{s=1}^n B_{rs}a_s
\right|^q
\le
\sum_{r=1}^n
\left(
\sum_{s=1}^n |B_{rs}||a_s|
\right)^q
\le
\sum_{r=1}^n
\left(
\sum_{s=1}^n |B_{rs}|
\right)^{q-1}
\sum_{s=1}^n |B_{rs}||a_s|^q
\nonumber\\
&\le
\|B\|_\infty^{q-1}
\sum_{s=1}^n |a_s|^q
\left(
\sum_{r=1}^n |B_{rs}|
\right)
\le
\|B\|_\infty^{q-1}
\|B\|_1
\sum_{s=1}^n |a_s|^q.\label{ineq_UB}
\end{align}
Thus,
\begin{align}
&\frac{1}{n}
\sum_{r=1}^n
\left|
\widehat{\mathbb{V}}_r
-
(\mathbb{Q}\mathbb{V})_r
\right|^q\nonumber\\
\le&
\|\mathbb{Q}\|_\infty^{q-1}\|\mathbb{Q}\|_1\frac{1}{n}\sum_{r=1}^n
\left|\left[
-(\widehat{\rho}^{\,1}-\rho_0)\mathbb{W}\mathbb{Y}
-
\mathbb{Z}(\widehat{\delta}^{\,1}-\delta_0)
\right]_r\right|^q\nonumber
\\
\le&
\|\mathbb{Q}\|_\infty^{q-1}\|\mathbb{Q}\|_1
(k+4)^{q-1}
\Bigg[
|\widehat{\rho}^{\,1}-\rho_0|^q
\frac{1}{n}
\sum_{r=1}^n
|(\mathbb{W}\mathbb{Y})_r|^q
+
\sum_{j=1}^{k+3}
|\widehat{\delta}^{\,1}_j-\delta_{0j}|^q
\frac{1}{n}
\sum_{r=1}^n
|\mathbb{Z}_{rj}|^q
\Bigg],\label{eq_V-QV}
\end{align}
where the second inequality holds because
\(
|x_1+\cdots+x_{k+4}|^q
\le
(k+4)^{q-1}\sum_{j=1}^{k+4}|x_j|^q
\).

By the moving-average representation derived in equation (3),
\(
Y_t=\mathcal E_t+\mathcal X_t\beta_0+\mathcal V_t.
\)
The deterministic component \(\mathcal E_t+\mathcal X_t\beta_0\) has uniformly bounded elements by Assumptions~4--6. Moreover, \(\mathcal V_t\) is an infinite historical-error accumulation of the same form as that considered in Lemma~17. Hence, by the same argument as in Lemma~17,
\(
\sup_{i,t,N}E|\mathcal V_{t,i}|^4=O(1).
\)
Therefore, applying Markov's inequality gives
\(
\frac{1}{n}\sum_{r=1}^n|\mathbb Y_r|^4=O_p(1).
\)
Since \(\mathbb W\) is UB, the equation (\ref{ineq_UB}) gives \[
\frac{1}{n}\sum_{r=1}^n|(\mathbb W\mathbb Y)_r|^q=O_p(1).
\]
The cases \(q=2,3\) follow from the fourth-moment bound.
Similarly, for each $j$, we obtain 
\[
\frac{1}{n}\sum_{r=1}^n|\mathbb Z_{rj}|^q=O_p(1).
\]
Together with \(\widehat{\rho}^{\,1}-\rho_0=o_p(1)\) and \(\widehat{\delta}^{\,1}_j-\delta_{0j}=o_p(1)\) for each \(j\), equation (\ref{eq_V-QV}) gives
\[
\frac{1}{n}
\sum_{r=1}^n
\left|
\widehat{\mathbb{V}}_r
-
(\mathbb{Q}\mathbb{V})_r
\right|^q
=
o_p(1),
\qquad q=2,3,4.
\]

Since $\widehat{\mathbb{V}}=\mathbb{Q}\mathbb{V}+\widehat{\mathbb{V}}-\mathbb{Q}\mathbb{V}$, the binomial expansion gives
\begin{align}
\frac{1}{n}\sum_{r=1}^n\left|
\widehat{\mathbb{V}}_r^q
-
(\mathbb{Q}\mathbb{V})_r^q
\right|
&\le
\frac{1}{n}\sum_{l=1}^q
\binom{q}{l}
\sum_{r=1}^n|(\mathbb{Q}\mathbb{V})_r|^{q-l}
\left|
\widehat{\mathbb{V}}_r-(\mathbb{Q}\mathbb{V})_r
\right|^l\nonumber\\
&\le \sum_{l=1}^q
\binom{q}{l}
\left[
\frac{1}{n}
\sum_{r=1}^n
|(\mathbb{Q}\mathbb{V})_r|^q
\right]^{(q-l)/q}
\left[
\frac{1}{n}
\sum_{r=1}^n
\left|
\widehat{\mathbb{V}}_r-(\mathbb{Q}\mathbb{V})_r
\right|^q
\right]^{l/q},\label{ineq_Vh-QV}
\end{align}
where the second inequality follows from H\"older's inequality. Since \(\mathbb{Q}\) is UB, equation \eqref{ineq_UB} gives
\[
\frac{1}{n}
\sum_{r=1}^n
|(\mathbb{Q}\mathbb{V})_r|^q
\le
\|\mathbb{Q}\|_\infty^{q-1}\|\mathbb{Q}\|_1
\frac{1}{n}
\sum_{r=1}^n
|\mathbb{V}_r|^q
=
O_p(1),\qquad q=2,3,4.
\]
Combining this with
\(
\frac{1}{n}
\sum_{r=1}^n
\left|
\widehat{\mathbb{V}}_r
-
(\mathbb{Q}\mathbb{V})_r
\right|^q
=
o_p(1),
\)
we obtain
\[
\frac{1}{n}
\sum_{r=1}^n
\widehat{\mathbb{V}}_r^q
-
\frac{1}{n}
\sum_{r=1}^n
(\mathbb{Q}\mathbb{V})_r^q
=
o_p(1),
\qquad q=2,3,4.
\]

Next, we show that the sample moments of \(\mathbb Q\mathbb V\) and \(\mathbb V\) are asymptotically equivalent. Since
\(
\mathbb Q\mathbb V
=
\mathbb V-\mathbb P\mathbb V,
\)
it suffices to control the average moments of \(\mathbb P\mathbb V\). Each element of \(\mathbb P\mathbb V\) is an individual-specific average of the errors over the observed periods. For observation \(r\) corresponding to individual \(i\),
\(
(\mathbb P\mathbb V)_r
=
\frac{1}{T_i}
\sum_{t:d_{it}=1}v_{it}.
\)
Since the errors are independent with mean zero, variance \(\sigma_0^2\), and fourth moment \(\mu_4\), we have
\[
\begin{aligned}
E|(\mathbb{P}\mathbb{V})_r|^2
=
E\left(
\frac{1}{T_i}
\sum_{t:d_{it}=1}v_{it}
\right)^2=
\frac{1}{T_i^2}
\sum_{t:d_{it}=1}E(v_{it}^2)
=
\frac{\sigma_0^2}{T_i}=O(T^{-1})
\end{aligned}
\]
Similarly,
\[
\begin{aligned}
E|(\mathbb{P}\mathbb{V})_r|^4
&=
E\left(
\frac{1}{T_i}
\sum_{t:d_{it}=1}v_{it}
\right)^4=
\frac{1}{T_i^4}
\left[
\sum_{t:d_{it}=1}E(v_{it}^4)
+
6
\sum_{\substack{s<t\\ d_{is}=d_{it}=1}}
E(v_{is}^2)E(v_{it}^2)
\right] \\
&=
\frac{1}{T_i^4}
\left[
T_i\mu_4
+
3T_i(T_i-1)\sigma_0^4
\right]
=
O(T^{-2}).
\end{aligned}
\]
Moreover, by H\"older's inequality,
\[
E|(\mathbb P\mathbb V)_r|^3
\le
\left(E|(\mathbb P\mathbb V)_r|^4\right)^{3/4}
=
O(T^{-3/2}).
\]
It follows that, for \(q=2,3,4\),
\(
E|(\mathbb P\mathbb V)_r|^q
=
O( T^{-q/2}).
\)
Therefore,
\[
\frac{1}{n}
\sum_{r=1}^n
|(\mathbb P\mathbb V)_r|^q
=
O_p(T^{-q/2})
=
o_p(1),
\]
by Markov's inequality.
Moreover, by Assumption~2,
\(
\frac{1}{n}
\sum_{r=1}^n
|\mathbb V_r|^q
=
O_p(1).
\)
Thus, similar to equation \eqref{ineq_Vh-QV}, we have
\[
\begin{aligned}
\frac{1}{n}
\sum_{r=1}^n
\left|
(\mathbb Q\mathbb V)_r^q-\mathbb V_r^q
\right|
&\le
\frac{1}{n}
\sum_{l=1}^q
\binom{q}{l}
\sum_{r=1}^n
|\mathbb V_r|^{q-l}
|(\mathbb P\mathbb V)_r|^l
\\
&\le
\sum_{l=1}^q
\binom{q}{l}
\left[
\frac{1}{n}
\sum_{r=1}^n
|\mathbb V_r|^q
\right]^{(q-l)/q}
\left[
\frac{1}{n}
\sum_{r=1}^n
|(\mathbb P\mathbb V)_r|^q
\right]^{l/q}=
o_p(1).
\end{aligned}
\]
Consequently,
\(
\frac{1}{n}
\sum_{r=1}^n
(\mathbb Q\mathbb V)_r^q
-
\frac{1}{n}
\sum_{r=1}^n
\mathbb V_r^q
=
o_p(1)\), for \(q=2,3,4\), which implies 
\[
\frac{1}{n}
\sum_{r=1}^n
\widehat{\mathbb{V}}_r^q
-
\frac{1}{n}
\sum_{r=1}^n
\mathbb V_r^q
=
o_p(1),
\qquad q=2,3,4.
\]
By the law of large numbers,
\[
\frac{1}{n}
\sum_{r=1}^n
\mathbb{V}_r^2
\xrightarrow{p}
\sigma_0^2,
\qquad
\frac{1}{n}
\sum_{r=1}^n
\mathbb{V}_r^3
\xrightarrow{p}
\mu_3,
\qquad
\frac{1}{n}
\sum_{r=1}^n
\mathbb{V}_r^4
\xrightarrow{p}
\mu_4.
\]
Therefore,
\[
\check{\sigma}^2\xrightarrow{p}\sigma_0^2,
\qquad
\check{\mu}_3\xrightarrow{p}\mu_3,
\qquad
\check{\mu}_4\xrightarrow{p}\mu_4,
\qquad
\check{c}
\xrightarrow{p}
\frac{\mu_4-3\sigma_0^4}{\sigma_0^4}.
\]
We next show that \(\widehat{\Sigma}\) is consistent.
By Theorem~3, we have \(\widehat{\theta}^{\,1}\xrightarrow{p}\theta_0\). In particular,
\(\widehat{\rho}^{\,1}-\rho_0=o_p(1)\).
Since
\(
\mathbb{S}(\widehat{\rho}^{\,1})
=
\mathbb{S}(\rho_0)
-
(\widehat{\rho}^{\,1}-\rho_0)\mathbb{W},
\)
applying \eqref{eq_inv_A+B} gives, for \(p=1,\infty\),
\[
\left\|
\mathbb{S}(\widehat{\rho}^{\,1})^{-1}
-
\mathbb{S}(\rho_0)^{-1}
\right\|_p
=
O_p(|\widehat{\rho}^{\,1}-\rho_0|)
=
o_p(1).
\]
Since \(\mathbb{W}\) is UB, for $p=1,\infty$, we have
\[
\begin{aligned}
\left\|
\widehat{\mathbb{G}}
-
\mathbb{G}_0
\right\|_p
=
\left\|
\mathbb{W}
\left[
\mathbb{S}(\widehat{\rho}^{\,1})^{-1}
-
\mathbb{S}(\rho_0)^{-1}
\right]
\right\|_p\le
\|\mathbb{W}\|_p
\left\|
\mathbb{S}(\widehat{\rho}^{\,1})^{-1}
-
\mathbb{S}(\rho_0)^{-1}
\right\|_p
=
o_p(1).
\end{aligned}
\]
Hence, \(\|\widehat{\mathbb{G}}-\mathbb{G}_0\|_1=o_p(1)\) and \(\|\widehat{\mathbb{G}}-\mathbb{G}_0\|_\infty=o_p(1)\).

From the definition of \(\widehat{\alpha}^{\,1}\), we have
\[
\begin{aligned}
\mathbb{D}\widehat{\alpha}^{\,1}-\mathbb{D}\alpha_0
=
\mathbb{P}
\left[
\mathbb{S}(\widehat{\rho}^{\,1})\mathbb{Y}
-
\mathbb{Z}\widehat{\delta}^{\,1}
\right]-\mathbb{D}\alpha_0=
\mathbb{P}\mathbb{V}
-
\mathbb{P}
\left[
(\widehat{\rho}^{\,1}-\rho_0)\mathbb{W}\mathbb{Y}
+
\mathbb{Z}(\widehat{\delta}^{\,1}-\delta_0)
\right].
\end{aligned}
\]
By Lemma~\ref{lm_VQV}, we have
\(
\frac{1}{n}
(\mathbb{P}\mathbb{V})'\mathbb{P}\mathbb{V}
=
o_p(1).
\)
Moreover, applying the same argument as in \eqref{eq_V-QV} gives
\(
\frac{1}{n}
\left\|
\mathbb{P}
\left[
(\widehat{\rho}^{\,1}-\rho_0)\mathbb{W}\mathbb{Y}
+
\mathbb{Z}(\widehat{\delta}^{\,1}-\delta_0)
\right]
\right\|^2
=
o_p(1).
\)
Combining the two bounds above,
we obtain
\[
\frac{1}{n}\|\mathbb{D}\widehat{\alpha}^{\,1}-\mathbb{D}\alpha_0\|^2=o_p(1).
\]
Since \(\widehat{\mathbb{G}}\) and \(\mathbb{G}_0\) are UB,
\(
\|\widehat{\mathbb{G}}-\mathbb{G}_0\|_1=o_p(1)\) and
\(\|\widehat{\mathbb{G}}-\mathbb{G}_0\|_\infty=o_p(1)\),
we have
\begin{align}
\frac{1}{n}
\left\|
\widehat{\mathbb{G}}\mathbb{Z}\widehat{\delta}^{\,1}
-
\mathbb{G}_0\mathbb{Z}\delta_0
\right\|^2
=o_p(1),
\qquad
\frac{1}{n}
\left\|
\widehat{\mathbb{G}}\mathbb{D}\widehat{\alpha}^{\,1}
-
\mathbb{G}_0\mathbb{D}\alpha_0
\right\|^2
=o_p(1).\label{eq_GZDGDA}
\end{align}
Consequently,
\[
\frac{1}{n}
\left\|
\widehat{\mathbb{G}}
\left(
\mathbb{Z}\widehat{\delta}^{\,1}
+
\mathbb{D}\widehat{\alpha}^{\,1}
\right)
-
\mathbb{G}_0
(\mathbb{Z}\delta_0+\mathbb{D}\alpha_0)
\right\|^2
=
o_p(1).
\]
Moreover,
\(
\frac{1}{n}
\left\|
\mathbb{G}_0
(\mathbb{Z}\delta_0+\mathbb{D}\alpha_0)
\right\|^2
=
O_p(1).
\)
Therefore,
\[
\begin{aligned}
&
\frac{1}{n}
\left[
\widehat{\mathbb{G}}
\left(
\mathbb{Z}\widehat{\delta}^{\,1}
+
\mathbb{D}\widehat{\alpha}^{\,1}
\right)
\right]'
\mathbb{Q}
\left[
\widehat{\mathbb{G}}
\left(
\mathbb{Z}\widehat{\delta}^{\,1}
+
\mathbb{D}\widehat{\alpha}^{\,1}
\right)
\right]
-
\frac{1}{n}
\left[
\mathbb{G}_0
\left(
\mathbb{Z}\delta_0+\mathbb{D}\alpha_0
\right)
\right]'
\mathbb{Q}
\left[
\mathbb{G}_0
\left(
\mathbb{Z}\delta_0+\mathbb{D}\alpha_0
\right)
\right]
\\
=&
\frac{1}{n}
\left[
\widehat{\mathbb{G}}
\left(
\mathbb{Z}\widehat{\delta}^{\,1}
+
\mathbb{D}\widehat{\alpha}^{\,1}
\right)
-
\mathbb{G}_0
(\mathbb{Z}\delta_0+\mathbb{D}\alpha_0)
\right]'
\mathbb{Q}
\left[
\widehat{\mathbb{G}}
\left(
\mathbb{Z}\widehat{\delta}^{\,1}
+
\mathbb{D}\widehat{\alpha}^{\,1}
\right)
-
\mathbb{G}_0
(\mathbb{Z}\delta_0+\mathbb{D}\alpha_0)
\right]
\\
&+
\frac{2}{n}
\left[
\mathbb{G}_0
(\mathbb{Z}\delta_0+\mathbb{D}\alpha_0)
\right]'
\mathbb{Q}
\left[
\widehat{\mathbb{G}}
\left(
\mathbb{Z}\widehat{\delta}^{\,1}
+
\mathbb{D}\widehat{\alpha}^{\,1}
\right)
-
\mathbb{G}_0
(\mathbb{Z}\delta_0+\mathbb{D}\alpha_0)
\right].
\end{aligned}
\]
For the first term above, we have
\[
\begin{aligned}
&\left|
\frac{1}{n}
\left[
\widehat{\mathbb{G}}
\left(
\mathbb{Z}\widehat{\delta}^{\,1}
+
\mathbb{D}\widehat{\alpha}^{\,1}
\right)
-
\mathbb{G}_0
(\mathbb{Z}\delta_0+\mathbb{D}\alpha_0)
\right]'
\mathbb{Q}
\left[
\widehat{\mathbb{G}}
\left(
\mathbb{Z}\widehat{\delta}^{\,1}
+
\mathbb{D}\widehat{\alpha}^{\,1}
\right)
-
\mathbb{G}_0
(\mathbb{Z}\delta_0+\mathbb{D}\alpha_0)
\right]
\right|
\\
\le&
\|\mathbb{Q}\|_\infty^{1/2}\|\mathbb{Q}\|_1^{1/2}
\frac{1}{n}
\left\|
\widehat{\mathbb{G}}
\left(
\mathbb{Z}\widehat{\delta}^{\,1}
+
\mathbb{D}\widehat{\alpha}^{\,1}
\right)
-
\mathbb{G}_0
(\mathbb{Z}\delta_0+\mathbb{D}\alpha_0)
\right\|^2
=
o_p(1).
\end{aligned}
\]
For the second term, by the Cauchy--Schwarz inequality,
\[
\begin{aligned}
&
\left|
\frac{2}{n}
\left[
\mathbb{G}_0
(\mathbb{Z}\delta_0+\mathbb{D}\alpha_0)
\right]'
\mathbb{Q}
\left[
\widehat{\mathbb{G}}
\left(
\mathbb{Z}\widehat{\delta}^{\,1}
+
\mathbb{D}\widehat{\alpha}^{\,1}
\right)
-
\mathbb{G}_0
(\mathbb{Z}\delta_0+\mathbb{D}\alpha_0)
\right]
\right|
\\
\le&
2
\left[
\frac{1}{n}
\left\|
\mathbb{G}_0
(\mathbb{Z}\delta_0+\mathbb{D}\alpha_0)
\right\|^2
\right]^{1/2}
\left[
\frac{1}{n}
\left\|
\mathbb{Q}
\left\{
\widehat{\mathbb{G}}
\left(
\mathbb{Z}\widehat{\delta}^{\,1}
+
\mathbb{D}\widehat{\alpha}^{\,1}
\right)
-
\mathbb{G}_0
(\mathbb{Z}\delta_0+\mathbb{D}\alpha_0)
\right\}
\right\|^2
\right]^{1/2}
\\
\le&
2
\|\mathbb{Q}\|_\infty^{1/2}\|\mathbb{Q}\|_1^{1/2}
\left[
\frac{1}{n}
\left\|
\mathbb{G}_0
(\mathbb{Z}\delta_0+\mathbb{D}\alpha_0)
\right\|^2
\right]^{1/2}
\left[
\frac{1}{n}
\left\|
\widehat{\mathbb{G}}
\left(
\mathbb{Z}\widehat{\delta}^{\,1}
+
\mathbb{D}\widehat{\alpha}^{\,1}
\right)
-
\mathbb{G}_0
(\mathbb{Z}\delta_0+\mathbb{D}\alpha_0)
\right\|^2
\right]^{1/2}
=
o_p(1).
\end{aligned}
\]
Hence,
\[
\begin{aligned}
\frac{1}{n}
\left[
\widehat{\mathbb{G}}
\left(
\mathbb{Z}\widehat{\delta}^{\,1}
+
\mathbb{D}\widehat{\alpha}^{\,1}
\right)
\right]'
\mathbb{Q}
\left[
\widehat{\mathbb{G}}
\left(
\mathbb{Z}\widehat{\delta}^{\,1}
+
\mathbb{D}\widehat{\alpha}^{\,1}
\right)
\right]-
\frac{1}{n}
\left[
\mathbb{G}_0
\left(
\mathbb{Z}\delta_0+\mathbb{D}\alpha_0
\right)
\right]'
\mathbb{Q}
\left[
\mathbb{G}_0
\left(
\mathbb{Z}\delta_0+\mathbb{D}\alpha_0
\right)
\right]
=
o_p(1).
\end{aligned}
\]
Similarly,
\[
\begin{aligned}
&
\frac{1}{n}
\widetilde{\Gamma}'\mathbb{Z}'\mathbb{Q}
\widehat{\mathbb{G}}
\left(
\mathbb{Z}\widehat{\delta}^{\,1}
+
\mathbb{D}\widehat{\alpha}^{\,1}
\right)
-
\frac{1}{n}
\widetilde{\Gamma}'\mathbb{Z}'\mathbb{Q}
\mathbb{G}_0
\left(
\mathbb{Z}\delta_0+\mathbb{D}\alpha_0
\right)
=
o_p(1).
\end{aligned}
\]

Since \(\|\widehat{\mathbb{G}}-\mathbb{G}_0\|_1=o_p(1)\), \(\|\widehat{\mathbb{G}}-\mathbb{G}_0\|_\infty=o_p(1)\), and $\widehat{\mathbb{G}}$ and $\mathbb{G}_0$ are UB, we have 
\begin{align}
\frac{1}{n}
\operatorname{tr}
\{
\widehat{\mathbb{G}}(\widehat{\mathbb{G}}+\widehat{\mathbb{G}}')
\}
=
\frac{1}{n}
\operatorname{tr}
\{
\mathbb{G}_0(\mathbb{G}_0+\mathbb{G}_0')
\}
+o_p(1).\label{eq_trGGG}
\end{align}

Since $\check{\sigma}^2\overset{p}{\rightarrow}\sigma_0^2$ and $\sigma_0^2$ is bounded away from zero, the results above imply
\(
\widehat{\Sigma}
-
\Gamma\Sigma_{\theta_0,n}\Gamma
=o_p(1).
\)
Since \(\lim_{n\rightarrow\infty}\Gamma\Sigma_{\theta_0,n}\Gamma=\Sigma_{\theta_0}\), we have
\[
\widehat{\Sigma}
\xrightarrow{p}
\Sigma_{\theta_0}.
\]

It remains to show the consistency of \(\widehat{\Omega}\). Similar to \eqref{eq_trGGG}, the same argument gives
\[
\frac{1}{n}\widehat{g}'\widehat{g}
=
\frac{1}{n}
\sum_{r=1}^n
(\mathbb{G}_0)_{rr}^2
+o_p(1),
\qquad
\frac{1}{n}
\operatorname{tr}(\widehat{\mathbb{G}})
=
\frac{1}{n}
\operatorname{tr}(\mathbb{G}_0)
+o_p(1),
\qquad
\frac{1}{n}
\widetilde{\Gamma}'\mathbb{Z}'\widehat{g}
=
\frac{1}{n}
\widetilde{\Gamma}'\mathbb{Z}'g_0
+o_p(1),
\]
where \(g_0=((\mathbb{G}_0)_{11},\ldots,(\mathbb{G}_0)_{nn})'\).

Moreover, according to equation \eqref{eq_GZDGDA},
\[
\begin{aligned}
&
\frac{1}{n}
\left\|
\widehat{\mathbb{G}}\mathbb{Z}\widehat{\delta}^{\,1}
+
\mathbb{Q}\widehat{\mathbb{G}}\mathbb{D}\widehat{\alpha}^{\,1}
-
\left(
\mathbb{G}_0\mathbb{Z}\delta_0
+
\mathbb{Q}\mathbb{G}_0\mathbb{D}\alpha_0
\right)
\right\|^2
\\
\le&
\frac{2}{n}
\left\|
\widehat{\mathbb{G}}\mathbb{Z}\widehat{\delta}^{\,1}
-
\mathbb{G}_0\mathbb{Z}\delta_0
\right\|^2
+
\frac{2}{n}
\left\|
\mathbb{Q}
\left(
\widehat{\mathbb{G}}\mathbb{D}\widehat{\alpha}^{\,1}
-
\mathbb{G}_0\mathbb{D}\alpha_0
\right)
\right\|^2
\\
\le&
\frac{2}{n}
\left\|
\widehat{\mathbb{G}}\mathbb{Z}\widehat{\delta}^{\,1}
-
\mathbb{G}_0\mathbb{Z}\delta_0
\right\|^2
+
2\|\mathbb{Q}\|_\infty\|\mathbb{Q}\|_1
\frac{1}{n}
\left\|
\widehat{\mathbb{G}}\mathbb{D}\widehat{\alpha}^{\,1}
-
\mathbb{G}_0\mathbb{D}\alpha_0
\right\|^2=
o_p(1).
\end{aligned}
\]
Since
\(
\frac{1}{n}\widehat{g}'\widehat{g}=O_p(1)
\), and \(
\frac{1}{n}g_0'g_0=O_p(1)
\),
the Cauchy--Schwarz inequality gives
\[
\begin{aligned}
&
\frac{1}{n}
\left[
\widehat{\mathbb{G}}\mathbb{Z}\widehat{\delta}^{\,1}
+
\mathbb{Q}\widehat{\mathbb{G}}\mathbb{D}\widehat{\alpha}^{\,1}
\right]'
\widehat{g}
-
\frac{1}{n}
\left[
\mathbb{G}_0\mathbb{Z}\delta_0
+
\mathbb{Q}\mathbb{G}_0\mathbb{D}\alpha_0
\right]'
g_0
=
o_p(1).
\end{aligned}
\]
Combining the above results with
\[
\check{\mu}_3\xrightarrow{p}\mu_3,
\qquad
\check{c}\xrightarrow{p}
\frac{\mu_4-3\sigma_0^4}{\sigma_0^4},
\qquad
\check{\sigma}^2\xrightarrow{p}\sigma_0^2,
\]
we obtain
\(
\widehat{\Omega}
-
\Gamma\Omega_{\theta_0,n}\Gamma
=
o_p(1).
\)
Since \(\lim_{n\rightarrow\infty}\Gamma\Omega_{\theta_0,n}\Gamma=\Omega_{\theta_0}\), it follows that
\[
\widehat{\Omega}
\xrightarrow{p}
\Omega_{\theta_0}.
\]

Finally, Assumption~9 implies that \(\Sigma_{\theta_0}\) is nonsingular. Therefore, \(\widehat{\Sigma}\) is nonsingular with probability approaching one, and by the continuous mapping theorem,
\[
\widehat{\Xi}
=
\widehat{\Sigma}^{-1}
(\widehat{\Sigma}+\widehat{\Omega})
\widehat{\Sigma}^{-1}
\xrightarrow{p}
\Sigma_{\theta_0}^{-1}
(\Sigma_{\theta_0}+\Omega_{\theta_0})
\Sigma_{\theta_0}^{-1}.
\]
Thus, \(n^{-1}\Gamma\widehat{\Xi}\Gamma\) is a consistent estimator of the asymptotic covariance matrix of \(\widehat{\theta}^{\,1}\). This completes the proof.
\end{proof}
\subsection{Proof of Theorem 5}
The proof follows the same route as Theorem 1. We first establish the consistency of
$\hat{\theta}^{(-\gamma)}$, and then prove the consistency of $\hat{\gamma}$ from its explicit
concentrated expression.

\textbf{Proof of uniform convergence}

The uniform convergence requires showing
\[
\sup_{\theta\in\Theta}
\left|
\frac{1}{n}\left(\widebar{L}_n^c(\theta)-\widebar{Q}_n(\theta)\right)
\right|
\overset{p}{\to}0,
\]
where $\widebar{Q}_n(\theta)\equiv E[\widebar{L}_n^c(\theta)]$. As in the proof of
Theorem 1, it suffices to show
\[
\sup_{\theta\in\Theta}
\frac{1}{n}
\left|
\widebar{\mathbb{V}}(\zeta)'\widebar{\mathbb{V}}(\zeta)
-
E\left[\widebar{\mathbb{V}}(\zeta)'\widebar{\mathbb{V}}(\zeta)\right]
\right|
\overset{p}{\to}0,
\]
where
\(
\widebar{\mathbb{V}}(\zeta)
\equiv
\widebar{\mathbb{Q}}
\left(\mathbb{S}(\rho)\mathbb{Y}-\mathbb{Z}\delta\right).
\)
Since
\(
\mathbb{Y}
=
\rho_0\mathbb{W}\mathbb{Y}
+
\mathbb{Z}\delta_0
+
\mathbb{D}\alpha_0
+
\mathbb{J}\tau_0
+
\mathbb{V},
\)
and $\widebar{\mathbb{Q}}\mathbb{D}=0$, $\widebar{\mathbb{Q}}\mathbb{J}=0$, we have
\(
\widebar{\mathbb{V}}(\zeta)
=
\widebar{\mathbb{Q}}
\left[
\mathbb{V}
-
(\rho-\rho_0)\mathbb{W}\mathbb{Y}
-
\mathbb{Z}(\delta-\delta_0)
\right].
\)
Thus,
\begin{align}
\frac{1}{n}\widebar{\mathbb{V}}(\zeta)'\widebar{\mathbb{V}}(\zeta)
=&
\frac{1}{n}\mathbb{V}'\widebar{\mathbb{Q}}\mathbb{V}
+
(\rho-\rho_0)^2
\frac{1}{n}\mathbb{Y}'\mathbb{W}'\widebar{\mathbb{Q}}\mathbb{W}\mathbb{Y}
+
\frac{1}{n}(\delta-\delta_0)'
\mathbb{Z}'\widebar{\mathbb{Q}}\mathbb{Z}
(\delta-\delta_0)
\nonumber\\
&-
2(\rho-\rho_0)
\frac{1}{n}\mathbb{Y}'\mathbb{W}'\widebar{\mathbb{Q}}\mathbb{V}
-
2\frac{1}{n}
(\delta-\delta_0)'\mathbb{Z}'\widebar{\mathbb{Q}}\mathbb{V}
+
2(\rho-\rho_0)
\frac{1}{n}
(\delta-\delta_0)'\mathbb{Z}'\widebar{\mathbb{Q}}\mathbb{W}\mathbb{Y}.
\label{eq_twfe_V'V}
\end{align}
Furthermore,
\[
\mathbb{W}\mathbb{Y}
=
\mathbb{G}_0
\left(
\mathbb{Z}\delta_0+\mathbb{D}\alpha_0+\mathbb{J}\tau_0+\mathbb{V}
\right).
\]
Compared with Theorem 1, the only additional component is
$\mathbb{G}_0\mathbb{J}\tau_0$. Since $\tau_{t0}$ is uniformly bounded by Assumption 4$'$ and
$\mathbb{G}_0$ is UB under Assumptions 3 and 5, this term can be treated in the same way as
$\mathbb{G}_0\mathbb{D}\alpha_0$. Therefore, applying Lemma~\ref{lem:twfe-ZV-ZZ} to each term in
\eqref{eq_twfe_V'V}, all polynomial coefficients in
$n^{-1}\widebar{\mathbb{V}}(\zeta)'\widebar{\mathbb{V}}(\zeta)$ converge uniformly to their expectations.
Since $\rho$ and $\delta$ are bounded in $\Theta$, we obtain
\[
\sup_{\theta\in\Theta}
\frac{1}{n}
\left|
\widebar{\mathbb{V}}(\zeta)'\widebar{\mathbb{V}}(\zeta)
-
E\left[
\widebar{\mathbb{V}}(\zeta)'\widebar{\mathbb{V}}(\zeta)
\right]
\right|
\overset{p}{\to}0.
\]
Because $\sigma^2$ is bounded away from zero, it follows that
\(
\sup_{\theta\in\Theta}
\left|
\frac{1}{n}\left(\widebar{L}_n^c(\theta)-\widebar{Q}_n(\theta)\right)
\right|
\overset{p}{\to}0.
\)

\textbf{Proof of equicontinuity}

Next we prove that $\frac{1}{n}\widebar{Q}_n(\theta)$ is uniformly equicontinuous for
$\theta\in\Theta$. The terms involving $-\frac{1}{2}\log\sigma^2$ and
$\frac{1}{n}\log|\mathbb{S}(\rho)|$ are identical to those in the proof of Theorem~1, and hence
are omitted. Therefore, it remains to verify that replacing the OWFE projection matrix by the
TWFE projection matrix $\widebar{\mathbb{Q}}$, together with the presence of the time effects
$\mathbb{J}\tau_0$, preserves the uniform equicontinuity of the normalized expectation term
\(
\frac{1}{n}
E\left[
\widebar{\mathbb{V}}(\zeta)'
\widebar{\mathbb{V}}(\zeta)
\right].
\)

Substituting
\(
\mathbb{Y}
=
\mathbb{S}_0^{-1}
\left(
\mathbb{Z}\delta_0+\mathbb{D}\alpha_0+\mathbb{J}\tau_0+\mathbb{V}
\right)
\)
into $\widebar{\mathbb{V}}(\zeta)$ gives
\begin{align*}
&E\left[
\widebar{\mathbb{V}}(\zeta)'
\widebar{\mathbb{V}}(\zeta)
\right]
\\
=&
E
\Bigl[
\Bigl\{
\mathbb{S}(\rho)\mathbb{S}_0^{-1}
\left(
\mathbb{Z}\delta_0+\mathbb{D}\alpha_0+\mathbb{J}\tau_0
\right)
-\mathbb{Z}\delta
\Bigr\}'
\widebar{\mathbb{Q}}
\Bigl\{
\mathbb{S}(\rho)\mathbb{S}_0^{-1}
\left(
\mathbb{Z}\delta_0+\mathbb{D}\alpha_0+\mathbb{J}\tau_0
\right)
-\mathbb{Z}\delta
\Bigr\}
\Bigr]
\\
&+
E
\Bigl[
\left(
\mathbb{S}(\rho)\mathbb{S}_0^{-1}\mathbb{V}
\right)'
\widebar{\mathbb{Q}}
\mathbb{S}(\rho)\mathbb{S}_0^{-1}\mathbb{V}
\Bigr]+
2E
\Bigl[
\Bigl\{
\mathbb{S}(\rho)\mathbb{S}_0^{-1}
\left(
\mathbb{Z}\delta_0+\mathbb{D}\alpha_0+\mathbb{J}\tau_0
\right)
-\mathbb{Z}\delta
\Bigr\}'
\widebar{\mathbb{Q}}
\mathbb{S}(\rho)\mathbb{S}_0^{-1}\mathbb{V}
\Bigr].
\end{align*}
By Lemma~\ref{lem:twfe-ZV-ZZ} and Assumption~1$'$, the last term is
$o(n)$. Moreover, \(
\mathbb{S}(\rho)\mathbb{S}_0^{-1}
=
I_n+(\rho_0-\rho)\mathbb{G}_0,
\)
and $\widebar{\mathbb{Q}}\mathbb{D}=0$, $\widebar{\mathbb{Q}}\mathbb{J}=0$. Therefore,
\begin{align}
&\frac{1}{n}
E\left[
\widebar{\mathbb{V}}(\zeta)'
\widebar{\mathbb{V}}(\zeta)
\right]\nonumber
\\
=&
\frac{(\rho_0-\rho)^2}{n}
E
\Bigl[
\left\{
\mathbb{G}_0
\left(
\mathbb{Z}\delta_0+\mathbb{D}\alpha_0+\mathbb{J}\tau_0
\right)
\right\}'
\widebar{\mathbb{Q}}
\left\{
\mathbb{G}_0
\left(
\mathbb{Z}\delta_0+\mathbb{D}\alpha_0+\mathbb{J}\tau_0
\right)
\right\}
\Bigr]
\nonumber\\
&+
\frac{1}{n}
E
\left[
(\delta_0-\delta)'
\mathbb{Z}'
\widebar{\mathbb{Q}}
\mathbb{Z}
(\delta_0-\delta)
\right]+
\frac{2(\rho_0-\rho)}{n}
E
\Bigl[
\left\{
\mathbb{G}_0
\left(
\mathbb{Z}\delta_0+\mathbb{D}\alpha_0+\mathbb{J}\tau_0
\right)
\right\}'
\widebar{\mathbb{Q}}
\mathbb{Z}
(\delta_0-\delta)
\Bigr]
+
\widebar{\sigma}_n^2
+
o(1),
\label{eq_twfe_equicontinuity_EVV}
\end{align}
where
\[
\begin{aligned}
\widebar{\sigma}_n^2
=
\frac{\sigma_0^2}{n}
\operatorname{tr}
\left[
\left(
\mathbb{S}(\rho)\mathbb{S}_0^{-1}
\right)'
\widebar{\mathbb{Q}}
\mathbb{S}(\rho)\mathbb{S}_0^{-1}
\right]=
\frac{\sigma_0^2}{n}
\operatorname{tr}
\left[
\left(
\mathbb{S}(\rho)\mathbb{S}_0^{-1}
\right)'
\mathbb{S}(\rho)\mathbb{S}_0^{-1}
\right]
-
\frac{\sigma_0^2}{n}
\operatorname{tr}
\left[
\left(
\mathbb{S}(\rho)\mathbb{S}_0^{-1}
\right)'
\widebar{\mathbb{P}}
\mathbb{S}(\rho)\mathbb{S}_0^{-1}
\right].
\end{aligned}
\]
It remains to show that the second term is negligible. Under Assumptions~3 and 5,
\(\mathbb{S}(\rho)\mathbb{S}_0^{-1}\) is UB uniformly in \(\rho\in\Lambda\). Since
\(\widebar{\mathbb{P}}\) is an orthogonal projection matrix with rank \(N+T-1\), we have
\[
\begin{aligned}
\left|
\operatorname{tr}
\left[
\left(
\mathbb{S}(\rho)\mathbb{S}_0^{-1}
\right)'
\widebar{\mathbb{P}}
\mathbb{S}(\rho)\mathbb{S}_0^{-1}
\right]
\right|
\leq&
\operatorname{rank}
\left[
\left(
\mathbb{S}(\rho)\mathbb{S}_0^{-1}
\right)'
\widebar{\mathbb{P}}
\mathbb{S}(\rho)\mathbb{S}_0^{-1}
\right]
\left\|
\left(
\mathbb{S}(\rho)\mathbb{S}_0^{-1}
\right)'
\widebar{\mathbb{P}}
\mathbb{S}(\rho)\mathbb{S}_0^{-1}
\right\|_2
\\
\leq&
\operatorname{rank}\left(\widebar{\mathbb{P}}\right)
\left\|
\mathbb{S}(\rho)\mathbb{S}_0^{-1}
\right\|_2^2
\leq
\operatorname{rank}\left(\widebar{\mathbb{P}}\right)
\left\|
\mathbb{S}(\rho)\mathbb{S}_0^{-1}
\right\|_1
\left\|
\mathbb{S}(\rho)\mathbb{S}_0^{-1}
\right\|_\infty
=
O(N+T).
\end{aligned}
\]
The first inequality uses the rank--spectral norm bound for the trace. The second inequality follows from
\(
\operatorname{rank}
\left[
\left(
\mathbb{S}(\rho)\mathbb{S}_0^{-1}
\right)'
\widebar{\mathbb{P}}
\mathbb{S}(\rho)\mathbb{S}_0^{-1}
\right]
\leq
\operatorname{rank}\left(\widebar{\mathbb{P}}\right)
\)
and \(\|\widebar{\mathbb{P}}\|_2=1\).
Consequently,
\[
\widebar{\sigma}_n^2
=
\frac{\sigma_0^2}{n}
\operatorname{tr}
\left[
\left(
\mathbb{S}(\rho)\mathbb{S}_0^{-1}
\right)'
\mathbb{S}(\rho)\mathbb{S}_0^{-1}
\right]
+
o(1),\]
uniformly in \(\rho\in\Lambda\). The leading term is the same as the corresponding term in
Theorem~1 and is therefore uniformly equicontinuous in \(\rho\). Hence,
\(\widebar{\sigma}_n^2\) is uniformly equicontinuous.

It remains to show that the first three terms on the right-hand side of
\eqref{eq_twfe_equicontinuity_EVV} are uniformly equicontinuous. These three terms can be
combined as
\(
\left(
\rho-\rho_0,\,
\delta'-\delta_0'
\right)
E\left[\widebar{\mathcal{H}}\right]
\left(
\rho-\rho_0,\,
\delta'-\delta_0'
\right)',
\)
where
\(
\widebar{\mathcal{H}}
\equiv\allowbreak
\frac{1}{n}
\left(
\mathbb{G}_0
\left(
\mathbb{Z}\delta_0+\mathbb{D}\alpha_0+\mathbb{J}\tau_0
\right),
\mathbb{Z}
\right)'\allowbreak
\widebar{\mathbb{Q}}
\left(
\mathbb{G}_0
\left(
\mathbb{Z}\delta_0+\mathbb{D}\alpha_0+\mathbb{J}\tau_0
\right),
\mathbb{Z}
\right).
\)
Under Assumption~4$'$, the elements of the regressors, \(\alpha_0\), and \(\tau_0\) are
uniformly bounded. Together with Assumptions~5--7 and Lemma~\ref{lem:twfe-ZV-ZZ}, we have
\(
E\left[\widebar{\mathcal{H}}\right]=O(1).
\)
Therefore,
\(
\left(
\rho-\rho_0,\,
\delta'-\delta_0'
\right)
E\left[\widebar{\mathcal{H}}\right]
\left(
\rho-\rho_0,\,
\delta'-\delta_0'
\right)'
\)
is a quadratic function of \((\rho,\delta)\) with uniformly bounded coefficients. Therefore it is uniformly equicontinuous.

Combining these results, we obtain that
\(
\frac{1}{n}
E\left[
\widebar{\mathbb{V}}(\zeta)'
\widebar{\mathbb{V}}(\zeta)
\right]
\)
is uniformly equicontinuous for \(\zeta\). Finally, since \(\sigma^2\) is bounded away from zero, multiplying the normalized expectation term by \(-1/(2\sigma^2)\) preserves
uniform equicontinuity. Together with the same treatment of
\(-\frac{1}{2}\log\sigma^2\) and \(\frac{1}{n}\log|\mathbb{S}(\rho)|\) as in Theorem~1, this
implies that
\(
\frac{1}{n}\widebar{Q}_n(\theta)
\)
is uniformly equicontinuous.

\textbf{Proof of global identification condition}

We now prove that for any $\theta^{(-\gamma)}\neq\theta_0^{(-\gamma)}$,
\[
\limsup_{n\to\infty}
\frac{1}{n}
\left[
\widebar{Q}_n(\theta)-\widebar{Q}_n(\theta_0)
\right]
<0.
\]
By \eqref{eq_twfe_equicontinuity_EVV}, as in equation \eqref{Q-Q0} in the proof of Theorem 1, we can write
\[
\frac{1}{n}
\left[
\widebar{Q}_n(\theta)-\widebar{Q}_n(\theta_0)
\right]
=
\widebar{L}_1(\rho,\sigma^2)
-
\frac{1}{2\sigma^2}
\widebar{L}_2(\rho,\delta)
+
o(1),
\]
where
\[
\widebar{L}_1(\rho,\sigma^2)
\equiv
-\frac{1}{2}\left(\ln\sigma^2-\ln\sigma_0^2\right)
+
\frac{1}{n}
\left(
\ln|\mathbb{S}(\rho)|-\ln|\mathbb{S}_0|
\right)
-
\frac{1}{2\sigma^2}
\left(
\widebar{\sigma}_n^2-\sigma^2
\right),
\]
and
\begin{align*}
\widebar{L}_2(\rho,\delta)
&\equiv
\frac{1}{n}E
\Bigl\{
\bigl[
(\rho_0-\rho)\mathbb{G}_0
(\mathbb{Z}\delta_0+\mathbb{D}\alpha_0+\mathbb{J}\tau_0)
+
\mathbb{Z}(\delta_0-\delta)
\bigr]'
\\&\qquad
\widebar{\mathbb{Q}}\bigl[
(\rho_0-\rho)\mathbb{G}_0
(\mathbb{Z}\delta_0+\mathbb{D}\alpha_0+\mathbb{J}\tau_0)
+
\mathbb{Z}(\delta_0-\delta)
\bigr]
\Bigr\}.
\end{align*}
By the information inequality, the same argument as in Theorem 1 gives
\(
\widebar{L}_1(\rho,\sigma^2)
\leq
\widebar{L}_1(\rho,\widebar{\sigma}_n^2)
\leq
0.
\)
Moreover, the terms involving $\gamma-\gamma_0$ in $\widebar{L}_2(\rho,\delta)$ are asymptotically
negligible after normalization by $n$, because $\mathbb{Z}^{(\gamma)}$ is a sparse binary vector with at
most $N$ nonzero elements. Hence
\[
\lim_{n\to\infty}
\widebar{L}_2(\rho,\delta)
=
\left(
\rho-\rho_0,\,
\delta^{(-\gamma)\prime}-\delta_0^{(-\gamma)\prime}
\right)
E\left[\widebar{\mathcal{H}}^{(-\gamma)}\right]
\left(
\rho-\rho_0,\,
\delta^{(-\gamma)\prime}-\delta_0^{(-\gamma)\prime}
\right)'.
\]

If Assumption 8$'$ (i) holds, then
$\lim_{n\to\infty}E\left[\widebar{\mathcal{H}}^{(-\gamma)}\right]$ exists and is nonsingular. Therefore,
$\widebar{L}_2(\rho,\delta)$ is strictly positive whenever
$(\rho,\delta^{(-\gamma)})\neq(\rho_0,\delta_0^{(-\gamma)})$, and
$(\rho_0,\delta_0^{(-\gamma)})$ is globally identified.
If Assumption 8$'$ (ii) holds, the information inequality implies
\(
\limsup_{n\to\infty}\widebar{L}_1(\rho,\sigma^2)<0
\)
whenever $\rho\neq\rho_0$. Thus $\rho_0$ is globally identified. Given $\rho=\rho_0$,
Assumption 4$'$ implies that
\(
\lim_{n\to\infty}
\frac{1}{n}
E\left(
\mathbb{Z}^{(-\gamma)\prime}
\widebar{\mathbb{Q}}
\mathbb{Z}^{(-\gamma)}
\right)
\)
exists and is nonsingular, so $\delta_0^{(-\gamma)}$ is globally identified. 

Once
$(\rho,\delta^{(-\gamma)})$ is globally identified, $\sigma_0^2$ is identified by the same scalar
information inequality as in Theorem 1. Hence $\hat{\theta}^{(-\gamma)}$ is consistent.

Finally, given $\hat{\theta}^{(-\gamma)}\overset{p}{\to}\theta_0^{(-\gamma)}$, the concentrated estimator
of $\gamma$ is
\begin{align*}
\hat{\gamma}
&=
\left(
\mathbb{Z}^{(\gamma)\prime}
\widebar{\mathbb{Q}}
\mathbb{Z}^{(\gamma)}
\right)^{-1}
\mathbb{Z}^{(\gamma)\prime}
\widebar{\mathbb{Q}}
\left(
\mathbb{S}(\hat{\rho})\mathbb{Y}
-
\mathbb{Z}^{(-\gamma)}\hat{\delta}^{(-\gamma)}
\right)
\\
&=
\gamma_0
+
\left(
\mathbb{Z}^{(\gamma)\prime}
\widebar{\mathbb{Q}}
\mathbb{Z}^{(\gamma)}
\right)^{-1}
\mathbb{Z}^{(\gamma)\prime}
\widebar{\mathbb{Q}}\mathbb{V}
+
o_p(1).
\end{align*}
The second equality follows from the consistency of $\hat{\theta}^{(-\gamma)}$ and the same order bounds
used above. As in the proof of Theorem~1, Assumptions~1$'$ and 4$'$ imply that
\(
\frac{1}{N}
\mathbb{Z}^{(\gamma)\prime}
\widebar{\mathbb{Q}}
\mathbb{Z}^{(\gamma)}
\)
is bounded away from zero and uniformly bounded from above.
Therefore,
\[
E\left[
\left(
\mathbb{Z}^{(\gamma)\prime}
\widebar{\mathbb{Q}}
\mathbb{Z}^{(\gamma)}
\right)^{-1}
\mathbb{Z}^{(\gamma)\prime}
\widebar{\mathbb{Q}}\mathbb{V}
\right]
=0,
\]
and
\[
\mathrm{Var}
\left[
\left(
\mathbb{Z}^{(\gamma)\prime}
\widebar{\mathbb{Q}}
\mathbb{Z}^{(\gamma)}
\right)^{-1}
\mathbb{Z}^{(\gamma)\prime}
\widebar{\mathbb{Q}}\mathbb{V}
\right]
=
\sigma_0^2
\left(
\mathbb{Z}^{(\gamma)\prime}
\widebar{\mathbb{Q}}
\mathbb{Z}^{(\gamma)}
\right)^{-1}
=
O\left(\frac{1}{N}\right).
\]
By Chebyshev's inequality,
\(
\left(
\mathbb{Z}^{(\gamma)\prime}
\widebar{\mathbb{Q}}
\mathbb{Z}^{(\gamma)}
\right)^{-1}
\mathbb{Z}^{(\gamma)\prime}
\widebar{\mathbb{Q}}\mathbb{V}
=
o_p(1).
\)
Thus $\hat{\gamma}\overset{p}{\to}\gamma_0$. Combining this with
$\hat{\theta}^{(-\gamma)}\overset{p}{\to}\theta_0^{(-\gamma)}$, we obtain
\(
\hat{\theta}\overset{p}{\to}\theta_0.
\)

\subsection{Proof of Theorem 6}

The proof follows the same Taylor expansion argument as in the proof of Theorem~2. We only present the parts that differ from the OWFE case.

By Theorem~5, \(\widehat{\theta}\xrightarrow{p}\theta_0\). From the first-order condition
\(\partial\widebar{L}_n^c(\widehat{\theta})/\partial\theta=0\), the Taylor expansion around
\(\theta_0\) gives
\[
0
=
\frac{\partial\widebar{L}_n^c(\theta_0)}{\partial\theta}
+
\frac{\partial^2\widebar{L}_n^c(\bar{\theta})}
{\partial\theta\partial\theta'}
(\widehat{\theta}-\theta_0),
\]
where \(\bar{\theta}\) lies between \(\widehat{\theta}\) and \(\theta_0\). Hence,
\[
\sqrt{n}\Gamma^{-1}(\widehat{\theta}-\theta_0)
=
\left(
-\frac{1}{n}
\Gamma
\frac{\partial^2\widebar{L}_n^c(\bar{\theta})}
{\partial\theta\partial\theta'}
\Gamma
\right)^{-1}
\left(
\frac{1}{\sqrt{n}}
\Gamma
\frac{\partial\widebar{L}_n^c(\theta_0)}{\partial\theta}
\right).
\]
By the TWFE score decomposition in Section~\ref{secB5},
\[
\frac{1}{\sqrt{n}}
\Gamma
\frac{\partial\widebar{L}_n^c(\theta_0)}{\partial\theta}
=
\frac{1}{\sqrt{n}}
\Gamma
\frac{\partial\widebar{L}_n^{c*}(\theta_0)}{\partial\theta}
-
\widebar{\Delta}_n.
\]
Moreover, by equation~(\ref{B.10}),
\[
\widebar{\Delta}_n
=
\sqrt{\frac{N}{c_uT}}
\varphi_n(\theta_0)
+
\sqrt{\frac{T}{c_uN}}
\psi_n(\theta_0)
+
O_p\left(\frac{1}{\sqrt{T}}\right)
+
O_p\left(\frac{1}{\sqrt{N}}\right),
\]
where \(c_u=n/(NT)\), \(\varphi_n(\theta_0)=O(1)\), and
\(\psi_n(\theta_0)=O(1)\). Here \(\varphi_n(\theta_0)\) is the OWFE incidental-parameter bias component, while \(\psi_n(\theta_0)\) is the additional bias component generated by the time fixed effects.

For the mean-zero score component, the same martingale difference CLT as in the proof of Theorem~2 applies, with \(\mathbb{Q}\) replaced by \(\widebar{\mathbb{Q}}\). Therefore,
\[
\frac{1}{\sqrt{n}}
\Gamma
\frac{\partial\widebar{L}_n^{c*}(\theta_0)}{\partial\theta}
\xrightarrow{d}
N
\left(
0,
\widebar{\Sigma}_{\theta_0}
+
\widebar{\Omega}_{\theta_0}
\right),
\]
where
\(
\widebar{\Sigma}_{\theta_0}
=
\lim_{n\to\infty}
\Gamma\widebar{\Sigma}_{\theta_0,n}\Gamma\),
\(
\widebar{\Omega}_{\theta_0}
=
\lim_{n\to\infty}
\Gamma\widebar{\Omega}_{\theta_0,n}\Gamma\), and $\widebar{\Sigma}_{\theta_0,n}$ and $\widebar{\Omega}_{\theta_0,n}$ are defined in equation (\ref{eq_TWFE_VAR}). By equation (\ref{eq_TWFE_VAR}), the corresponding second moment satisfies
\[
E
\left[
\frac{1}{\sqrt{n}}
\Gamma
\frac{\partial\widebar{L}_n^{c*}(\theta_0)}{\partial\theta}
\frac{\partial\widebar{L}_n^{c*}(\theta_0)}{\partial\theta'}
\Gamma
\frac{1}{\sqrt{n}}
\right]
=
\Gamma\widebar{\Sigma}_{\theta_0,n}\Gamma
+
\Gamma\widebar{\Omega}_{\theta_0,n}\Gamma
+
O\left(\frac{1}{T}\right)
+
O\left(\frac{1}{N}\right).
\]

For the Hessian matrix, equations~(\ref{B.7}) and (\ref{B.8}) imply
\[
-\frac{1}{n}
\Gamma
\frac{\partial^2\widebar{L}_n^c(\bar{\theta})}
{\partial\theta\partial\theta'}
\Gamma
=
\Gamma\widebar{\Sigma}_{\theta_0,n}\Gamma
+
O_p\left(\frac{1}{\sqrt{NT}}\right)
+
\left\|
\Gamma^{-1}(\bar{\theta}-\theta_0)
\right\|_2O_p(1).
\]
Since \(\widehat{\theta}\xrightarrow{p}\theta_0\), we first have
\[
-\frac{1}{n}
\Gamma
\frac{\partial^2\widebar{L}_n^c(\bar{\theta})}
{\partial\theta\partial\theta'}
\Gamma
=
\Gamma\widebar{\Sigma}_{\theta_0,n}\Gamma
+
o_p(1).
\]
Under Assumption~9$'\), \(\Gamma\widebar{\Sigma}_{\theta_0,n}\Gamma\) is nonsingular for sufficiently large \(n\), which has been proved in Section \ref{secB5}. Hence,
\[
\sqrt{n}\Gamma^{-1}(\widehat{\theta}-\theta_0)
=
\left[
\Gamma\widebar{\Sigma}_{\theta_0,n}\Gamma
+
o_p(1)
\right]^{-1}
\left[
\frac{1}{\sqrt{n}}
\Gamma
\frac{\partial\widebar{L}_n^{c*}(\theta_0)}{\partial\theta}
-
\widebar{\Delta}_n
\right].
\]
Since the mean-zero score component is \(O_p(1)\) and
\(
\widebar{\Delta}_n
=
O_p\left(\sqrt{\frac{N}{T}}\right)
+
O_p\left(\sqrt{\frac{T}{N}}\right),
\)
we obtain
\[
\Gamma^{-1}(\widehat{\theta}-\theta_0)
=
O_p
\left(
\max
\left\{
\frac{1}{T},
\frac{1}{N}
\right\}
\right).
\]
Therefore, by (\ref{B.7}) and (\ref{B.8}),
\[
-\frac{1}{n}
\Gamma
\frac{\partial^2\widebar{L}_n^c(\bar{\theta})}
{\partial\theta\partial\theta'}
\Gamma
=
\Gamma\widebar{\Sigma}_{\theta_0,n}\Gamma
+
O_p
\left(
\max
\left\{
\frac{1}{T},
\frac{1}{N}
\right\}
\right).
\]
Using the standard inverse expansion, we have
\begin{align}
\left(
-\frac{1}{n}
\Gamma
\frac{\partial^2\widebar{L}_n^c(\bar{\theta})}
{\partial\theta\partial\theta'}
\Gamma
\right)^{-1}
=
\left(
\Gamma\widebar{\Sigma}_{\theta_0,n}\Gamma
\right)^{-1}
+
O_p
\left(
\max
\left\{
\frac{1}{T},
\frac{1}{N}
\right\}
\right).\label{eq_invH_}
\end{align}
Combining the above results gives
\[
\begin{aligned}
\sqrt{n}\Gamma^{-1}(\widehat{\theta}-\theta_0)
=&
\left[
\left(
\Gamma\widebar{\Sigma}_{\theta_0,n}\Gamma
\right)^{-1}
+
O_p
\left(
\max
\left\{
\frac{1}{T},
\frac{1}{N}
\right\}
\right)
\right]
\frac{1}{\sqrt{n}}
\Gamma
\frac{\partial\widebar{L}_n^{c*}(\theta_0)}{\partial\theta}
\\
&-
\left[
\left(
\Gamma\widebar{\Sigma}_{\theta_0,n}\Gamma
\right)^{-1}
+
O_p
\left(
\max
\left\{
\frac{1}{T},
\frac{1}{N}
\right\}
\right)
\right]
\widebar{\Delta}_n.
\end{aligned}
\]
Since
\[
\widebar{\Delta}_n
=
\sqrt{\frac{N}{c_uT}}
\varphi_n(\theta_0)
+
\sqrt{\frac{T}{c_uN}}
\psi_n(\theta_0)
+
O_p\left(\frac{1}{\sqrt{T}}\right)
+
O_p\left(\frac{1}{\sqrt{N}}\right),
\]
we have
\[
\begin{aligned}
&\sqrt{n}\Gamma^{-1}(\widehat{\theta}-\theta_0)
+
\sqrt{\frac{N}{c_uT}}
\left(
\Gamma\widebar{\Sigma}_{\theta_0,n}\Gamma
\right)^{-1}
\varphi_n(\theta_0)
+
\sqrt{\frac{T}{c_uN}}
\left(
\Gamma\widebar{\Sigma}_{\theta_0,n}\Gamma
\right)^{-1}
\psi_n(\theta_0)
\\
=&
\left[
\left(
\Gamma\widebar{\Sigma}_{\theta_0,n}\Gamma
\right)^{-1}
+
o_p(1)
\right]
\frac{1}{\sqrt{n}}
\Gamma
\frac{\partial\widebar{L}_n^{c*}(\theta_0)}{\partial\theta}
+
O_p
\left(
\max
\left\{
\frac{1}{\sqrt{T}},
\frac{1}{\sqrt{N}},
\sqrt{\frac{N}{T^3}},
\sqrt{\frac{T}{N^3}}
\right\}
\right).
\end{aligned}
\]
Because
\(\lim_{n\rightarrow\infty}
\Gamma\widebar{\Sigma}_{\theta_0,n}\Gamma
=
\widebar{\Sigma}_{\theta_0},
\)
and the mean-zero score component converges to
\(
N\left(
0,
\widebar{\Sigma}_{\theta_0}
+
\widebar{\Omega}_{\theta_0}
\right),
\)
we obtain
\begin{align*}
\sqrt{n}\Gamma^{-1}(\widehat{\theta}-\theta_0)
&+
\sqrt{\frac{N}{c_uT}}
\left(
\Gamma\widebar{\Sigma}_{\theta_0,n}\Gamma
\right)^{-1}
\varphi_n(\theta_0)
+
\sqrt{\frac{T}{c_uN}}
\left(
\Gamma\widebar{\Sigma}_{\theta_0,n}\Gamma
\right)^{-1}
\psi_n(\theta_0)
\\
&+
O_p
\left(
\max
\left\{
\frac{1}{\sqrt{T}},
\frac{1}{\sqrt{N}},
\sqrt{\frac{N}{T^3}},
\sqrt{\frac{T}{N^3}}
\right\}
\right)
\xrightarrow{d}
N
\left(
0,
\widebar{\Sigma}_{\theta_0}^{-1}
\left(
\widebar{\Sigma}_{\theta_0}
+
\widebar{\Omega}_{\theta_0}
\right)
\widebar{\Sigma}_{\theta_0}^{-1}
\right).
\end{align*}
Since the $\gamma$-components of both $\varphi_n(\theta_0)$ and $\psi_n(\theta_0)$ are zero, and $\Gamma$ only differs from the identity matrix in the $\gamma$-component, we have
\(
\Gamma^{-1}\varphi_n(\theta_0)=\varphi_n(\theta_0)
\) and \(
\Gamma^{-1}\psi_n(\theta_0)=\psi_n(\theta_0).
\)
So equivalently,
\begin{align*}
\sqrt{n}\Gamma^{-1}
\left[
\widehat{\theta}
-
\left(
\theta_0
-
\frac{\widebar{\Sigma}_{\theta_0,n}^{-1}\varphi_n(\theta_0)}{c_uT}
-
\frac{\widebar{\Sigma}_{\theta_0,n}^{-1}\psi_n(\theta_0)}{c_uN}
\right)
\right]
+
O_p
\left(
\max
\left\{
\frac{1}{\sqrt{T}},
\frac{1}{\sqrt{N}},
\sqrt{\frac{N}{T^3}},
\sqrt{\frac{T}{N^3}}
\right\}
\right)\\
\xrightarrow{d}
N
\left(
0,
\widebar{\Sigma}_{\theta_0}^{-1}
\left(
\widebar{\Sigma}_{\theta_0}
+
\widebar{\Omega}_{\theta_0}
\right)
\widebar{\Sigma}_{\theta_0}^{-1}
\right).
\end{align*}
Furthermore, by Section~\ref{secB5}, \(
\widebar{\Sigma}_{\theta_0,n}^{-1}\varphi_n(\theta_0)=O_p(1)\), and
\(\widebar{\Sigma}_{\theta_0,n}^{-1}\psi_n(\theta_0)=O_p(1).
\)
If \(v_{it}\)'s are normally distributed, then
\(\widebar{\Omega}_{\theta_0}=0\), and the asymptotic variance reduces to
\(\widebar{\Sigma}_{\theta_0}^{-1}\).
This completes the proof.

\subsection{Proof of Theorem 7}

The proof follows directly from Theorem~6 and the same plug-in argument as in the proof of
Theorem~3. Define the bias-corrected estimator as
\[
\widehat{\theta}^{\,1}
=
\widehat{\theta}
+
\left[
\left(
-\frac{1}{n}
\frac{\partial^2\widebar{L}_n^c(\theta)}
{\partial\theta\partial\theta'}
\right)^{-1}
\left\{
\frac{\varphi_n(\theta)}{c_uT}
+
\frac{\psi_n(\theta)}{c_uN}
\right\}
\right]_{\theta=\widehat{\theta}}.
\]
By \eqref{eq_invH_}, we have
\[
\left(
-\frac{1}{n}
\Gamma
\frac{\partial^2\widebar{L}_n^c(\widehat{\theta})}
{\partial\theta\partial\theta'}
\Gamma
\right)^{-1}
=
\left(
\Gamma\widebar{\Sigma}_{\theta_0,n}\Gamma
\right)^{-1}
+
O_p
\left(
\max
\left\{
\frac{1}{T},
\frac{1}{N}
\right\}
\right).
\]
For the plug-in bias correction term, the proof is the same as in the OWFE case. The continuity of \(\varphi_n(\theta)\) follows from Section~\ref{B.4}, and the continuity of
\(\psi_n(\theta)\) follows similarly. Therefore, by the same mean-value argument as in the OWFE case,
\[
\varphi_n(\widehat{\theta})-\varphi_n(\theta_0)
=
\left\|
\Gamma^{-1}(\widehat{\theta}-\theta_0)
\right\|_2O_p(1),
\qquad
\psi_n(\widehat{\theta})-\psi_n(\theta_0)
=
\left\|
\Gamma^{-1}(\widehat{\theta}-\theta_0)
\right\|_2O_p(1).
\]
By Theorem~6,
\(
\left\|
\Gamma^{-1}(\widehat{\theta}-\theta_0)
\right\|_2
=
O_p
\left(
\max
\left\{
\frac{1}{T},
\frac{1}{N}
\right\}
\right).
\)
Therefore,
\[
\begin{aligned}
&
\widehat{\theta}^{\,1}
-
\widehat{\theta}
-
\frac{\widebar{\Sigma}_{\theta_0,n}^{-1}\varphi_n(\theta_0)}{c_uT}
-
\frac{\widebar{\Sigma}_{\theta_0,n}^{-1}\psi_n(\theta_0)}{c_uN}
\\
=&
\left\{
\left[
\left(
-\frac{1}{n}
\frac{\partial^2\widebar{L}_n^c(\theta)}
{\partial\theta\partial\theta'}
\right)^{-1}
\right]_{\theta=\widehat{\theta}}
-
\widebar{\Sigma}_{\theta_0,n}^{-1}
\right\}
\left[
\frac{\varphi_n(\theta_0)}{c_uT}
+
\frac{\psi_n(\theta_0)}{c_uN}
\right]
\\
&+
\left[
\left(
-\frac{1}{n}
\frac{\partial^2\widebar{L}_n^c(\theta)}
{\partial\theta\partial\theta'}
\right)^{-1}
\right]_{\theta=\widehat{\theta}}
\left[
\frac{\varphi_n(\widehat{\theta})-\varphi_n(\theta_0)}{c_uT}
+
\frac{\psi_n(\widehat{\theta})-\psi_n(\theta_0)}{c_uN}
\right]
\\
=&
O_p
\left[
\left(
\frac{1}{T}
+
\frac{1}{N}
\right)
\max
\left\{
\frac{1}{T},
\frac{1}{N}
\right\}
\right].
\end{aligned}
\]
Multiplying both sides by \(\sqrt{n}\Gamma^{-1}\), we obtain
\[
\begin{aligned}
\sqrt{n}\Gamma^{-1}
\left[
\widehat{\theta}^{\,1}
-
\widehat{\theta}
-
\frac{\widebar{\Sigma}_{\theta_0,n}^{-1}\varphi_n(\theta_0)}{c_uT}
-
\frac{\widebar{\Sigma}_{\theta_0,n}^{-1}\psi_n(\theta_0)}{c_uN}
\right]
=&
O_p
\left[
\sqrt{NT}
\left(
\frac{1}{T}
+
\frac{1}{N}
\right)
\max
\left\{
\frac{1}{T},
\frac{1}{N}
\right\}
\right]
\\
=&
O_p
\left(
\frac{1}{\sqrt{NT}}
+
\sqrt{\frac{N}{T^3}}
+
\sqrt{\frac{T}{N^3}}
\right)
=
o_p(1),
\end{aligned}
\]
where the last equality follows from \(N/T^3\to0\) and \(T/N^3\to0\).
Therefore,
\[
\begin{aligned}
\sqrt{n}\Gamma^{-1}
\left(
\widehat{\theta}^{\,1}-\theta_0
\right)
=&
\sqrt{n}\Gamma^{-1}
\left[
\widehat{\theta}
-
\left(
\theta_0
-
\frac{\widebar{\Sigma}_{\theta_0,n}^{-1}\varphi_n(\theta_0)}{c_uT}
-
\frac{\widebar{\Sigma}_{\theta_0,n}^{-1}\psi_n(\theta_0)}{c_uN}
\right)
\right]
+
o_p(1).
\end{aligned}
\]
By Theorem~6, if \(N/T^3\to0\) and \(T/N^3\to0\), then
\[
\sqrt{n}\Gamma^{-1}
\left[
\widehat{\theta}
-
\left(
\theta_0
-
\frac{\widebar{\Sigma}_{\theta_0,n}^{-1}\varphi_n(\theta_0)}{c_uT}
-
\frac{\widebar{\Sigma}_{\theta_0,n}^{-1}\psi_n(\theta_0)}{c_uN}
\right)
\right]
\xrightarrow{d}
N
\left(
0,
\widebar{\Sigma}_{\theta_0}^{-1}
\left(
\widebar{\Sigma}_{\theta_0}
+
\widebar{\Omega}_{\theta_0}
\right)
\widebar{\Sigma}_{\theta_0}^{-1}
\right).
\]
Hence,
\[
\sqrt{n}\Gamma^{-1}
\left(
\widehat{\theta}^{\,1}-\theta_0
\right)
\xrightarrow{d}
N
\left(
0,
\widebar{\Sigma}_{\theta_0}^{-1}
\left(
\widebar{\Sigma}_{\theta_0}
+
\widebar{\Omega}_{\theta_0}
\right)
\widebar{\Sigma}_{\theta_0}^{-1}
\right).
\]
If \(v_{it}\)'s are normally distributed, then
\(\widebar{\Omega}_{\theta_0}=0\), and the asymptotic variance reduces to
\(\widebar{\Sigma}_{\theta_0}^{-1}\).
This completes the proof.

\subsection{Proof of Theorem 8}

Let \(\widehat{\theta}^{\,1}=(\widehat{\rho}^{\,1},
\widehat{\delta}^{\,1\prime},\widehat{\sigma}^{2,1})'\) be the TWFE
bias-corrected estimator. Define
\[(\widehat\alpha^{\,1\prime},\widehat\tau^{\,1\prime})'
=
[(\mathbb D,\mathbb J)'(\mathbb D,\mathbb J)]^{-1}
(\mathbb D,\mathbb J)'
\{\mathbb S(\widehat\rho^{\,1})\mathbb Y-\mathbb Z\widehat\delta^{\,1}\},\]
and
\(\widehat{\mathbb V}
=
\widebar{\mathbb Q}
\{\mathbb S(\widehat\rho^{\,1})\mathbb Y-\mathbb Z\widehat\delta^{\,1}\}\).
Let \(\check\sigma^2=n^{-1}\widehat{\mathbb V}'\widehat{\mathbb V}\),
\(\check\mu_3=n^{-1}\sum_{r=1}^n\widehat{\mathbb V}_r^3\),
\(\check\mu_4=n^{-1}\sum_{r=1}^n\widehat{\mathbb V}_r^4\), and
\(\check c=(\check\mu_4-3\check\sigma^4)/\check\sigma^4\), where
\(\check\sigma^4=(\check\sigma^2)^2\). Let
\(\widehat{\mathbb G}=\mathbb W\mathbb S(\widehat\rho^{\,1})^{-1}\) and
\(\widehat g=(\widehat{\mathbb G}_{11},\ldots,\widehat{\mathbb G}_{nn})'\).

Define
\[
\widehat{\widebar{\Sigma}}
=
\begin{pmatrix}
\left[\begin{array}{cc}
\displaystyle\frac{1}{n\check\sigma^2}
\{\widehat{\mathbb G}(\mathbb Z\widehat\delta^{\,1}
+\mathbb D\widehat\alpha^{\,1}
+\mathbb J\widehat\tau^{\,1})\}'
\widebar{\mathbb Q}
\{\widehat{\mathbb G}(\mathbb Z\widehat\delta^{\,1}
+\mathbb D\widehat\alpha^{\,1}
+\mathbb J\widehat\tau^{\,1})\}
\\\displaystyle+
\frac{1}{n}\operatorname{tr}
\{\widehat{\mathbb G}(\widehat{\mathbb G}+\widehat{\mathbb G}')\}
\end{array}\right]
& * & *
\\[2.2ex]
\displaystyle
\frac{1}{n\check\sigma^2}
\widetilde{\Gamma}'\mathbb Z'\widebar{\mathbb Q}
\widehat{\mathbb G}
(\mathbb Z\widehat\delta^{\,1}
+\mathbb D\widehat\alpha^{\,1}
+\mathbb J\widehat\tau^{\,1})
&
\displaystyle
\frac{1}{n\check\sigma^2}
\widetilde{\Gamma}'\mathbb Z'\widebar{\mathbb Q}\mathbb Z\widetilde{\Gamma}
& *
\\[2.2ex]
\displaystyle
\frac{1}{n\check\sigma^2}\operatorname{tr}(\widehat{\mathbb G})
&
\mathbf{0}_{1\times(k+3)}
&
\displaystyle
\frac{1}{2\check\sigma^4}
\end{pmatrix}.
\]
Next, define
\[
\widehat{\widebar{\Omega}}
=
\begin{pmatrix}
\displaystyle
\frac{\check c}{n}\widehat g'\widehat g
+
\frac{2\check\mu_3}{n\check\sigma^4}
\left[
\widehat{\mathbb G}\mathbb Z\widehat\delta^{\,1}
+
\widebar{\mathbb Q}\widehat{\mathbb G}
(\mathbb D\widehat\alpha^{\,1}
+\mathbb J\widehat\tau^{\,1})
\right]'
\widehat g
& * & *
\\[2.2ex]
\displaystyle
\frac{\check\mu_3}{n\check\sigma^4}
\widetilde{\Gamma}'\mathbb Z'\widehat g
&
\mathbf{0}_{(k+3)\times(k+3)}
& *
\\[2.2ex]
\displaystyle
\frac{\check c}{2n\check\sigma^2}
\operatorname{tr}(\widehat{\mathbb G})
&
\mathbf{0}_{1\times(k+3)}
&
\displaystyle
\frac{\check c}{4\check\sigma^4}
\end{pmatrix}.
\]
Set
\[
\widehat{\widebar{\Xi}}
=
\widehat{\widebar{\Sigma}}^{-1}
(\widehat{\widebar{\Sigma}}+\widehat{\widebar{\Omega}})
\widehat{\widebar{\Sigma}}^{-1}.
\]

\begin{proof}
We first show the consistency of the residual moment estimators. Since
\[
\widehat{\mathbb{V}}
-
\widebar{\mathbb{Q}}\mathbb{V}
=
\widebar{\mathbb{Q}}
\left[
-(\widehat{\rho}^{\,1}-\rho_0)\mathbb{W}\mathbb{Y}
-
\mathbb{Z}(\widehat{\delta}^{\,1}-\delta_0)
\right],
\]
the same argument as in the proof of Theorem~4, with \(\mathbb{Q}\) replaced by
\(\widebar{\mathbb{Q}}\), gives
\[
\frac{1}{n}
\sum_{r=1}^n
\left|
\widehat{\mathbb{V}}_r
-
(\widebar{\mathbb{Q}}\mathbb{V})_r
\right|^q
=
o_p(1),
\qquad q=2,3,4.
\]
Here we use that \(\widebar{\mathbb{Q}}\) is UB, and the moment bounds for
\(\mathbb{W}\mathbb{Y}\) and the columns of \(\mathbb{Z}\) are the same as those used in
Theorem~4. Therefore, by the binomial expansion and H\"older's inequality,
\[
\frac{1}{n}
\sum_{r=1}^n
\widehat{\mathbb{V}}_r^q
-
\frac{1}{n}
\sum_{r=1}^n
(\widebar{\mathbb{Q}}\mathbb{V})_r^q
=
o_p(1),
\qquad q=2,3,4.
\]

It remains to compare the sample moments of \(\widebar{\mathbb Q}\mathbb V\)
and \(\mathbb V\). Since
\(\widebar{\mathbb Q}\mathbb V=\mathbb V-\widebar{\mathbb P}\mathbb V\),
it suffices to control the average moments of \(\widebar{\mathbb P}\mathbb V\).
Since \(\widebar{\mathbb P}\) is an orthogonal projection matrix with
\(\mathrm{rank}(\widebar{\mathbb P})=N+T-1\), for each observation \(r\),
\[
E\left|(\widebar{\mathbb P}\mathbb V)_r\right|^2
=
\sigma_0^2(\widebar{\mathbb P})_{rr}.
\]
Moreover, since
\((\widebar{\mathbb P}\mathbb V)_r
=
\sum_{s=1}^n(\widebar{\mathbb P})_{rs}\mathbb V_s\), the fourth-moment
formula for a linear combination of independent errors gives
\[
\begin{aligned}
E\left|(\widebar{\mathbb P}\mathbb V)_r\right|^4
=
(\mu_4-3\sigma_0^4)
\sum_{s=1}^n
(\widebar{\mathbb P})_{rs}^4
+
3\sigma_0^4
\left[
\sum_{s=1}^n
(\widebar{\mathbb P})_{rs}^2
\right]^2\le
c_4
\left[
\sum_{s=1}^n
(\widebar{\mathbb P})_{rs}^2
\right]^2
=
c_4(\widebar{\mathbb P})_{rr}^2,
\end{aligned}
\]
where \(c_4=|\mu_4-3\sigma_0^4|+3\sigma_0^4\). The last equality holds because \(\widebar{\mathbb P}\) is symmetric and idempotent,
\[
\sum_{s=1}^n
(\widebar{\mathbb P})_{rs}^2
=
(\widebar{\mathbb P}\widebar{\mathbb P})_{rr}
=
(\widebar{\mathbb P})_{rr}.
\]
By H\"older's inequality,
\[
E\left|(\widebar{\mathbb P}\mathbb V)_r\right|^3
\le
\left\{
E\left|(\widebar{\mathbb P}\mathbb V)_r\right|^4
\right\}^{3/4}
\le
c_4^{3/4}(\widebar{\mathbb P})_{rr}^{3/2}.
\]
Since \(0\le(\widebar{\mathbb P})_{rr}\le 1\) and
\[
\sum_{r=1}^n
(\widebar{\mathbb P})_{rr}
=
\mathrm{tr}(\widebar{\mathbb P})
=
N+T-1,
\]
we obtain, for \(q=2,3,4\),
\[
E\left[
\frac{1}{n}
\sum_{r=1}^n
|(\widebar{\mathbb P}\mathbb V)_r|^q
\right]
\le
c_q\frac{N+T-1}{n}
=
O\left(\frac{1}{T}\right)
+
O\left(\frac{1}{N}\right),
\]
where \(c_2=\sigma_0^2\), \(c_3=c_4^{3/4}\), and
\(c_4=|\mu_4-3\sigma_0^4|+3\sigma_0^4\).
Therefore,
\[
\frac{1}{n}
\sum_{r=1}^n
|(\widebar{\mathbb P}\mathbb V)_r|^q
=
o_p(1),
\qquad q=2,3,4.
\]
Applying the same binomial expansion as in the proof of Theorem~4, we obtain
\[
\frac{1}{n}
\sum_{r=1}^n
(\widebar{\mathbb Q}\mathbb V)_r^q
-
\frac{1}{n}
\sum_{r=1}^n
\mathbb V_r^q
=
o_p(1),
\qquad q=2,3,4.
\]
Consequently,
\[
\frac{1}{n}
\sum_{r=1}^n
\widehat{\mathbb{V}}_r^q
-
\frac{1}{n}
\sum_{r=1}^n
\mathbb{V}_r^q
=
o_p(1),
\qquad q=2,3,4.
\]
By the law of large numbers, it follows that
\[
\check{\sigma}^2\xrightarrow{p}\sigma_0^2,
\qquad
\check{\mu}_3\xrightarrow{p}\mu_3,
\qquad
\check{\mu}_4\xrightarrow{p}\mu_4,
\qquad
\check{c}
\xrightarrow{p}
\frac{\mu_4-3\sigma_0^4}{\sigma_0^4}.
\]

We next show that \(\widehat{\widebar{\Sigma}}\) is consistent. By the consistency of the TWFE bias-corrected estimator, \(\widehat{\theta}^{\,1}\xrightarrow{p}\theta_0\). In particular, \(\widehat{\rho}^{\,1}-\rho_0=o_p(1)\). As shown in the proof of Theorem~4, for \(p=1,\infty\),
\[
\left\|
\mathbb{S}(\widehat{\rho}^{\,1})^{-1}
-
\mathbb{S}(\rho_0)^{-1}
\right\|_p
=
o_p(1),
\qquad
\left\|
\widehat{\mathbb{G}}-
\mathbb{G}_0
\right\|_p
=
o_p(1).
\]
From the definition of \((\widehat{\alpha}^{\,1\prime},\widehat{\tau}^{\,1\prime})'\),
\[
\begin{aligned}
\mathbb{D}\widehat{\alpha}^{\,1}
+
\mathbb{J}\widehat{\tau}^{\,1}
-
(\mathbb{D}\alpha_0+
\mathbb{J}\tau_0)
=&
\widebar{\mathbb{P}}
\left[
\mathbb{S}(\widehat{\rho}^{\,1})\mathbb{Y}
-
\mathbb{Z}\widehat{\delta}^{\,1}
\right]
-
(\mathbb{D}\alpha_0+
\mathbb{J}\tau_0)
\nonumber\\
=&
\widebar{\mathbb{P}}\mathbb{V}
-
\widebar{\mathbb{P}}
\left[
(\widehat{\rho}^{\,1}-\rho_0)\mathbb{W}\mathbb{Y}
+
\mathbb{Z}(\widehat{\delta}^{\,1}-\delta_0)
\right].
\end{aligned}
\]
As in the proof of Theorem~4, with \(\mathbb P\) replaced by
\(\widebar{\mathbb P}\), we have
\[
\frac{1}{n}
\left\|
\mathbb{D}\widehat{\alpha}^{\,1}
+
\mathbb{J}\widehat{\tau}^{\,1}
-
(\mathbb{D}\alpha_0+
\mathbb{J}\tau_0)
\right\|^2
=
o_p(1).
\]
Since \(\widehat{\mathbb{G}}\) and \(\mathbb{G}_0\) are UB, and
\(\|\widehat{\mathbb{G}}-\mathbb{G}_0\|_1=o_p(1)\) and
\(\|\widehat{\mathbb{G}}-\mathbb{G}_0\|_\infty=o_p(1)\), we have
\begin{align}
\frac{1}{n}
\left\|
\widehat{\mathbb{G}}\mathbb{Z}\widehat{\delta}^{\,1}
-
\mathbb{G}_0\mathbb{Z}\delta_0
\right\|^2
&=
o_p(1),
\nonumber\\
\frac{1}{n}
\left\|
\widehat{\mathbb{G}}
(\mathbb{D}\widehat{\alpha}^{\,1}+\mathbb{J}\widehat{\tau}^{\,1})
-
\mathbb{G}_0
(\mathbb{D}\alpha_0+\mathbb{J}\tau_0)
\right\|^2
&=
o_p(1).
\label{eq:twfe-GZ-fixed-vcov}
\end{align}
Consequently,
\[
\frac{1}{n}
\left\|
\widehat{\mathbb{G}}
(\mathbb{Z}\widehat{\delta}^{\,1}
+
\mathbb{D}\widehat{\alpha}^{\,1}
+
\mathbb{J}\widehat{\tau}^{\,1})
-
\mathbb{G}_0
(\mathbb{Z}\delta_0+\mathbb{D}\alpha_0+
\mathbb{J}\tau_0)
\right\|^2
=
o_p(1).
\]
Moreover,
\(
 n^{-1}\|\mathbb{G}_0(\mathbb{Z}\delta_0+\mathbb{D}\alpha_0+
\mathbb{J}\tau_0)\|^2=O_p(1)
\), because the additional term \(\mathbb{G}_0\mathbb{J}\tau_0\) is handled in the same way as
\(\mathbb{G}_0\mathbb{D}\alpha_0\). Therefore, using the Cauchy--Schwarz inequality and the UB property of
\(\widebar{\mathbb{Q}}\),
\[
\begin{aligned}
&\frac{1}{n}
\left[
\widehat{\mathbb{G}}
(\mathbb{Z}\widehat{\delta}^{\,1}
+
\mathbb{D}\widehat{\alpha}^{\,1}
+
\mathbb{J}\widehat{\tau}^{\,1})
\right]'
\widebar{\mathbb{Q}}
\left[
\widehat{\mathbb{G}}
(\mathbb{Z}\widehat{\delta}^{\,1}
+
\mathbb{D}\widehat{\alpha}^{\,1}
+
\mathbb{J}\widehat{\tau}^{\,1})
\right]
\nonumber\\
&-
\frac{1}{n}
\left[
\mathbb{G}_0
(\mathbb{Z}\delta_0+
\mathbb{D}\alpha_0+
\mathbb{J}\tau_0)
\right]'
\widebar{\mathbb{Q}}
\left[
\mathbb{G}_0
(\mathbb{Z}\delta_0+
\mathbb{D}\alpha_0+
\mathbb{J}\tau_0)
\right]
=
o_p(1).
\end{aligned}
\]
Similarly,
\[
\begin{aligned}
&
\frac{1}{n}
\widetilde{\Gamma}'\mathbb{Z}'\widebar{\mathbb{Q}}
\widehat{\mathbb{G}}
(\mathbb{Z}\widehat{\delta}^{\,1}
+
\mathbb{D}\widehat{\alpha}^{\,1}
+
\mathbb{J}\widehat{\tau}^{\,1})-
\frac{1}{n}
\widetilde{\Gamma}'\mathbb{Z}'\widebar{\mathbb{Q}}
\mathbb{G}_0
(\mathbb{Z}\delta_0+
\mathbb{D}\alpha_0+
\mathbb{J}\tau_0)
=
o_p(1).
\end{aligned}
\]
Also, the trace term satisfies
\begin{align}
\frac{1}{n}
\operatorname{tr}
\{
\widehat{\mathbb{G}}(\widehat{\mathbb{G}}+
\widehat{\mathbb{G}}')
\}
=
\frac{1}{n}
\operatorname{tr}
\{
\mathbb{G}_0(\mathbb{G}_0+
\mathbb{G}_0')
\}
+
o_p(1).
\label{eq:twfe-trGGG-vcov}
\end{align}
Similarly, the plug-in argument used in the proof of Theorem~4 gives
\[
\widehat{\widebar{\Sigma}}
-
\Gamma\widebar{\Sigma}_{\theta_0,n}\Gamma
=
o_p(1).
\]
Since \(\lim_{n\to\infty}\Gamma\widebar{\Sigma}_{\theta_0,n}\Gamma=\widebar{\Sigma}_{\theta_0}\), we obtain
\[
\widehat{\widebar{\Sigma}}
\xrightarrow{p}
\widebar{\Sigma}_{\theta_0}.
\]

It remains to show the consistency of \(\widehat{\widebar{\Omega}}\). Similar to
\eqref{eq:twfe-trGGG-vcov},
\[
\frac{1}{n}\widehat{g}'\widehat{g}
=
\frac{1}{n}
\sum_{r=1}^n
(\mathbb{G}_0)_{rr}^2
+
o_p(1),
\qquad
\frac{1}{n}\operatorname{tr}(\widehat{\mathbb{G}})
=
\frac{1}{n}\operatorname{tr}(\mathbb{G}_0)+o_p(1),
\qquad
\frac{1}{n}\widetilde{\Gamma}'\mathbb{Z}'\widehat{g}
=
\frac{1}{n}\widetilde{\Gamma}'\mathbb{Z}'g_0+o_p(1),
\]
where \(g_0=((\mathbb{G}_0)_{11},\ldots,(\mathbb{G}_0)_{nn})'\). Moreover, by
\eqref{eq:twfe-GZ-fixed-vcov},
\[
\begin{aligned}
&
\frac{1}{n}
\left\|
\widehat{\mathbb{G}}\mathbb{Z}\widehat{\delta}^{\,1}
+
\widebar{\mathbb{Q}}\widehat{\mathbb{G}}
(\mathbb{D}\widehat{\alpha}^{\,1}+
\mathbb{J}\widehat{\tau}^{\,1})
-
\left[
\mathbb{G}_0\mathbb{Z}\delta_0
+
\widebar{\mathbb{Q}}\mathbb{G}_0
(\mathbb{D}\alpha_0+
\mathbb{J}\tau_0)
\right]
\right\|^2
\nonumber\\
\le&
\frac{2}{n}
\left\|
\widehat{\mathbb{G}}\mathbb{Z}\widehat{\delta}^{\,1}
-
\mathbb{G}_0\mathbb{Z}\delta_0
\right\|^2
+
2\|\widebar{\mathbb{Q}}\|_\infty\|\widebar{\mathbb{Q}}\|_1
\frac{1}{n}
\left\|
\widehat{\mathbb{G}}
(\mathbb{D}\widehat{\alpha}^{\,1}+
\mathbb{J}\widehat{\tau}^{\,1})
-
\mathbb{G}_0
(\mathbb{D}\alpha_0+
\mathbb{J}\tau_0)
\right\|^2
=
o_p(1).
\end{aligned}
\]
Since \(n^{-1}\widehat{g}'\widehat{g}=O_p(1)\) and \(n^{-1}g_0'g_0=O_p(1)\), the Cauchy--Schwarz inequality gives
\[
\begin{aligned}
\frac{1}{n}
\left[
\widehat{\mathbb{G}}\mathbb{Z}\widehat{\delta}^{\,1}
+
\widebar{\mathbb{Q}}\widehat{\mathbb{G}}
(\mathbb{D}\widehat{\alpha}^{\,1}+
\mathbb{J}\widehat{\tau}^{\,1})
\right]'
\widehat{g}
-
\frac{1}{n}
\left[
\mathbb{G}_0\mathbb{Z}\delta_0
+
\widebar{\mathbb{Q}}\mathbb{G}_0
(\mathbb{D}\alpha_0+
\mathbb{J}\tau_0)
\right]'
g_0
=
o_p(1).
\end{aligned}
\]
Combining the above results with the consistency of \(\check{\mu}_3\), \(\check{c}\), and \(\check{\sigma}^2\), we obtain
\[
\widehat{\widebar{\Omega}}
-
\Gamma\widebar{\Omega}_{\theta_0,n}\Gamma
=
o_p(1).
\]
Since \(\lim_{n\to\infty}\Gamma\widebar{\Omega}_{\theta_0,n}\Gamma=\widebar{\Omega}_{\theta_0}\), it follows that
\[
\widehat{\widebar{\Omega}}
\xrightarrow{p}
\widebar{\Omega}_{\theta_0}.
\]

Finally, Assumption~9$'$ implies that \(\widebar{\Sigma}_{\theta_0}\) is nonsingular. Therefore, \(\widehat{\widebar{\Sigma}}\) is nonsingular with probability approaching one, and by the continuous mapping theorem,
\[
\widehat{\widebar{\Xi}}
=
\widehat{\widebar{\Sigma}}^{-1}
(\widehat{\widebar{\Sigma}}+
\widehat{\widebar{\Omega}})
\widehat{\widebar{\Sigma}}^{-1}
\xrightarrow{p}
\widebar{\Sigma}_{\theta_0}^{-1}
(\widebar{\Sigma}_{\theta_0}+
\widebar{\Omega}_{\theta_0})
\widebar{\Sigma}_{\theta_0}^{-1}.
\]
Thus, \(n^{-1}\Gamma\widehat{\widebar{\Xi}}\Gamma\) is a consistent estimator of the asymptotic covariance matrix of \(\widehat{\theta}^{\,1}\). This completes the proof.
\end{proof}

\section{Detailed Simulation Results}
\subsection{Supplemental Tables}\label{tabresult}
Tables \ref{Tab_N230}--\ref{Tab_L160} report supplementary simulation results for the baseline OWFE estimator. Tables \ref{Tab_TWFE_N30}--\ref{Tab_TWFE_L60} report supplementary simulation results for the TWFE estimator.
\begin{landscape}
\begin{table}[h]
\caption{Results of OWFE estimator with Normal error, UP=30\%, $\theta_0=(0.1, 0.3,0.5,1,1,1)$.}
\label{Tab_N230}
\centering
\begin{threeparttable}
\resizebox{!}{205pt}{
\usebox{\mytableA}
}
\begin{minipage}[t]{1.33\textwidth}
\footnotesize
Notes: 
$\mathrm{UP}=1-\frac{n}{NT}$. In the table, \textbf{bold} numbers indicate that the bias-corrected results have improved, while \textit{italicized} numbers indicate that the bias-corrected results perform equally well as before. The 95\%CP and 90\%CP  indicate the frequencies that the true parameter value is contained within the 95\% and 90\% confidence intervals of the estimator, respectively.
\end{minipage}
\end{threeparttable}
\end{table}
\end{landscape}

\begin{landscape}
\begin{table}[h]
\caption{Results of OWFE estimator with Normal error, UP=60\%, $\theta_0=(0.1, 0.3,0.5,1,1,1)$.}
\label{Tab_N260}
\centering
\begin{threeparttable}
\resizebox{!}{205pt}{
\usebox{\mytableB}
}

\begin{minipage}[t]{1.33\textwidth}
\footnotesize
Notes: $\mathrm{UP}=1-\frac{n}{NT}$. In the table, \textbf{bold} numbers indicate that the bias-corrected results have improved, while \textit{italicized} numbers indicate that the bias-corrected results perform equally well as before. The 95\%CP and 90\%CP  indicate the frequencies that the true parameter value is contained within the 95\% and 90\% confidence intervals of the estimator, respectively.
\end{minipage}
\end{threeparttable}
\end{table}
\end{landscape}

\begin{landscape}
\begin{table}[h]
\caption{Results of OWFE estimator with Exponential error, UP=30\%, $\theta_0=(0.5,0.2,0.1,1,1,1)$.}
\label{Tab_E130}
\centering
\begin{threeparttable}
\resizebox{!}{205pt}{
\usebox{\mytableC}
}

\begin{minipage}[t]{1.33\textwidth}
\footnotesize
Notes: $\mathrm{UP}=1-\frac{n}{NT}$. In the table, \textbf{bold} numbers indicate that the bias-corrected results have improved, while \textit{italicized} numbers indicate that the bias-corrected results perform equally well as before. The 95\%CP and 90\%CP  indicate the frequencies that the true parameter value is contained within the 95\% and 90\% confidence intervals of the estimator, respectively.
\end{minipage}
\end{threeparttable}
\end{table}
\end{landscape}

\begin{landscape}
\begin{table}[h]
\caption{Results of OWFE estimator with Exponential error, UP=60\%, $\theta_0=(0.5,0.2,0.1,1,1,1)$.}
\label{Tab_E160}
\centering
\begin{threeparttable}
\resizebox{!}{205pt}{
\usebox{\mytableD}
}

\begin{minipage}[t]{1.33\textwidth}
\footnotesize
Notes: $\mathrm{UP}=1-\frac{n}{NT}$. In the table, \textbf{bold} numbers indicate that the bias-corrected results have improved, while \textit{italicized} numbers indicate that the bias-corrected results perform equally well as before. The 95\%CP and 90\%CP  indicate the frequencies that the true parameter value is contained within the 95\% and 90\% confidence intervals of the estimator, respectively.
\end{minipage}
\end{threeparttable}
\end{table}
\end{landscape}

\begin{landscape}
\begin{table}[h]
\caption{Results of OWFE estimator with Laplace error, UP=30\%, $\theta_0=(0.5,0.2,0.1,1,1,1)$.}
\label{Tab_L130}
\centering
\begin{threeparttable}
\resizebox{!}{205pt}{
\usebox{\mytableE}
}

\begin{minipage}[t]{1.33\textwidth}
\footnotesize
Notes: $\mathrm{UP}=1-\frac{n}{NT}$. In the table, \textbf{bold} numbers indicate that the bias-corrected results have improved, while \textit{italicized} numbers indicate that the bias-corrected results perform equally well as before. The 95\%CP and 90\%CP  indicate the frequencies that the true parameter value is contained within the 95\% and 90\% confidence intervals of the estimator, respectively.
\end{minipage}
\end{threeparttable}
\end{table}
\end{landscape}

\begin{landscape}
\begin{table}[h]
\caption{Results of OWFE estimator with Laplace error, UP=60\%, $\theta_0=(0.5,0.2,0.1,1,1,1)$.}
\label{Tab_L160}
\centering
\begin{threeparttable}
\resizebox{!}{205pt}{
\usebox{\mytableF}
}

\begin{minipage}[t]{1.33\textwidth}
\footnotesize
Notes: $\mathrm{UP}=1-\frac{n}{NT}$. In the table, \textbf{bold} numbers indicate that the bias-corrected results have improved, while \textit{italicized} numbers indicate that the bias-corrected results perform equally well as before. The 95\%CP and 90\%CP  indicate the frequencies that the true parameter value is contained within the 95\% and 90\% confidence intervals of the estimator, respectively.
\end{minipage}

\end{threeparttable}
\end{table}
\end{landscape}

\begin{landscape}
\begin{table}[h]
\caption{Results of TWFE estimator with Normal error, UP=30\%, $\theta_0=(0.5,0.2,0.1,1,1,1)$.}
\label{Tab_TWFE_N30}
\centering
\begin{threeparttable}
\resizebox{!}{205pt}{
\usebox{\mytableG}
}
\begin{minipage}[t]{1.33\textwidth}
\footnotesize
Notes: $\mathrm{UP}=1-\frac{n}{NT}$. In the table, \textbf{bold} numbers indicate that the bias-corrected results have improved, while \textit{italicized} numbers indicate that the bias-corrected results perform equally well as before. The 95\%CP and 90\%CP indicate the frequencies that the true parameter value is contained within the 95\% and 90\% confidence intervals of the estimator, respectively.
\end{minipage}
\end{threeparttable}
\end{table}
\end{landscape}

\begin{landscape}
\begin{table}[h]
\caption{Results of TWFE estimator with Normal error, UP=60\%, $\theta_0=(0.5,0.2,0.1,1,1,1)$.}
\label{Tab_TWFE_N60}
\centering
\begin{threeparttable}
\resizebox{!}{205pt}{
\usebox{\mytableH}
}
\begin{minipage}[t]{1.33\textwidth}
\footnotesize
Notes: $\mathrm{UP}=1-\frac{n}{NT}$. In the table, \textbf{bold} numbers indicate that the bias-corrected results have improved, while \textit{italicized} numbers indicate that the bias-corrected results perform equally well as before. The 95\%CP and 90\%CP indicate the frequencies that the true parameter value is contained within the 95\% and 90\% confidence intervals of the estimator, respectively.
\end{minipage}
\end{threeparttable}
\end{table}
\end{landscape}

\begin{landscape}
\begin{table}[h]
\caption{Results of TWFE estimator with Exponential error, UP=30\%, $\theta_0=(0.5,0.2,0.1,1,1,1)$.}
\label{Tab_TWFE_E30}
\centering
\begin{threeparttable}
\resizebox{!}{205pt}{
\usebox{\mytableI}
}
\begin{minipage}[t]{1.33\textwidth}
\footnotesize
Notes: $\mathrm{UP}=1-\frac{n}{NT}$. In the table, \textbf{bold} numbers indicate that the bias-corrected results have improved, while \textit{italicized} numbers indicate that the bias-corrected results perform equally well as before. The 95\%CP and 90\%CP indicate the frequencies that the true parameter value is contained within the 95\% and 90\% confidence intervals of the estimator, respectively.
\end{minipage}
\end{threeparttable}
\end{table}
\end{landscape}

\begin{landscape}
\begin{table}[h]
\caption{Results of TWFE estimator with Exponential error, UP=60\%, $\theta_0=(0.5,0.2,0.1,1,1,1)$.}
\label{Tab_TWFE_E60}
\centering
\begin{threeparttable}
\resizebox{!}{205pt}{
\usebox{\mytableJ}
}
\begin{minipage}[t]{1.33\textwidth}
\footnotesize
Notes: $\mathrm{UP}=1-\frac{n}{NT}$. In the table, \textbf{bold} numbers indicate that the bias-corrected results have improved, while \textit{italicized} numbers indicate that the bias-corrected results perform equally well as before. The 95\%CP and 90\%CP indicate the frequencies that the true parameter value is contained within the 95\% and 90\% confidence intervals of the estimator, respectively.
\end{minipage}
\end{threeparttable}
\end{table}
\end{landscape}

\begin{landscape}
\begin{table}[h]
\caption{Results of TWFE estimator with Laplace error, UP=30\%, $\theta_0=(0.5,0.2,0.1,1,1,1)$.}
\label{Tab_TWFE_L30}
\centering
\begin{threeparttable}
\resizebox{!}{205pt}{
\usebox{\mytableK}
}
\begin{minipage}[t]{1.33\textwidth}
\footnotesize
Notes: $\mathrm{UP}=1-\frac{n}{NT}$. In the table, \textbf{bold} numbers indicate that the bias-corrected results have improved, while \textit{italicized} numbers indicate that the bias-corrected results perform equally well as before. The 95\%CP and 90\%CP indicate the frequencies that the true parameter value is contained within the 95\% and 90\% confidence intervals of the estimator, respectively.
\end{minipage}
\end{threeparttable}
\end{table}
\end{landscape}

\begin{landscape}
\begin{table}[h]
\caption{Results of TWFE estimator with Laplace error, UP=60\%, $\theta_0=(0.5,0.2,0.1,1,1,1)$.}
\label{Tab_TWFE_L60}
\centering
\begin{threeparttable}
\resizebox{!}{205pt}{
\usebox{\mytableL}
}
\begin{minipage}[t]{1.33\textwidth}
\footnotesize
Notes: $\mathrm{UP}=1-\frac{n}{NT}$. In the table, \textbf{bold} numbers indicate that the bias-corrected results have improved, while \textit{italicized} numbers indicate that the bias-corrected results perform equally well as before. The 95\%CP and 90\%CP indicate the frequencies that the true parameter value is contained within the 95\% and 90\% confidence intervals of the estimator, respectively.
\end{minipage}
\end{threeparttable}
\end{table}
\end{landscape}

\subsection{Supplemental Figures}\label{D.2}
\begin{figure}[H]
\centering
\includegraphics[width=\textwidth]{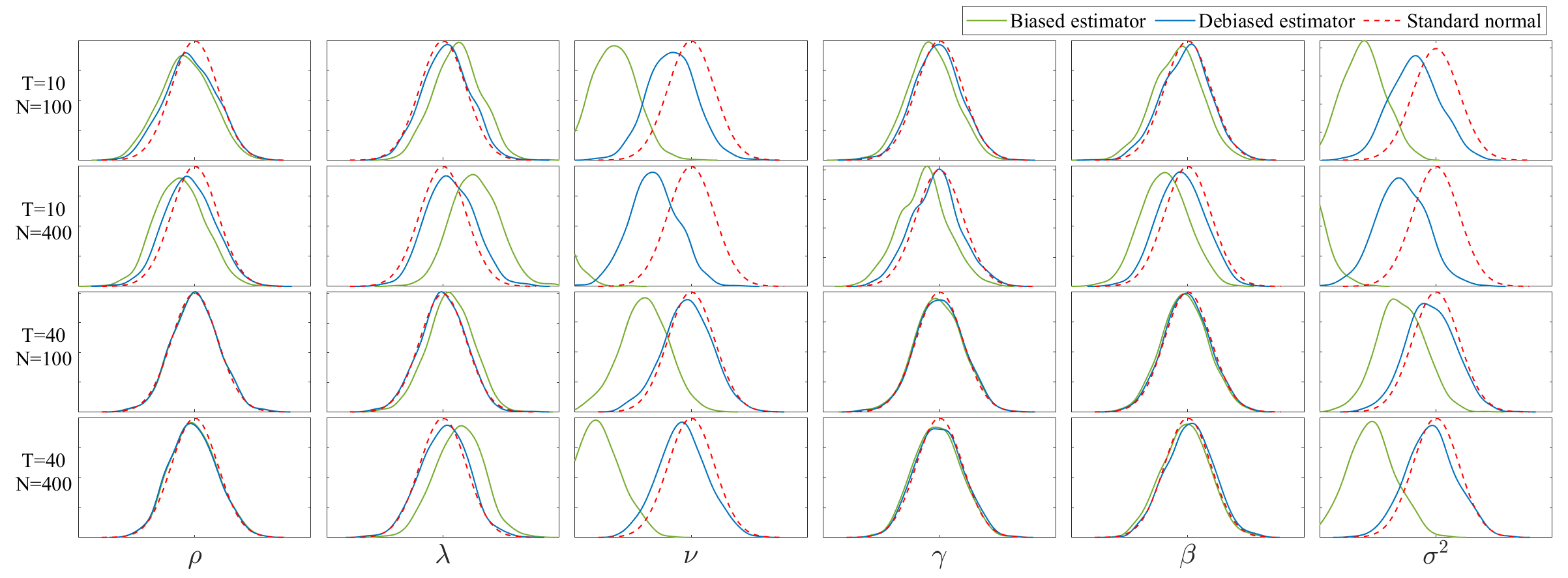}
\caption{PDF of OWFE standardized estimator with normal error, UP=30\%, $\theta_0=(0.1,0.3,0.5,1,1,1)$}
\label{Fig_N230}
\end{figure}

\begin{figure}[H]
\centering
\includegraphics[width=\textwidth]{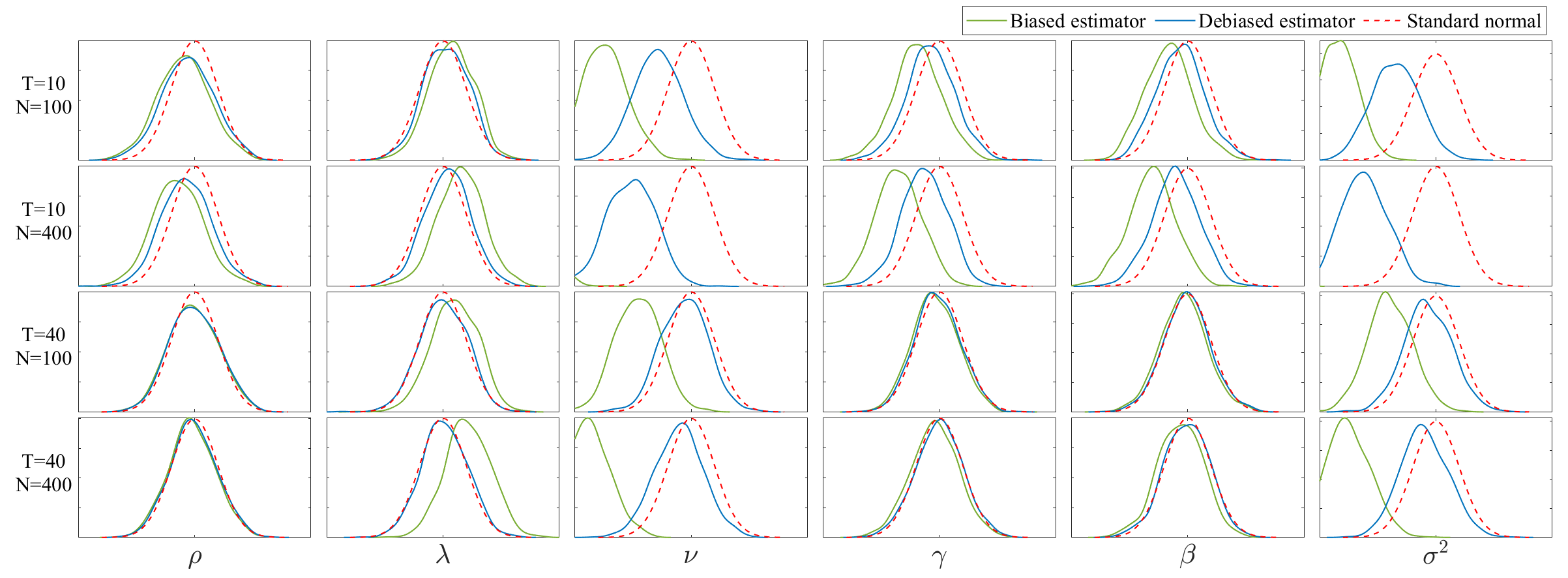}
\caption{PDF of OWFE standardized estimator with normal error, UP=60\%, $\theta_0=(0.1,0.3,0.5,1,1,1)$}
\label{Fig_N260}
\end{figure}

\begin{figure}[H]
\centering
\includegraphics[width=\textwidth]{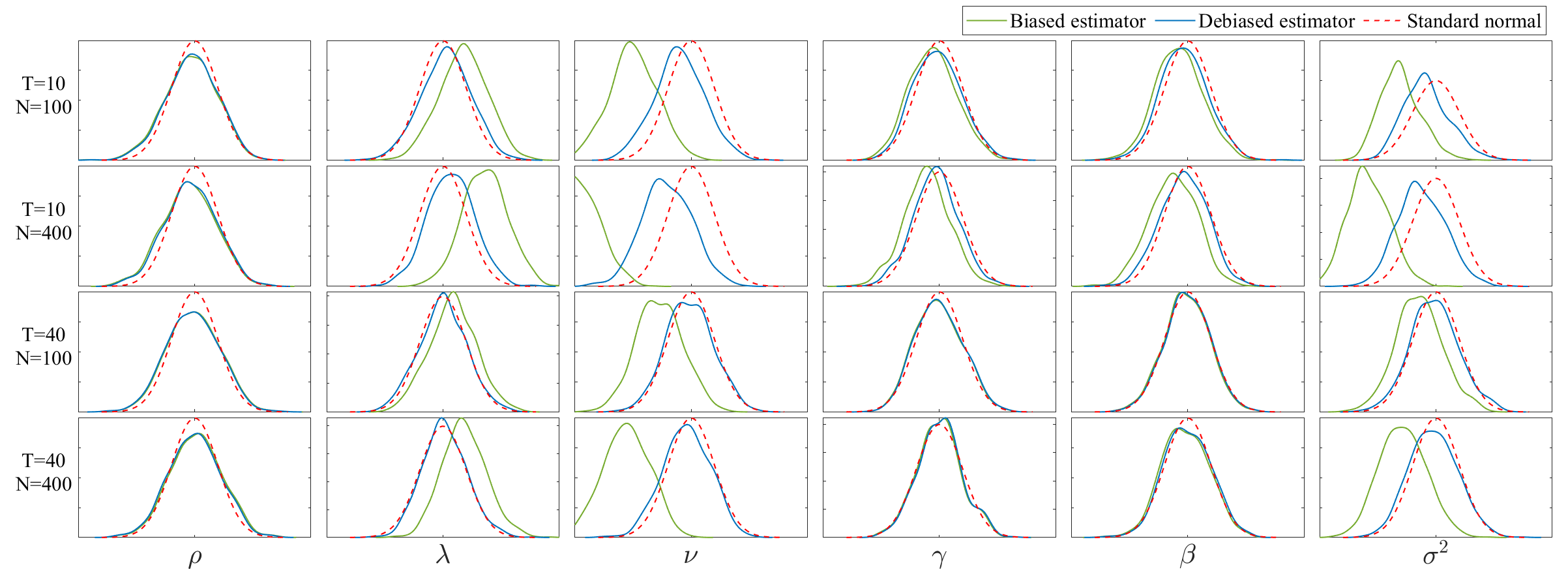}
\caption{PDF of OWFE standardized estimator with Exponential error, UP=30\%, $\theta_0=(0.5,0.2,0.1,1,1,1)$}
\label{Fig_E130}
\end{figure}

\begin{figure}[H]
\centering
\includegraphics[width=\textwidth]{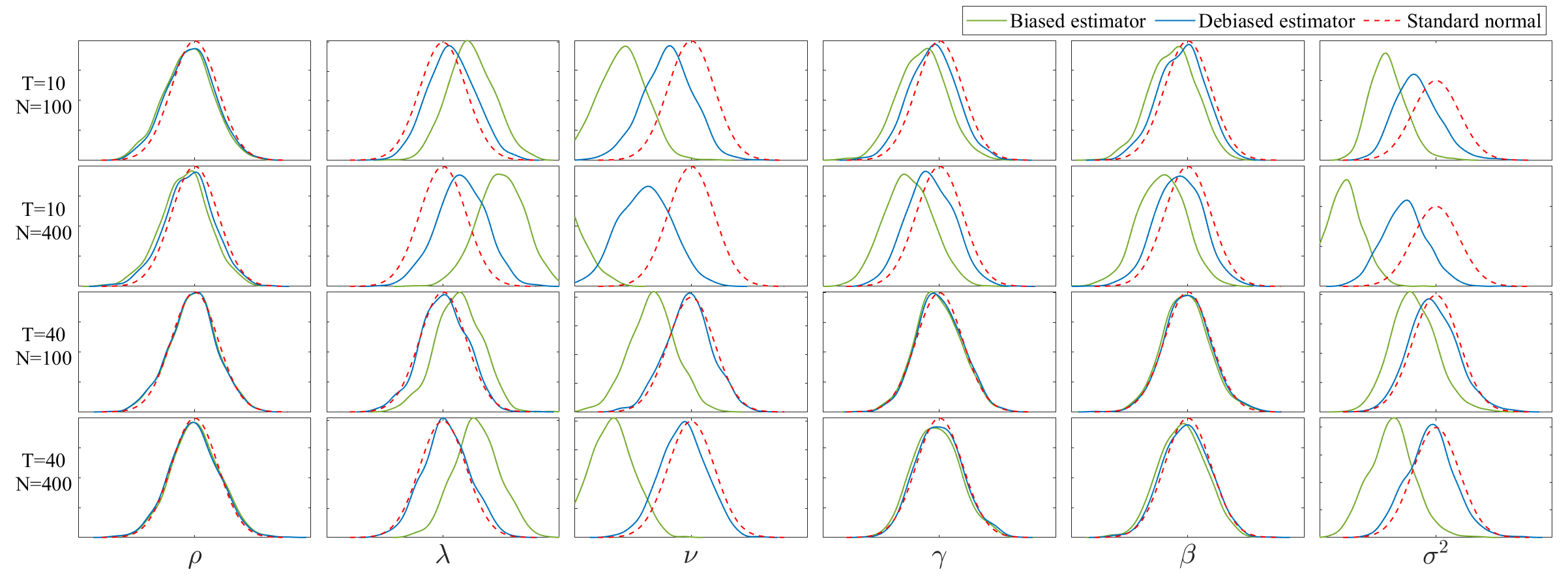}
\caption{PDF of OWFE standardized estimator with Exponential error, UP=60\%, $\theta_0=(0.5,0.2,0.1,1,1,1)$}
\label{Fig_E160}
\end{figure}

\begin{figure}[H]
\centering
\includegraphics[width=\textwidth]{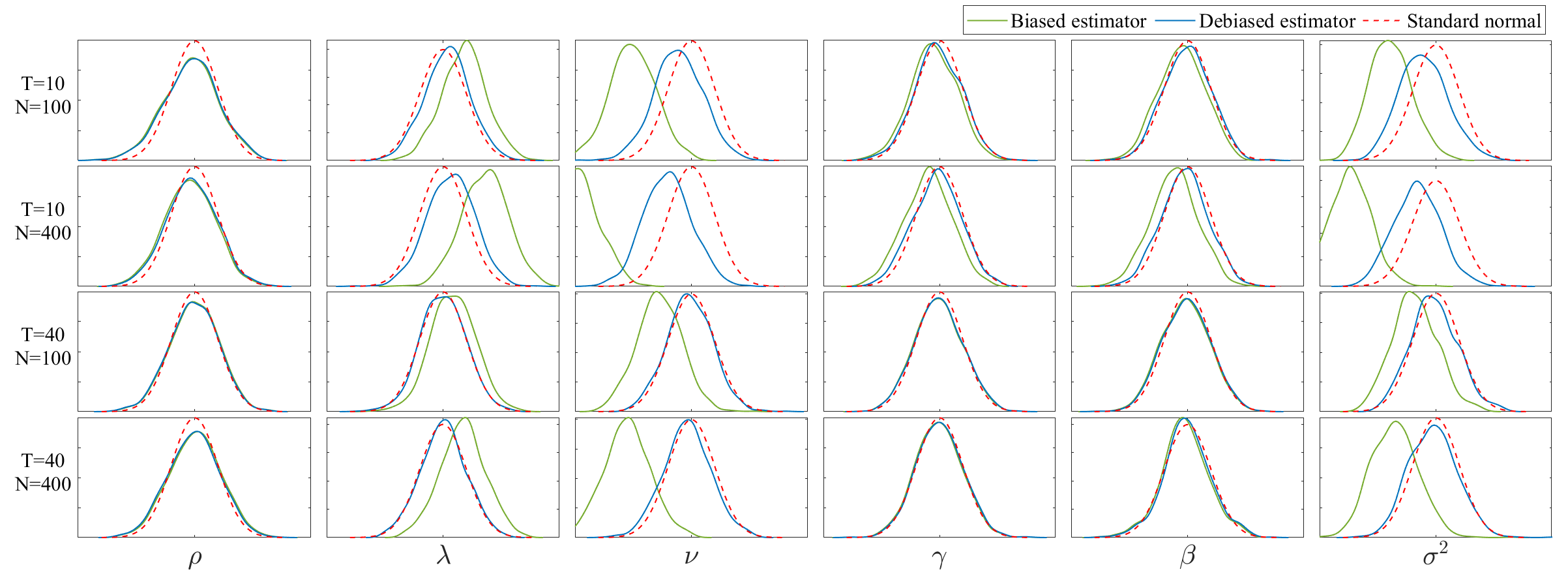}
\caption{PDF of OWFE standardized estimator with Laplace error, UP=30\%, $\theta_0=(0.5,0.2,0.1,1,1,1)$}
\label{Fig_L130}
\end{figure}

\begin{figure}[H]
\centering
\includegraphics[width=\textwidth]{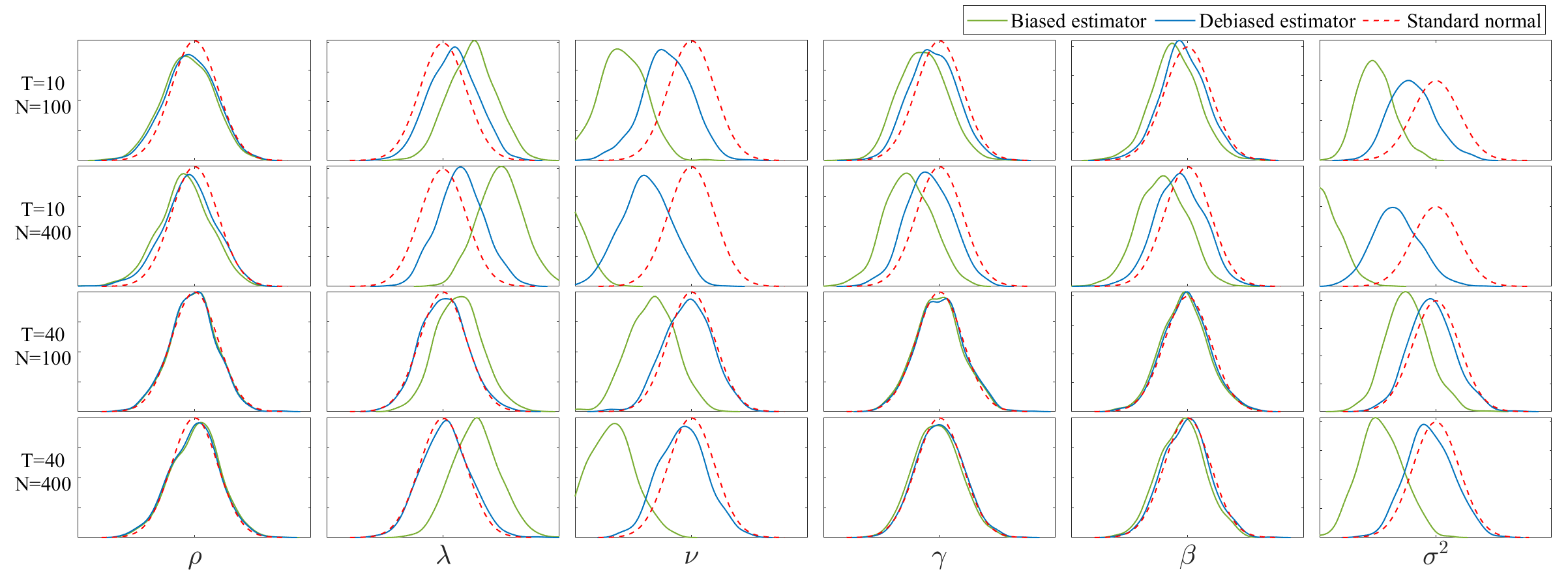}
\caption{PDF of OWFE standardized estimator with Laplace error, UP=60\%, $\theta_0=(0.5,0.2,0.1,1,1,1)$}
\label{Fig_L160}
\end{figure}
\newpage
\section{Detailed Empirical Results}
\subsection{Descriptive Statistics of the Data}
\begin{table}[h]
\caption{Descriptive Statistics}\label{Tab_DS}
\resizebox{0.99\textwidth}{!}{
\usebox{\mytableDS}
}
\end{table}

\subsection{Estimation Results}\label{realdata}
\begin{table}[H]
\centering
\caption{Estimation Results for Auckland Dataset}
\label{Tab_F1}
\begin{threeparttable}
\usebox{\mytableAuckV}
\begin{tablenotes}
\footnotesize
\item *In this dataset, $N=5896$, $T=30$, $n=80716$, $\mathrm{UP}=1-\frac{n}{NT}=54.37\%$. Estimates are presented with their standard errors in parentheses. Statistical significance is indicated by stars: *** for $p<0.01$, ** for $p<0.05$, and * for $p<0.1$.
\end{tablenotes}
\end{threeparttable}
\end{table}

\begin{table}[h]
\caption{Estimation Results for Queenstown Dataset}
\label{Tab_E_Queenstown}
\centering
\begin{threeparttable}
\usebox{\mytableAuckv}
\begin{tablenotes}
\footnotesize
\item *In this dataset, $N=6095$, $T=30$, $n=95455$, $\mathrm{UP}=1-\frac{n}{NT}=47.80\%$. Estimates are presented with their standard errors in parentheses. Statistical significance is indicated by stars: *** for $p<0.01$, ** for $p<0.05$, and * for $p<0.1$.
\end{tablenotes}
\end{threeparttable}
\end{table}

\begin{table}[h]
\caption{Estimation Results for Wellington Dataset}
\label{Tab_E_Wellington}
\centering
\begin{threeparttable}
\usebox{\mytableQueenV}
\begin{tablenotes}
\footnotesize
\item *In this dataset, $N=1549$, $T=30$, $n=18723$, $\mathrm{UP}=1-\frac{n}{NT}=59.71\%$. Estimates are presented with their standard errors in parentheses. Statistical significance is indicated by stars: *** for $p<0.01$, ** for $p<0.05$, and * for $p<0.1$.
\end{tablenotes}
\end{threeparttable}
\end{table}

\begin{table}[H]
\caption{Estimation Results for Williamsburg Dataset}
\label{Tab_E_Williamsburg}
\centering
\begin{threeparttable}
\usebox{\mytableQueenv}
\begin{tablenotes}
\footnotesize
\item *In this dataset, $N=1566$, $T=17$, $n=14567$, $\mathrm{UP}=1-\frac{n}{NT}=45.28\%$. Estimates are presented with their standard errors in parentheses. Statistical significance is indicated by stars: *** for $p<0.01$, ** for $p<0.05$, and * for $p<0.1$.
\end{tablenotes}
\end{threeparttable}
\end{table}

\begin{table}[H]
\caption{Estimation Results for Midtown Dataset}
\label{Tab_E_Midtown}
\centering
\begin{threeparttable}
\usebox{\mytableNZV}
\begin{tablenotes}
\footnotesize
\item *In this dataset, $N=2390$, $T=17$, $n=23672$, $\mathrm{UP}=1-\frac{n}{NT}=41.74\%$. Estimates are presented with their standard errors in parentheses. Statistical significance is indicated by stars: *** for $p<0.01$, ** for $p<0.05$, and * for $p<0.1$.
\end{tablenotes}
\end{threeparttable}
\end{table}

\begin{table}[H]
\caption{Estimation Results for Hell's Kitchen Dataset}
\label{Tab_E_HellsKitchen}
\centering
\begin{threeparttable}
\usebox{\mytableNZv}
\begin{tablenotes}
\footnotesize
\item *In this dataset, $N=1943$, $T=17$, $n=16862$, $\mathrm{UP}=1-\frac{n}{NT}=48.95\%$. Estimates are presented with their standard errors in parentheses. Statistical significance is indicated by stars: *** for $p<0.01$, ** for $p<0.05$, and * for $p<0.1$.
\end{tablenotes}
\end{threeparttable}
\end{table}